\documentclass{ws-ijmpd}

\usepackage[super,compress]{cite}

\newcommand{\be}{\begin{equation}}
\newcommand{\ee}{\end{equation}}
\newcommand{\bea}{\begin{eqnarray}}
\newcommand{\eea}{\end{eqnarray}}
\newcommand{\der}{\partial}
\newcommand{\vphi}{\varphi}

\begin{document}

\markboth{Israel Quiros}{Selected topics in scalar-tensor theories}


\catchline{}{}{}{}{}


\title{SELECTED TOPICS IN SCALAR-TENSOR THEORIES AND BEYOND}

\author{ISRAEL QUIROS}

\address{Departamento de Ingenier\'ia Civil, Divisi\'on de Ingenier\'ias, Universidad de Guanajuato\\
Guanajuato, C.P. 36000, M\'exico\\
iquiros@fisica.ugto.mx}

\maketitle

\begin{history}
\received{Day Month Year}
\revised{Day Month Year}
\end{history}

\begin{abstract}
Scalar fields have played an important role in the development of the fundamental theories of physics as well as in other branches of physics such as gravitation and cosmology. For a long time these escaped detection until 2012 year when the Higgs boson was observed for the first time. Since then alternatives to the general theory of relativity like the Brans-Dicke theory, scalar-tensor theories of gravity and their higher derivative generalizations -- collectively known as Horndeski theories -- have acquired renewed interest. In the present review we discuss on several selected topics regarding these theories, mainly from the theoretical perspective but with due mention of the observational aspect. Among the topics covered in this review we pay special attention to the following: 1) the asymptotic dynamics of cosmological models based in the Brans-Dicke, scalar-tensor and Horndeski theories, 2) inflationary models, extended quintessence and the Galileons, with emphasis in causality and stability issues, 3) the chameleon and Vainshtein screening mechanisms that may allow the elusive scalar field to evade the tight observational constraints implied by the solar system experiments, 4) the conformal frames conundrum with a brief discussion on the disformal transformations and 5) the role of Weyl symmetry and scale invariance in the gravitation theories. The review is aimed at specialists as well as at non-specialists in the subject, including postgraduate students.\end{abstract}


\ccode{PACS numbers: 04.20.Fy; 04.20.Ha; 04.50.-h; 04.50.Kd; 98.80.-k; 98.80.Jk}


\tableofcontents



\section{Introductory notes}\label{sect-intro}	

The search for the elusive scalar field in terrestrial experiments seems to have given a positive result with the discovery of the Higgs particle\cite{higgs_prl_1964, higgs_phys_lett_1964, englert_prl_1964, guralnik_prl_1964, higgs_discov_1, higgs_discov_2, higgs_discov_3, higgs_discov_4}. Being a tensor of order zero it is the simplest of the tensor fields. Despite of its important role in the theory of particle physics as a way to provide the masses of other gauge fields and particles, there are not fundamental forces carried by the scalar field. Unlike this there are vector fields which are the carriers of the electroweak and strong interactions, while a tensor field: the graviton, is the carrier of the gravitational interactions. Its appearance as a product of the compactification of the extra-dimensions in higher-dimensional Kaluza-Klein type theories like the string theory, in addition to the frustrated desire to find a place to the scalar field as a major player in the fundamental laws of nature, has fueled its repeated use in the search for solutions to unsolved problems in different areas of the theory of gravity, such as in cosmology and in astrophysics. 

There is, however, a more theoretically-motivated origin of the use of scalar fields in the gravitational theories. According to the famous theorem by Lovelock\cite{lovelock-1971, lovelock-1972} -- see also the related Refs. \citen{zanelli-lov-1991, lov-cai-prd-2002, zegers-lov-2005, cai-lov-2006, exiri-lov-2008, lov-padman-rpp-2010, lov-li-ctp-2011, navarro-lov-2011, acoleyen-lov-2011, clifton-phys-rept-2012, dolan-lov-2014, reall-lov-2014}, among many others -- the unique metric higher-derivative theory, $$S=\int d^4x\sqrt{|g|}{\cal L}\left(g_{\mu\nu},\der_\sigma g_{\mu\nu},\der_\sigma\der_\lambda g_{\mu\nu}\right),$$ that gives rise to second-order field equations for all metric components, is based in the Lagrangian density, $${\cal L}=\sum_{n=0}^Kc_n{\cal L}_{(n)},$$ where $c_n$ are arbitrary constants and the ${\cal L}_{(n)}$ are the $2n$-dimensional Euler densities that are given by $${\cal L}_{(n)}=\frac{1}{2^n}\delta^{a_1b_1\cdots a_nb_n}_{c_1d_1\cdots c_nd_n}R^{c_1d_1}_{a_1b_1}\cdots R^{c_nd_n}_{a_nb_n},$$ where $\delta^{a_1b_1\cdots a_nb_n}_{c_1d_1\cdots c_nd_n}$ are the (totally antisymmetric) generalized Kronecker delta function. In four dimensions the only non-vanishing Euler densities are ${\cal L}_{(0)}\propto 1$, ${\cal L}_{(1)}\propto R$ and ${\cal L}_{(2)}\propto{\cal G}$, where $${\cal G}=R^2-4R_{\mu\nu}R^{\mu\nu}+R_{\mu\nu\sigma\lambda}R^{\mu\nu\sigma\lambda},$$ is the Gauss-Bonnet term. This latter term, however, does not contribute towards the equations of motion since it amounts to a total derivative, i. e., it is a topological term in four dimensions.\footnote{In order to have cosmological implications in four dimensions, the Gauss-Bonnet term may be coupled to a scalar field\cite{nojiri-gb-prd-2005}.} Hence, in four dimensions the only action that gives rise to second-order motion equations is just the Einstein-Hilbert action (including a cosmological constant). The resulting motion equations are just the Einstein's equations of general relativity (GR). Lovelock's theorem entails that if one wants to construct metric theories of gravity with field equations that differ from those of GR, one is left with a few options\cite{clifton-phys-rept-2012}: Either i) accept higher than second derivatives of the metric in the field equations, or ii) adopt higer-dimensional spacetimes, or iii) consider other fields beyond the metric tensor, among other exotic possibilities. The latter option, precisely, opens up the door to scalar fields as a feasible modification of Einstein's theory.

In this review we shall give an -- as comprehensive as possible -- exposition on the scalar field as a major player in the description of the laws of gravity, including its role in cosmology. When we refer to it as ``a major player in the description of the laws of gravity'', we mean that, in addition to the graviton, the scalar field is also a carrier of the gravitational interactions. Hence, here we have to differentiate its use as an additional -- perhaps exotic -- matter field in general relativity, from its use as one of the carriers of the gravitational interactions of matter itself. 

In order to make our point clear, let us consider the Einstein-Hilbert (EH) action complemented with a matter piece in the form of a self-interacting scalar field $\vphi$:\footnote{The mostly positive signature of the metric is assumed: $(-,+,+,+)$. Hence, for instance, the determinant of the metric: $g<0$, is always a negative quantity so that we write $\sqrt{|g|}$ in order to avoid imaginary values of the volume density measure.}

\bea S=\frac{1}{16\pi G_N}\int d^4x\sqrt{|g|}\left(R-2\Lambda\right)+\int d^4x\sqrt{|g}\,{\cal L}_\vphi,\label{eh-vphi-act}\eea where $\sqrt{|g|}$ is a scalar density of weight $+1$ ($|g|$ is the absolute value of the determinant of the metric $g_{\mu\nu}$), so that $d^4x\sqrt{|g|}$ is an invariant measure, $R$ is the curvature scalar and $\Lambda$ is the cosmological constant. In the matter piece of the action -- the second term in the right-hand-side (RHS) of \eqref{eh-vphi-act} -- ${\cal L}_\vphi=-(\der\vphi)^2/2-V(\vphi)$ is the Lagrangian density of the scalar field, with $V(\vphi)$ -- the self-interaction potential for $\vphi$, while $(\der\vphi)^2\equiv g^{\mu\nu}\nabla_\mu\vphi\nabla_\nu\vphi$ is (twice) its kinetic term. Although along the text -- and unless otherwise stated -- we shall use a simplified system of units where $8\pi G_\text{N}=M^{-2}_\text{Pl}=c=\hbar=1$, in this introductory section we use the units' system where the (reduced) Planck mass $M_\text{Pl}$ and the Newton's constant $G_\text{N}$ are in the following relationship ($c=\hbar=1$): $M^2_\text{Pl}=(8\pi G_\text{N})^{-1}$, with $M_\text{Pl}\simeq 1.22\times 10^{19}$ GeV. The Einstein's field equations resulting from \eqref{eh-vphi-act}:

\bea G_{\mu\nu}+\Lambda g_{\mu\nu}=8\pi G_\text{N} T^{(\vphi)}_{\mu\nu},\;\;T^{(\vphi)}_{\mu\nu}=\der_\mu\vphi\der_\nu\vphi-\frac{1}{2}\,(\der\vphi)^2g_{\mu\nu}-Vg_{\mu\nu},\label{eh-feq}\eea where $G_{\mu\nu}\equiv R_{\mu\nu}-R g_{\mu\nu}/2$ is the Einstein's tensor, reveal what is already clear from \eqref{eh-vphi-act}: that this is just Einstein's GR with a matter piece in the form of a perfect fluid with stress-energy tensor\cite{pimentel_cqg_1989}:

\bea T^{(\vphi)}_{\mu\nu}=\left(\rho_\vphi+p_\vphi\right)u_\mu u_\nu+p_\vphi g_{\mu\nu},\label{vphi-set}\eea where $u_\mu:=\nabla_\mu\vphi/\sqrt{(\der\vphi)^2}$, with the energy density and pressure of the scalar field defined in the following way:

\bea \rho_\vphi:=-\frac{1}{2}\left(\nabla\vphi\right)^2+V(\vphi),\;\;p_\vphi:=-\frac{1}{2}\left(\nabla\vphi\right)^2-V(\vphi).\label{vphi-rho-p}\eea 

The action \eqref{eh-vphi-act} describes the gravitational interactions of a self-interacting scalar field in a de Sitter (or anti-de Sitter, depending on the sign of $\Lambda$) background within the GR setting. Several cosmological models intended to describe the dark energy (DE) and the dark matter (DM) in a unified framework rest, precisely, in this theory (see, for instance, Ref. \refcite{matos_quiros_prd_2009} and also Ref. \refcite{quiros_prd_2005} for a higher dimensional alternative). Other unified models of DE and DM are based in \eqref{eh-vphi-act} with the replacement ${\cal L}_\vphi\rightarrow{\cal L}_\vphi+{\cal L}_\psi$, where ${\cal L}_\psi=-(\nabla\psi)^2/2-V(\psi)$ is the Lagrangian density of an additional scalar field $\psi$ (the cosmological constant term may be relaxed). The different unified models \cite{matos_urena_cqg_2000, matos_urena_prd_2001, padmanabhan_prd_2002} rely on the different ways in which the self-interaction potentials $V(\vphi)$ and $V(\psi)$ are specified. 

But what if allow for a non-minimal coupling between the scalar field $\vphi$ and the curvature in the form of, for instance: $\propto f(\vphi) R$, where $f$ is a continuous non-vanishing function of the scalar field? As we shall see below, this would entail that the strength of the gravitational interactions is a point-dependent quantity that is controlled by $f^{-1}$. As a consequence gravity would not be exclusively a curvature effect any more, but it would be contributed also by the non-geometric scalar field. Theories of this type are called as scalar-tensor theories (STT-s) of gravity since both the metric tensor and the scalar field carry the gravitational interactions.

Hence, what to understand by an scalar-tensor theory of gravity? According to most common understanding scalar-tensor gravity is a non-fully geometrical, metric theory of gravity where the scalar-field is non-minimally coupled to the curvature\cite{fujii_book_2004}. As already said, this means that the gravitational phenomena are partly due to the curvature of spacetime and partly due to the scalar field that sets the strength of the gravitational interactions at each point in spacetime. As we shall see, if allow for higher-order derivatives of the scalar field in the Lagrangian density\cite{horndeski_gal, nicolis_gal, deffayet_vikman_gal, deffayet_deser_gal, deffayet_prd_2011, fab_4_prl_2012, deffayet-rev, tsujikawa_lect_not, kazuya_rpp_2016}, there are other subtle ways in which a scalar field can act as a co-carrier of the gravitational interactions. But, even if consider the latter non-trivial ways, a good measure to classify a given theory as a scalar-tensor one is the effective gravitational coupling -- the one measured in Cavendish experiments -- being a point-depending quantity. For further discussion on what to understand by a STT see the subsection \ref{subsect-horn-sst}.

One should be careful with such kind of classifications since, our understanding of what a STT entails and why it is different from GR with a scalar field as a matter source of the Einstein's equations, may be correct only if ignore the quantum effects of matter. When these effects are considered, even if we start with GR with a scalar field among the matter degrees of freedom, the quantum interactions of matter at the level of first loop corrections may induce a non-minimal coupling of the scalar field with the curvature\cite{callan_ann_phys_1970}, so that we end up with a STT. In this review we relay exclusively on the classical description of the gravitational phenomena so that possible quantum interactions of matter are ignored. 

The review has been designed to be self-contained, however there are very good reviews and books covering several of the subjects included here, so that the focus will be mainly in those issues that either have not been (adequately) covered or have to be updated. Among these we mention the following: 1) the Horndeski theories as a generalization of the scalar-tensor theories (section \ref{sect-horn}), 2) the study of the asymptotic dynamics of Brans-Dicke and scalar-tensor theories (section \ref{sect-dsyst-bd}), 3) the issue of causality and the related Laplacian instability that are of importance in the check of generalizations of STT-s (section \ref{sect-c2s}), 4) the screening mechanisms that explain how the scalar field has escaped detection in local experiments (section \ref{sect-screen}), 5) the conformal frames issue, perhaps one of the oldest problems in the discussion on scalar-tensor theories (section \ref{sect-cf}), the Weyl symmetry and scale invariance, a subject that is intimately related with the (possible) existence of scalar fields in nature (section \ref{sect-scale-inv}), among others. Due to lack of space, other topics such as, for instance, the gravitational waves in STT-s, have not been included and the reader is submitted to the corresponding bibliography. In this regard we recommend the book \refcite{faraoni-book} -- chapter 3 -- that contains a compact and complete introduction to this particular issue (see also Refs. \citen{stt_grav_w_1, stt_grav_w_2, stt_grav_w_3, stt_grav_w_4, stt_grav_w_5, stt_grav_w_6, stt_grav_w_7, stt_grav_w_8, stt_grav_w_9, stt_grav_w_10}). Another subject that is not covered in the present review is the discussion on exact cosmological solutions that is covered in detail in the book \refcite{faraoni-book} as well. Here we prefer to investigate the qualitative aspects of the cosmological dynamics instead. This means that we prefer to discuss on general classes of asymptotic solutions that represent generic cosmological behavior. The cosmological aspect of the scalar-tensor theories is studied in two separate sections \ref{sect-cosmo} and \ref{sect-dsyst-bd} respectively. In the former we discuss on scalar-tensor models of inflation, extended quintessence and on generalized scalar-tensor theories, while in the latter we expose a detailed dynamical systems study of the Brans-Dicke theory and of sensible cases within the Horndeski theories. 

One of the issues that represents more difficulty to those who do research on the STT subject, is the one on the conformal frames equivalence\cite{faraoni_rev_1997, faraoni_ijtp_1999, faraoni_prd_2007, sarkar_mpla_2007, deruelle_veiled_2011, deruelle_nordstrom_2011, quiros_grg_2013, hyun_arxiv, rondeau_prd_2017, majhi_prd_2017}. It is related to the fact that under a specific redefinition of the field variables of the theory (the metric, the scalar field and the matter fields), called as conformal transformations, a given STT may be transformed into a -- in principle infinite -- number of conformal frames. The different conformal frames are usually regarded as different representations of the given theory. The question then arises on whether or not to consider the different conformal frames as equivalent physical representations of the theory. In case these were not equivalent, which one of the conformal representations is the physical one? Here we want to avoid, as much as we can, the conformal frames issue, so that along the review we present the scalar-tensor theories in the so called Jordan frame (known also as string frame under an appropriate redefinition of the scalar field) exclusively. This was the original formulation of the Brans-Dicke theory\cite{bd-1961, brans_phd_thesis} and of their subsequent generalization, collectively known as scalar-tensor theories\cite{nordvedt_astroph_j_1970, wagoner_prd_1970}. The conformal frames issue, including the formulation of the Brans-Dicke and STT-s in the Einstein's frame, is discussed in detail in section \ref{sect-cf}.

This review is intended, mainly, for those who want to start doing research on the issue of alternative theories of gravity, but may be useful also for those researchers who are active in the field and want to keep themselves updated.


\section{Scalar fields and the fundamental theories of physics}\label{sect-fund-theor-phys}

The first appearance of scalar fields in the fundamental theories of physics can be traced back to 1913 year\cite{nordstrom_ann_phys_1913}. However it was not until the pioneering works by Pascual Jordan\cite{jordan_1948, jordan_z_phys_1959} and by Carl Brans and Robert Dicke\cite{bd-1961, brans_phd_thesis}, that a systematic investigation of the role that scalar fields may play in the gravitation theory was undertaken (for a more exact and complete history of the advent of STT-s we recommend Ref. \refcite{goenner_grg_2012}). Almost by the same time a scalar field was invoked as a way to allow for spontaneous break down of an internal Lie symmetry\cite{weinberg_salam_phys_rev_1962, higgs_prl_1964, higgs_phys_lett_1964, englert_prl_1964, guralnik_prl_1964}. It happens that the scalar field-based mechanism of spontaneous symmetry breaking\cite{coleman_weinberg_prd_1973, coleman_jackiw_prd_1974} provides masses not only for vectors but also for leptons and quarks\cite{susskind_prd_1979}. As a matter of fact the Higgs boson provides the masses for the gauge fields of the standard model of particles (SMP)\cite{weinberg_prl_1967, salam_1968, coleman_weinberg_prd_1973, coleman_jackiw_prd_1974, susskind_prd_1979}. A theory of gravity was also proposed where a scalar field-mediated break down of symmetry is incorporated\cite{zee_prl_1979}. Since then the scalar field has repeatedly appeared in several branches of theoretical physics including cosmology under the name of the inflaton field that drives the early inflationary stage of the cosmic expansion\cite{linde_rpp_1979, kazanas_astrophys_j_1980, guth_prd_1981, linde_plb_1982, dolgov_plb_1982, steinhardt_prl_1982, starobinsky_plb_1982, guth_prl_1982, linde_plb_1983, vilenkin_prd_1983, linde_rpp_1984, matarrese_prd_1985, brandenberger(infl_rev)_rpp_1985, linde_prd_1994, lidsey(infl_rev)_rmp_1997, lyth_phys_rep_1999, bassett(infl_rev)_rmp_2006, linde_lnp_2008}, or in the form of the quintessence\cite{carroll_prl_1998, zlatev_prl_1999, amendola_prd_2000, wang_atrophys_j_2000, barreiro_prd_2000, sahni_prd_2000, pavon_plb_2001, chimento_prd_2003, caldwell_prl_2005, copeland_rev_ijmpd_2006} or K-essence\cite{picon_prl_2000, picon_prd_2001, chiba_prd_2002, malquarti_prd_2003, scherrer_prl_2004, babichev_jhep_2008} fields, that are potential candidates for the dark energy, since these are able to explain the present inflationary stage. It is found also within another fundamental theory of physics: string theory, where it is acknowledged as the dilaton\cite{green_book_1987, polchinski_book_1998, copeland-wands-rev}, an unavoidable result of the compactification of the extra-dimensions.  

In this section, as an illustration of the role of the scalar fields in the fundamental theories of physics, we shall briefly discuss on the basis of spontaneous symmetry breaking -- including the Zee's gravitational theory\cite{zee_prl_1979} -- and of inflationary models only. A discussion of the inevitable appearance of scalar fields in the Kaluza-Klein scheme and in string theory as the result of the compactification of the extra-space, can be found in Refs. \citen{copeland-wands-rev, coley-bd-kk, wesson-bd-kk-rev} (see also Ref. \refcite{fujii_book_2004}).


\subsection{Spontaneous symmetry breaking in gauge theories}\label{subsect-ssb}

Spontaneous symmetry breaking is the mechanism through which the gauge fields and particles of the SMP acquire masses. Here, in order to illustrate how the mechanism works, we shall consider a very simple classic Lagrangian ${\cal L}$, composed of\cite{linde_rpp_1984}: i) a Lagrangian of a self-interacting scalar field ${\cal L}_\vphi$, ii) a Lagrangian of a fermion field ${\cal L}_\psi$ and iii) a Lagrangian for the interaction of the scalar and the fermion fields ${\cal L}_\text{int}$:

\bea {\cal L}_\vphi=-\frac{1}{2}\left(\der\vphi\right)^2-V(\vphi),\;{\cal L}_\psi=\bar\psi\gamma^\mu\der_\mu\psi,\;{\cal L}_\text{int}=q\bar\psi\vphi\psi,\label{ssb-lag}\eea where we chose the following self-interaction potential:

\bea V(\vphi)=-\frac{\mu^2}{2}\,\vphi^2+\frac{\lambda}{4}\,\vphi^4.\label{ssb-pot}\eea In the above equations $\mu^2$ amounts to the mass squared of the scalar field, while $\lambda$ and $q$ are coupling constants. Here, since for simplicity we are considering classic field effects only, $V(\vphi)$ in \eqref{ssb-pot} is just an effective potential which does not receive quantum corrections. The effective potential $V(\vphi)$ is a local maximum at $\vphi=0$ where $\der^2_\vphi V|_{\vphi=0}=-\mu^2<0$, while it is a minimum at $\vphi=\pm v\equiv\pm\sqrt{\mu^2/\lambda}$ where $\der^2_\vphi V|_{\vphi=\pm v}=2\mu^2>0$. Notice that the Lagrangian ${\cal L}_\vphi$ of the scalar field preserves the symmetry: $\vphi\rightarrow-\vphi$. However, once the scalar field has established at one of the stable minima $\pm v$, the above reflection symmetry breaks down.

In order to see how the particle spectrum of the theory changes after symmetry breaking, let us expand the full Lagrangian ${\cal L}$ around the minima of the potential. Recall that before symmetry breaking we have a scalar field with negative mass squared $-\mu^2<0$ that interacts with a massless fermion through the term $\propto\bar\psi\vphi\psi$. After expanding around the minima: $\vphi\rightarrow\sqrt{\mu^2/\lambda}+\sigma$ one gets: $${\cal L}=-\frac{1}{2}(\der\sigma)^2-\mu^2\sigma^2+\sqrt{\lambda\mu^2}\sigma^3-\frac{\lambda}{4}\sigma^4+\frac{\mu^4}{4\lambda}+\bar\psi\left(\gamma^\mu\der_\mu+q\sqrt\frac{\mu^2}{\lambda}\right)\psi+q\bar\psi\sigma\psi.$$ This Lagrangian corresponds to a self-interacting scalar field $\sigma$ with positive mass squared $2\mu^2>0$, which interacts with a fermion field with mass $m=q\sqrt{\mu^2/\lambda}$, through the term $\propto\bar\psi\sigma\psi$. In Minkowski backgrounds the constant term $\mu^4/4\lambda$ does not contribute towards the equations of motion and may be safely ignored, however, in curved backgrounds this latter term should be interpreted as a cosmological constant. As seen we started with a tachyon scalar (negative mass squared) interacting with a massless fermion and, after symmetry breaking, what we obtained was a massive scalar field interacting with a fermion field with the mass $m$. By the same mechanism we can give mass to massless vector fields $A_\mu$ interacting with the scalar field\cite{higgs_prl_1964, higgs_phys_lett_1964, englert_prl_1964, guralnik_prl_1964}. In this latter case it is useful to consider a complex scalar field $\vphi=(\vphi_1+i\vphi_2)/2$, besides the full Lagrangian is invariant under the $U(1)$ group of gauge transformations.

Embedding of the symmetry breaking mechanism into the cosmological framework requires considering a temperature dependent self-interaction potential (free energy). For instance, in Ref. \refcite{linde_rpp_1984} the following example is investigated:

\bea V(\vphi,T)=-\frac{\mu^2}{2}\left(1-\frac{\lambda T^2}{4\mu^2}\right)\vphi^2+\frac{\lambda}{4}\vphi^4-\frac{\pi^2T^4}{90}-\frac{\mu^2 T^2}{24},\label{t-dep-pot}\eea where $T$ is the temperature of the cosmic background. The extrema of this potential are at $$\vphi=0,\;\vphi=\pm\sqrt{\frac{\mu^2}{\lambda}\left(1-\frac{\lambda T^2}{4\mu^2}\right)}.$$ For temperatures below some critical temperature: $T<T_c$, where $T_c=2\sqrt{\mu^2/\lambda}$, there is a local maximum at $\vphi=0$ while the other extrema are local minima. For temperatures $T>T_c$ there is only a global minimum at $\vphi=0$. Hence, at the very high temperatures existing immediately after the bigbang, the reflection symmetry $\vphi\rightarrow-\vphi$ of the Lagrangian \eqref{ssb-lag} is an actual symmetry of the laws of physics -- recall that this is a very simple example that does not depict any realistic situation -- but as the background temperature decreases below the critical temperature $T_c$ with the course of the cosmic expansion, eventually a break down of the reflection symmetry occurs. This picture can be improved by considering more realistic Lagrangians including gauge vector fields, etc.


\subsubsection{Gravity theory with spontaneous symmetry breaking}\label{subsect-zee-theory}

Zee's theory of gravity was motivated by the hope to attribute the smallness of the Newton's constant $G_N$ to the massiveness of some scalar particle. The proposed action\cite{zee_prl_1979}:

\bea S=\int d^4x\sqrt{|g|}\left[\frac{1}{2}\,\xi\vphi^2R-\frac{1}{2}(\der\vphi)^2-V(\vphi)+{\cal L}_m\right],\label{zee-action}\eea where $\xi$ is a dimensionless coupling constant and ${\cal L}_m$ is the Lagrangian density of the matter degrees of freedom, can be recast into the form of the Brans-Dicke action (see section \ref{sect-bd} below) with coupling parameter $\omega_\text{BD}=1/4\xi$, if make the following scalar field redefinition: $\phi=\xi\vphi^2$. If the potential $V$ in \eqref{zee-action} is minimized when $\vphi=v$ -- in the quantum theory $v$ may be considered as the vacuum expectation value (VEV) of the scalar field -- then the above action reduces to Einstein-Hilbert action with the Newton's constant $8\pi G_N=\xi^{-1}v^{-2}$ and with the cosmological constant $V(v)$. Hence, assuming that $\xi\sim 1$, since $$v=\frac{1}{\sqrt{8\pi G_N}}=M_\text{Pl}\sim 10^{19}\text{GeV},$$ where $M_\text{Pl}$ is the reduced Planck mass, an extraordinarily large scalar field VEV is required. It is not surprising that the scale at which the $SU(5)$ symmetry of the strong-electroweak grand unification theory (GUT)\cite{georgi_glashow_prl_1974, georgi_prl_1974, minkowski_ann_phys_1975, weinberg_prd_1976} breaks down to $SU(3)\otimes SU(2)\otimes U(1)$ -- the symmetry of the unified electroweak (EW) scheme -- is about $10^{14}$GeV, which is close to the Planck scale. The idea of Ref. \refcite{zee_prl_1979} is that the scalar field $\vphi$ in \eqref{zee-action} is precisely the Higgs field\cite{higgs_prl_1964, higgs_phys_lett_1964}. In other words, the suggestion of Ref. \refcite{zee_prl_1979} is that a unified mechanism is responsible for the mass scale of gravity and for the symmetry breaking $SU(5)\rightarrow SU(3)\otimes SU(2)\otimes U(1)$. A similar idea was retaken in Ref. \citen{accetta_prd_1985, kaiser_prd_1995, cervantes_prd_1995, cervantes_npb_1995, barvinsky_jcap_2008, bezrukov_plb_2008, simone_plb_2009, bezrukov_jhep_2009, bezrukov_jhep_2011, bezrukov_jhep_2012} in order to explain the primordial inflation as driven by the Higgs field (see subsection \ref{subsect-higgs-infl}).


\subsection{Primordial inflation}\label{subsect-prim-infl}

The inflationary scenario\cite{kazanas_astrophys_j_1980, guth_prd_1981, linde_plb_1982, dolgov_plb_1982, steinhardt_prl_1982, starobinsky_plb_1982, guth_prl_1982, linde_plb_1983, vilenkin_prd_1983, linde_rpp_1984, matarrese_prd_1985, brandenberger(infl_rev)_rpp_1985, linde_prd_1994, lidsey(infl_rev)_rmp_1997, lyth_phys_rep_1999, bassett(infl_rev)_rmp_2006, linde_lnp_2008} was proposed in order to solve several problems of the standard cosmological model\cite{guth_prd_1981, linde_plb_1982}: i) horizon, ii) flatness, iii) homogeneity and isotropy, and iv) primordial monopole (and other relics like cosmic strings and topological defects) problems. The main ingredient of the inflationary models is a scalar field $\vphi$ which, for obvious reasons is called as ``inflaton''. In what follows we briefly explain the basic model in order to show how a single scalar field can drive an inflationary stage of the cosmological expansion (see below). For other details of the inflationary models such as: understanding of their physical basis as well as the problems they solve, the small quantum fluctuations and the associated cosmological perturbations that produced the cosmic structure we see today, etc., see, for instance, the reviews Refs. \citen{brandenberger(infl_rev)_rpp_1985, lidsey(infl_rev)_rmp_1997, lyth_phys_rep_1999, bassett(infl_rev)_rmp_2006, linde_lnp_2008}. For a discussion on the status of the subject after the data recorded by the Planck satellite released in 2013 year\cite{planck_2013} we recommend Ref. \refcite{infl_plack_2013}.  

We start with the basic equations: the Einstein's field equations sourced by scalar field matter fluid plus the Klein-Gordon equation for the inflaton $\vphi$:

\bea &&G_{\mu\nu}=T_{\mu\nu}^{(\vphi)}=\der_\mu\vphi\der_\nu\vphi-\frac{1}{2}g_{\mu\nu}(\der\vphi)^2-g_{\mu\nu}V(\vphi),\nonumber\\
&&\Box\vphi=\der_\vphi V.\label{efe-dil-eq}\eea In a cosmological context it is useful to consider as a good description of the expanding universe, the Friedmann-Robetson-Walker (FRW) metric which in spherical coordinates is given by the line element:

\bea ds^2=-dt^2+a^2(t)\left(\frac{dr^2}{1-kr^2}+r^2d\Omega^2\right),\label{frw-metric}\eea where $a(t)$ is the scale factor of the universe, $d\Omega^2\equiv d\theta^2+\sin^2\theta d\phi^2$, and the normalized spatial curvature $k=+1,0,-1$, depending on whether we consider closed, flat or open universes, respectively. Inserting this metric into the field equations \eqref{efe-dil-eq} yields to the following set of cosmological motion equations:\footnote{We did not write the Raychaudhuri equation since it is not independent of the Friedmann and Klein-Gordon equations in \eqref{cosmo-dil-eq}: it can be obtained by differentiating the Friedmann equation and substituting $\ddot\vphi$ from the Klein-Gordon equation.}

\bea &&3H^2+\frac{3k}{a^2}=\rho_\vphi\equiv\frac{1}{2}\dot\vphi^2+V(\vphi),\nonumber\\
&&\ddot\vphi+3H\dot\vphi=-\der_\vphi V,\label{cosmo-dil-eq}\eea where $H\equiv\dot a/a$ is the Hubble parameter. Notice that the second equation above -- the Klein-Gordon equation -- is the motion equation for a damped oscillator if $V(\vphi)=\mu^2\vphi^2/2$. The second term in the left-hand-side (LHS) of \eqref{cosmo-dil-eq}: $3H\dot\vphi$, plays the role of the damping term. 

The physical idea behind the inflationary behavior of the universe described by \eqref{cosmo-dil-eq} is quite simple (here we drop the spatial curvature term): Suppose that near of the initial value $\vphi(t_0)=\vphi_0$ the potential is very flat: $\der_\vphi^2V\ll 1$. Then the field very slowly moves along its potential, so that its kinetic energy may be neglected as compared with the potential energy: $\dot\vphi^2\ll V$ $\Rightarrow\,3H^2\approx V_0$, where $V_0=V(\vphi_0)$. As a consequence the universe exponentially expands: $a(t)\sim\exp{(\sqrt{V_0/3}\;t)}$. With the expansion, eventually the field moves away from the initial condition and the Hubble parameter starts decreasing. The damping term also decreases and at some (sufficiently long) period of time after the beginning of the inflationary expansion, the field $\vphi$ starts oscillating around the minimum of the potential and the universe becomes hot. Let us develop this idea further. The Friedmann equation -- first equation in \eqref{cosmo-dil-eq} -- can be rewritten in the following form: 

\bea \Omega_\vphi-1=\frac{k}{a^2H^2},\label{flat-eq}\eea where $\Omega_\vphi\equiv\rho_\vphi/3H^2$ is the Hubble-normalized (dimensionless) energy density of matter (in this case of the scalar field). For usual decelerated expansion $\ddot a<0$, the quantity $a^2H^2$ decreases with the expansion of the universe. Actually, consider that the source of the Einstein's equations is not a scalar field but a barotropic fluid with energy density $\rho_m$ and pressure $p_m$, satisfying the equation of state: $p_m=\omega\rho_m$, where the constant $\omega$ is usually called as 'equation of state' (EOS) parameter properly. Then the Friedmann equations can be written as: 

\bea a^2H^2=\frac{M^2}{3a^{3\omega+1}}-k,\label{a2h2-eq}\eea where $M^2$ is an integration constant. Then, for known forms of matter: dust ($\omega=0$), radiation ($\omega=1/3$) or stiff matter ($\omega=1$), the quantity $a^2H^2$ decreases with the expansion as said. According to \eqref{flat-eq} this would entail that any departure from spatial flatness at the bigbang will inevitably grow up with the expansion. This is to be contrasted with the observational evidence on almost spatial flatness of our present universe. This is what is known as the flatness problem. Here, for illustration, we shall address this problem, but the reader should understand that the other puzzles mentioned above also find resolution within the inflationary paradigm\cite{lidsey(infl_rev)_rmp_1997, lyth_phys_rep_1999, bassett(infl_rev)_rmp_2006, linde_lnp_2008}. 

As seen from \eqref{a2h2-eq}, in order to solve the flatness problem one needs is a universe filled with a fluid with EOS: $\omega<-1/3$, since if the latter condition is fulfilled then $a^2H^2$ will grow up with the expansion, thus erasing any initial departure from spatial flatness. Although known forms of matter do not meet the required bound on the EOS parameter, the scalar field comes to rescue. The inflationary scenario based on the inflaton field works only if the effective potential of the inflaton $V(\vphi)$ is very flat near the initial condition $\vphi=\vphi_0$ and if assume chaotic initial conditions\cite{linde_plb_1983, linde_lnp_2008}. As a matter of fact the chaotic inflation may occur in any theory where the potential has a sufficiently flat region, which allows the existence of the slow-roll regime: 

\bea |\ddot\vphi|\ll 3H|\dot\phi|,\;\frac{\dot\vphi^2}{2}\ll V,\label{slow-roll-cond}\eea besides, if the spatial curvature is taken into account, to the above conditions one must add: $H^2\gg k/a^2$. The slow-roll conditions \eqref{slow-roll-cond} can be written in an alternative way if introduce the slow-roll parameters:

\bea \epsilon=\frac{1}{2}\left(\frac{\der_\vphi V}{V}\right)^2,\;\eta=\frac{\der_\vphi^2V}{V},\;\zeta^2=\frac{\der_\vphi V\der_\vphi^3V}{V^2}.\label{slow-roll-param}\eea In terms of the latter parameters the slow-roll conditions read: $\epsilon\ll 1$, and $|\eta|\ll 1$. These conditions must hold for a prolonged time period in order to get the necessary amount of inflation. The inflationary phase ends when $\epsilon\sim|\eta|\sim 1$. A useful quantity to describe the amount of inflation is the number of e-foldings\cite{bassett(infl_rev)_rmp_2006}: $$N\equiv\ln\frac{a_f}{a}=\int_t^{t_f}Hdt\approx\int_{\vphi_f}^\vphi\frac{V}{\der_\vphi V}\,d\vphi,$$ where the subscript $f$ means evaluation of the given quantity at the end of inflation. In order to solve the flatness problem we should have $N\geq 60$.


\subsubsection{Classification of inflationary models}

A useful classification of the different models of single-field inflation, in connection with the observations, may be the one in which the inflationary models are divided into three groups, according to the region occupied in the ($\epsilon$,$\eta$)-plane by a given inflationary potential\cite{lyth_phys_rep_1999, dodelson_prd_1997}:

\begin{itemize}

\item{\it Small field.} In this case $\eta<-\epsilon$. An example of a small-field inflationary potential is $$V(\phi)=M^4\left[1-\left(\frac{\phi}{\mu}\right)^p\right].$$

\item{\it Large field (chaotic).} For these models $0<\eta<2\epsilon$, and typical examples are the power-law potential: $$V(\phi)=M^4\left(\frac{\phi}{\mu}\right)^p,$$ and the exponential one: $$V(\phi)=M^4\exp\left(\frac{\phi}{\mu}\right).$$

\item{\it Hybrid.} For potentials that drive hybrid inflation $0<2\epsilon<\eta$. An example can be: $$V(\phi)=M^4\left[1+\left(\frac{\phi}{\mu}\right)^p\right].$$

\end{itemize}


\subsubsection{Primordial non-gaussianity in the density perturbations}

The primordial non-gaussianity in the density perturbations are thought to be at the origin of structures in the universe. Non-gaussianity represents an important observable to discriminate among the competing scenarios\cite{non_g_phys_rep_2004, non_g_2000, non_g_2005, non_g_2007, non_g_2010_1, non_g_2010_2}. 

According to the inflationary paradigm the observable universe today should be flat, i.e., $|\Omega_k|\ll 1$, where $\Omega_k\equiv 1-\Omega$, there should exist primordial curvature perturbations whose power spectrum ${\cal P}_R(k)\sim k^{n_s-1}$, has a slightly tilted spectral index, $|n_s-1|\ll 1$, and, unless the inflaton potential or the initial conditions are fine tuned, the primordial perturbations should be gaussian. It has been demonstrated conclusively\cite{non_g_2010_2, maldacena_non_g, acquaviva_npb_2003} that slow-roll models where the density perturbations are produced by fluctuations of the inflaton itself, predict negligible non-gaussianity (for a pedagogical review on primordial non-Gaussianities from inflation see the Ref. \refcite{non_g_2010_2}). A detection of a degree of non-gaussianity by the next generation of experiments would therefore favor either an exotic inflationary model, or a model where density perturbations are generated by other dynamics\cite{non_g_2010_2}. Up to the present, from the optimal analysis of the WMAP 5-year data, no evidence of non-Gaussianity has been found\cite{non_g_2010_1}.


\section{Brans-Dicke theory}\label{sect-bd}

In this review we shall discuss on the scalar field as a major player in the fundamental laws of physics, in particular in the laws of gravity. Hence, here we focus in theories where the scalar field is non-minimally coupled to the curvature, i. e., in scalar-tensor theories. We start by the prototype and simplest in the class: the Brans-Dicke (BD) theory of gravity\cite{bd-1961}. It is thought to embody the Mach's principle\cite{dicke_am_sci_1959, rothman_am_sci_2017}. 

Mathematically the BD theory is expressed by the following action principle:\footnote{For the physical principles which the BD theory is based on we recommend Ref. \refcite{fujii_book_2004}.}

\bea S_\text{BD}=\int d^4x\sqrt{|g|}\left[\phi R-\frac{\omega_\text{BD}}{\phi}(\der\phi)^2-2V(\phi)+2{\cal L}_m\right],\label{bd-action}\eea where $\phi$ is the BD scalar field, $V(\phi)$ is the self-interaction potential for $\phi$, and $\omega_\text{BD}$ is a free constant -- the only free parameter of the theory -- called as the BD parameter. It should be noticed that in the original formulation of the BD theory\cite{bd-1961} the scalar field's self-interaction potential was not considered, a case usually called as massless BD theory. In the form depicted by the action \eqref{bd-action}, the BD theory is said to be given in the Jordan frame (JF). The action \eqref{bd-action}, with perhaps a quite different aspect and by ignoring the scalar field's self-interaction term, was first given by other scientists including Pascual Jordan\cite{jordan_1948, jordan_z_phys_1959, brans_phd_thesis, goenner_grg_2012}, this is why the BD theory is sometimes called as Jordan-Brans-Dicke (JBD) theory. For a nice historical account of the development of the JBD theory we recommend Ref. \refcite{goenner_grg_2012}.

In the BD theory \eqref{bd-action} the scalar field plays the role of the point-dependent gravitational coupling (not the same as the measured Newton's constant): 

\bea \phi=\frac{1}{8\pi G(x)}=M^2_\text{Pl}(x),\label{grav-coup}\eea where $M_\text{Pl}(x)$ is the point-dependent reduced Planck mass. The BD scalar field sets the strength of the gravitational interactions at each point in spacetime. In consequence, this is not a completely geometrical theory of gravity since the gravitational effects are not only encoded in the curvature of the spacetime but, also, in the interaction with the propagating scalar field degree of freedom.


\subsection{Brans-Dicke equations of motion}\label{subsect-bd-moteq}

From \eqref{bd-action}, by varying with respect to the metric, the Einstein-Brans-Dicke (EBD) equations of motion can be derived (see Ref. \refcite{ejp-2016} for the details of the derivation):

\bea G_{\mu\nu}=\frac{1}{\phi}\left[T^{(\phi)}_{\mu\nu}+T^{(m)}_{\mu\nu}\right]+\frac{1}{\phi}(\nabla_\mu\nabla_\nu-g_{\mu\nu}\Box)\phi,\label{bd-feq}\eea where $G_{\mu\nu}\equiv R_{\mu\nu}-g_{\mu\nu}R/2$ is the Einstein's tensor,

\bea T^{(\phi)}_{\mu\nu}\equiv\frac{\omega_\text{BD}}{\phi}\left[\der_\mu\phi\der_\nu\phi-\frac{1}{2}g_{\mu\nu}\left(\der\phi\right)^2\right]-g_{\mu\nu}V(\phi),\label{def-1}\eea is the stress-energy tensor of the BD-field, and 

\bea T^{(m)}_{\mu\nu}:=-\frac{2}{\sqrt{|g|}}\frac{\delta\left(\sqrt{|g|}{\cal L}_m\right)}{\delta g^{\mu\nu}},\label{mat-set}\eea is the stress-energy tensor of the matter degrees of freedom. By taking variations of \eqref{bd-action} with respect to the BD field, the following ``Klein-Gordon-Brans-Dicke'' (KGBD) equation of motion is obtained (see Ref. \refcite{ejp-2016} for details):

\bea 2\omega_\text{BD}\frac{\Box\phi}{\phi}-\omega_\text{BD}\left(\frac{\der\phi}{\phi}\right)^2+R=2\der_\phi V,\label{kgbd}\eea or, equivalently:

\bea (3+2\omega_\text{BD})\Box\phi=2\phi\der_\phi V-4V+T^{(m)}.\label{kgbd-eq}\eea

Using the field equations \eqref{bd-feq} and \eqref{kgbd}, the relationship:

\bea \Box(\nabla_\mu\phi)-\nabla_\mu(\Box\phi)=R_{\mu\nu}\nabla^\nu\phi,\label{usef-rel}\eea and the Bianchi identity $\nabla^\nu G_{\mu\nu}=0$, the standard conservation equation for the stress-energy tensor of the matter fields is obtained:

\bea \nabla^\nu T^{(m)}_{\nu\mu}=0.\label{cons-eq}\eea This entails that the matter fields respond only to the metric $g_{\mu\nu}$, i. e., these follow geodesics of that metric. 

Hence, what is the role of the scalar field in the gravitational interactions of matter? As seen from equations \eqref{bd-feq} and \eqref{kgbd-eq} above, the matter acts as a source of the metric and of the scalar fields and, then the metric says back the matter how it should move. The scalar field just modulates the strength of the interactions of matter with the metric field through the effective gravitational coupling.


\subsection{Alternative presentation of the BD theory}\label{subsect-alt-bd}

Under the following replacement: $$\phi=\frac{1}{2}\xi\vphi^2,\;\epsilon\xi^{-1}=4\omega_\text{BD}\;(\epsilon=\text{Sign}\,\omega_\text{BD}=\pm 1),$$ the action \eqref{bd-action} can be written in an alternative way where the kinetic term for the scalar field gets its standard form \cite{fujii_book_2004}: 

\bea S_\text{BD}=\int d^4x\sqrt{|g|}\left[\frac{1}{2}\xi\vphi^2R-\frac{\epsilon}{2}(\der\vphi)^2-\bar V(\vphi)+{\cal\bar L}_m\right],\label{bd-action'}\eea where the free parameter $\xi$ is a dimensionless constant and, in order to match the standard symbology in the bibliography (see, for instance, Ref. \refcite{fujii_book_2004}), we have replaced $V(\phi)$ and ${\cal L}_m$ in \eqref{bd-action} by: $$V(\phi)\rightarrow\bar V(\vphi)=2V(\phi(\vphi)),\;{\cal L}_m\rightarrow{\cal\bar L}_m=2{\cal L}_m,$$ respectively. This entails, in turn, that $$T^{(m)}_{\mu\nu}=\frac{\bar T^{(m)}_{\mu\nu}}{2}\Rightarrow T^{(m)}=\frac{\bar T^{(m)}}{2}.$$ The corresponding BD field equations read:

\bea &&2\phi G_{\mu\nu}=\bar T^{(m)}_{\mu\nu}+T^{(\vphi)}_{\mu\nu}+2\left(\nabla_\mu\nabla_\nu-g_{\mu\nu}\Box\right)\phi,\nonumber\\
&&\Box\phi=\frac{1}{\zeta^2}\,\bar T^{(m)}+\frac{1}{4\zeta^2}\left(\vphi\der_\vphi\bar V-4\bar V\right),\;\nabla^\nu\bar T^{(m)}_{\nu\mu}=0,\label{vphi-bd-feq}\eea where\cite{fujii_book_2004} $$\zeta^{-2}=6+\epsilon\xi^{-1}=2(3+2\omega_\text{BD}),$$ and

\bea T^{(\vphi)}_{\mu\nu}=\epsilon\left[\nabla_\mu\vphi\nabla_\nu\vphi-\frac{1}{2}\,g_{\mu\nu}\left(\nabla\vphi\right)^2\right]-g_{\mu\nu}\bar V.\label{vphi-bd-set}\eea Take care of the mixing of fields $\phi$ and $\vphi$, required for maximum simplicity of the above equations.

In this review we shall use the standard presentation of the BD theory as it was published in the seminal paper Ref. \refcite{bd-1961}, i. e., the one based in the action \eqref{bd-action} -- and the derived equations of motion -- as well as in its associated conformal representation\cite{dicke-1962} (see subsection \ref{subsect-bd-jf-ef}). For a detailed discussion on STT-s given in the form \eqref{bd-action'} we recommend the book \refcite{fujii_book_2004}.


\subsection{Weak-field limit of BD theory}\label{subect-bd-weakf}

This is an approximate solution to equations \eqref{bd-feq} and \eqref{kgbd-eq}, which is first order in the matter density. For simplicity we start considering the case with vanishing potential $V(\phi)=0$, i. e., a massless BD field and then we shall generalize to the case with non-vanishing potential. In subsection \ref{subsect-horn-geff} we shall discuss again on this issue for more general scalar-tensor theories that are based on Lagrangians that are higher order in the derivatives of the scalar field. 

Mathematically what we do is to expand up to linear terms in the metric and scalar field perturbations:

\bea g_{\mu\nu}&=&\eta_{\mu\nu}+h_{\mu\nu}(x),\nonumber\\
\phi&=&\phi_0+\sigma(x),\label{perts}\eea where $\eta_{\mu\nu}=\text{diag}(-1,1,1,1)$ is the Minkowski metric and $h_{\mu\nu}(x)$, $\sigma(x)$ are small point-dependent metric and scalar perturbations, respectively. The linearized Einstein's tensor reads:

\bea G^L_{\mu\nu}=\frac{1}{2}\left[-\Box h_{\mu\nu}+\der_\mu\der_\lambda h^\lambda_\nu+\der_\nu\der_\lambda h^\lambda_\mu-\der_\mu\der_\nu h-\eta_{\mu\nu}\left(\der_\lambda\der_\tau h^{\lambda\tau}-\Box h\right)\right],\label{lin-etensor}\eea where $h=h^\mu_\mu$ and the tensorial indexes are raised and lowered by means of $\eta_{\mu\nu}$. If substitute the linearized Einstein's tensor \eqref{lin-etensor} back to \eqref{bd-feq} and take into account the linear expansion \eqref{perts} in its RHS, up to the first order in the perturbations we get:

\bea G^L_{\mu\nu}=\frac{1}{\phi_0}\,T^{(m)}_{\mu\nu}+\frac{1}{\phi_0}\left(\der_\mu\der_\nu-\eta_{\mu\nu}\Box\right)\sigma.\label{lin-bd-feq}\eea In order to avoid the mixing between the perturbed fields $h_{\mu\nu}$ and $\sigma$ in this equation it will be useful to diagonalize it by introducing a new field:

\bea \psi_{\mu\nu}=h_{\mu\nu}-\frac{1}{2}\,\eta_{\mu\nu}h-\frac{1}{\phi_0}\,\eta_{\mu\nu}\sigma,\label{lin-diag}\eea and the four coordinate conditions: $\der_\nu\psi^\nu_\mu=0$, that we can safely declare thanks to the four degrees of freedom available to make diffeomorphisms. The diagonalized linearized Einstein-BD equations then read:

\bea \Box\psi_{\mu\nu}=-\frac{2}{\phi_0}\,T^{(m)}_{\mu\nu},\label{lin-ebd-eq}\eea whose retarded-time solution is given by:

\bea \psi_{\mu\nu}=\frac{4}{\phi_0}\int d^3x\,\frac{T^{(m)}_{\mu\nu}}{r}.\label{lin-bd-eq-sol}\eea 

On the other hand, the linearized KGBD equation of motion \eqref{kgbd-eq}:

\bea \Box\sigma=\frac{T^{(m)}}{3+2\omega_\text{BD}},\label{lin-kgbd-eq}\eea has the following retarded-time solution:

\bea \sigma=-\frac{2}{3+2\omega_\text{BD}}\int d^3x\,\frac{T^{(m)}}{r}.\label{lin-kgbd-eq-sol}\eea One can invert \eqref{lin-diag} to get:

\bea h_{\mu\nu}=\psi_{\mu\nu}-\frac{1}{2}\,\eta_{\mu\nu}\psi-\frac{1}{\phi_0}\,\eta_{\mu\nu}\sigma.\label{lin-hmn-sol}\eea Then, taking into account that in the Newtonian limit (matter objects at rest): $T^{(m)}=-\rho$ ($\rho$ is the energy density of matter), since $g_{\mu\nu}=\eta_{\mu\nu}+h_{\mu\nu}$:

\bea &&g_{\mu\nu}=\eta_{\mu\nu}+\frac{4}{\phi_0}\left[\int d^3x\,\frac{T^{(m)}_{\mu\nu}}{r}-\frac{1+\omega_\text{BD}}{3+2\omega_\text{BD}}\int d^3x\frac{T^{(m)}\eta_{\mu\nu}}{r}\right],\nonumber\\
&&g_{00}=-1+\frac{2}{\phi_0}\left(\frac{4+2\omega_\text{BD}}{3+2\omega_\text{BD}}\right)\int d^3x\,\frac{\rho}{r},\nonumber\\
&&g_{ij}=\delta_{ij}\left[1+\frac{4}{\phi_0}\left(\frac{1+\omega_\text{BD}}{3+2\omega_\text{BD}}\right)\int d^3x\,\frac{\rho}{r}\right].\label{lin-metric-comp}\eea If compare the Newtonian potential of a point particle of mass $M$: $U=G_N M/r$, obtained in the weak-field -- and low velocities -- limit of general relativity: $g_{00}=-1+2U$, with the one above, one gets the following relationship:

\bea 8\pi G_\text{eff}=\frac{1}{\phi_0}\left(\frac{4+2\omega_\text{BD}}{3+2\omega_\text{BD}}\right).\label{bd-eff-g}\eea This means that the factor $\phi\approx \phi_0$ that multiplies the curvature scalar in \eqref{bd-action} -- the one that sets the strength of the gravitational interactions point by point -- is just the gravitational coupling associated with the tensor part of the gravitational interaction. Meanwhile, the effective gravitational coupling constant that is measured in Cavendish-type experiments is $G_\text{eff}$ in \eqref{bd-eff-g}. This is also contributed by the scalar piece of the gravitational interactions, the one that originates the strange factor $(4+2\omega_\text{BD})/(3+2\omega_\text{BD})$ in \eqref{bd-eff-g}. 

The null-null component of the metric: $g_{00}$, determines the gravitational weight of the body and also the redshift. Then, since the factor $(4+2\omega_\text{BD})/(3+2\omega_\text{BD})$ is being absorbed in the definition of the measured gravitational constant $G_\text{eff}$ in \eqref{bd-eff-g}, there is no difference in the results of the gravitational redshift experiments within the BD theory as compared with general relativity. Nonetheless, the deflection of light experiments within the BD theory lead to results that differ from those within GR. This is due to the fact that the deflection of light is influenced by the ratio $g_{ii}/g_{00}$ instead. It is obtained that\cite{bd-1961}: $$\delta\theta=\frac{4G_\text{eff} M}{r_*}\left[\frac{3+2\omega_\text{BD}}{4+2\omega_\text{BD}}\right],$$ where $r_*$ is the closest approach distance to the astrophysical object -- the Sun, for instance -- by the light ray. From the latter equation and equations \eqref{lin-kgbd-eq}, \eqref{lin-metric-comp} and \eqref{bd-eff-g}, it is evident how the GR limit can be recovered from the BD theory: just take the $\omega_\text{BD}\rightarrow\infty$ limit. In this limit from \eqref{bd-eff-g} it follows that $8\pi G_\text{eff}=\phi_0^{-1}$, while from \eqref{lin-metric-comp}, for a stationary mass point of mass M we get that: $$g_{00}=-1+\frac{2G_\text{eff} M}{r},\;g_{ij}=\delta_{ij}\left(1+\frac{2G_\text{eff} M}{r}\right).$$ The fact that in the (weak-field) $\omega_\text{BD}\rightarrow\infty$ limit the measured gravitational constant $8\pi G_\text{eff}=1/\phi_0$, means that the strength of the gravitational interaction in this limit is entirely due to the metric tensor field, i. e., that the BD scalar field is decoupled from the gravitational field. This is why GR is recovered in this limit. See, however, Ref. \refcite{bd_faraoni_prd_1999}, where by means of the conformal transformations tool the author shows that the known result of Brans-Dicke theory reducing to general relativity when $\omega_\text{BD}\rightarrow\infty$, is false if the trace of the matter energy-momentum tensor vanishes.

\subsubsection{Brans-Dicke theory with non-vanishing potential}

For the general case with $V\neq 0$, following a procedure similar to the one applied above\cite{stabile_weak_f_1}, we obtain the following expression for the effective (measured) gravitational coupling in the Brans-Dicke theory\cite{stabile_weak_f_1, stabile_weak_f_2, stabile_weak_f_3, salgado_wek_f}:

\bea 8\pi G_\text{eff}=\frac{1}{\phi_0}\left[\frac{3+2\omega_\text{BD}+e^{-M_0r}}{3+2\omega_\text{BD}}\right],\label{bd-v-eff-g}\eea where $\phi_0$ is the value of the field around which the perturbations \eqref{perts} are performed, while the mass (squared) of the propagating scalar perturbation is given by:

\bea M^2_0=\frac{\phi_0 V''_0}{3+2\omega_\text{BD}},\label{bd-mass-eff-g}\eea with $V_0=V(\phi_0)$, $V'_0=\der_\phi V|_{\phi_0}$, $V''_0=\der^2_\phi V|_{\phi_0}$, etc. It is seen that in the formal limit $M_0\rightarrow\infty$, i. e., when the propagating scalar degree of freedom decouples from the rest of the field spectrum of the theory, we recover general relativity with $8\pi G_\text{eff}=1$ (the choice $\phi_0=1$ is implicit). Meanwhile, in the limit of a light scalar field $M_0\rightarrow 0$ we retrieve the expression \eqref{bd-eff-g} for the measured gravitational coupling in the original formulation of the BD theory\cite{bd-1961}.


\subsection{The PPN approximation}\label{subsect-bd-ppn}

In order to discuss on another of the well-known tests of metric gravitational theories: the perihelion shift of Mercury's orbit, we have to go to, at least, the second approximation in the expansion of the metric coefficients. This is where the parametrized post-Newtonian (PPN) formalism\cite{will-lrr-2014} enters the scene: the comparison of metric theories of gravity with experiment becomes a quite simple task when the slow-motion, weak-field limit is considered. The PPN approximation is sufficiently accurate in view of present and future solar-system tests. Following the PPN formalism, the spacetime metric can be expanded around the Minkowski metric $\eta_{\mu\nu}$ in terms of dimensionless gravitational potentials potentials of a varying degree of smallness: 

\bea U({\bf x},t)=\int\frac{d^3x'\rho({\bf x}',t)}{|{\bf x}-{\bf x}'|},\label{grav-pot-will}\eea where the degree of smallness is set according to the following rules\cite{will-lrr-2014}: $U\sim v^2\sim\Pi\sim p/\rho\sim\epsilon$, $v^i\sim\epsilon^{1/2}$, etc. Here we have considered that $v^i=dx^i/dt$ is the coordinate velocity, $\Pi$ is the internal energy per unit rest mass, and $p$, $\rho$ are the pressure and the density of rest mass, respectively, as measured in the comoving frame. A consistent post-Newtonian limit requires determination of $g_{00}$ up to ${\cal O}(\epsilon^2)$, while $g_{ij}$ is to be computed up to ${\cal O}(\epsilon)$. Different metric theories lead to differing coefficients in front of the mentioned gravitational potentials. The PPN formalism inserts parameters in place of the mentioned coefficients. For the PPN metric in general relativity and in scalar-tensor theories we have\cite{will-lrr-2014}:

\bea &&g_{00}=-1+2U-2\beta U^2+2(\gamma+1)\Phi_1+2(1+3\gamma-2\beta)\Phi_2+2\Phi_3+6\gamma\Phi_4+{\cal O}(\epsilon^3),\nonumber\\
&&g_{ij}=(1+2\gamma U)\delta_{ij}+{\cal O}(\epsilon^3),\label{metric-c-ppn}\eea where

\bea &&U=\int\frac{d^3x'\rho'}{|{\bf x}-{\bf x}'|},\;\Phi_1=\int\frac{d^3x'\rho'v'^2}{|{\bf x}-{\bf x}'|}.\;\Phi_2=\int\frac{d^3x'\rho' U'}{|{\bf x}-{\bf x}'|},\nonumber\\
&&\Phi_3=\int\frac{d^3x'\rho'\Pi'}{|{\bf x}-{\bf x}'|},\;\Phi_4=\int\frac{d^3x'p'}{|{\bf x}-{\bf x}'|}.\label{potentials-ppn}\eea In the above equations the PPN parameters $\gamma$ and $\beta$ represent how much space-curvature is produced by unit rest mass and how much nonlinearity is in the superposition law for gravity, respectively. These are, in a sense, the most important parameters in metric theories and are the only non-vanishing PPN parameters in GR and in STT-s. For GR $\gamma=\beta=1$, while fo the BD theory:

\bea \gamma=\frac{1+\omega_\text{BD}}{2+\omega_\text{BD}},\;\beta=1.\label{gamma-bd}\eea 

The parameter $\gamma$ enters the equations for the determination of the deflection of light, the time delay of light, etc. For the deflection of light by the Sun (consider for definiteness a grazing ray) the deflection angle is computed as: $$\delta\theta\approx\frac{1}{2}(1+\gamma)\,1.''7505,$$ while for the time delay of a light ray that passes close to the Sun one gets: $$\delta t\approx\frac{1}{2}(1+\gamma)\left[240-20\ln\left(\frac{d^2}{r}\right)\right]\,\mu s,$$ where $d$ is the distance of closest approach of the light ray to the Sun in solar radii and $r$ is the distance from the Sun in astronomical units. Measurements of the above quantities in different solar system experiments place constraints on $\gamma$ and, hence, on the BD coupling parameter: $$\omega_\text{BD}=\frac{2\gamma-1}{1-\gamma}.$$ The Cassini experiment leads to the constraint\cite{will-lrr-2014, alonso_prd_2017}: $\omega_\text{BD}>40000$, so that, at least in the solar system, the BD theory must not differ appreciably from GR. However, the latter constraint is valid only for the massless BD theory. If allow for the BD scalar to be self-interacting, i. e., if there is a non-vanishing self-interacting potential for the scalar field, then the PPN formalism does not apply and it may happen that, thanks to the chameleon screening mechanism, the BD field is screened in the solar system scale, so that it escapes detection. For the chameleon screening mechanism within the BD theory see section \ref{subsect-cham}.


\subsection{Exact solutions: Brans-Dicke wormholes without exotic matter}\label{subsect-bd-wh}

In this section we shall briefly (and exclusively) discuss on static spherically symmetric solutions of the Brans-Dicke equations \eqref{bd-feq}, \eqref{kgbd-eq}. For a good and detailed account of this subject see Refs. \citen{brans_phd_thesis, agnese_la_camera_prd_1995, torres_wh_prd_1997, nandi_prd_1998, lobo_prd_2010, faraoni_sols_prd_2016, belknap_prd_2017}. Regarding the exact cosmological solutions, there can be found in the bibliography very nice expositions\cite{faraoni-book, elizalde_nojiri_prd_2004, faraoni_sols_ann_phys_2018}. In this review we prefer to look for the more general and fundamental qualitative properties of the cosmological dynamics (see section \ref{sect-dsyst-bd}). In this case the tools of the dynamical systems offer a 'bird eye' view on the whole space of relevant solutions. These are correlated with critical points in certain equivalent state space or phase space (see Ref. \refcite{ejp-2015} for a brief introduction to this subject).

As it was stated in Ref. \refcite{agnese_la_camera_prd_1995} -- and recently confirmed in Ref. \refcite{faraoni_sols_prd_2016} -- all static spherically symmetric solutions of the Brans-Dicke equations \eqref{bd-feq}, \eqref{kgbd-eq}, depending on the value of the PPN parameter \eqref{gamma-bd}: $$\gamma=\frac{1+\omega_\text{BD}}{2+\omega_\text{BD}},$$ are either naked singularities if $\gamma<1$ or wormholes if $\gamma>1$. Black hole solutions are not found in this case. 

There was an explosion of interest in the past on the physics of wormholes fueled, mainly, by the classical analysis of traversable wormholes\cite{morris_thorne_am_j_phys_1988, visser_npb_1989, kar_prd_1994, hochberg_prl_1998} and the possibility of constructing time machines\cite{morris_prl_1988}. The latter possibility was inevitably correlated with the existence of matter violating the weak energy condition\cite{wald_gr_1984} (WEC) and this posed a serious problem to the hypothetical construction of such machines. Although the interest in the wormholes has decayed with time, there remains a useful exercise to show how these geometrical structures with non-trivial topology may be obtained without the need to resort to exotic WEC-violating matter. Brans-Dicke theory -- and the more general scalar-tensor theories -- offer an instructive illustration of the latter possibility\cite{accetta_npb_1990, agnese_la_camera_prd_1995, torres_wh_prd_1997, nandi_prd_1998, lobo_prd_2010, sushkov_kozyrev_prd_2011, montelongo_mpla_2011, bronnikov_gc_2010, kar_prd_2016, dogru_ijmpd_2015}. The dynamical BD wormhole was studied in Ref. \refcite{accetta_npb_1990}, while the static case was investigated in Refs. \citen{agnese_la_camera_prd_1995, torres_wh_prd_1997, nandi_prd_1998, lobo_prd_2010, sushkov_kozyrev_prd_2011, montelongo_mpla_2011}. Here, for sake of simplicity, we shall expose this subject by focusing in static BD wormholes exclusively.


\subsubsection{WEC violation and exotic matter}

The weak energy condition\cite{wald_gr_1984} establishes that the energy density of matter as measured by an observer with 4-velocity $u^\mu$ should be non-negative. More generally; given a time-like vector $\xi^\mu$, the following condition must hold: $T_{\mu\nu}\xi^\mu\xi^\nu\geq 0$, where the $T_{\mu\nu}$ are the components of the stress-energy tensor of the matter degrees of freedom. Within the frame of general relativity the above condition can be written also in the following equivalent form: 

\bea G_{\mu\nu}\xi^\mu\xi^\nu\geq 0.\label{wec}\eea According to Ref. \refcite{morris_thorne_am_j_phys_1988}, for static wormhole configurations within GR, observers moving sufficiently fast through the wormhole's throat will see negative energy density of matter, i. e., the WEC \eqref{wec} will be violated. The authors of this reference called the type of matter -- populating the throat of the wormhole -- with that property as ``exotic matter''. As a matter of fact not only static spherically symmetric wormholes require WEC violating matter at their throats; any traversible non-spherical non-static wormhole would require exotic matter at its throat as well. The geometrical argument given in Ref. \refcite{morris_thorne_am_j_phys_1988} is quite simple: the cross-sectional area of the bundle of light rays entering the wormhole at one mouth and emerging from the other, must be initially decreasing and then increasing, so that the energy density of matter through which the bundle of light rays passes, should be negative in order to provide the required gravitational repulsion. 

One of the basic assumptions on which the work of Ref. \refcite{morris_thorne_am_j_phys_1988} is based, is that the gravitational phenomena are adequately described by general relativity. Here we replace GR by the Brans-Dicke theory in order to look for a possible relaxation of the requirement of exotic matter. Our reasoning line is based in the following idea. According to the BD equations of motion \eqref{bd-feq}: $$G_{\mu\nu}\xi^\mu\xi^\nu=\frac{1}{\phi}\left[\rho_{(m)}+\tilde{\rho}_{(\phi)}\right]+\frac{1}{\phi}\,\xi^\mu\xi^\nu\left(\nabla_\mu\nabla_\nu-g_{\mu\nu}\Box\right)\phi,$$ where we have assumed the matter in the form of a perfect fluid with 4-velocity $\xi^\mu$ (a time-like vector), $\rho_{(m)}$ is the matter density as seen by an observer which is co-moving with the perfect fluid and $\tilde{\rho}_{(\phi)}=\xi^\mu\xi^\nu T_{\mu\nu}^{(\phi)}$, with $T_{\mu\nu}^{(\phi)}$ defined in \eqref{def-1}. While for reasonable assumptions on non-negativity of the BD coupling parameter and of the self-interaction potential the sum within square brackets in the RHS of the above equation is always non-negative: $\rho_{(m)}+\tilde{\rho}_{(\phi)}\geq 0$, the sign of the term $\propto\xi^\mu\xi^\nu\left(\nabla_\mu\nabla_\nu-g_{\mu\nu}\Box\right)\phi$ is non-definite, so that, in principle this latter term may be negative, in which case it may happen that $G_{\mu\nu}\xi^\mu\xi^\nu<0$, i. e., the WEC is violated even if the measured matter density is non-negative. A similar situation occurs if consider, for instance, the null energy condition (NEC). In this case one replaces the time-like vector $\xi^\mu$ by the null vector $k^\mu$, so that, from the BD motion equations \eqref{bd-feq} one obtains: $$G_{\mu\nu}k^\mu k^\nu=\frac{1}{\phi}T^{(m)}_{\mu\nu}k^\mu k^\nu+\frac{\omega_\text{BD}}{\phi^2}(k^\mu\der_\mu\phi)^2+\frac{1}{\phi}(\nabla_\mu\nabla_\nu\phi)k^\mu k^\nu.$$ Assuming that the matter does not violate the NEC, there is yet room for violation of the energy condition by the BD gravitational field if either: i) $\omega_\text{BD}<0$, $\der_\mu\phi\neq 0$, or ii) $\nabla_\mu\nabla_\nu\phi<0$. The above arguments specially apply in the vacuum case since the gravitational scalar field can be the source of the violation of the energy conditions even in the absence of matter. Below we shall illustrate the present discussion with concrete examples.


\subsubsection{Vacuum Brans-Dicke wormholes} 

Here we shall discuss the static spherically symmetric solutions to the vacuum BD equations of motion about a point mass of mass $M$:

\bea G_{\mu\nu}&=&\frac{\omega_\text{BD}}{\phi^2}\left[\der_\mu\phi\der_\nu\phi-\frac{1}{2}g_{\mu\nu}\left(\der\phi\right)^2\right]+\frac{1}{\phi}\left(\nabla_\mu\nabla_\nu-g_{\mu\nu}\Box\right)\phi,\nonumber\\
\Box\phi&=&0,\label{vac-bd-feq}\eea where, for simplicity, we consider a vanishing self-interaction potential: $V=0$. The most general static spherically symmetric line element can be written as:

\bea ds^2=-e^{\nu(r)}dt^2+e^{\mu(r)}dr^2+e^{\lambda(r)}r^2 d\Omega^2,\label{vac-bd-met}\eea where, as usual: $d\Omega^2\equiv d\theta^2+\sin^2\theta d\vphi^2$, while $\nu$, $\mu$ and $\lambda$ are arbitrary functions of the radial coordinate $r$. Here, following Ref. \refcite{agnese_la_camera_prd_1995} we use the gauge where $$\lambda-\mu=\ln\left(1-\frac{2\eta}{r}\right).$$ Inserting this metric into the vacuum BD field equations the following solutions are obtained\cite{agnese_la_camera_prd_1995, krori_jmp_1982}:

\bea &&e^{\nu(r)}=\left(1-\frac{2\eta}{r}\right)^A,\;e^{\mu(r)}=\left(1-\frac{2\eta}{r}\right)^B,\nonumber\\
&&e^{\lambda(r)}=\left(1-\frac{2\eta}{r}\right)^{1+B},\;\phi(r)=\phi_0\left(1-\frac{2\eta}{r}\right)^{-\frac{A+B}{2}},\label{agnese_bd_wh}\eea where $\eta$, $A$, and $B$ are integration constants and, according to \eqref{bd-eff-g}: $$\phi_0=\frac{1}{8\pi G_\text{eff}}\left(\frac{4+2\omega_\text{BD}}{3+2\omega_\text{BD}}\right)=\frac{2}{G_\text{eff}}\frac{1}{1+\gamma},$$ with $\gamma$ -- the only PPN parameter in \eqref{gamma-bd} that differs from the corresponding GR one. This latter parameter and the constants $A$, $B$ obey the following constraint:

\bea 1-\gamma=\frac{(A+B)^2}{2(1+AB)}.\label{gamma-const}\eea By going into the PPN limit, the constants $\eta$, $A$ and $B$ can be written in terms of the mass $M$ and of the PPN parameter $\gamma$:

\bea A=\sqrt\frac{2}{1+\gamma},\;B=-\gamma\sqrt\frac{2}{1+\gamma},\;\eta=M\sqrt\frac{1+\gamma}{2}.\label{const-vac-bd-met}\eea It is seen that in the limit $\omega_\text{BD}\rightarrow\infty$ $\Rightarrow\gamma\rightarrow 1$, the Schwarzschild GR solution is recovered.

The question now is whether the above static vacuum BD spacetime can support a wormhole geometry. In order to answer to this question it is demanded that we write the metric \eqref{vac-bd-met} in the canonical Morris-Thorne wormhole metric\cite{morris_thorne_am_j_phys_1988}:

\bea ds^2=-e^{2\Phi(R)}dt^2+\frac{dR^2}{1-b(R)/R}+R^2d\Omega^2,\label{wh-met}\eea where the functions $\Phi=\Phi(R)$ and $b=b(R)$ are known as the redshift and shape function, respectively. The former determines the gravitational redshift while the latter determines the spatial shape of the wormhole. Where the standard radial coordinate $R$ is related with the radial coordinate $r$ above in the following way: 

\bea R(r)=r\left(1-\frac{2\eta}{r}\right)^{[1-\gamma\sqrt{2/(1+\gamma)}]/2}.\label{R-r}\eea A wormhole spacetime arises whenever $\gamma>1$, in which case there is a minimum allowed value of $r$: $r\geq r_*$, where $$r_*=\eta\left(1+\gamma\sqrt\frac{2}{1+\gamma}\right),$$ which means, in turn, that $R\geq R_*$ ($R_*=R(r_*)$). From \eqref{R-r} it follows that $r_*>2\eta$, so that $R_*>0$. We have:

\bea &&\Phi(R)=\sqrt\frac{2}{1+\gamma}\ln\sqrt{1-\frac{2\eta}{r(R)}},\nonumber\\
&&\frac{b(R)}{R}=1-\frac{\left[1-\frac{\eta}{r(R)}\left(1+\gamma\sqrt\frac{2}{1+\gamma}\right)\right]^2}{1-2\eta/r(R)},\label{wh-met-fuct}\eea where $r(R)$ means taking the inverse of \eqref{R-r}. Notice that in the limit $\gamma\rightarrow 1$ ($\omega_\text{BD}\rightarrow\infty$), an event horizon arises at $R=2M$ and the above solution depicts a static spherically symmetric black hole.

Here we shall discuss the case with $\gamma>1$, which is the one leading to a static wormhole as we shall see. Actually, in this case, the redshift function $\Phi(R)$ is finite everywhere and since $R_*>0$, there is no event horizon. Meanwhile, the shape function meets the following bound: $b(R)/R\leq 1$, and also the limits $$\lim_{R\rightarrow R_*}\frac{b(R)}{R}=1,\;\lim_{R\rightarrow\infty}\frac{b(R)}{R}=0.$$ Hence, the functions $\Phi(R)$ and $b(R)$ meet the requirements necessary for the description of a wormhole geometry\cite{morris_thorne_am_j_phys_1988}. For more details on the geometry of the static spherically symmetric vacuum BD wormhole see, for instance, Refs. \citen{agnese_la_camera_prd_1995, nandi_prd_1998, lobo_prd_2010}.

It remains to check the WEC for this wormhole. By inserting the above wormhole solutions into the vacuum BD equations \eqref{vac-bd-feq} one obtains that: $$G_{00}=\frac{(1-\gamma)(1+2\gamma)}{(1+\gamma)}\frac{\eta^2}{r^4(R)}\left[1-\frac{2\eta}{r(R)}\right]^{2\left[\sqrt\frac{2}{1+\gamma}-1\right]},$$ which, since for the static spherically symmetric wormhole $\gamma>1$, is a negative quantity: $G_{00}<0$. This, in turn, implies that the WEC is violated, as required for the occurrence of a wormhole. The interesting fact here is that the gravitational repulsion which is necessary at the throat of the wormhole in order to warrant conversion of the cross-sectional area of the bundle of light rays entering at one mouth from decreasing into increasing cross-sectional area, is provided by the BD scalar field which is a part of the gravitational field itself and, by no means can be considered as matter (not even exotic). For wormholes within GR -- see, for instance, Refs. \citen{morris_thorne_am_j_phys_1988, visser_npb_1989, kar_prd_1994, morris_prl_1988} -- the gravitational repulsion at the throat can be supplied only by the exotic matter instead. 

We want to make a necessary comment to the above wormhole solution within BD theory: as shown in Ref. \refcite{nandi_prd_1998}, the static spherically symmetric vacuum BD solution supporting wormhole geometry arises only in the narrow interval of the BD coupling: $-3/2<\omega_\text{BD}<-4/3$, that is ruled out by the solar system experiments. This, however, does not mean the end of the story: we may allow for self-interacting BD scalar field (non-vanishing self-interaction potential) that supports the wormhole geometry, and then we may check whether the chameleon screening mechanism works appropriately as to hide the BD field from solar system experiments (see the section \ref{sect-screen}). Besides, the above stringent bound is due to a specific choice of an integration constant. In Ref. \refcite{lobo_prd_2010} the general study was performed and the $\omega_\text{BD}$-interval was improved to include, in particular, the case $\omega_\text{BD}=0$.


\subsubsection{Brans-Dicke wormholes in the presence of background matter}

For sake of completeness let us briefly discuss on static spherically symmetric Brans-Dicke wormholes in the presence of ordinary matter\cite{torres_wh_prd_1997}. Here we follow Ref. \refcite{torres_wh_prd_1997} where the stress-energy tensor of matter was chosen in such a way that its non-vanishing components are: $$T^0_0=-\rho(r),\;T^r_r=-\tau(r),\;T^\theta_\theta=T^\vphi_\vphi=p(r),$$ with the matter obeying the following equation of state: $-\tau+2p=\epsilon\rho$ ($\epsilon$ is a unspecified constant). Here we choose the gauge: $\lambda(r)=0$, in the line-element \eqref{vac-bd-met}. We get:

\bea &&\phi(r)=\phi_0\,e^{k\nu(r)/2},\;\nu(r)=-\frac{\alpha}{r},\nonumber\\
&&e^{-\mu(r)}=\frac{\alpha\,e^\frac{2\alpha B}{Ar}}{r}\left(1-\frac{\alpha A}{4r}\right)^{-(8l+1)}\left(I+C\right),\label{bd-mat-sol}\eea where $\alpha$ is a positive constant, $C$ is an integrations constant, $$k=\frac{\epsilon-1}{2\omega_\text{BD}+3+(\omega_\text{BD}+1)(\epsilon-1)},\;l=-\frac{B}{A^2},$$ and the constants $A$, $B$ above are given by:

\bea &&A=-2\frac{\epsilon+2(1+\omega_\text{BD})}{2+\omega_\text{BD}+\epsilon(1+\omega_\text{BD})},\nonumber\\
&&B=-\frac{8(1+\epsilon)+\epsilon^2(\omega_\text{BD}+2)+4\omega_\text{BD}^2(1+\epsilon)+\omega_\text{BD}(11+12\epsilon)}{\left[2+\omega_\text{BD}+\epsilon(1+\omega_\text{BD})\right]^2}.\nonumber\eea The integral $I$ in \eqref{bd-mat-sol} is given by\cite{torres_wh_prd_1997}: $$I=\frac{r}{\alpha}+2lA\ln\left(\frac{r}{\alpha}\right)+\frac{r}{\alpha}\sum_{n=2}^\infty\frac{(-1)^n}{n!(n-1)}\left(\frac{2\alpha B}{Ar}\right)^n\,_3F_1\left(-n,8l,b;b;-\frac{A^2}{8B}\right),$$ where $_3F_1$ is the hypergeometric function. In order to fix the constant $C$ a value of the throat radius $r_*$ is to be chosen so that the ``flaring out'' condition\cite{morris_thorne_am_j_phys_1988}: $$\lim_{r\rightarrow r_*^+}e^{-\mu(r)}=0^+,$$ is satisfied. For non-negative $A\geq 0$ the flaring out condition holds for all values of the BD coupling constant $\omega_\text{BD}$ but for $\omega_\text{BD}=-(2+\epsilon)/(1+\epsilon)$, in which case the constant $A$ is diverging.

The metric \eqref{vac-bd-met}, \eqref{bd-mat-sol} describes two asymptotically flat spacetimes joined by a throat, i. e., a wormhole. Since at the throat $\exp\mu\rightarrow\infty$, then for the terms contributing towards the null-null component of the stress-energy tensor of matter we get:

\bea \tau_*=\frac{\phi_0\,e^{-k\alpha/2r_*}}{r_*^2},\;\rho_*=\tau_*\frac{k+1+r_*/\alpha}{1-\epsilon r_*/\alpha},\;p_*=\frac{\tau_*}{2}\frac{\epsilon(k+1)+1}{1-\epsilon r_*/\alpha}.\label{set-throat}\eea Since we want standard matter at the throat, hence: 

\bea \rho_*\geq 0,\;\rho_*-\tau_*\geq 0,\;\rho_*+p_*\geq 0.\label{e-cond}\eea But, at the same time the WEC should be violated at the throat, i. e., $$\frac{2(\omega_\text{BD}+1)+\epsilon}{2\omega_\text{BD}+3}\rho_*<0,$$ besides, $r_*\geq\alpha|A/4|$, i. e.\cite{torres_wh_prd_1997}: $$r_*\geq\alpha\left|\frac{2+\omega_\text{BD}}{4+3\omega_\text{BD}}\right|.$$ Take, for instance, the particular case with $\epsilon=2$. In this case the WEC is violated if $-2\leq\omega_\text{BD}\leq-3/2$ and, for $\omega_\text{BD}=-1.75$, for instance, the matter fulfills the conditions \eqref{e-cond}. This case offers an example where we have normal matter at the throat of the wormhole while the WEC is violated by the BD field, so that there is no need for the exotic matter.


\section{Scalar-tensor theories}\label{sect-stt}

Scalar-tensor theories of gravity\cite{faraoni-book, nordvedt_astroph_j_1970, wagoner_prd_1970, bergmann_ijtp_1968, reasenberg-1979, damour_prl_1993, damour_prd_1993, damour_prd_1996, damour-epj-1998, farese_polarski_prd_2001, anderson_yunes_prd_2017, massaeli_epjc_2017, belinchon_ijmpd_2017, africanos_ijmpd_2018, brazilenos_rastal_epjc_2014, rincon_epjc_2018} are a generalization of the Brans-Dicke theory to allow the BD coupling to be a function of the scalar field: $\omega_\text{BD}\rightarrow\omega(\phi)$, i. e., to be a varying parameter:

\bea S_\text{ST}=\int d^4x\sqrt{|g|}\left[\phi R-\frac{\omega(\phi)}{\phi}(\der\phi)^2-2V(\phi)+2{\cal L}_m\right].\label{stt-action}\eea The above action can be found also in an alternative presentation (see, for instance, Refs. \citen{wagoner_prd_1970, faraoni-book}): 

\bea S_\text{ST}=\int d^4x\sqrt{|g|}\left[f(\phi)R-\omega(\phi)(\der\phi)^2-2V(\phi)+2{\cal L}_m\right].\label{stt-alt-action}\eea However, it is not difficult to prove that this latter action is transformed into \eqref{stt-action} by a simple redefinition of the scalar field and of the coupling function: $$f(\phi)\rightarrow\phi,\;\;\omega(\phi)\rightarrow\frac{f(\phi)}{(\der_\phi f)^2}\,\omega(\phi).$$ In this review we shall use both presentations indistinctly, however, below we shall write only the motion equations that can be derived from the action \eqref{stt-action} and the interested reader is encouraged to make the transformation to the variables of \eqref{stt-alt-action} if desired.

The field equations that are derivable from \eqref{stt-action} are very similar to the BD field equations \eqref{bd-feq} and \eqref{kgbd-eq} with the replacement $\omega_\text{BD}\rightarrow\omega=\omega(\phi)$:

\bea &&G_{\mu\nu}=\frac{1}{\phi}\,T^{(m)}_{\mu\nu}+\frac{\omega}{\phi^2}\left[\der_\mu\phi\der_\nu\phi-\frac{1}{2}g_{\mu\nu}\left(\nabla\phi\right)^2\right]-g_{\mu\nu}\frac{V(\phi)}{\phi}+\frac{1}{\phi}(\nabla_\mu\nabla_\nu-g_{\mu\nu}\Box)\phi,\nonumber\\
&&\Box\phi+\frac{\der_\phi\omega}{3+2\omega}(\der\phi)^2=\frac{T^{(m)}}{3+2\omega}+\frac{2}{3+2\omega}\left(\phi\der_\phi V-2V\right),\label{stt-feq}\eea but for the second term in the LHS of the Klein-Gordon (KG) equation in \eqref{stt-feq}.

The need for a generalization of BD theory, besides its heuristic potential, is rooted in the tight constraints on the BD coupling parameter $\omega_\text{BD}$ that the solar system experiments have established. If one allows for the possibility of a varying coupling: $\omega_\text{BD}\rightarrow\omega(\phi)$, the latter experimental constraints may be avoided or, at least, alleviated.


\subsection{PPN formalism}\label{subsect-stt-ppn}

In order to better understand the above mentioned possibility, let us to go to the PPN approximation that, for solar system experiments, is enough. The PPN parameters for STT-s given by \eqref{stt-action} with vanishing potential are\cite{will-lrr-2014, brazilenos_rastal_epjc_2014, deng_xie(ppn)_prd_2016}:

\bea \gamma=\frac{1+\omega_0}{2+\omega_0},\;\beta=1+\frac{\lambda}{4+2\omega_0},\label{stt-ppn-par}\eea where 

\bea \left.\omega_0\equiv\omega(\phi_0),\;\lambda\equiv\frac{\phi d\omega/d\phi}{(3+2\omega)(4+2\omega)}\right|_{\phi_0}.\label{stt-def-par}\eea In the above equations the scalar field is evaluated today ($\phi_0$) and it is determined by appropriate cosmological boundary conditions given far from the system of interest. In a similar way the effective gravitational constant $G_\text{eff}$ -- the one measured in Cavendish type experiments -- can be expressed through $\phi_0$ and $\omega_0$ (here we include the contribution from a non-vanishing potential):

\bea 8\pi G_\text{eff}=\frac{1}{\phi_0}\left[\frac{3+2\omega_0+e^{M_0 r}}{3+2\omega_0}\right],\label{stt-gn}\eea where the mass $M_0$ of the scalar perturbation is given by \eqref{bd-mass-eff-g} with following the replacement $\omega_\text{BD}\rightarrow\omega_0$. The following formal limits (we consider the choice $\phi_0=1$): $M_0\rightarrow\infty$ and $M_0\rightarrow 0$, lead to general relativity and to massless STT, respectively.

In order to avoid the tight constraints from the solar system experiments, the function $\omega=\omega(\phi)$ and the parameter $\lambda$ could have the property that, at present (today) and in weak-field environments, the value of the scalar field $\phi_0$ is such that $\omega_0\rightarrow\infty$, while $\lambda\rightarrow 0$. This property entails that in the mentioned situations GR is closely approached. However, it could happen that in the past/future and in strong field environments the related values of the function $\omega$ and of the parameter $\lambda$ are far from the GR values, making the corresponding STT to appreciably differ from general relativity. A warning is to be placed here: not every possible choice of the coupling function $\omega=\omega(\phi)$ leads to recovering of general relativity at present and in weak field environments, as required by solar system experiments\cite{will-lrr-2014, damour-epj-1998, reasenberg-1979, alonso_prd_2017}. Actually, since it is mandatory that both conditions: 

\bea \omega_0\rightarrow\infty,\;\lambda\rightarrow 0,\label{stt-gr-cond}\eea be simultaneously satisfied, there are theories for which the first limit above is satisfied while the second one is not. Take as an example the choice\cite{faraoni-book}: $$\omega(\phi)=\frac{\alpha}{\left(1-\frac{\phi}{\phi_0}\right)^\beta},$$ where $\alpha$ and $\beta$ are arbitrary non-negative constants. In this case we have that $$\lambda=\frac{\alpha\beta\phi}{\left[3\left(1-\frac{\phi}{\phi_0}\right)^\beta+2\alpha\right]\left[4\left(1-\frac{\phi}{\phi_0}\right)^\beta+2\alpha\right]\left(1-\frac{\phi}{\phi_0}\right)},$$ so that $\omega_0\rightarrow\infty$, $\lambda\rightarrow\infty$.


\subsection{Wormholes}\label{subsect-stt-wh}

As it was for the Brans-Dicke theory -- see the discussion in subsection \ref{subsect-bd-wh} -- for the STT \eqref{stt-action}, due to the term with the second derivatives of the scalar field in the RHS of the Einstein-BD (EBD) equation in \eqref{stt-feq}, it may happen that even for matter with positive energy density, the quantity: $G_{\mu\nu}\xi^\mu\xi^\nu<0$ ($\xi^\mu$ are the coordinates of an arbitrary time-like vector) is negative, so that the WEC is violated. This, in turn, means that wormhole configurations may arise. A nice illustration can be found in Ref. \refcite{kar_prd_2016}, where the authors obtained a large class of static, spherically symmetric wormhole spacetimes in the scalar-tensor gravity \eqref{stt-action}, \eqref{stt-feq} with vanishing potential $V=0$, for which the (standard) matter satisfies the weak energy condition. Under the assumption of traceless matter, these wormholes are characterized by having vanishing Ricci curvature: $R=0$. This latter condition appreciably simplifies the BD equations of motion. Actually, if combine the trace of the EBD equation in \eqref{stt-feq} with the Klein-Gordon motion equation for the scalar field (second line equation in \eqref{stt-feq}), one obtains:

\bea R=-\frac{2\omega T^{(m)}}{(3+2\omega)\phi}-\frac{(\der\phi)^2}{(3+2\omega)\phi^2}\left[3\phi\der_\phi\omega-\omega(2\omega+3)\right].\nonumber\eea If further require vanishing of the second term in the RHS of the above equation:

\bea \frac{d\omega}{d\phi}-\frac{\omega(2\omega+3)}{3\phi}=0\;\Rightarrow\;\omega(\phi)=\frac{3C_0\phi}{1-2C_0\phi},\label{stt-wh-cond}\eea where $C_0$ is an arbitrary integration constant. In Ref. \refcite{kar_prd_2016} the authors made the choice $C_0=-1/2$. Assuming the static spherically symmetric line element in the standard wormhole form: $$ds^2=-e^{2\Phi(r)}dt^2+\frac{dr^2}{1-b(r)/r}+r^2d\Omega^2,$$ the authors found the following class of solutions:

\bea &&\Phi(r)=\ln\left[C_1\left(m+\frac{\beta}{r}\right)+C_2\sqrt{1-\frac{2m}{r}-\frac{\beta}{r^2}}\right],\nonumber\\
&&b(r)=2m+\frac{\beta}{r},\label{wh-class-sol}\eea where $m$ and $\beta$ are free parameters and $C_1$, $C_2$ are integration constants. The above redshift function is everywhere finite and non-vanishing. It was shown that for the following constraint on the parameter space: $m>0$, $\beta>0$, $C_2/C_1m>-1$, the obtained spacetimes represent static spherically symmetric wormholes. Unfortunately, for the choice of the coupling function in Ref. \refcite{kar_prd_2016}: $$\omega(\phi)=-\frac{3\phi}{2(1+\phi)}\;\Rightarrow\;\lambda=-\frac{3}{2}\left(\frac{\phi_0}{4+\phi_0}\right),$$ the (weak field) general relativity limit does not exist. For recent work on wormholes supported by STT-s see Ref. \citen{kar_prd_2016, chew_prd_2018, dogru_ijmpd_2015}.


\subsection{Compact astrophysical objects in the STT}

Among the compact objects\cite{gerosa_cqg_2016} that have been studied within the frame of the STT-s we can mention the neutron\cite{sotani_prd_2017, kokkotas_prd_2017, sotani_kokkotas_prd_2017, mendes_ortiz_prd_2016} and boson stars. Boson stars\cite{bos-strs-rev-1, bos-strs-rev-2, bos-strs-rev-3, bos-strs-ruffini, bos-strs-4, bos-strs-5, bos-strs-6, bos-strs-7, bos-strs-8, bos-strs-8-0, bos-strs-8-1, bos-strs-8-2, bos-strs-9, bos-strs-10, bos-strs-11, bos-strs-12, bos-strs-12-1, bos-strs-13, bos-strs-14, bos-strs-15, bos-strs-16, bos-strs-17, bos-strs-18, bos-strs-19, bos-strs-20, bos-strs-21, bos-strs-22, bos-strs-23, bos-strs-24, bos-strs-25, bos-strs-26, bos-strs-27, bos-strs-28, bos-strs-29, bos-strs-30, bos-strs-31} are a gravitationally bound macroscopic state made up of scalar bosons. Unlike neutron stars, whose pressure support derives from the Pauli exclusion principle, for boson stars this is replaced by Heisenberg's uncertainty principle. Bosons are incorporated into GR by considering the Lagrangian density of a non-interacting complex, massive scalar field\cite{bos-strs-ruffini}: $${\cal L}_\text{bos}=-\frac{1}{2}|\der\psi|^2-\frac{1}{2}m^2|\psi|^2-\frac{1}{4}\lambda|\psi|^4,$$ where $|\der\psi|^2\equiv g^{\mu\nu}\der_\mu\psi^*\der_\nu\psi$ and $|\psi|^2\equiv\psi^*\psi$. This leads to the following stress-energy tensor for the bosonic matter:

\bea T_{\mu\nu}^{(m)}=\frac{1}{2}\left(\der_\mu\psi^*\der_\nu\psi+\der_\mu\psi\der_\nu\psi^*\right)-\frac{1}{2} g_{\mu\nu}\left(|\der\psi|^2+m^2|\psi|^2+\frac{1}{2}\lambda|\psi|^4\right).\label{boson-set}\eea By working with the resulting Einstein's motion equation plus the Klein-Gordon equation for the boson, the boson star solution is obtained\cite{bos-strs-ruffini}. It was found that the mass of one such boson star is of the order $M\simeq M^2_\text{Pl}/m$, where $m$ is the boson mass. This mass can be increased by a factor of $M_\text{Pl}/m$ if consider a self-interaction potential for the scalar field\cite{bos-strs-4}, yielding a mass of the order of the Chandrasekhar mass.\footnote{The method of Ref. \refcite{bos-strs-ruffini} has been improved in \refcite{bos-strs-27}, where the authors showed how to go beyond the Ruffini-Bonazzola ansatz towards an exact solution of the interacting operator Klein-Gordon equation, which can be solved iteratively to ever higher precision.} More recent works\cite{bos-strs-14, bos-strs-15} show that the bosonic compact objects can act as the dark matter. These have been invoked, also, as a possible candidate for the very compact (supermassive) object at the center of the Galaxy, Sgr A*\cite{bos-strs-13}. Millimeter very long baseline interferometry will soon produce accurate images of the closest surroundings of this supermassive compact object. These images may reveal the existence of a central faint region, the so-called shadow, which may be interpreted as the observable footprint of the event horizon of a black hole. According to Ref. \refcite{bos-strs-13}, the computed images of an accretion torus around Sgr A*, assuming this compact object is a boson star with no event horizon, show that very relativistic rotating boson stars produce images extremely similar to Kerr black holes, showing in particular a shadow-like structure. However, in Ref. \refcite{bos-strs-24}, the authors discuss how these horizonless ultra-compact objects are actually distinct from black holes, both phenomenologically and dynamically. The possibility that self-interacting bosonic dark matter forms star-like objects has been investigated in Ref. \refcite{bos-strs-14}, while in Ref. \refcite{bos-strs-15} a detailed analysis of how bosonic dark matter ``condensates'' interact with compact stars was provided. The possibility of weakening the black hole uniqueness theorem for rotating configurations and soliton-type collisions of excited boson stars has been reviewed in Ref. \refcite{bos-strs-16}. A very interesting result has been presented in Ref. \refcite{bos-strs-29} in the form of a theorem, where it was proved that self-gravitating static scalar fields whose self-interaction potential is a monotonically increasing function of its argument, cannot form spherically symmetric (asymptotically flat) bound matter configurations. This theorem rules out, in particular, the existence of spatially regular static boson stars made of nonlinear massive scalar fields.

Given the simplicity of the boson star, it is natural to examine boson star solutions in theories of gravity other than GR to look for new phenomena. The simplest modification of GR is given by the scalar-tensor theories (see section \ref{sect-stt}), among which the BD theory is the prototype. Brans-Dicke boson stars were first examined in Ref. \refcite{bd-bos-strs-1} for a particular value of the BD coupling constant $\omega_\text{BD}=6$, and then, in Ref. \refcite{bd-bos-strs-2} these were generalized to other values of $\omega_\text{BD}$ and to other scalar-tensor theories as well. Below we shall briefly sketch the formalism used to investigate the properties of the boson stars within the STT. 

The formalism relies on the action \eqref{stt-action} for a STT with vanishing self-interaction potential $V(\phi)=0$, where the stress-energy tensor of matter is given by \eqref{boson-set}. We recall that, in order for the given STT to avoid the tight constraints coming from local experiments in the Solar system, the following limits should be jointly verified: $$\omega_0\rightarrow\infty,\;\lambda\rightarrow 0,$$ with $\omega_0=\omega(\phi_0)$ and $\lambda$ given by equations in \eqref{stt-def-par} where the scalar field is evaluated today ($\phi_0$) and it is determined by appropriate cosmological boundary conditions given far away from the astrophysical object. The Klein-Gordon equation derived from ${\cal L}_\text{bos}$ is: $$\Box\psi-m^2\psi-\lambda|\psi|^2\psi^*=0.$$ Further one assumes spherical symmetry so that the background metric is given by the line element: $ds^2=-B(r)dt^2+A(r)dr^2+r^2d\Omega^2$. Due, precisely, to the spherical symmetry for the boson one adopts: $\psi(t,r)=\chi(r)\exp(-iwt)$, where the function $\chi$ is to be expanded in creation and annihilation operators. In this regard, semi-classically, one may imagine $T_{\mu\nu}^{(m)}$ as an expectation value in a given configuration with a large number of bosons\cite{bd-bos-strs-2}. With the spherically symmetric metric and the boson field $\psi(t,r)$ in the form given above, substituted back into the STT motion equations -- with the stress-energy tensor of matter \eqref{boson-set} -- one obtains the equations for the structure of the star. The obtained equations of motion reduce to those of BD theory if set $\omega(\phi)$ to a constant, while if set $\phi=\phi_0$-- a constant, then the GR motion equations are obtained. 

The next step is to set adequate boundary conditions such as: i) finite mass, which implies that $\chi(\infty)=0$, ii) non-singularity at the origin, i. e., $\chi(0)=\chi_0$-- finite, and $d\chi/dr|_0=0$, iii) asymptotic flatness, which means that $B(\infty)=1$ and $A(\infty)=1$, iv) $\phi(\infty)$ should obey appropriate cosmological boundary conditions at the time of stellar formation (this includes that $d\phi/dr|_\infty\rightarrow 0$). Finally one chooses an appropriate integration method (Runge-Kutta for instance) in order to numerically integrate the system of equations from the center of the star outwards. For precise details we recommend, for instance, the Refs. \citen{bos-strs-8, bd-bos-strs-2}. It has been demonstrated that STT-based boson stars can be stable at any time of cosmic history (equilibrium stars are denser in the past) and that the radius corresponding to the maximal boson star mass remains roughly the same during cosmological evolution\cite{bos-strs-8}. It has been shown, also, that the phenomenon of spontaneous scalarization predicted in neutron stars within the framework of scalar-tensor theories of gravity,\footnote{In Refs. \citen{damour_prd_1996, damour-scalariz-1} the authors discovered that neutron star models within STT may undergo a phase transition that consists in the appearance of a non-trivial configuration of the scalar field in the absence of sources with vanishing asymptotic value. Such configurations are endowed with a new global quantity termed scalar charge and the process was called as spontaneous scalarization.} also takes place in STT boson stars with vanishing self-interaction term for the boson field (other than the mass term)\cite{bos-strs-8-1, bos-strs-8-2}. In general, other parameters like the STT boson stars masses, have values that do not differ too much from the GR-based model\cite{sotani_prd_2017}.


\section{Scalar-tensor and $f(R)$-theories}\label{sect-fdr}

There are indications that including higher order terms into the gravitational action makes the given theory of gravity more compatible with quantum (renormalizable) variants\cite{stelle-prd} whose predictions can be trusted back enough into the past. One example is the addition of four-order terms like $R_{\mu\nu\tau\rho}R^{\mu\nu\tau\rho}$, $R_{\mu\nu}R^{\mu\nu}$ and $R^2$ into the Einstein-Hilbert action that gives a class of multimass models of gravity\cite{stelle-grg} where, in addition to the usual massless excitations of the fields, there are massive scalar and spin-2 excitations with a total of 8 degrees of freedom. The unwanted (yet tractable) property of this theory is that the massive spin-2 mode is ghost-like\cite{ovrut-prd-1, ovrut-prd-2}. 

Stability issues are central in the study of higher-order modifications of general relativity because they are plagued by several kinds of instabilities, some of which are catastrophic, leading to subsequent ruling out of the corresponding theories. Amongst others is the fundamental Ostrogradsky instability, based on the powerful no-go theorem of the same name\cite{ostro-theor}: ``There is a linear instability in the Hamiltonians associated with Lagrangians which depend upon more than one time derivative in such a way that the dependence cannot be eliminated by partial integration.'' This result is general and can be extended to higher-order derivatives in general. As a consequence, the only Ostrogradsky-stable higher-order modifications of Einstein-Hilbert action are those in the form of a function of the curvature scalar\cite{woodard_2007}, called as $f(R)$ theories\cite{capozziello(fdr)_ijmpd_2002, capozziello(fdr)_2003, vollick_prd_2003, chiba_plb_2003, carroll_duvvuri_prd_2004, nojiri(fdr)_grg_2004, carroll_defelice_prd_2005, sotiriou_cqg_2006, chiba(fdr)_prd_2007, nojiri_odintsov(fdr_rev)_2007, nojiri_odintsov(fdr)_plb_2007, nojiri_odintsov(fdr)_prd_2008, cognola_prd_2008, shaw_prd_2008, quiros_prd_2009, sotiriou_rmp_2010, defelice_tsujikawa_lrr_2010, nojiri_odintsov_phys_rep_2011, ruf_prd_2018, ohta_ptep_2018, nojiri_plb_2018}. These theories are given by a straightforward generalization of the Lagrangian in the Einstein-Hilbert action: $$S_\text{EH}=\frac{1}{2}\int d^4x\sqrt{|g|}R\;\;\rightarrow\;\;S=\frac{1}{2}\int d^4x\sqrt{|g|}f(R).$$ In general, including the matter degrees of freedom, the action for the $f(R)$ theories and the corresponding motion equations look like (here we consider, specifically, the metric formalism\cite{nojiri_odintsov(fdr_rev)_2007}, for the metric-affine formalism see Ref. \refcite{sotiriou_liberati_ann_phys_2007} and for the metric, metric-affine and the Palatini formalisms we recommend Refs. \citen{sotiriou_rmp_2010, capozziello_rev_grg_2008}):

\bea S=\frac{1}{2}\int d^4x\sqrt{|g|}f(R)+\int d^4x\sqrt{|g|}{\cal L}_m,\label{fdr-action}\eea and 

\bea R_{\mu\nu}-\frac{f(R)}{2\der_R f}\,g_{\mu\nu}-\frac{1}{\der_R f}\left(\nabla_\mu\nabla_\nu-g_{\mu\nu}\Box\right)(\der_R f)=\frac{1}{\der_R f}\,T_{\mu\nu},\label{fdr-feq}\eea respectively. Taking the divergence of both sides of \eqref{fdr-feq} yields to the standard conservation equation: $\nabla^\mu T_{\mu\nu}^{(m)}=0$. Notice, also, that the trace of \eqref{fdr-feq}:

\bea (\der_R f)R-2f(R)+3\Box(\der_R f)=T,\label{fdr-trace}\eea is a differential equation relating $R$ and $T$, and not an algebraic equation as it is customary.


\subsection{Accelerating expansion with the $f(R)$ theories} 

The resurgence of the modifications of Einstein's theory of gravity such as the $f(R)$ theories has been fueled by their ability to produce acceleration of the cosmic expansion without the need for an unknown and exotic form of antigravitating matter or dark energy\cite{defelice_tsujikawa_lrr_2010, nojiri_odintsov_phys_rep_2011, fdr_cosmo_1, fdr_cosmo_2, fdr_cosmo_3, fdr_cosmo_4, fdr_cosmo_5, fdr_cosmo_6, fdr_cosmo_7, fdr_cosmo_8, fdr_cosmo_9, fdr_cosmo_10, fdr_cosmo_11, fdr_stt_const}. Here we shall briefly show how this acceleration of the expansion comes about.

Let us assume that the spacetime metric is adequately described by the flat FRW metric \eqref{frw-metric}, the equations of motion \eqref{fdr-feq} can be written in the following form:

\bea &&3H^2=\frac{1}{f_{,R}}\left(\rho_m+\rho_\text{eff}\right),\nonumber\\
&&2\dot H=-\frac{1}{f_{,R}}\left(\rho_m+p_m+\rho_\text{eff}+p_\text{eff}\right),\label{fdr-feq-frw}\eea where $f_{,R}\equiv\der_R f$, $f_{,RR}\equiv\der^2_R f$, etc., while the effective energy density and pressure are defined as,

\bea &&\rho_\text{eff}=\frac{1}{2}\left(Rf_{,R}-f\right)-3H\dot R f_{,RR},\nonumber\\
&&p_\text{eff}=f_{,RR}\left(\ddot R+2H\dot R+\frac{f_{,RRR}}{f_{,RR}}\dot R^2\right)+\frac{1}{2}\left(f-Rf_{,R}\right),\label{fdr-p-rho-eff}\eea respectively. Notice that, since in a flat FRW spacetime $R=6(\dot H+2H^2)$ $\Rightarrow\dot R=6(\ddot H+4H\dot H)$, $\ddot R=6(\dddot H+4\dot H^2+4H\ddot H)$, the motion equations \eqref{fdr-feq-frw} contain, in general, up to four-order derivatives of the scale factor. In order for the the effective gravitational coupling to be positive it is required that $f_{,R}> 0$, while $f_{,RR}>0$ to avoid the Dolgov-Kawasaki instability\cite{quiros_prd_2009, dolgov_kawasaki, faraoni(instab_fdr)_prd_2006}.

In order to simplify the analysis let us further assume the vacuum situation, i. e., in the above equations we set $\rho_m=p_m=0$. From equations \eqref{fdr-feq-frw}, \eqref{fdr-p-rho-eff} it follows that,

\bea \frac{\ddot a}{a}=-\frac{1}{6f_{,R}}\left(\rho_\text{eff}+3p_\text{eff}\right),\label{fdr-ddot-a}\eea so that, in order to get accelerated expansion, it is required that $\rho_\text{eff}+3p_\text{eff}<0$, i. e., $$3f_{,RR}\left(\ddot R+H\dot R\right)+3f_{,RRR}\dot R^2+f-Rf_{,R}<0.$$ In particular, the $f(R)$ model mimics a cosmological constant if $p_\text{eff}=-\rho_\text{eff}$, i. e., $$f_{,RR}(\ddot R-H\dot R)+f_{,RRR}\dot R^2=0.$$


\subsection{Equivalence between $f(R)$ and scalar-tensor theories}

Without entering into the discussion of what is meant by dynamical equivalence of given theories (we refer the reader, for instance, to Ref. \refcite{sotiriou_etall_ijmpd_2008} for a detailed discussion on this issue), a subject that will be abundantly discussed in section \ref{sect-cf} in what regards to the conformal transformations of the metric, here we shall sketch the demonstration of the dynamical equivalence of $f(R)$ theories with STT-s, in particular with Brans-Dicke theory\cite{chiba_plb_2003} (see also Ref. \refcite{teyssandier-jmp-1983}) .

Let us write the following Einstein-Hilbert action that has been modified with the introduction of the field $\psi$:

\bea S=\frac{1}{2}\int d^4x\sqrt{|g|}\left[f(\psi)+\der_\psi f(R-\psi)\right].\label{modif-eh-action}\eea Variation of \eqref{modif-eh-action} with respect to the field $\psi$ gives: $\der^2_\psi f(R-\psi)=0,$ so that, provided that $\der^2_\psi f\neq 0$, one gets $\psi=R$, and substituting back into \eqref{modif-eh-action} one gets the action for the $f(R)$ theories. Hence, the action \eqref{modif-eh-action} and the gravitational part of \eqref{fdr-action} are dynamically equivalent. Now one redefines the field $\psi$ through $\phi=\der_\psi f$, and write $$V(\phi)=\psi(\phi)\phi-f(\psi(\phi)).$$ The action \eqref{modif-eh-action} can then be written in the form of a BD action: 

\bea S=\frac{1}{2}\int d^4x\sqrt{|g|}\left[\phi R-V(\phi)\right],\label{fdr-bd-action}\eea without the kinetic term, i. e., with vanishing coupling, $\omega_\text{BD}=0$. Take, for instance the following theory\cite{carroll_duvvuri_prd_2004}: $$f(R)=R-\frac{\mu^4}{R},$$ where $\mu$ is a constant parameter with dimension of mass. In this case the dynamically equivalent action can be written in the BD form \eqref{fdr-bd-action} with potential: $V(\phi)=2\mu^2\sqrt{\phi-1}$. The following motion equations are obtained from the action \eqref{fdr-bd-action}:

\bea &&G_{\mu\nu}=\frac{1}{\phi}\,T^{(m)}_{\mu\nu}-\frac{V(\phi)}{2\phi}\,g_{\mu\nu}+\frac{1}{\phi}\left(\nabla_\mu\nabla_\nu-g_{\mu\nu}\Box\right)\phi,\nonumber\\
&&3\Box\phi=T^{(m)}+\phi\der_\phi V-2V,\label{fdr-bd-feq}\eea where, in order to obtain the BDKG equation above, we have previously substituted the trace of the Einstein's field equation into the constraint $R=\der_\phi V$, that is obtained by varying the action \eqref{fdr-bd-action} with respect to $\phi$. 

Above we have sketched the demonstration of the dynamical equivalence of $f(R)$ theories with the BD theory in the metric formalism. If follow, instead, the alternative Palatini formalism it is shown that the equivalence with the Brans-Dicke theory still holds true, but this time what one obtains is BD theory with coupling parameter\cite{flanagan_prl_2004} $\omega_\text{BD}=-3/2$.

Given the dynamical equivalence between both these theories\cite{ohta_ptep_2018} the question is which one to choose\cite{sotiriou_rmp_2010} in order to describe the gravitational phenomena? From the theoretical point of view there is no reason to prefer one theory over the other, meanwhile, from the practical viewpoint the answer depends on one's physical motivation. In a more particle physics motivated work the natural choice would be the BD theory, while in a general relativity motivated work one would prefer the $f(R)$ theory. Nonetheless, one thing should be clear: although there is a large amount of work done on the Brans-Dicke theory, the particular cases with $\omega_\text{BD}=0$ and $\omega_\text{BD}=-3/2$, have not been studied in any detail since both are ruled-out by the solar system experiments\cite{will-lrr-2014} and, besides, both are outstanding: in the former case the scalar field is not a propagating degree of freedom as a consequence of the absence of the kinetic term, while in the latter case the BD field is a ghost due to the wrong sign of the kinetic term. Besides, the combination $3+2\omega_\text{BD}$ appears in the BDKG equation \eqref{kgbd-eq}, and the particular value $\omega_\text{BD}=-3/2$ leads to the following constraint equation: $$4V-2\phi\der_\phi V=T^{(m)}.$$ Hence, the choice of the self-interaction potential can not be made independent of the matter content of the theory. Besides, the Einstein's frame representation of the BD theory with this choice of the coupling constant is a theory where the scalar field is devoid of its kinetic term. This means that, indeed, the study of the $f(R)$ theories offers useful information on the BD theory with the mentioned values of the BD coupling constant.


\section{Extended theories of gravity}\label{sect-etg}

Possible modifications of Einstein's GR, including scalar tensor and/or $f(R)$ theories, can be investigated under the standards of the so called extended theories of gravity (ETG)\cite{capoz-phys-rept-2011, gottlober-cqg-1990, schmidt-cqg-1990, wands-cqg-1994, capoz-etg-grg-2000, nojiri-ijgmp-2007, capozziello_rev_grg_2008, capoz-etg-prd-2015}. These are understood as generalizations of GR that contain corrections and enlargements of the Einstein theory\cite{capoz-phys-rept-2011} such as the addition of higher-order curvature invariants and/or non-minimally (and also minimally) coupled scalar fields into the gravitational action:\footnote{Here, as through the whole review, we follow the metric formalism. For both, metric, metric affine and Palatini formalisms considerations we recommend the review papers Refs. \citen{sotiriou_rmp_2010, capoz-phys-rept-2011}.}

\bea S=\int d^4x\sqrt{|g|}f\left(R,R_{\mu\nu}R^{\mu\nu},R_{\mu\nu\sigma\lambda}R^{\mu\nu\sigma\lambda},\Box R,\Box^2R,\ldots\Box^kR,\phi\right),\label{etg-action}\eea where $f$ is an arbitrary function of the curvature invariants and of the scalar field. The particular case when $f(R)$, i. e., the so called $f(R)$ theories, has been discussed in the former section, while the case when $f=f(R,R_{\mu\nu}R^{\mu\nu},R_{\mu\nu\sigma\lambda}R^{\mu\nu\sigma\lambda})$ has been investigated, for instance, in Ref. \refcite{chiba-jcap-2005} (see also Ref. \citen{stelle-prd, stelle-grg, ovrut-prd-1, ovrut-prd-2} for a related quantum mechanical and perturbative exploration of this case), where it was shown that the corresponding action:

\bea S=\int d^4x\sqrt{|g|}\;f\left(R,R_{\mu\nu}R^{\mu\nu},R_{\mu\nu\sigma\lambda}R^{\mu\nu\sigma\lambda}\right),\label{etg-action-1}\eea is equivalent to multi-scalar–tensor gravity theory with four-derivative terms\footnote{As mentioned in the introduction to this review, Lovelock's theorem\cite{lovelock-1971, lovelock-1972} entails, precisely, that if one wants to construct metric theories of gravity with field equations that differ from those of GR, one of the possibilities is to accept derivatives of the metric higher than second order in the field equations.} if introduce the auxiliary scalar fields $\vphi_i$ ($i=1,2,3$). The action \eqref{etg-action-1} is then replaced by the following, 

\bea &&S=\int d^4x\sqrt{|g|}\left[f(\vphi_1,\vphi_2,\vphi_3)+\frac{\der f}{\der\vphi_1}\left(R-\vphi_1\right)+\frac{\der f}{\der\vphi_2}\left(R_{\mu\nu}R^{\mu\nu}-\vphi_2\right)\right.\nonumber\\
&&\left.\;\;\;\;\;\;\;\;\;\;\;\;\;\;\;\;\;\;\;\;\;\;\;\;\;\;\;\;\;\;\;\;\;\;\;\;\;\;\;\;\;\;\;\;\;\;\;\;\;\;\;\;\;\;\;\;\;\;\;\;\;\;\;\;\;\;\;\;+\frac{\der f}{\der\vphi_3}\left(R_{\mu\nu\sigma\lambda}R^{\mu\nu\sigma\lambda}-\vphi_3\right)\right].\label{action-xx}\eea Variation of the above action with respect to the $\vphi_i$-s yields: 

\bea &&\delta S=\int d^4x\sqrt{|g|}\delta\vphi_j\left[\frac{\der^2f}{\der\vphi_j\der\vphi_1}\left(R-\vphi_1\right)+\frac{\der^2f}{\der\vphi_j\der\vphi_2}\left(R_{\mu\nu}R^{\mu\nu}-\vphi_2\right)\right.\nonumber\\
&&\left.\;\;\;\;\;\;\;\;\;\;\;\;\;\;\;\;\;\;\;\;\;\;\;\;\;\;\;\;\;\;\;\;\;\;\;\;\;\;\;\;\;\;\;\;\;\;\;\;\;\;\;\;\;\;\;\;\;\;\;\;\;\;+\frac{\der^2f}{\der\vphi_j\der\vphi_3}\left(R_{\mu\nu\sigma\lambda}R^{\mu\nu\sigma\lambda}-\vphi_3\right)\right],\nonumber\eea so that, given that the matrix $\der^2f/\der\vphi_j\der\vphi_i$ is non-degenerate, the corresponding motion equations amount to: $\vphi_1=R$, $\vphi_2=R_{\mu\nu}R^{\mu\nu}$ and $\vphi_3=R_{\mu\nu\sigma\lambda}R^{\mu\nu\sigma\lambda}$, respectively. If substitute these relationships back into the action \eqref{action-xx} we obtain the starting action \eqref{etg-action-1}. It was demonstrated in Refs. \citen{stelle-prd, stelle-grg, ovrut-prd-1, ovrut-prd-2} and also in Ref. \refcite{chiba-jcap-2005}, that if the action \eqref{etg-action-1} is expanded around a vacuum spacetime, there appear massive spin-2 ghost excitations that render the vacuum unstable. Another particular case of interest will be briefly discussed below. For a detailed exposition on ETG-s we recommend the review Ref. \refcite{capoz-phys-rept-2011}.


\subsection{Case where $f\left(R,\Box R,\Box^2R,\ldots\Box^kR\right)$}

Here we shall discuss on one particular case of interest for cosmology and astrophysics\cite{capoz-phys-rept-2011, capozziello_rev_grg_2008}. The corresponding ETG-s are given by the action\cite{schmidt-cqg-1990, wands-cqg-1994}:

\bea S=\int d^4x\sqrt{|g|}\;f\left(R,\Box R,\Box^2R,\ldots\Box^kR\right)+S_m,\label{etg-action-2}\eea where $S_m$ stands for the action of the matter fields (including minimally coupled scalar fields). This is the action for $(2k+4)$-order gravity. Following Ref. \refcite{wands-cqg-1994} here we will show how the above action can be rewritten as well in the form of a multi-scalar-tensor theory. For simplicity we shall omit the matter piece of the action. The first step is to introduce new variables: $\vphi_i=\vphi_0,\vphi_1,\cdots,\vphi_k$. Then, the function $f(\Box^iR)$ is written as $f(\vphi_i)$, so that from the dynamically equivalent action the derived $\vphi_i$-motion equations: $\vphi_i=\Box^iR$, are anticipated. The resulting action that is dynamically equivalent to \eqref{etg-action-2} looks like:

\bea S=\int d^4x\sqrt{|g|}\left[f(\vphi_i)+\sum_{j=0}^k\frac{\der f}{\der\vphi_j}\left(\Box^jR-\vphi_j\right)\right],\label{etg-2-equiv}\eea  so that the $\vphi_i$-s motion equations are: $$\sum_{i=0}^kF_{ij}\left(\Box^iR-\vphi_i\right)=0.$$ Hence, if the matrix $F_{ij}=\der^2f/\der\vphi_j\der\vphi_i$ is non-degenerate we obtain: $\Box^iR=\vphi_i$, as required. The derivatives of the Ricci scalar in the action can be reduced to terms linear in $R$ by integration by parts: $$\int d^4x\sqrt{|g|}\frac{\der f}{\der\vphi_ i}\Box R=\int d^4x\sqrt{|g|}\left[\Box\left(\frac{\der f}{\der\vphi_i}\right)\right]\,R,$$ where the boundary term has been omitted. The following action is obtained:

\bea S=\int d^4x\sqrt{|g|}\left[\left(\sum_{j=0}^k\Box^j\frac{\der f}{\der\vphi_j}\right)\,R+f(\vphi_i)-\sum_{j=0}^k\vphi_ j\frac{\der f}{\der\vphi_j}\right].\label{quasi-final}\eea Next we identify the scalar functional that is multiplying the Ricci scalar in the above action with a new scalar field: $$\phi=\sum_{j=0}^k\Box^j\frac{\der f}{\der\vphi_j}.$$ The $\vphi_k$ variable can be eliminated by writing it as a functional of the scalar field $\phi$ and of the remaining $\vphi_i$ ($i\neq k$). As a consequence, the $2k+4$-order gravity given by \eqref{etg-action-2} can be written as a second-order scalar-tensor theory with $k+1$ scalar fields. 

In order to see how this formalism works, let us apply it to a concrete example that has been developed in Ref. \refcite{wands-cqg-1994}. Let us choose the sixth-order gravity given by $f=R+\alpha R\Box R$, where $\alpha$ is a free constant parameter. We have that $f(\vphi_ i)=\vphi_0(1+\alpha\vphi_1)$, so that the action \eqref{quasi-final} is written as: $$S=\int d^4x\sqrt{|g|}\left[\left(1+\alpha\vphi_1+\alpha\Box\vphi_0\right)R-\alpha\vphi_0\vphi_1\right],$$ where $$\phi=\frac{\der f}{\der\vphi_0}+\Box\frac{\der f}{\der\vphi_12}=1+\alpha\vphi_1+\alpha\Box\vphi_0.$$ Substituting this scalar field back into the above action and, writing the auxiliary scalar field $\vphi_1$ as a function of $\phi$ and of $\vphi_0$: $$\alpha\vphi_1=\phi-1-\alpha\Box\vphi_0,$$ we can write the action as one for a scalar-tensor gravity:

\bea S=\int d^4x\sqrt{|g|}\left[\phi R-\frac{\xi}{\sqrt{2\alpha}}\left(\phi-1\right)-\frac{1}{2}(\der\xi)^2\right],\label{etg-2-stt}\eea where we have redefined $\xi=\sqrt{2\alpha}\vphi_0$, and we have taken into account that, up to a boundary term, $$\int d^4x\sqrt{|g|}\xi\Box\xi=-\int d^4x\sqrt{|g|}(\der\xi)^2.$$ Notice that \eqref{etg-2-stt} depicts Brans-Dicke theory with vanishing coupling parameter $\omega_\text{BD}=0$, for a BD scalar field $\phi$, and with an additional canonical scalar field $\xi$, as matter source.


\section{Extended scalar-tensor theories: Horndeski theories}\label{sect-horn}

Horndeski theories\cite{horndeski_gal, nicolis_gal, deffayet_vikman_gal, deffayet_deser_gal, deffayet_prd_2011, fab_4_prl_2012, deffayet-rev, tsujikawa_lect_not, kazuya_rpp_2016} represent the most general higher derivatives extension of STT-s whose dynamics is governed by second-order motion equations. The recent history of these theories is quite peculiar. Inspired by the five-dimensional Dvali-Gabadadze-Porratti (DGP) model\cite{dgp_plb_2000, deffayet(dgp)_prd_2002, luty_porrati_rattazzi_jhep_2003, nicolis_rattazzi_jhep_2004, lue(dgp)_prd_2004, koyama_maartens_jcap_2006, roy-rev}, in Ref. \refcite{nicolis_gal} the authors derived the five Lagrangians that lead to field equations invariant under the Galilean symmetry $\der_\mu\phi\rightarrow\der_\mu\phi+b_\mu$ in the Minkowski space-time. The scalar field that respects the Galilean symmetry was dubbed ``Galileon''. Each of the five Lagrangians leads to second-order differential equations, keeping the theory free from unstable spin-2 ghosts, and from the corresponding instability of the resulting theory. If the analysis in Ref. \refcite{nicolis_gal} is generalized to the curved spacetime, then these Lagrangians need to be promoted to their covariant forms. This was done in Ref. \citen{deffayet_vikman_gal, deffayet_deser_gal} where the authors derived the covariant Lagrangians ${\cal L}_i$ ($i=1,..., 5$) that keep the field equations up to second-order. In Ref. \refcite{kobayashi} it was shown that these Lagrangians are equivalent to the ones discovered by Horndeski\cite{horndeski_gal}.

According to Refs. \citen{deffayet_vikman_gal, deffayet_deser_gal}, the most general 4-dimensional scalar-tensor theories having second-order motion equations are described by the linear combinations of the following Lagrangians (${\cal L}_1=M^3\phi$, where the constant $M$ has the dimension of mass):

\bea &&{\cal L}_2=K,\;{\cal L}_3 =-G_3(\Box\phi),\;{\cal L}_4=G_4 R+G_{4,X}\left[(\Box\phi)^2-(\nabla_\mu\nabla_\nu\phi)^2\right],\nonumber\\
&&{\cal L}_5=G_5 G_{\mu\nu}\nabla^{\mu}\nabla^\nu\phi-\frac{1}{6}G_{5,X}\left[(\Box\phi)^3-3\Box\phi(\nabla_\mu\nabla_\nu\phi)^2+2(\nabla_\mu\nabla_\nu\phi)^3\right],\label{horn-lags}\eea where $K=K(\phi,X)$ and $G_i=G_i(\phi,X)$ ($i=3,4,5$), are functions of the scalar field $\phi$ and its kinetic energy density $X=-(\der\phi)^2/2$, while $G_{i,\phi}$ and $G_{i,X}$, represent the derivatives of the functions $G_i$ with respect to $\phi$ and $X$, respectively. In the Lagrangian ${\cal L}_5$ above, for compactness of writing, we have adopted the same definitions used in Ref. \refcite{kobayashi}:
 
\bea (\nabla_\mu\nabla_\nu\phi)^2&:=\nabla_\mu\nabla_\nu\phi\nabla^\mu\nabla^\nu\phi,\;(\nabla_\mu\nabla_\nu\phi)^3&:=\nabla^\mu\nabla_\alpha\phi\nabla^\alpha\nabla_\beta\phi\nabla^\beta\nabla_\mu\phi.\label{def}\eea

The general action for the Horndeski theories:

\bea S_\text{Horn}=\int d^4x\sqrt{|g|}\left({\cal L}_2+{\cal L}_3+{\cal L}_4+{\cal L}_5+{\cal L}_m\right),\label{horn-action}\eea where the ${\cal L}_i$ are given by \eqref{horn-lags} and ${\cal L}_m$ stands for the Lagrangian of the matter degrees of freedom, comprises several well-known particular cases\cite{tsujikawa_lect_not}:

\begin{itemize}

\item{\it General relativity with a minimally coupled scalar field.} This is given by the following choice of the relevant functions in \eqref{horn-lags}:
 
\bea G_4=\frac{1}{2},\;G_3=G_5=0\;\Rightarrow\;S=\int d^4x\sqrt{|g|}\left[\frac{1}{2}\,R+K(\phi,X)+{\cal L}_m\right].\label{k-ess-class}\eea This choice comprises quintessence; $K(\phi,X)=X-V$, and k-essence, for instance, $K(\phi,X)=f(\phi)g(X)$, where $f$ and $g$ are arbitrary functions of their arguments.

\item{\it Brans-Dicke theory.} The following choice corresponds to the BD theory that is the prototype STT:

\bea &&K(\phi,X)=\frac{\omega_\text{BD}}{\phi}\,X-V(\phi),\;G_3=G_5=0,\;G_4=\frac{\phi}{2},\nonumber\\
&&\;\;\;\;\;\;\;\;\;\;\;\;\;\;\;\Rightarrow S=\frac{1}{2}\int d^4x\sqrt{|g|}\left[\phi R-\frac{\omega_\text{BD}}{\phi}(\der\phi)^2-2V\right].\label{bd-class}\eea

\item{\it $f(R)$-theory.} In this case we have that:

\bea &&K=-\frac{1}{2}\left(\der_R fR-f\right),\;G_4=\frac{1}{2}\der_R f,\;G_3=G_5=0,\nonumber\\
&&\;\;\;\;\;\;\;\;\;\;\;\Rightarrow S_{f(R)}=\frac{1}{2}\int d^4x\sqrt{|g|}f(R).\label{fdr-class}\eea Notice that under the replacement $\phi=\der_R f$, $V=(\der_R f-f)/2$, in \eqref{fdr-class} leads to BD theory \eqref{bd-class} with vanishing coupling $\omega_\text{BD}=0$.

\item{\it Non-minimal coupling (NMC) theory.} This is described by the functions:

\bea &&K=\omega(\phi)X-V(\phi),\;G_4=\frac{1-\xi\phi^2}{2},\;G_3=G_5=0,\nonumber\\
&&\;\;\;\;\;\;\;\;\;\;\;\Rightarrow S_\text{nmc}=\int dx^4\sqrt{|g|}\left[\frac{1-\xi\phi^2}{2}R-\frac{\omega(\phi)}{2}(\der\phi)^2-V(\phi)\right].\label{nmc-class}\eea Higgs inflation (see subsection \ref{subsect-higgs-infl} below) corresponds to the choice $\omega(\phi)=1$, $V(\phi)=\lambda(\phi^2-v^2)^2/4$.

\item{\it Covariant Galileons.} For vanishing potential the covariant Galileon model\cite{nicolis_gal, deffayet_vikman_gal, deffayet_deser_gal, deffayet_prd_2011, fab_4_prl_2012, deffayet-rev} is recovered from the general Horndeski action \eqref{horn-action} by the following choice of the functions\cite{tsujikawa_lect_not}:

\bea K=-2c_2,\;G_3=-2c_3 X,\;G_4=\frac{1}{2}-4c_4 X^2,\;G_5=-4c_5 X^2,\label{cov-gal-class}\eea where the $c_i$-s are constants. For this case we do not include the resulting action since it is a quite complex expression. 

\item{\it Cubic Galileon.} For this particular case that will be the subject of the subsection \ref{subsect-gal}, in the functions in \eqref{horn-lags} one sets: $$K=\frac{2\omega_\text{BD}}{\phi}X-2\Lambda\phi,\;G_3=-2f(\phi)X,\;G_4=\phi,\;G_5=0,$$ and the resulting action reads\cite{kazuya_gal}: 

\bea S_\text{cub}=\int d^4x\sqrt{|g|}\left[\phi R-\frac{\omega_\text{BD}}{\phi}(\der\phi)^2-2\Lambda\phi+f(\phi)\Box\phi(\der\phi)^2\right].\label{cub-gal-action}\eea

\item{\it Kinetic coupling to the Einstein's tensor.} This is another particular and very interesting case within the class of the Horndeski theories (see subsection \ref{sect-c2s} where causality and Laplacian instability issues are discussed for this particular case). It corresponds to the following choice: $$K=X-V,\;G_3=0,\;G_4=\frac{1}{2},\;G_5=-\frac{\alpha}{2}\phi,$$ that leads to the action: 

\bea S_\text{kc}=\frac{1}{2}\int d^4x\sqrt{|g|}\left[R+2(X-V)+\alpha G_{\mu\nu}\der^\mu\phi\der^\nu\phi\right],\label{k-coup-action}\eea where we have taken into account that integration by parts of the term $-\alpha\phi G_{\mu\nu}\nabla^\mu\nabla^\nu\phi$ amounts to $-\alpha G_{\mu\nu}\der^\mu\phi\der^\nu\phi$.

\end{itemize} 

One of the advantages of the Galileons as introduced in Ref. \refcite{nicolis_gal} is that it is possible to obtain the equivalent of the DGP self-accelerating phase without the unwanted ghost instability\cite{kazuya_gal}. Galileon models have been applied to reproduce the present speed-up of the cosmic expansion\cite{kazuya_gal, pujolas_jcap_2010, chow_gal, kobayashi_prd_2010} and, also, the primordial inflation\cite{japan_gal_prl_2010, burrage_jcap_2011} (see subsection \ref{subsect-infl}). The implications of these models for the non-gaussianity issue\cite{ngauss_1, ngauss_2, ngauss_3, ngauss_4, ngauss_5, ngauss_6, ngauss_7}, as well as for gravitational wave emission in the context of the Vainshtein screening have been studied in Ref. \citen{derham_prd_2013, chu_prd_2013}  (see subsection \ref{subsect-vain}).


\subsection{Cosmological perturbations and effective gravitational couplings in the Horndeski theories}\label{subsect-horn-geff}

The linear perturbations about the flat FRW metric: $$ds^2=-(1+2\psi)dt^2-2\der_i\chi dtdx^i+a^2(t)(1+2\Phi)\delta_{ij}dx^idx^j,$$ where $\psi$, $\chi$, and $\Phi$ are the scalar metric perturbations, in the theory given by the action \eqref{horn-action}, were studied in Ref. \refcite{defelice(horn_perts)_plb_2011}. The spatial gauge where the $g_{ij}$ is diagonal is assumed. The scalar field as well as the matter fields, are also perturbed: $\phi(t)\rightarrow\phi(t)+\delta\phi(t,{\bf x})$, $\rho_m\rightarrow\rho_m+\delta\rho_m$. Following Ref. \refcite{defelice(horn_perts)_plb_2011}, for compactness of writing, let us to introduce the following useful quantities:

\bea &&{\cal F}_T\equiv 2\left[G_4-X\left(\ddot\phi G_{5,X}+G_{5,\phi}\right)\right],\nonumber\\
&&{\cal G}_T\equiv 2\left[G_4-2XG_{4,X}-X\left(H\dot\phi G_{5,X}-G_{5,\phi}\right)\right],\label{horn-perts-usef-quant}\eea and also, the expansion:

\bea &&\Theta=-\dot\phi XG_{3,X}+2H\left(G_4-4XG_{4,X}-4X^2G_{4,XX}\right)+\dot\phi\left(G_{4,\phi}+2XG_{4,\phi X}\right)\nonumber\\
&&\;\;\;\;\;\;\;\;\;\;\;\;\;\;\;\;\;\;\;\;\;\;\;-H^2\dot\phi\left(5XG_{5,X}+2X^2G_{5,XX}\right)+2HX\left(3G_{5,\phi}+2XG_{5,\phi X}\right).\label{horn-expansion}\eea 

For the discussion on the evolution of matter perturbations relevant to large-scale structure, the modes deep inside the Hubble radius ($k^2/a^2\gg H^2$) are the ones that play the most important role. In the quasi-static approximation on sub-horizon scales\footnote{The range of validity of the quasi-static approximation may be very limited in theories where the sound speed $c_s\ll 1$.}, so that the dominant contributions in the perturbation equations are those including $k^2/a^2$ and $\delta$ -- the density contrast of matter, the following Poisson equation on $\psi$ is obtained\cite{defelice(horn_perts)_plb_2011}: $$\frac{k^2}{a^2}\,\psi\simeq-4\pi G_\text{eff}\delta\rho_m,$$ where the effective gravitational coupling $G_\text{eff}$, is the one measured in local experiments. It is given by the following expression (recall that we are working in the units system where $8\pi G_N=M_\text{pl}^{-2}=1$):

\bea 8\pi G_\text{eff}=\frac{2\left(B_6D_9-B_7^2\right)(k/a)^2-2B_6M^2}{\left(B_8^2D_9+A_6^2B_6-2A_6B_7B_8\right)(k/a)^2-B_8^2M^2},\label{horn-eff-g}\eea where

\bea &&A_6=2(\Theta-H{\cal G}_T)/\dot\phi,\;B_6=2{\cal F}_T,\nonumber\\
&&B_7=2\left[\dot{\cal G}_T+H\left({\cal G}_T-{\cal F}_T\right)\right]/\dot\phi,\;B_8=2{\cal G}_T,\nonumber\\
&&D_9=\left[2(\dot\Theta+H\Theta)-4H\dot{\cal G}_T+2H^2({\cal F}_T-2{\cal G}_T)+\rho_m\right]/\dot\phi^2.\label{horn-coeff}\eea The coefficient $M^2$ is related with the mass squared of the field $\delta\phi$ and it is given by:

\bea M^2=-K_{,\phi\phi}+K_{,\phi X}(\ddot\phi+3H\dot\phi)+2XK_{,\phi\phi X}+2XK_{,\phi XX}\ddot\phi+...,\label{M2}\eea where the ellipsis stands for terms containing second, third and fourth-order derivatives of the functions $G_i$ on the variables $\phi$ and $X$. For the full expression of $M^2$ see Eq. (35) of Ref. \refcite{defelice(horn_perts)_plb_2011}.


\subsubsection{Are the Horndeski theories a generalization of just scalar-tensor theories or of something else?}\label{subsect-horn-sst}

In the bibliography one usually finds the statement that the Horndeski theories are a generalization -- or an extension -- of the scalar-tensor theories. But, what really means that a given theory of gravity is a scalar-tensor theory? In this review, as already discussed, such a statement entails that the gravitational phenomena are not completely due to the curvature of spacetime but, that these are partly a result of the curvature and partly due to an additional scalar field degree of freedom. Take as an example a scalar field with the typical non-minimal coupling to the curvature of the form, ${\cal L}_\text{nmc}\propto f(\phi)R$. In this case the gravitational coupling $\propto f^{-1}(\phi)$, so that it sets the strength of the gravitational interactions at each point in spacetime. This is the most obvious way in which the tensor part of the gravitational interactions is modified by the scalar field. In addition, the measured (effective) gravitational coupling is modified in a non-trivial way by the scalar part of the gravitational interaction. For instance, if the scalar field possesses a standard kinetic term,\footnote{For vanishing kinetic term without the potential the scalar field is a non-propagating degree of freedom, so that the resulting theory coincides with general relativity. But if the scalar field's potential is non-vanishing, it could happen that for vanishing kinetic term the theory is a scalar-tensor one, as it is, for instance, for $f(R)$-theories.} $-(\der\phi)^2/2$, the above non-minimal coupling implies that the measured gravitational constant is given by (see section \ref{sect-stt}): 

\bea 8\pi G_\text{eff}=\frac{1}{f(\phi)}\left[\frac{4+2f/(\der_\phi f)^2}{3+2f/(\der_\phi f)^2}\right],\label{nmc-eff-g}\eea where we are assuming vanishing self-interaction potential. The factor that multiplies $f^{-1}(\phi)$ in \eqref{nmc-eff-g} comes from the scalar degree of freedom that, together with the two polarizations of the graviton, carry the gravitational interactions. For the particular case when the STT is given by \eqref{stt-action}, in the above equation one have to replace, $f(\phi)\rightarrow\phi$ and $f/(\der_\phi f)^2\rightarrow\omega(\phi)$. 

Under a conformal transformation of the metric, the given STT can be formulated in terms of the Einstein's frame variables (see subsection \ref{subsect-jf-ef}). In particular, the non-minimal coupling of the scalar field with the curvature, $f(\phi) R$, can be removed at the cost of the appearance of  non-minimal coupling between the matter Lagrangian and the scalar field, $f^{-2}(\phi){\cal L}_m$. In this latter case what we have is not properly a STT but standard GR -- where the gravitational effects are completely due to the curvature of spacetime -- with the presence of an additional universal fifth-force acting on the matter fields. Hence, the measured gravitational coupling is just the Newton's constant $G_N$ (in our units $G_N=1/8\pi$).\footnote{No matter how trivial the differences between GR and the STT may look, certain confusion may arise due to the presence of higher derivatives of the scalar field and to complicated self-couplings. To make things worse, additional confusion may be related with the issue on the physical equivalence between the different conformal frames in which a given scalar-tensor theory can be formulated, also known as the 'conformal transformations issue' \cite{dicke-1962, faraoni_rev_1997, faraoni_ijtp_1999, faraoni_prd_2007, sarkar_mpla_2007, sotiriou_etall_ijmpd_2008, deruelle_veiled_2011, deruelle_nordstrom_2011, quiros_grg_2013}. According to several authors \cite{dicke-1962, faraoni_prd_2007, sotiriou_etall_ijmpd_2008} a given STT is physically equivalent to GR with a scalar field that is non-minimally coupled to the matter degrees of freedom. If this point of view were correct then there would not be physical distinction between GR with an additional universal fifth force and the STT.} 

A good indicator that the given theory is a STT is that its corresponding effective gravitational coupling be a function of the scalar field, i. e., that it could be expressible in the form of \eqref{nmc-eff-g} through, possibly, a redefinition of the scalar field. After the Horndeski generalizations of the scalar-tensor theories, one should require that, not only the scalar field but also its higher order derivatives and mixed (non-linear) terms where curvature quantities are multiplied by these elements, can modify the effective coupling that is measured in Cavendish-like experiments \eqref{horn-eff-g}. For the Brans-Dicke theory, as stated above: $$K(\phi,X)=\frac{\omega_\text{BD}}{\phi}\,X-V(\phi),\;G_3=G_5=0,\;G_4=\frac{\phi}{2}.$$ The corresponding effective gravitational coupling \eqref{horn-eff-g} is given by:

\bea 8\pi G_\text{eff}=\frac{1}{\phi}\left[\frac{4+2\omega_\text{BD}+2\phi(Ma/k)^2}{3+2\omega_\text{BD}+2\phi(Ma/k)^2}\right],\label{horn-bd-eff-g}\eea where, neglecting terms ${\cal O}(H^2\phi)$, $M^2\simeq\der^2_\phi V+\der_\phi V/\phi$. In the limit $M^2\rightarrow 0$, i. e., when the scalar field is massless as in the original BD theory without the potential, we recover the result for $G_\text{eff}$ in \eqref{bd-eff-g}. Meanwhile, in the limit $M^2\rightarrow\infty$, i. e., when the scalar field decouples from the rest of the matter degrees of freedom of the theory -- also when $\omega_\text{BD}\rightarrow\infty$ -- the GR behavior is reproduced.

But, what about other theories included in the Horndeski class \eqref{horn-action}? Take, for instance, the class determined by the choice \eqref{k-ess-class}. Looking at the resulting action, for an arbitrary function $K(\phi,X)$, one immediately recognizes the so called k-essence theories (these include the quintessence models for the particular choice $K(\phi,X)=X-V(\phi)$). In this case, since $G_4=1/2$, ${\cal F}_T={\cal G}_T=1$, and given that $G_3=G_5=0$, one gets that $\Theta=H$, and consequently, $A_6=B_7=0$, $B_6=B_8=2$. Hence, for the effective gravitational coupling \eqref{horn-eff-g} one obtains $8\pi G_\text{eff}=1$, which means that k-essence is just general relativity plus a scalar field -- with a perhaps exotic kinetic energy term --  as matter source of the Einstein's equations. 

For the choice:

\bea G_4=\frac{1}{2},\;G_5=0,\;G_3=G_3(\phi,X)\neq 0,\label{qbic-class}\eea that includes the cubic Galileon model, ${\cal F}_T={\cal G}_T=1$, while $\Theta=H-\dot\phi XG_{3,X}$, and $$A_6=2(\Theta-H)/\dot\phi,\;B_6=B_8=2,\;D_9=[2\dot\Theta+2H(\Theta-H)+\rho_m]/\dot\phi^2,$$ so that

\bea 8\pi G_\text{eff}=\frac{[2\dot\Theta-2H\dot\phi XG_{3,X}+\rho_m](k/a)^2-M^2\dot\phi^2}{[2\dot\Theta-2H\dot\phi XG_{3,X}+4X^3G^2_{3,X}+\rho_m](k/a)^2-M^2\dot\phi^2}.\label{qbic-class-eff-g}\eea Notice that if, $G_3=G_3(\phi)$, is a function of the scalar field alone, the resulting theory is equivalent to GR. In order for the above choice to represent a STT it is required that $G_3$ be an explicit function of the kinetic term: $G_3\propto f(X)$.

The cubic Galileon represents an example where the scalar-tensor character of a given theory may be very subtle. Actually, for the choice \eqref{k-ess-class} it is clear why the resulting theory is general relativity with a scalar field as matter source: there is no direct coupling of the scalar field (or of its derivatives) to the curvature. These couplings are explicit in the terms: $$G_4(\phi,X)R,\;\;G_5(\phi,X)G_{\mu\nu}\nabla^\mu\nabla^\nu\phi,$$ but as long as $G_4=\text{const}=1/2$ -- i. e., $G_4\neq G_4(\phi,X)$ -- and $G_5=0$, there is no (explicit) direct coupling between the scalar field a the curvature. 

The interesting thing is that according to the choice \eqref{qbic-class}, $G_4=1/2$, $G_5=0$, as in \eqref{k-ess-class}, so that one should expect that the resulting theory should be general relativity as well. However, if take a closer look at \eqref{qbic-class-eff-g}, it is seen that thanks to the term $4X^3G^2_{3,X}$ in the denominator, $G_\text{eff}\propto f(\phi,\dot\phi,\ddot\phi,X)$, so that this is not general relativity: The choice \eqref{qbic-class} is a scalar-tensor theory! We may explain this result in the following way. For simplicity let us assume that $G_3=G_3(X)$ is an explicit function of the kinetic energy of the scalar field alone. Variation of the Lagrangian ${\cal L}_3$ in \eqref{horn-lags} with respect to the scalar field can be written as: $$\delta{\cal L}_3=-G_{3,X}\delta X(\Box\phi)-G_3\Box(\delta\phi),$$ where $\delta X=\nabla^\mu\phi\nabla_\mu(\delta\phi)$. After further modification, up to a divergence, $\nabla_\mu V^\nu$, where $$V^\mu=G_{3,X}\nabla^\mu\phi(\Box\phi)\delta\phi+G_3\nabla^\mu(\delta\phi)-G_{3,X}\nabla^\mu X\delta\phi,$$ the variation of the Lagrangian can be put into the following form: 

\bea &&\delta{\cal L}_3=\left[G_{3,XX}\nabla_\mu X\nabla^\mu\phi(\Box\phi)+G_{3,X}(\Box\phi)^2+G_{3,X}\nabla^\mu\phi\nabla_\mu(\Box\phi)\right.\nonumber\\
&&\left.\;\;\;\;\;\;\;\;\;\;\;\;\;\;\;\;\;\;\;\;\;\;\;\;\;\;\;\;\;\;\;\;\;\;\;\;\;\;\;\;\;\;\;\;\;\;-G_{3,XX}\nabla_\mu X\nabla^\mu X-G_{3,X}\Box X\right]\delta\phi,\nonumber\eea where the presence of third-order derivatives is evident. According to the relationship, 

\bea \Box(\nabla_\mu\phi)-\nabla_\mu(\Box\phi)=R_{\mu\nu}\nabla^\mu\phi\nabla^\nu\phi,\label{curv-der-rel}\eea we have that (see the definitions \eqref{def}): 

\bea \Box X=(\nabla_\mu\phi\nabla_\nu\phi)^2+\nabla^\mu\phi\Box(\nabla_\mu\phi)=(\nabla_\mu\phi\nabla_\nu\phi)^2+\nabla^\mu\phi\nabla_\mu(\Box\phi)+R_{\mu\nu}\nabla^\mu\phi\nabla^\nu\phi,\nonumber\eea so that the variation of the cubic Lagrangian can be rewritten into the form where it contains derivatives no higher than the 2nd order:

\bea &&\delta{\cal L}_3=\left\{G_{3,XX}\nabla_\mu X\left[\nabla^\mu\phi(\Box\phi)-\nabla^\mu X\right]+G_{3,X}\left[(\Box\phi)^2-(\nabla_\mu\phi\nabla_\nu\phi)^2\right]\right.\nonumber\\
&&\left.\;\;\;\;\;\;\;\;\;\;\;\;\;\;\;\;\;\;\;\;\;\;\;\;\;\;\;\;\;\;\;\;\;\;\;\;\;\;\;\;\;\;\;\;\;\;\;\;\;\;\;\;\;\;\;\;\;\;\;\;\;\;\;\;\;\;\;\;\;\;-G_{3,X}R_{\mu\nu}\nabla_\mu\phi\nabla_\nu\phi\right\}\delta\phi.\label{delta-lag}\eea This has been achieved at the cost of introducing a term (last term above) where the Ricci curvature tensor is coupled to the derivatives of the scalar field. In this form, it is evident that any first-order variation of the cubic Lagrangian induces a derivative coupling of the scalar field to the curvature, thus making explicit the scalar-tensor character of the cubic Galileon theory.


\subsubsection{Limitations of the present analysis}

We want to (markedly) underline that our analysis in this section -- and throughout the whole review -- is valid until quantum effects can not be ignored. The quantum effects of the interaction of the matter fields may induce a non-minimal coupling with the curvature\cite{callan_ann_phys_1970}. Here we give a brief account of the demonstration given in Ref. \refcite{callan_ann_phys_1970}. Let us consider the simplest renormalizable quantum field theory that is given by the Lagrangian:

\bea {\cal L}_\phi=-\frac{1}{2}(\der\phi)^2-\frac{1}{2}\mu_0^2\phi^2-\lambda_0\phi^4.\label{renorm-lag}\eea The conventional stress-energy tensor $$T_{\mu\nu}^{(\phi)}=\der_\mu\phi\der_\nu\phi+g_{\mu\nu}{\cal L}_\phi,$$ does not have finite matrix elements, however, the modified tensor:

\bea \Theta_{\mu\nu}^{(\phi)}=T_{\mu\nu}^{(\phi)}+\frac{1}{6}\left(\nabla_\mu\nabla_\nu-g_{\mu\nu}\Box\right)\phi^2,\label{mod-set}\eea has finite matrix elements to all orders in $\lambda$. When we take into account the gravitational interactions, if we want the gravitational effects to be finite in to lowest order in the gravitational coupling and to all orders in all the other couplings, then, in the RHS of the Einstein's (GR) motion equations: $$G_{\mu\nu}=\frac{1}{M^2_\text{Pl}}T^{(\phi)}_{\mu\nu},$$ one has to make the replacement: $T^{(\phi)}_{\mu\nu}\rightarrow\Theta_{\mu\nu}^{(\phi)}$. This means, in turn, that the Einstein-Hilbert action principle: $$S_\text{EH}=\int d^4x\sqrt{|g|}\left[\frac{M^2_\text{Pl}}{2}R+{\cal L}_\phi\right],$$ is to be replaced by the STT action: $$S_*=\int d^4x\sqrt{|g|}\left[f(\phi)R+{\cal L}_\phi\right],$$ where $f(\phi)=M^2_\text{Pl}/2-\phi^2/12.$ In order to have observable gravitational effects when calculating, for instance, the amplitude of the scattering of a graviton in an external field, one has to rely on the stress-energy tensor that has finite matrix elements, i. e., on $\Theta_{\mu\nu}^{(\phi)}$. This, in turn, requires of a STT from the start. 

Hence, the classification of gravity theories into scalar-tensor theories and/or other metric theories according to the present (fully classical) discussion, is correct given that the quantum effects are ignored.


\subsection{Beyond Horndeski}\label{subsect-beyond-horn}

It has been shown that the Horndeski framework can be extended to include other Lagrangians\cite{bhorn-1, bhorn-2, bhorn-3, bhorn-4, bhorn-5, langlois_noui_jcap_2016_1, bhorn(vainsh), mancarella_jcap_2017}. The main idea behind those extensions of Horndeski Lagrangians is that in order to avoid the Ostrogradsky instability\cite{woodard_2007, ostro-theor}, it is sufficient but not necessary that the motion equations be second order in the derivatives\cite{bhorn-ostrog-1, bhorn-ostrog-2, crisostomi_jhep_2016, chagoya_tasinato_jhep_2017}. These extensions have been coined as ``beyond Horndeski'' theories and are based in the Horndeski Lagrangians \eqref{horn-lags} with the addition of the pieces\cite{bhorn-1, bhorn-2}:

\bea &&{\cal L}^\text{bhorn}_4=F_4(\phi,X)\epsilon^{\mu\nu\rho}_{\;\;\;\;\;\;\sigma}\epsilon^{\mu'\nu'\rho'\sigma}\nabla_\mu\phi\nabla_{\mu'}\phi(\nabla_\nu\nabla_{\nu'}\phi)(\nabla_\rho\nabla_{\rho'}\phi),\nonumber\\
&&{\cal L}^\text{bhorn}_5=F_5(\phi,X)\epsilon^{\mu\nu\rho\sigma}\epsilon^{\mu'\nu'\rho'\sigma'}\nabla_\mu\phi\nabla_{\mu'}\phi(\nabla_\nu\nabla_{\nu'}\phi)(\nabla_\rho\nabla_{\rho'}\phi)(\nabla_\sigma\nabla_{\sigma'}\phi),\label{bhorn-lags}\eea to the Lagrangians ${\cal L}_4$ and ${\cal L}_5$ in \eqref{horn-lags}, respectively. In \eqref{bhorn-lags}, $\epsilon_{\mu\nu\rho\sigma}$ is the totally antisymmetric Levi-Civita tensor density. The condition that $$F_4(\phi,X)=0,\;\;F_5(\phi,X)=0,$$ ensures that the motion equations are second order in the derivatives of the scalar field, which brings us back to the Horndeski theories. 

The motion equations resulting from the beyond-Horndeski Lagrangians in \eqref{bhorn-lags} involve up to third-order derivatives but this does not imply that we have extra degrees of freedom (DOF) when compared with the Horndeski dynamics. According to the Ostrogradsky theorem\cite{ostro-theor} the higer-derivative theories are pathological due to the propagating extra DOF that behave like ghosts. However, as shown in Ref. \refcite{bhorn-1} by means of the Hamiltonian formalism, the simple counting of DOF within beyond Horndeski theories leads to three propagating degrees of freedom like in scalar-tensor/Horndeski theories: the two polarizations of the tensor (graviton) field and the helicity-0 scalar field. Further work on the beyond Horndeski theories\cite{bhorn-4} has shown that the combination of Horndeski and beyond Horndeski Lagrangians of the same order, say $a{\cal L}_4+b{\cal L}^\text{bhorn}_4$, leads to Horndeski theory, so that the resulting mixed Lagrangian propagates three DOF, which means that the theory is free of Ostrogradsky ghosts. Meanwhile, if combine Horndeski and beyond Horndeski Lagrangians of different order, say $a{\cal L}_5+b{\cal L}^\text{bhorn}_4$ the primary constraint arising in isolated beyond Horndeski framework, that allows to remove the extra DOF, is lost\cite{bhorn-4}. In other words: beyond Horndeski is a healthy but isolated theory: combined with Horndeski, it either becomes Horndeski, or likely propagates a ghost.

In Ref. \refcite{bhorn(vainsh)}, working in the quasistatic approximation in the Newtonian gauge, the authors investigated the nonlinear effect of the scalar-field fluctuations that can screen the fifth force inside the Vainshtein radius within the frame of the beyond Horndeski theories. This effect is known as the Vainshtein screening (see subsection \ref{subsect-vain} below). All the nonlinear terms which could be relevant on small scales were taken into account. It has been shown in that reference that one of the solutions outside and near the source reproduces the standard behavior, $\Phi=-\psi\propto r^{-1}$.\footnote{Here we use the definitions for the linear perturbations in Ref. \refcite{defelice(horn_perts)_plb_2011} -- same as the ones in subsection \ref{subsect-horn-geff} above -- and not the ones in \refcite{bhorn(vainsh)}, where, for instance, the metric potentials $\Phi$ and $\psi$ are of the same sign.} However, the new non-linear interactions beyond Horndeski change the behavior of the gravitational potentials inside of the matter overdensity in a fundamental way: The strength of the gravitational interaction depends not only on the enclosed mass but also on the local matter energy density. As a result $\Phi$ and $\psi$ no longer coincide, implying that GR is not recovered inside the source.


\section{Scalar-tensor cosmological models}\label{sect-cosmo}

Among the major challenges in the description of the cosmological history of our universe is understanding and explaining the following issues: i) horizon, ii) flatness, iii) homogeneity and isotropy, and iv) primordial monopole (and other relics like cosmic strings and topological defects) problems, among others.\footnote{For other details of the problems the inflationary paradigm solves as well as on the small quantum fluctuations and the associated cosmological perturbations that produced the cosmic structure we see toady, etc., see, for instance, the reviews in Refs. \citen{brandenberger(infl_rev)_rpp_1985, lidsey(infl_rev)_rmp_1997, lyth_phys_rep_1999, bassett(infl_rev)_rmp_2006, linde_lnp_2008}.} The inflationary scenario\cite{kazanas_astrophys_j_1980, guth_prd_1981, linde_plb_1982, dolgov_plb_1982, steinhardt_prl_1982, starobinsky_plb_1982, guth_prl_1982, linde_plb_1983, vilenkin_prd_1983, linde_rpp_1984, matarrese_prd_1985, brandenberger(infl_rev)_rpp_1985, linde_prd_1994, lidsey(infl_rev)_rmp_1997, lyth_phys_rep_1999, bassett(infl_rev)_rmp_2006, linde_lnp_2008} was invoked, precisely, in order to solve these issues within the standard cosmological model. The main ingredient of the inflationary models is a scalar field $\phi$, called inflaton, that enters the RHS of the Einstein's equations as an exotic matter field (see subsection \ref{subsect-prim-infl}). In a similar way another scalar field, called quintessence, is supposed to solve another of the major open questions of modern cosmology: understanding the nature of the dark energy that is causing the observed accelerated expansion in the universe. 

That the universe is expanding at an accelerating peace at present -- and since nont long ago -- is supported by cosmological observations from type Ia supernovae sample from panSTARRS\cite{riess, perlmutter, snia, snia-1, snia-2}, combined with the baryon acoustic oscillations (BAO)\cite{bao, bao-1, bao-2, bao-3, bao-4, bao-5} and the cosmic microwave background (CMB)\cite{cmb, cmb-1, cmb-2, cmb-3, cmb-4}, large scale structure (LSS)\cite{lss, lss-1, lss-2, lss-3}, weak lensing\cite{lensing, lensing-1, lensing-2, lensing-3, lensing-4}, and the integrated Sachs-Wolfe effect\cite{swe, swe-1, swe-2, swe-3, swe-4, swe-5, swe-6}. In the context of general relativity (GR), the cosmological constant $\Lambda$, which can be interpreted as the energy of the vacuum\cite{lambda, lambda-1, lambda-2}, provides the simplest explanation of this alien component of the cosmic budget\cite{peebles}. In the $\Lambda$CDM model, the cosmological constant accounts approximately for the 70 \% of the total energy content of the universe, meanwhile the cold dark matter (CDM) component amounts to around 25 \%. The baryonic matter and the radiation complete the cosmic inventory. Although this model provides a good fit to a big range of independent observations, there is some tension with several observations as it has been pointed out in Refs. \citen{ref1, ref2}. In particular, there is tension between measurements of the amplitude of the power spectrum of density perturbations inferred using the CMB and directly measured by LSS on smaller scales. However, one of the major drawbacks of this model is related with the fact that there is not yet a satisfactory theoretical explanation for the very small value of $\Lambda$. Furthermore, the $\Lambda$CDM model suffers from a fine tuning or ``coincidence problem''\cite{zlatev_prl_1999, ccp, ccp-1}: Why is the dark matter density comparable to the vacuum energy density now, given that their time evolution is so different? Further improvement of the $\Lambda$CDM model implies the possibility that the dark energy is not a constant but evolves with time\cite{ratra, ratra-1}. As mentioned above, the simplest model for an evolving dark energy are light scalar fields known as quintessence, where a scalar field is postulated as the would be explanation of the observed accelerating rate of the expansion of the universe\cite{carroll_prl_1998, zlatev_prl_1999, amendola_prd_2000}. Quintessence differs from the cosmological constant in that, while the former is dynamic, that is, it changes over time, the latter remains a constant during the cosmic history. Many models of quintessence have a tracker behavior that partly solves the cosmological constant problem\cite{paul}. In these models, the quintessence field has a density which closely tracks (but is less than) the radiation density until matter-radiation equality, which triggers quintessence to start having characteristics similar to dark energy, eventually dominating the universe.

Both the inflaton and the quintessence field above mentioned, are assumed in the form of self-interacting scalar fields with energy density $\rho_\phi=\dot\phi^2/2+V$ and parametric pressure $p_\phi=\dot\phi^2/2-V$ that are added in the RHS of the Einstein's motion equations. In both cases the EOS parameter $$\omega_\phi=\frac{p_\phi}{\rho_\phi}=\frac{\dot\phi^2-2V}{\dot\phi^2+2V}\;\;\Rightarrow\;\;-1\leq\omega_\phi\leq 1.$$ Besides, at present $\omega_\phi\simeq-1$, in order to fit to the existing observational evidence (similar for primordial inflation). This means that we have to add an exotic matter component to the Einstein's equations in order to explain the mentioned stages of the cosmic evolution which, in turn implies a real challenge for the SMP since it should provide the particles that should act as adequate candidates for the inflaton and for the dark energy. 

An interesting alternative that at first sigh does not imply any challenge for the SMP, is to assume not Einstein theory but the scalar-tensor theory as the correct gravity theory. In such a case the primordial inflation as well as the present stage of accelerated expansion may be explained by the gravitational interactions themselves and no exotic form of matter is required.


\subsection{Models of inflation}\label{subsect-infl}

The inflationary paradigm sketched in subsection \ref{subsect-prim-infl} relies in the Einstein's GR theory, i. e., the inflaton $\vphi$ is a matter field. Unlike this, in the Zee's theory of induced gravity briefly exposed in subsection \ref{subsect-zee-theory}, being a BD theory with a symmetry breaking potential, (previous to symmetry breaking) the scalar field plays a fundamental role in the gravitational interactions due to the non-minimal coupling $\sim\xi\vphi^2 R$ in \eqref{zee-action}. In Refs. \citen{accetta_prd_1985, kaiser_prd_1995, cervantes_prd_1995, cervantes_npb_1995, barvinsky_jcap_2008, bezrukov_plb_2008, simone_plb_2009, bezrukov_jhep_2009, bezrukov_jhep_2011, bezrukov_jhep_2012} the Zee's idea was retaken in order to explain the primordial inflation.


\subsubsection{Induced gravity inflation}\label{subsect-indgrav-infl} 

Take, as illustration, the model developed in Ref. \refcite{accetta_prd_1985}. It is based in the Zee's action \eqref{zee-action} with self-interacting, symmetry breaking potential: $$V(\vphi)=\frac{\lambda}{8}\left(\vphi^2-v^2\right)^2,$$ where $v$ is the VEV of the scalar field $\vphi$. Under appropriate bound on the coupling $\lambda$ (typically $\lambda\leq 10^{-12}$), this model exhibit slow-roll inflation that leads to an acceptable magnitude of the density perturbations. For definiteness we shall consider here chaotic initial conditions leading to chaotic inflation\cite{linde_plb_1983}, so that the scalar field evolves from $\vphi\gg v$ to $\vphi=v$. Lets assume the FRW line element \eqref{frw-metric}, then, the motion equations derivable from \eqref{zee-action} are:

\bea &&3H^2+6H\frac{\dot\vphi}{\vphi}=\frac{1}{\xi\vphi^2}\left(\frac{1}{2}\dot\vphi^2+V\right)-\frac{3k}{a^2},\nonumber\\
&&\ddot\vphi+3H\dot\vphi+\frac{\dot\vphi^2}{\vphi}=\frac{-\vphi V'+4V}{(1+6\xi)\vphi},\label{zee-cosmo-feq}\eea where the overdot means derivative with respect to the cosmic time $t$, while the prime denotes derivative with respect to the $\vphi$-field. In order to solve the motion equations \eqref{zee-cosmo-feq} we shall assume slow-roll conditions: 

\bea \left|\frac{\dot\vphi}{\vphi}\right|\ll H,\;\dot\vphi^2\ll V,\;\ddot\vphi\ll H\dot\vphi.\label{slwr-cond}\eea In this friction term-dominated regime, the above motion equations can be written in the following simplified form:

\bea &&H^2=\frac{V(\vphi)}{3\xi\vphi^2}-\frac{k}{a^2},\nonumber\\
&&3H\dot\vphi=\frac{4V-\vphi V'}{(1+6\xi)\vphi}.\label{zee-slwr-feq}\eea The curvature term in the first equation in \eqref{zee-slwr-feq} can be neglected since it will become quickly negligible as compared with the term $\propto V/\xi\vphi^2$. The slow-roll conditions above \eqref{slwr-cond} are satisfied when $|\vphi-v|\geq\sqrt\xi v$, and $\lambda,\xi\ll 1$. The slow-roll motion equations \eqref{zee-slwr-feq} can be combined to get: $\dot\vphi=-\sqrt{2\lambda\xi/3}\,v^2$, or, after integration:

\bea \vphi(t)=\vphi_0-\sqrt{\frac{2}{3}\lambda\xi}\,v^2t,\label{zee-slwr-sol}\eea where $\vphi_0$ is the initial value of the inflaton field and we are considering chaotic inflation, so that $\vphi_0>v$. Taking into account that $dt=-\sqrt{3/2\lambda\xi}d\vphi/v^2$, the Friedmann equation in \eqref{zee-slwr-feq} can be readily integrated to yield:

\bea \frac{a(\vphi)}{a_0}=\left(\frac{\vphi}{\vphi_0}\right)^\frac{1}{4\xi}\exp\left(\frac{\vphi_0^2-\vphi^2}{8\xi v^2}\right).\label{slwr-scale-f}\eea During the stage where $\vphi\approx\vphi_0\gg v$, the scale factor grows exponentially: $$a(t)\approx a_0\exp\left(\sqrt\frac{\lambda}{24\xi}\,\vphi_0 t\right).$$ During the time interval required for the scalar field to go from $\vphi_0$ to $v$, the $\ln$ of the scale factor grows by: $$\ln\frac{a_v}{a_0}=\frac{1}{8\xi}\left[\frac{\vphi^2_0}{v^2}-1-2\ln\left(\frac{\vphi_0}{v}\right)\right],$$ which for small enough $\xi\ll 1$ can be large enough to solve the flatness and horizon puzzles.

When $|\vphi-v|\sim{\cal O}(\sqrt\xi v)$, $\vphi$ starts oscillating about $\vphi=v$ with frequency $m_\vphi=\sqrt\lambda v$. Due to the eventual coupling of $\vphi$ to other matter fields, a term $\sim\Gamma\dot\phi$ will arise\cite{turner(coh_osc)_prd_1983} which originates the decay of the coherent oscillations. For further evolution within $t_v\leq t\leq\Gamma^{-1}$, the inflaton $\vphi$ will oscillate with frequency $\approx m_\vphi$, and the energy density $\rho_\vphi\propto a^{-3}$ will decay as dust matter. At $t\approx\Gamma^{-1}$ these oscillations will decay and reheat the universe to a temperature $T_\text{reh}\approx\sqrt{M_\text{Pl}\Gamma}$.

The most serious problem with the above scenario is that in order to successfully implement the inflationary paradigm, it is required a very small -- and unnatural -- coupling constant $\lambda<10^{-12}$ and this, in turn, makes very difficult an acceptable reheating.


\subsubsection{Higgs inflation}\label{subsect-higgs-infl} 

The above mentioned problem can be avoided in the induced gravity inflation model of Refs. \citen{cervantes_prd_1995, cervantes_npb_1995}. In Ref. \refcite{cervantes_prd_1995} the authors investigate the cosmological consequences of a theory of induced gravity in which the BD scalar field is identified with the Higgs field of the first symmetry breaking of a minimal $SU(5)$ GUT, meanwhile, in Ref. \refcite{cervantes_npb_1995} the coupling of the induced gravity theory is to the minimal standard model of the internal gauge group $SU(3)\otimes SU(2)\otimes U(1)$ with the $SU(2)\otimes U(1)$ Higgs field $\vphi$. This latter model requires of unnaturally large values of the Higgs mass in order to get successful cosmology. 

A very popular model of Higgs inflation was developed in Ref. \citen{bezrukov_plb_2008, bezrukov_jhep_2009} (see also Ref. \refcite{kaiser_prd_1995}). The main achievement of the latter work -- as well as of the related works -- was to show that the SMP can give rise to inflation. In order to explain the main idea of Ref. \refcite{bezrukov_plb_2008}, we start with the SMP Lagrangian non-minimally coupled to gravity: $${\cal L}_\text{tot}=\left(\frac{M}{2}+\xi H^\dag H\right)R+{\cal L}_\text{smp},$$ where $M$ is some mass parameter, the coupling $\xi$ is a free constant, $H$ is the Higgs field and ${\cal L}_\text{smp}$ is the SMP part of the Lagrangian. If $\xi=0$, then $M=M_\text{Pl}$ is the Planck mass. Although this model yields to the right particle phenomenology, it produces large matter fluctuations, many orders of magnitude larger than observed, thus leading to incorrect cosmological consequences. If $M=0$, then we recover the induced gravity models of Refs. \citen{zee_prl_1979, cervantes_prd_1995, cervantes_npb_1995}. In this case the Higgs field mass is very large which is in conflict with the experiment\cite{higgs_discov_1, higgs_discov_2} that gives a mass $\approx 125$ GeV. The model proposed in Ref. \refcite{bezrukov_plb_2008} presupposes the existence of an intermediate choice of $M$ and $\xi$ which is good for particle physics and for cosmology at the same time. 

The model of Higgs inflation in Ref. \refcite{bezrukov_plb_2008} is based in the following action (we consider only the scalar sector of the SMP):

\bea S=\int d^4x\sqrt{|g|}\left[\frac{M^2+\xi h^2}{2}\,R-\frac{1}{2}\,(\der h)^2-\frac{\lambda}{4}\left(h^2-v^2\right)^2\right],\label{bezrukov-action}\eea where the unitary gauge $H=h/\sqrt{2}$ ($H^\dag H=h^2/2$) is being considered and, for simplicity, all of the gauge interactions are neglected. Besides, it is assumed that $1\ll\sqrt\xi\lll 10^{17}$, so that $M=M_\text{Pl}$ with good accuracy. We want to point out that in this subsection we are considering $M_\text{Pl}=1/\sqrt{8\pi G_N}\neq 1$.

It is possible to eliminate the non-minimal coupling $\propto h^2R$ by means of a conformal transformation of the metric together with a redefinition of the Higgs field  (see section \ref{sect-cf}):

\bea \hat g_{\mu\nu}=\left(1+\frac{\xi h^2}{M^2_\text{Pl}}\right)g_{\mu\nu},\;d\chi=\frac{\sqrt{1+(1+6\xi)\xi h^2/M^2_\text{Pl}}}{1+\xi h^2/M^2_\text{Pl}}\,dh.\label{bezrukov_conf_t}\eea Under the above transformations the Jordan frame (JF) action \eqref{bezrukov-action} is mapped into the Einstein frame (EF) action:

\bea S_\text{EF}=\int d^4x\sqrt{|\hat g|}\left[\frac{M^2_\text{Pl}}{2}\,\hat R-\frac{1}{2}(\der\chi)^2-U(\chi)\right],\label{bezrukov-ef-action}\eea where the curvature quantities with the hat: $\hat R_{\alpha\beta\mu\nu}$ -- the Riemann-Christoffel tensor, $\hat R_{\mu\nu}$ -- the Ricci tensor, $\hat R$ -- the curvature scalar, etc., are defined as usual but in terms of the metric with the hat $\hat g_{\mu\nu}$, and

\bea U(\chi)=\frac{\lambda}{4\Omega^4(\chi)}\left[h^2(\chi)-v^2\right]^2,\;\Omega^2(\chi)=1+\frac{\xi h^2(\chi)}{M^2_\text{Pl}}.\label{bezrukov-ef-pot}\eea In the weak field limit $h\simeq\chi$ $\Rightarrow\Omega^2\simeq 1$, so that the potential $U$ coincides with the JF potential $V$. On the contrary, for large field values: $h\gg M_\text{Pl}/\sqrt\xi$ $\Rightarrow\chi\gg\sqrt{6}M_\text{Pl}$, the situation is drastically different. In this limit: $$h\simeq\frac{M_\text{Pl}}{\sqrt\xi}\exp\left(\frac{\chi}{\sqrt{6}M_\text{Pl}}\right).$$ This means, in turn, that the EF Higgs potential: $$U(\chi)\simeq\frac{\lambda M^2_\text{Pl}}{4\xi^2}\left[1+\exp\left(-\frac{2\chi}{\sqrt{6}M_\text{Pl}}\right)\right]^{-2},$$ is exponentially flat. The flatness of this potential at $\chi\gg M_\text{Pl}$ is, precisely, what makes possible a succesful chaotic inflation. The obtained slow-roll parameters are:

\bea &&\epsilon=\frac{M_\text{Pl}}{2}\left(\frac{\der_\chi U}{U}\right)^2\simeq\frac{4M^2_\text{Pl}}{3\xi^2h^4},\nonumber\\
&&\eta=M^2_\text{Pl}\frac{\der^2_\chi U}{U}\simeq-\frac{4M^2_\text{Pl}}{2\xi h^2},\nonumber\\
&&\zeta^2=M^4_\text{Pl}\frac{\der_\chi U\der_\chi^3 U}{U}\simeq\frac{16M^4_\text{Pl}}{9\xi^2h^4}.\nonumber\eea Slow roll ends at $\epsilon\simeq 1$ $\Rightarrow$: $$h_{\epsilon\simeq 1}\simeq\left(\frac{4}{3}\right)^{1/4}\frac{M_\text{Pl}}{\sqrt\xi}.$$ The number of $e$-foldings between $h=h_0$ and $h=h_{\epsilon\simeq 1}$ is given by: $$N=\frac{1}{M^2_\text{Pl}}\int_{h_{\epsilon\simeq 1}}^{h_0}\frac{U}{\der_h U}\left(\frac{d\chi}{dh}\right)^2dh\simeq\frac{3\xi}{4}\frac{h_0^2-h^2_{\epsilon\simeq 1}}{M^2_\text{Pl}}.$$

For values of $\sqrt\xi\lll 10^{17}$, since the SMP mass scale $v$ does not appear in the relevant formulae, the inflationary physics does not depend on $v$. After the end of inflation, given that the interactions of the SMP fields with the Higgs are strong, the reheating happens just after the slow-roll stage ends up, with $T_\text{reh}\simeq(2\lambda/\pi^2g)^{1/4}M_\text{Pl}/\sqrt\xi\simeq 10^{15}$ GeV, where $g=106.75$ is the number of degrees of freedom of the SMP.

According to Ref. \refcite{bezrukov_plb_2008}, the radiative corrections do not spoil the flatness of the potential in the region $h\sim 10M_\text{Pl}/\sqrt\xi$, or $\chi\sim 6M_\text{Pl}$, which is essential for chaotic inflation to happen. However, in Ref. \refcite{bezrukov_jhep_2009} the analysis of \refcite{bezrukov_plb_2008} was extended to account for two-loop radiative corrections. As a result, the interval for allowed Higgs masses was somewhat modified, exceeding the region in which the Standard Model can be considered as a viable effective field theory. There are other works\cite{barbon_prd_2009, mark_jhep_2010, salvio_plb_2013} where the validity of the Higgs inflation is profoundly challenged.


\subsubsection{G-inflation}\label{subsect-g-infl} 

Given that STT offer an interesting arena where to look for inflationary behavior, it is then for sure that the Horndeski theories, being their higher-derivatives generalizations, may offer fruitful alternatives for the explanation of the primordial inflation also. Below we shall discuss on two of these possible alternatives: i) the Galileon-driven inflation and ii) inflation driven by the kinetic coupling to the Einstein's tensor.

A new class of inflation model, called as ``G-inflation'', was proposed in Ref. \citen{g-infl-1, g-infl-2, g-infl-3}. This class of inflationary models is specified by a Galileon-like nonlinear derivative interaction of the form $G_3(\phi,X)\Box\phi$ in the Lagrangian with the resulting equations of motion being of second order. The G-inflation is then based on a Horndeski action \eqref{horn-action} with the following choice of the functions: $G_4=1/2$, $G_5=0$ (recall that in this review $X\equiv-(\der\phi)^2/2$);

\bea S_\text{g.inf}=\int d^4x\sqrt{|g|}\left[\frac{1}{2}R+K(\phi,X)-G_3(\phi,X)\Box\phi\right].\label{g-infl-action}\eea In a homogeneous and isotropic cosmological background with FRW metric (flat spatial sections) given by the line-element \eqref{frw-metric}, the motion equations derivable from \eqref{g-infl-action} read:

\bea &&\;\;3H^2=2K_{,X}X-K+3G_{3,X}H\dot\phi^3-2G_{3,\phi}X,\nonumber\\
&&-2\dot H=2K_{,X}X+3G_{3,X}H\dot\phi^3-2\left(2G_{3,\phi}+G_{3,X}\ddot\phi\right)X,\nonumber\\
&&{\cal D}\left(\ddot\phi+3H\dot\phi\right)=-2K_{,X\phi}X+K_{,\phi}+2G_{3,\phi\phi}X-6G_{3,XX}HX\dot X\nonumber\\
&&\;\;\;\;\;\;\;\;\;\;\;\;\;\;\;\;\;\;-6G_{3,X}\left[\dot HX+H\dot X+3H^2X\right]+6\left(K_{,XX}-2G_{3,X\phi}\right)HX\dot\phi,\label{g-infl-feqs}\eea where $${\cal D}\equiv K_{,X}-2\left(G_{3,\phi}-G_{3,X\phi}X\right)+2XK_{,XX}-4G_{3,X\phi}X.$$ The terms with the Hubble parameter (and with its time derivative) in the above equations are the curvature associated terms, so that these are ones that violate the Galileon symmetry. Let us discuss on particular simple models.

\begin{itemize}

\item{\it Kinetic-driven G-inflation.} One simple choice of the functions $K$ and $G_3$ is the following: $K(\phi,X)=K(X)$, $G_3(\phi,X)=g_0 X$, where $g_0$ is some constant with the dimensions of inverse mass squared. In this case the motion equations \eqref{g-infl-feqs} greatly simplify;

\bea &&\;\;3H^2=2K_{,X}X-K+3g_0H\dot\phi^3,\nonumber\\
&&-2\dot H=2K_{,X}X+3g_0H-2g_0X\ddot\phi,\nonumber\\
&&\left(K_{,X}+2XK_{,XX}\right)\ddot\phi=-3K_{,X}H\dot \phi-6g_0\left[\dot HX+H\dot X+3H^2X\right].\label{g-infl-feqs'}\eea Due to our choice of $K(\phi,X)=K(X)$, the obtained inflation is kinetically driven. Notice that, since we have chosen $G_3(\phi,X)=g_0X$, the action \eqref{g-infl-action} has a shift symmetry: $\phi\rightarrow\phi+c$. An exactly de Sitter solution with constant $\dot\phi=\alpha_0$, exist in this case. It is given by\cite{g-infl-1}:

\bea 3H^2=-K,\;\;2K_{,X}X+3g_0H\dot\phi^3=2X\left(K_{,X}+3g_0H\dot\phi\right)=0.\label{g-infl-dsitt}\eea An example is provided by the following choice of the function $K=K(X)$:

\bea K(X)=-X+\frac{g_0}{2\mu}X^2,\label{g-infl-eje}\eea where $\mu$ is a constant parameter with dimension of mass. Under the above choice one have that $X=\alpha^2_0/2$, so that, from the right-hand and left-hand equations in \eqref{g-infl-dsitt} it follows that, $$H=\frac{1-\frac{g_0\alpha_0^2}{2\mu}}{3g_0\alpha_0},\;\;3H^2=\frac{\alpha^2_0}{2}\left(1-\frac{g_0\alpha^2_0}{4\mu}\right),$$ respectively. Comparing both equations for the Hubble rate, for sake of consistency, we get the following constraint on the constants $\alpha_0$, $g_0$ and $\mu$: $$\frac{1-\frac{g_0\alpha_0^2}{2\mu}}{\frac{g_0\alpha_0^2}{2\mu}\sqrt{1-\frac{g_0\alpha_0^2}{4\mu}}}=\sqrt{6}\mu.$$ It is seen that for this de Sitter solution to exist, the time derivative of the Galileon $\dot\phi=\alpha_0$ should be a bounded quantity: $-\sqrt{4\mu/g_0}\leq\alpha_0\leq\sqrt{4\mu/g}$, or, in terms of its kinetic energy density, $X\leq 2\mu/g_0$. The numerical investigation of the model with the following refinement (necessary to look for a change of the sign of the linear kinetic term): $K(X)=-A(\phi)X+g_0X^2/2\mu$, shows that soon after $\phi=\phi_\text{end}$, to change the sign of $A=A(\phi)$, all of the higher derivative terms become negligible and the Galileon behaves as a massless canonical scalar field, so that its energy density quickly dilutes as $\rho_\phi\propto a^{-6}$. Since the shift symmetry of the original Lagrangian prevents direct interaction between the Galileon and standard-model particles, reheating proceeds only through gravitational
particle production\cite{ford-reheat}. The estimated reheating temperature, $T_R\simeq 10^{-2}H_\text{end}^2/M_\text{Pl}$, where $H_\text{end}$ is the Hubble parameter at the end of inflation. For more details on how this toy model produces the primordial inflation see Ref. \refcite{g-infl-1}.

The fact that the above exact de Sitter solution exists, does not mean that it is a generic solution of the field equations \eqref{g-infl-feqs'}. In other words: the exact solution may exist but it may not necessarily be structurally stable, so that it may be attained only under very specific initial conditions. The generic solutions of \eqref{g-infl-feqs'} with arbitrary function $K(\phi,X)$ may be found after a dynamical systems study in an equivalent phase space. This will be the subject of subsection \ref{subsect-gal}.  

\bigskip
\item{\it Potential-driven G-inflation.} Here in \eqref{g-infl-feqs} we set: $$K(\phi,X)=X-V(\phi),\;\;G_3(\phi,X)=-g(\phi)X,$$ so that the motion equations become,

\bea &&\;\;3H^2=X\left[1-(6-\alpha)gH\dot\phi\right]+V,\nonumber\\
&&-\dot H=X\left[1-(3+\eta-\alpha)gH\dot\phi\right],\nonumber\\
&&(3-\eta)H\dot\phi-(9-3\epsilon-6\eta+2\alpha\eta)g(H\dot\phi)^2=-(1+2\beta)\der_\phi V,\label{v-g-infl-feqs}\eea where the following slow-roll parameters have been defined: $$\epsilon:=-\frac{\dot H}{H^2},\;\eta:=-\frac{\ddot\phi}{H\dot\phi},\;\alpha:=\left(\frac{\der_\phi g}{g}\right)\frac{\dot\phi}{H},\;\beta:=\frac{\der_\phi^2g X^2}{\der_\phi V}.$$ It is assumed that all these quantities are small: $\epsilon\sim|\eta|\sim|\alpha|\sim|\beta|\ll 1$. Then, it follows that $g=g(\phi)$ should be a slowly varying function of $\phi$. Besides: $$X\ll V,\;\;|gH\dot\phi X|\ll V,$$ so that the slow-roll motion equations read: $$3H^2\simeq V,\;\;3H\dot\phi(1-3gH\dot\phi)\simeq-\der_\phi V.$$ Two regimes can be differentiated: i) $|gH\dot\phi|\ll 1$, and ii) $|gH\dot\phi|\gg 1$. The former case is standard slow-roll inflation, while the latter leads to modifications of the standard inflation due to the cubic (derivative) self-interaction. In this latter case one have that, $$9H^2\dot\phi^2\simeq\frac{\der_\phi V}{g},$$ where, necessarily, $\der_\phi V/g>0$. The solution of the above slow-roll equations in this second case leads to: $\dot\phi\simeq\sqrt{\der_\phi V/3gV}$, where the scalar field rolls down the potential (ghost instability is avoided provided that $g\dot\phi<0$). Compared with standard slow-roll inflation (case i above), in the Galileon-mediated inflation (case ii) the speed of inflation is suppressed by the factor $1/\sqrt{g\der_\phi V}$ ($g\der_\phi V\gg 1$). In other words, the cubic self-interaction of the Galileon flattens the potential by the same factor, which means that inflation takes place for a wider range of potentials\cite{higgs-g-infl}.

\end{itemize}

Higgs-G-inflation models\cite{higgs-g-infl, kamada(higgs_g_infl)_prd_2011} represent a further modification of G-inflation where the scalar field $\phi$ is identified with the Higgs field ${\cal H}$. In this case in \eqref{g-infl-action} we set: $$K(\phi,X)=X-\frac{\lambda}{4}\phi^4,\;\;G_3(\phi,X)=-\frac{\phi}{M^4}X,$$ where $M$ is some mass parameter. As in the discussion above, the cubic self-interaction affects the inflationary stage if $g\der_\phi V\gg 1$, i. e., if $\phi\gg M_\text{Pl}M/M_c,$ where $M_c:=\lambda^{1/4}M_\text{Pl}$ is some additional mass scale. If $M\ll M_c$, Higgs G-inflation proceeds even if standard Higgs inflation would otherwise be impossible. For more details see Ref. \refcite{higgs-g-infl}. The generated primordial density perturbation has been shown to be consistent with the present observational data\cite{kamada(higgs_g_infl)_prd_2011}.


\subsubsection{Inflation driven by the kinetic coupling to the Einstein's tensor}\label{subsect-k-infl} 

Among the Horndeski generalizations of scalar-tensor theories we find the so called theories with kinetic coupling to the Einstein's tensor\cite{sushkov, saridakis-sushkov, sushkov-a, k-coup-skugoreva, matsumoto, granda, gao, germani-prl, germani}, that are given by the action \eqref{k-coup-action}: $$S_\text{kc}=\frac{1}{2}\int d^4x\sqrt{|g|}\left[R+2(X-V)+\alpha G_{\mu\nu}\der^\mu\phi\der^\nu\phi\right].$$ In Ref. \refcite{sushkov} it was shown that a cosmological model with nonminimal derivative coupling of the form $G_{\mu\nu}\der^\mu\phi\der^\nu\phi$ is able to explain in a unified way both a quasi-de Sitter phase and an exit from it without any fine-tuned potential, while in Ref. \refcite{saridakis-sushkov} it was found that, depending on the coupling parameter, the universe transits from one de Sitter stage to another. Hence the kinetic coupling provides an essentially new inflationary mechanism\cite{sushkov-a} where, in contrast to the standard inflationary scenario, the dynamics of the inflation does not depend on the scalar field potential and is only determined by the coupling parameter $\alpha$\cite{k-coup-skugoreva}. Below we shall briefly outline the most salient features of this new inflationary scenario.

Let us consider the background FRW metric with flat spatial sections that is given by \eqref{frw-metric}. If substitute this metric back into the field equations that are derived from the action \eqref{k-coup-action} -- under the assumption of vanishing potential -- one obtains the following cosmological equations\cite{sushkov}:

\bea 3H^2&=&\frac{X}{1-3\alpha X}\;\;\Rightarrow\;\;X=\frac{3H^2}{1+3\alpha H^2},\nonumber\\
-\left(2\dot H+3H^2\right)&=&\frac{X-2\alpha H\dot X}{1-\alpha X},\nonumber\\
\left(1+3\alpha H^2\right)\dot X&=&-6HX-6\alpha H\left(2\dot H+3H^2\right)X,\label{k-coup-feqs}\eea where we have considered that $a(t)=\exp[\alpha(t)]$ and, as before: $X\equiv\dot\phi^2/2$. Notice from the Friedmann equation above (top left-hand equation) that, for positive coupling, $\alpha>0$, the kinetic energy density is a bounded quantity: $0\leq X<1/3\alpha$. In this case the Hubble parameter is unbounded and a cosmological singularity -- like the bigbang singularity, for instance -- may arise. Meanwhile, from the right-hand Friedmann equation in \eqref{k-coup-feqs}, it follows that for negative coupling, $\alpha<0$, the Hubble parameter (squared) is bounded instead: $0\leq H^2\leq -1/3\alpha$. In this last case, whenever $X$ is a monotonically decreasing function of the cosmic time, asymptotically into the past: $$\lim_{t\rightarrow-\infty} H^2=-\frac{1}{3\alpha},$$ a de Sitter stage with $a(t)\propto\exp(\sqrt{-1/3\alpha}\;t)$, is approached. This early de Sitter stage is what is assumed to provide an alternative description of the primordial inflation\cite{sushkov}. Other alternatives arise if consider non-vanishing self-interaction potentials like in Ref. \refcite{saridakis-sushkov}. In particular the universe may transit from one de Sitter solution to another. A realistic inflationary scenario based in the action \eqref{k-coup-action} (see above) with the constant potential, $V(\phi)=\Lambda$-- the measured value of the cosmological constant, was investigated in Ref. \refcite{sushkov-a}. In that scenario the primordial inflationary epoch driven by non-minimal kinetic coupling comes to the end at $t_f\simeq 10^{-35}$ sec. Later on the universe enters into the matter-dominated epoch which lasts approximately for $\sim 10^{18}$ sec. Finally the cosmological term (the constant potential) comes into play, and the universe enters into the secondary inflationary epoch with $a(t)\propto\exp(\sqrt{\Lambda/3}\;t)$. We recommend Ref. \refcite{sushkov-a} for details.


\subsection{Extended quintessence}\label{subsect-quint}

In order to explain the missing dark energy within the context of general relativity, it is necessary to put -- by hand -- a mysterious matter component with negative pressure in the RHS of the Einstein's field equations. A simple model that does the work is the one based in a self-interacting (very light), slowly-rolling scalar field, which is generically called as quintessence\cite{matos_urena_cqg_2000, carroll_prl_1998, zlatev_prl_1999, amendola_prd_2000, wang_atrophys_j_2000, barreiro_prd_2000, sahni_prd_2000, pavon_plb_2001, chimento_prd_2003, caldwell_prl_2005, copeland_rev_ijmpd_2006, peebles, lcdm, cosh-pot, urena_prd_2000, quiros_cqg_2003}. If one were interested in the search for an alternative explanation to the present stage of accelerated cosmic expansion, not relying in Einstein's GR, the scalar-tensor theories offer a not much more complex possibility. In this latter case the scalar field is not put by hand but it is a part of the gravitational interactions themselves. The price to pay is a non-minimal interaction between the curvature and the scalar field, which is inherent in STT of gravity, and does not arise in standard quintessence models.  

Extended quintessence\cite{uzan(ext-quint)_prd_1999, amendola(ext-quint)_prd_1999, chiba(ext-quint)_prd_1999, ext-quint, ext-quint-1, ext-quint-2, ext-quint-3, boisseau_prl_2000, torres_prd_2002, ext-quint-4, ext-quint-5, ext-quint-exp-pot, ext-quint-faraoni, ext-quint-nesseris, ext-quint-slwr, new-ext-quint, ext-quint-nbody, ext-quint-hrycyna, ext-quint-fan, lymperis_prd_2017, ext-quint-li, ext-quint-dsyst} is the collective name under which scalar-tensor theories of gravity (with non-vanishing self-interacting scalar field) are known in a cosmological context where the energy density of the scalar field provides most of the cosmic energy today and is the responsible for the present inflationary stage of the cosmic expansion. Such cosmic scenarios have the appealing feature that the same BD-type field that causes the gravitational coupling to vary from point to point in spacetime, is the origin of the spacetime variations of the cosmological 'constant'. 

The action for the extended quintessence is given as it follows\cite{ext-quint}:

\bea S_\text{EQ}=\int d^4x\sqrt{|g|}\left[\frac{1}{2}f(\phi,R)-\frac{1}{2}\omega(\phi)(\der\phi)^2-V(\phi)+\beta{\cal L}_{(m)}\right],\label{ext-quint-action}\eea where $\beta$ is a constant that allows to fix units and ${\cal L}_{(m)}$ is the Lagrangian of the matter degrees of freedom. The related equations of motion read:

\bea &&G_{\mu\nu}=\frac{\beta}{\der_R f}\,T^{(m)}_{\mu\nu}+\frac{\omega}{\der_R f}\left[\der_\mu\phi\der_\nu\phi-\frac{1}{2}g_{\mu\nu}(\der\phi)^2\right]\nonumber\\
&&\;\;\;\;\;\;\;\;\;\;\;\;\;\;\;\;\;\;\;\;\;\;\;\;\;\;\;+g_{\mu\nu}\frac{f-(\der_R f) R-2V}{2\der_R f}+\frac{1}{\der_R f}\left(\nabla_\mu\nabla_\nu-g_{\mu\nu}\Box\right)(\der_R f),\nonumber\\
&&\omega\Box\phi+\der_\phi\omega(\der\phi)^2+\frac{1}{2}\,\der_\phi f=\der_\phi V.\label{ext-quint-feq}\eea 

Here we shall focus in particular cases when $f(\phi,R)=F(\phi)R$. In these cases $\der_R f=F(\phi)$ $\Rightarrow f-(\der_R f)R=0$, so that the above motion equations get simplified:

\bea &&G_{\mu\nu}=\frac{\beta}{F}\,T^{(m)}_{\mu\nu}+\frac{\omega}{F}\left[\der_\mu\phi\der_\nu\phi-\frac{1}{2}g_{\mu\nu}(\der\phi)^2\right]-\frac{V}{F}g_{\mu\nu}\nonumber\\
&&\;\;\;\;\;\;\;\;\;\;\;\;\;\;\;\;\;\;\;\;\;\;\;\;+\frac{F''}{F}\left[\der_\mu\phi\der_\nu\phi-g_{\mu\nu}(\der\phi)^2\right]+\frac{F'}{F}\left(\nabla_\mu\nabla_\nu-g_{\mu\nu}\Box\right)\phi,\nonumber\\
&&\omega\Box\phi+\omega'(\der\phi)^2+\frac{F'}{2}\,R=V',\label{ext-quint-moteq}\eea where the prime denotes derivative with respect to the scalar field $\phi$. It is to be recalled that under the simultaneous replacement: $$F(\phi)\rightarrow\phi,\;\frac{F}{(F')^2}\rightarrow\omega_\text{BD},$$ the latter motion equations transform into the field equations of the BD theory.

In order to illustrate the dynamics of the extended quintessence scenario depicted by \eqref{ext-quint-moteq}, here we shall choose the following relevant coupling functions (we set $\beta=1$):

\bea F(\phi)=1-\xi\left(\phi^2-\phi^2_0\right),\;\omega(\phi)=1,\label{ext-quint-choice}\eea where $\xi$ is the non-minimal coupling constant (the choice $\xi=1/6$ is usually known as conformal coupling) while $\phi_0$ is the magnitude of the scalar field today. This value has been chosen in such a way as to set the present value $F_0=F|_{\phi=\phi_0}=1$, as required (recall that we work in the units where $8\pi G_N=c=1$). We further assume a homogeneous and isotropic FRW metric \eqref{frw-metric} as an adequate model for the background geometry. The cosmological dynamics is then governed by the following equations:

\bea &&3H^2=\frac{1}{F}\left(\rho_m+\frac{1}{2}\dot\phi^2+V\right)-3\frac{F'}{F}H\dot\phi,\nonumber\\
&&\dot H=-\frac{1}{2F}\left(\rho_m+p_m+\dot\phi^2\right)+\frac{F'}{2F}\left(H\dot\phi-\ddot\phi\right)-\frac{F''}{2F}\dot\phi^2,\nonumber\\
&&\ddot\phi+3H\dot\phi=3F'\left(\dot H+2H^2\right)-V',\label{ext-quint-cosmo-eq}\eea where the overdot means derivative with respect to the cosmic time $t$ while $F$ is given by \eqref{ext-quint-choice}. In \eqref{ext-quint-cosmo-eq} we have assumed the matter degrees of freedom in the form of a perfect (barotropic) fluid with energy density $\rho_m$ and pressure $p_m$. From these equations one can obtain the following expression for the deceleration parameter $q\equiv-\ddot a/aH^2$:

\bea q=\frac{\rho_m+3p_m+2(\dot\phi^2-V)}{6\left(F+3F'^2/2\right)H^2}+\frac{F''\dot\phi^2-2F'H\dot\phi-F'V'}{2\left(F+3F'^2/2\right)H^2}+\frac{3F'^2}{2\left(F+3F'^2/2\right)},\label{ext-quint-q}\eea where, according to \eqref{ext-quint-choice}, $F'=-2\xi\phi$, $F''=-2\xi$. This is to be contrasted with the expression for the deceleration parameter for standard quintessence (minimal coupling of the scalar field):

\bea q=\frac{\rho_m+3p_m+2(\dot\phi^2-V)}{6H^2}.\label{quint-q}\eea In the latter case the condition for accelerated expansion can be written as: $V>\dot\phi^2+(\rho_m+3p_m)/2$. In the extended quintessence scenario, despite that the terms $\xi\dot\phi^2$ and $-\xi\phi V'$ in \eqref{ext-quint-q} may also contribute towards accelerated expansion, since these are first-order in the coupling parameter $\xi$ -- which is assumed to be small -- one should expect that the standard quintessence condition for accelerated expansion could be only slightly modified. 

There are two important experimental constraints any STT should meet\cite{will-lrr-2014, damour-epj-1998, reasenberg-1979}:

\begin{enumerate}

\item From solar system experiments it follows that: $$\left|\frac{\dot G_\text{eff}}{G_\text{eff}}\right|=\left|\frac{\dot F_0}{F_0}\right|\leq 10^{-11}\;\text{per year},$$

\item while, from effects induced on photon trajectories: $$\omega_\text{BD}=\frac{F_0}{F'^2_0}\geq 4\times 10^4.$$

\end{enumerate} These constraints lead to: $$\xi\leq\frac{10^{-11}}{2\phi_0\dot\phi_0},\;\xi\leq\frac{10^{-2}}{4\phi_0},$$ respectively. The second constraint does not depend on the rate of change of the scalar field but only on its present value, so that it is the most convenient one in order to establish bounds on the coupling $\xi$. In Ref. \refcite{chiba(ext-quint)_prd_1999}, for instance, by fixing the potential to be of tracker-type: $V(\phi)=M^4(\phi/M)^{-\alpha}$ with $\alpha=4$, the following constraints on the coupling parameter: $-10^{-2}\leq\xi\leq 10^{-2},$ were obtained while searching for a dynamical behavior in accordance with the existing at the time observational evidence. This limit on the coupling constant is not as strong as existing limits on the couplings to ordinary matter\cite{carroll_prl_1998} such as to the electromagnetic field ($\leq 10^{-6}$) and to QCD ($\leq 10^{-4}$).

The qualitative behavior of the extended quintessence driven by the motion equations \eqref{ext-quint-cosmo-eq} can be summarized as follows. At sufficiently early times when the curvature is high enough, the piece of the effective potential in the KG equation \eqref{ext-quint-cosmo-eq}: $V_\text{eff}=V+3\xi(\dot H+2H^2)\phi^2$, that comes from the non-minimal coupling $\propto\xi(\dot H+2H^2)\phi^2$, dominates over the pure self-interaction potential $V$. Then the field $\phi$ settles down to a slow-roll regime where the friction term $3H\dot\phi$ balances the term $6\xi(\dot H+2H^2)\phi$. In this regime, which lasts until $V$ becomes significant, $\phi$ is nearly a constant. After that $\phi$ starts to roll fast and the term $\propto\xi(\dot H+2H^2)\phi$ can be neglected so that the field behaves as a minimally coupled tracker field with the inverse power-law potential $V\propto\phi^{-4}$. 

The non-minimal coupling in the extended quintessence models modifies the estimations for the integrated Sachs-Wolfe effect: $\delta C_{l\simeq 10}/C_{l\simeq 10}\simeq 6[1-F(\phi_\text{dec})]\simeq 12\xi\phi_0^2$ (the $C_l$ are the expansion coefficients of the two-point correlation function into Legendre polynomials\cite{two_p_coeff} and $\phi_\text{dec}$ stands for the value of the field at decoupling), and also for the positions of the acoustic peak multipoles that shift in the following amount: $\delta l/l\simeq[F(\phi_\text{dec})-1]/8\simeq\xi\phi_0^2/8$. These effects can be $10-30\%$ with respect to standard quintessence\cite{ext-quint}. 

Another interesting consequence of non-minimally coupled quintessence models is that, under some circumstances, these models can lead to a reduction in the primordial helium abundance\cite{ext-quint-3}. This reduction is a desirable effect given the tension between estimates of the primordial helium abundance predicted by the standard bigbang nucleosynthesis, $Y_P=0.248\pm 0.001$, and the likely lower actual helium abundance: $Y_P=0.234\pm 0.003$ according to Ref. \refcite{olive_1995} and $Y_P=0.244\pm 0.002$ according to Ref. \refcite{izotov_1998}.


\subsubsection{Slow-rolling extended thawing quintessence}

In this subsection, in order to get more insight into the models with non-minimally coupled quintessence, we discuss on a model of extended quintessence called as extended ``thawing'' quintessence\cite{ext-quint-slwr}. Quintessence models which in the past are described by an almost constant scalar field and begin to roll down the potential recently, are called ``thawing'' models\cite{thaw_quint_chiba}. Of particular interest are the slow-roll conditions for these models. The usual slow-roll conditions: $$\epsilon=\frac{1}{2}\left(\frac{V'}{V}\right)^2\ll 1,\;\eta=\left|\frac{V''}{V}\right|\ll 1,$$ that are required in order to get the correct primordial inflation, are not expected to hold in the case of thawing quintessence since, in general, the term $\ddot\phi$ is not necessarily small compared with the friction term $3H\dot\phi$ in the KG equation of motion. The correct slow-roll conditions in this case were derived in Ref. \refcite{thaw_quint_chiba}.

In the case of standard (minimally coupled or uncoupled) quintessence the equations of motion in the flat FRW metric read:

\bea &&3H^2=\rho_m+\rho_\phi,\nonumber\\
&&\dot H=-\frac{1}{2}\left(\rho_m+p_m+\rho_\phi+p_\phi\right),\nonumber\\
&&\ddot\phi+3H\dot\phi=-V',\label{thaw-quint-feq}\eea where $\rho_\phi=\dot\phi^2/2+V$, $p_\phi=\dot\phi^2/2-V$, $\rho_m$ and $p_m$ are the energy density and pressure of the background fluid (matter/radiation) with the following equation of state: $p_m=\omega_m\rho_m$ ($\omega_m$ is the EOS parameter of the matter/radiation).

By slow-roll quintessence it is meant a model of quintessence whose kinetic energy density is much smaller than its potential, $\dot\phi^2\ll 2V$. But, unlike primordial inflation, since the Hubble rate $H$ in the Friedmann equation in \eqref{thaw-quint-feq} is not determined by the energy density of the quintessence alone but, also, by the energy density of matter/radiation, it is not required that $\ddot\phi$ be smaller than the friction term $3H\dot\phi$ in the KG motion equation in \eqref{thaw-quint-feq}. Actually, slowly rolling thawing quintessence models have the approximate equations of state $p_\phi+\rho_\phi\simeq 0$ $\Rightarrow\omega_\phi=p_\phi/\rho_\phi\simeq -1$, so that the friction term $3H(\rho_\phi+p_\phi)/\dot\phi$ coming from the conservation equation for the quintessence: $\dot\rho_\phi+3H(\rho_\phi+p_\phi)=0$, is not effective and $\ddot\phi$ is not necessarily small compared with $3H\dot\phi$. It is useful to introduce the parameter $\beta$ that determines the size of the acceleration of the field relative to the damping term\cite{thaw_quint_linder, thaw_quint_crittenden}:

\bea \beta=\frac{\ddot\phi}{3H\dot\phi}=-1-\frac{V'}{3H\dot\phi}\;\;\Rightarrow\;\;\dot\phi=-\frac{V'}{3(\beta+1)H}.\label{beta-eq}\eea For thawing models $\beta$ is approximately a constant in the sense that $|\dot\beta|\ll H|\beta|$. In particular, while the Universe remains matter-dominated, $\beta=1/2$ and then begins to decline towards zero when the quintessence starts dominating\cite{thaw_quint_crittenden}.

Taking into account the slow-roll condition $\dot\phi^2\ll 2V$ and the RHS equation in \eqref{beta-eq}, one gets: $$\frac{V'^2}{18(1+\beta)^2H^2V}\ll 1,$$ or, in terms of the slow-roll parameter $\epsilon$:

\bea \epsilon:=\frac{V'^2}{6H^2V}\ll 1,\label{thaw-eps-slwr}\eea where we have omitted the (almost) constant term $1+\beta$ and the factor of $1/6$ is kept in order for the above definition of the $\epsilon$-slow-roll parameter to coincide with the one in primordial inflation when $3H^2\simeq V$. This equation is the quintessence counterpart of the primordial inflation slow-roll condition $(V'/V)^2\ll 1$.

In order to get the second slow-roll condition, let us to take the time derivative of the RHS equation in \eqref{beta-eq}: 

\bea \ddot\phi=\left[\frac{V''}{9(1+\beta)H^2}-\frac{1+\omega_m}{2}+\frac{\dot\beta}{3(1+\beta)H}\right]\frac{V'}{1+\beta},\label{ddotphi-eq}\eea where we have taken into account that, assuming matter/radiation domination, from the Friedmann and Raychaydhuri equation in \eqref{thaw-quint-feq} it follows that, $$\frac{\dot H}{H^2}\simeq-\frac{3}{2}\,(1+\omega_m).$$ On the other hand, since according to \eqref{beta-eq}: $$\ddot\phi=3\beta H\dot\phi=-\frac{\beta V'}{1+\beta},$$ then \eqref{ddotphi-eq} can be rewritten in the following form: $$\beta=\frac{1+\omega_m}{2}-\frac{\dot\beta}{3(1+\beta)H}-\frac{V''}{9(1+\beta)H^2},$$ or, since $\dot\beta\ll H\beta$:

\bea \beta\simeq\frac{1+\omega_m}{2}-\frac{V''}{9(1+\beta)H^2}.\label{thaw-beta-eq}\eea Given that $\beta$ is almost a constant during matter/radiation domination, then the following slow-roll condition must hold: $V''/9(1+\beta)H^2\ll 1$, or in terms of the corresponding slow-roll parameter;

\bea \eta:=\frac{V''}{3H^2}\;\;\Rightarrow\;\;|\eta|\ll 1.\label{thaw-eta-slwr}\eea Hence, $\beta=(\omega_m+1)/2$.

Equations \eqref{thaw-eps-slwr} and \eqref{thaw-eta-slwr} constitute the slow-roll conditions for thawing quintessence during the matter/radiation epoch\cite{thaw_quint_chiba}. If follow the same procedure with a thawing quintessence field but, in place of the motion equations \eqref{thaw-quint-feq}, take into account equations \eqref{ext-quint-cosmo-eq} for extended quintessence, the result is quite different. Actually, taking into account the above conditions for slow-rolling thawing quintessence one obtains: $$\beta=\frac{\ddot\phi}{3H\dot\phi}=\frac{\omega_m-1}{2}.$$ The slow-roll condition \eqref{thaw-eps-slwr} is transformed into the following:

\bea \epsilon:=\frac{V'^2_\text{eff}}{6H^2V_*},\label{ext-thaw-eps-slwr}\eea where $V_\text{eff}=V+3FH^2(1-3\omega_m)$ and $V_*=V-3F'H\dot\phi$, while equation \eqref{ddotphi-eq} is transformed into the following equation: $$\ddot\phi=-\frac{\dot H}{H}\dot\phi-\frac{V''\dot\phi}{3(1+\beta)H}-\frac{F''H(1-3\omega_m)\dot\phi}{1+\beta}+\frac{3F'H^2(1-3\omega_m)}{1+\beta}-\frac{\dot\beta\dot\phi}{1+\beta},$$ with $\dot\phi=-V'_\text{eff}/3(1+\beta)H$. Hence, if take into account that $|\dot\beta|\ll H|\beta|$, $$\beta=\frac{\omega_m-1}{2}-\frac{V''}{9(1+\beta)H^2}-\frac{F''(1-3\omega_m)}{3(1+\beta)}+\frac{V'}{V'_\text{eff}}.$$ From this latter equation, since $\beta$ is assumed to be a constant during the matter/radiation dominated stage of the cosmic evolution, $\beta=(\omega_m-1)/2$, the following slow-roll conditions follow ($\eta:=V''/3H^2$):

\bea |\eta|\ll 1,\;|F''(1-3\omega_m)|\ll 1,\;\left|\frac{V'}{V'_\text{eff}}\right|\ll 1.\label{ext-thaw-other-slwr}\eea The latter slow-roll conditions for extended thawing quintessence are the counterpart of the slow-roll condition, $|V''|/V\ll 1$, for primordial inflation.


\subsubsection{The limits of extended quintessence}

In Ref. \refcite{ext-quint-nesseris} the limits of extended quintessence were explored in details. The authors used a low redshift expansion of the cosmological equations of extended quintessence to divide the observable Hubble history parameter space in four sectors: A forbidden sector I where the scalar field of the theory becomes imaginary (the kinetic term becomes negative), a forbidden sector II where the scalar field rolls up (instead of down) its potential, an allowed freezing quintessence sector III where the scalar field is currently decelerating down its potential towards freezing and an allowed thawing sector IV where the scalar field is currently accelerating down its potential.

In the extended quintessence model where the motion equations \eqref{ext-quint-cosmo-eq} drive the dynamics, the measured effective Newton's constant is defined as:

\bea G_\text{eff}=\frac{1}{F}\left(\frac{1+2F'^2/F}{1+3F'^2/2F}\right).\label{ext-quint-measured-G}\eea From local experiments in the solar system it follows that\cite{will-lrr-2014}: $F'^2/F<2.5\times 10^{-5}$, hence, the measured gravitational coupling, $$G_\text{eff}\simeq\frac{1}{F}.$$ The field equations \eqref{ext-quint-cosmo-eq} and the relevant parameters can be written in terms of the redshift $z$ by recalling that, $$a(t)=\frac{a_0}{1+z(t)}\;\;\Rightarrow\;\;dt=-\frac{dz}{(1+z)H(z)}.$$ Combining the Friedmann and the Raychaydhuri equations in \eqref{ext-quint-cosmo-eq} we can get:

\bea &&\phi'^2=-F''-\left(\frac{H'}{H}+\frac{2}{1+z}\right)F+\frac{2FH'}{(1+z)H}-3(1+z)\Omega_{0m}\left(\frac{H_0}{H}\right)^2F_0,\nonumber\\
&&V=\frac{(1+z)^2H^2}{2}\left\{F''+\left(\frac{H'}{H}-\frac{4}{1+z}\right)F'+\frac{6F}{(1+z)^2}\left[1-\frac{(1+z)H'}{3H}\right]\right.\nonumber\\
&&\left.\;\;\;\;\;\;\;\;\;\;\;\;\;\;\;\;\;\;\;\;\;\;\;\;\;\;\;\;\;\;\;\;\;\;\;\;\;\;\;\;\;\;\;\;\;\;\;\;\;\;\;\;\;\;\;\;\;\;\;\;\;\;\;\;\;\;-3(1+z)\Omega_{0m}\left(\frac{H_0}{H}\right)^2F_0\right\},\label{limits-eq}\eea where the prime denotes derivative with respect to the redshift $z$, $H_0\simeq 10^{-10}h$ yrs$^{-1}$ and $\Omega_{0m}$ is the current normalized matter density. The authors of \refcite{ext-quint-nesseris} then explore the observational consequences that emerge from the following generic inequalities: 

\bea \phi'^2(z)>0,\;V'(z)>0,\;(\phi'^2(z))'>0,\;(\phi'^2(z))'<0,\label{limits-ineq}\eea where, from left to the right the above inequalities entail: a real scalar field, that rolls down (not up) its potential, represents freezing and thawing quintessence, respectively. Since the observational consequences of extended quintessence at low redshifts is our interest here, the relevant functions and parameters are to be expanded around $z=0$:

\bea &&\phi(z)=1+\phi_1 z+\phi_2 z^2+\cdots,\nonumber\\
&&V(z)=1+V_1 z+V_2 z^2+\cdots,\nonumber\\
&&H^2(z)=1+h_1 z+h_2 z^2+\cdots,\nonumber\\
&&F(z)=1+F_1 z+F_2 z^2+\cdots,\label{expansion-low-z}\eea where $$F_1=g_1\equiv\frac{\dot G_0}{G_0H_0},\;F_2=g_1\left(g_1-\frac{h_1}{4}-\frac{1}{2}\right)-\frac{g_2}{2},$$ with $g_n\equiv G_0^{(n)}/G_0H_0^n$. These expansions are to be substituted into \eqref{limits-eq}, equating terms order by order in $z$ and ignoring the coefficient $g_1$ due to local experiments that set $|g_1|<10^{-13}H_0^{-1}\;\text{yrs}^{-1}\simeq 10^{-3}h^{-1}\ll 1$. For the zeroth and first order in $z$ it is found that

\bea &&h_1-3\Omega_{0m}+g_2=\phi_1^2>0,\nonumber\\
&&-h_1(1+h_1)+2h_2-3\Omega_{0m}(1-h_1)-g_2(1+h_1)-g_3=(\phi'^2)'|_{z=0}.\label{bounding}\eea The second equation above along with the conditions for freezing or thawing quintessence divide the allowed $(h_1,h_2)$ parameter space into a freezing ($(\phi'^2(z))'>0$) and a thawing ($(\phi'^2(z))'<0$) sector, respectively, for each set of $(g_2,g_3)$. Unfortunately, due to the way in which the existing codes parametrize the measured gravitational coupling $G(t)$, the local experiments provide bounds on $g_1$ but not on the $g_i$ ($i\geq 2$). Nevertheless, rough order estimates can be made on $g_2$. Actually, it follows from solar system experiments that $|\dot G_0/G_0|<10^{-13}$ yrs$^{-1}$, hence, assuming that the total variation $\Delta G/G_0$ over the time scale $\Delta t$, $$\left|\frac{\Delta G}{G_0}\right|\simeq\left|\frac{\dot G_0}{G_0}\right|\Delta t\simeq\left|\frac{\ddot G_0}{G_0}\right|(\Delta t)^2<10^{-11},$$ we get that $$\left|\frac{\ddot G_0}{G_0}\right|<10^{-15}\;\text{yrs}^{-2}\;\;\Rightarrow\;\;|g_2|<10^5h^{-2}.$$ Generalizing the above argument to any order in $\Delta t$ one gets\cite{ext-quint-nesseris}: $|g_n|<10^{8n-11}h^{-n}$. In Ref. \refcite{ext-quint-nesseris}, by using the SNIa data and the Chevalier-Polarski-Linder parametrization\cite{chevalier_polarski_2001, linder_prl_2003}: $$H^2(z)=H^2_0\left[\Omega_{0m}(1+z)^3+(1-\omega_{0m})(1+z)^{3(1+w_0+w_1)}e^{-\frac{3w_1z}{1+z}}\right],$$ to the Supernova Legacy Survey (SNLS) data set\cite{snls} with prior $\Omega_{0m}=0.24$, where $$w(z)=w_0+w_1\frac{z}{1+z},$$ that leads to: 

\bea &&h_1=3(1+w_0-\Omega_{0m}w_0),\nonumber\\
&&h_2=\frac{3}{2}\left[2+5w_0(1-\Omega_{0m})+(3w_0^2+w_1)(1-\Omega_{0m})\right],\nonumber\eea the authors obtained the following improved bound on $g_2$: $$g_2=\frac{\ddot G_0}{G_0 H_0^2}>-1.91\;\;\Rightarrow\;\;\frac{\ddot G_0}{G_0}>-1.91 H_0^2\simeq-2\times 10^{-20}h^2\;\text{yrs}^{-2},$$ valid at $2\sigma$ level.


\section{Asymptotic dynamics of Brans-Dicke and scalar-tensor theories}\label{sect-dsyst-bd}

In addition to the well-known fact that in the $\omega_\text{BD}\rightarrow\infty$ limit of Brans-Dicke theory GR is recovered (but for matter with traceless stress-energy tensor), several works\cite{damour_prl_1993, serna_cqg_2002} have been dedicated to show that indeed general relativity can be recovered in certain asymptotic limit of BD theory. In Ref. \refcite{damour_prl_1993} STT-s of gravity were shown to generically contain an attractor mechanism toward general relativity, with the redshift at the beginning of the matter-dominated era providing the measure for the present level of deviation from general relativity. Meanwhile, in Ref. \refcite{serna_cqg_2002} the authors investigated the conditions for convergence toward GR of STT gravity defined by an arbitrary coupling function $\alpha$ in the Einstein frame. They showed that, in general, the evolution of the scalar field is governed by two opposite mechanisms: an attraction mechanism which tends to drive scalar-tensor model toward Einstein's GR, and a repulsion mechanism which has the contrary effect.

However, when asymptotic dynamics as the above mentioned, are implied, it is very useful to rely on the tools of the dynamical systems\cite{ejp-2015, copeland-wands-rev, cosmology-books, cosmology-books-1, copeland, luis-mayra, kolitch_ann_phys_1996, quiros-cqg-2010, quiros-plb-2009, quiros-plb-2009-1, quiros_cqg_2006, lazkoz_genly_plb_2006, genly_lazkoz_quiros_plb_2007, genly, paliatha, paliatha-1, granda_ijmpd_2017, uggla-ref, uggla-ref-1}. There are several works in the bibliography on the study of the asymptotic dynamics of the BD-theory and also of scalar-tensor theories\cite{ds-bd, ds-bd-1, ds-bd-2, holden_wands_cqg_1998, olga, indios, faraoni, iranies, genly-1, genly-1-1, genly-1-2, gregory_ann_phys_1997}. In the work of Ref. \refcite{gregory_ann_phys_1997} the authors studied the BD-theory with the quadratic potential $V\propto\phi^2$, while in Ref. \refcite{holden_wands_cqg_1998} the Brans-Dicke theory with vanishing potential was investigated. Other previous works on the asymptotic dynamics of the BD-theory include the Refs. \citen{carloni-cqg-2008, hrycyna, hrycyna-1, hrycyna-2, quiros-prd-2015-1}. Here we concentrate in the latter -- most recent -- work and we shall show that, in general, the general relativity de Sitter solution does not arise in the BD theory but for the quadratic self-interaction potential. We discuss also on the rich structure of the phase space of the BD-theory, including the existence of a BD-de Sitter phase that shares no similitude with the standard GR-de Sitter solution.

The Brans-Dicke theory (in the Jordan frame) is depicted by the action \eqref{bd-action}. In this section, for convenience, we rescale the BD scalar field and also redefine the self-interaction potential:

\bea \phi=e^\vphi,\;V(\phi)=e^\vphi\,U(\vphi),\label{vphi}\eea so that, the action \eqref{bd-action} is transformed into the dilatonic BD action:

\bea S=\int d^4x\sqrt{|g|}e^\vphi\left\{R-\omega_\textsc{bd}(\der\vphi)^2-2U+2e^{-\vphi}{\cal L}_m\right\}.\label{dbd-action}\eea Within the context of the low-energy effective string theory the latter action is meant to represent the so called 'string frame' representation of the theory \cite{copeland-wands-rev}. Here we prefer, for the moment, to keep talking about dilatonic JF-BD theory instead of string-frame effective action. The following motion equations are obtained from (\ref{dbd-action}):

\bea &&G_{\mu\nu}=\left(\omega_\textsc{bd}+1\right)\left[\der_\mu\vphi\der_\nu\vphi-\frac{1}{2}g_{\mu\nu}(\der\vphi)^2\right]-g_{\mu\nu}\left[\frac{1}{2}(\der\vphi)^2+U(\vphi)\right]\nonumber\\
&&\;\;\;\;\;\;\;\;\;\;\;\;\;\;\;\;\;\;\;\;\;\;\;\;\;\;\;\;\;\;\;\;\;\;\;\;\;\;\;\;\;\;\;\;\;\;\;\;\;\;\;\;\;\;+\nabla_\mu\der_\nu\vphi-g_{\mu\nu}\Box\vphi+e^{-\vphi}T^{(m)}_{\mu\nu},\nonumber\\
&&\Box\vphi+(\der\vphi)^2=\frac{2}{3+2\omega_\textsc{bd}}\left(\der_\vphi U-U\right)+\frac{e^{-\vphi}}{3+2\omega_\textsc{bd}}\,T^{(m)},\label{feq}\eea where $T^{(m)}_{\mu\nu}$ is the stress-energy tensor of the matter degrees of freedom. Let us assume FRW spacetimes with flat spatial sections ($k=0$), with the line-element \eqref{frw-metric}: $ds^2=-dt^2+a^2(t)\delta_{ij}dx^idx^j$, $i,j=1,2,3$. We assume the matter content of the Universe in the form of a cosmological perfect fluid, which is characterized by the state equation $p_m=w_m\rho_m$, relating the barotropic pressure $p_m$ and the energy density $\rho_m$ of the fluid ($w_m$ is the EOS parameter of the matter fluid). Under these assumptions the cosmological equations (\ref{feq}) are written as it follows:

\bea &&3H^2=\frac{\omega_\textsc{bd}}{2}\,\dot\vphi^2-3H\dot\vphi+U+e^{-\vphi}\rho_m,\nonumber\\
&&\dot H=-\frac{\omega_\textsc{bd}}{2}\,\dot\vphi^2+2H\dot\vphi+\frac{\der_\vphi U-U}{3+2\omega_\textsc{bd}}-\frac{2+\omega_\textsc{bd}\left(1+w_m\right)}{3+2\omega_\textsc{bd}}\,e^{-\vphi}\rho_m,\nonumber\\
&&\ddot\vphi+3H\dot\vphi+\dot\vphi^2=2\frac{U-\der_\vphi U}{3+2\omega_\textsc{bd}}+\frac{1-3w_m}{3+2\omega_\textsc{bd}}\,e^{-\vphi}\rho_m,\nonumber\\
&&\dot\rho_m+3H\left(w_m+1\right)\rho_m=0,\label{efe}\eea where, as before, $H\equiv\dot a/a$ is the Hubble parameter. Since we are interested here in the asymptotic dynamics of the theory, we shall apply the dynamical systems tools in order to get related useful information. A very compact and basic introduction to the application of the dynamical systems in cosmological settings with scalar fields can be found in the references \citen{copeland_rev_ijmpd_2006, copeland-wands-dsyst, coley, luis, bohmer-rev, ejp-2015}.


\subsection{Dynamical systems}\label{subsect-dsyst}

Usually, when one deals with the asymptotic dynamics of BD cosmological models it is customary to choose the following variables of the phase space\cite{gregory_ann_phys_1997, holden_wands_cqg_1998, hrycyna, hrycyna-1, hrycyna-2}:

\bea x\equiv\frac{\dot\vphi}{\sqrt{6}H}=\frac{\vphi'}{\sqrt 6},\;y\equiv\frac{\sqrt{U}}{\sqrt{3}H},\;\xi\equiv 1-\frac{\der_\vphi U}{U},\label{vars}\eea where the tilde means derivative with respect to the variable $\tau\equiv\ln a$ -- the number of e-foldings. As a matter of fact $x$ and $y$ in Eq. \eqref{vars}, are the same variables which are usually considered in similar dynamical systems studies of FRW cosmology, within the frame of Einstein's general relativity with a scalar field matter source\cite{copeland-wands-dsyst}. In terms of the above variables the Friedmann constraint in Eq. \eqref{efe} can be written as

\bea \Omega^\text{eff}_m\equiv\frac{e^{-\vphi}\rho_m}{3H^2}=1+\sqrt{6}x-\omega_\textsc{bd}\,x^2-y^2\geq 0.\label{friedmann-c}\eea Notice that one might define a dimensionless potential energy density and an ``effective kinetic'' energy density

\bea \Omega_U=\frac{U}{3H^2}=y^2,\;\Omega^\text{eff}_K=x\left(\omega_\textsc{bd}x-\sqrt{6}\right),\label{omega-u-k}\eea respectively, so that the Friedmann constraint can be re-written in the following compact form: $$\Omega^\text{eff}_K+\Omega_U+\Omega^\text{eff}_m=1.$$ 

The definition for the dimensionless effective kinetic energy density $\Omega^\text{eff}_K$ has not the same meaning as in GR with a scalar field: It may be a negative quantity without challenging the known laws of physics. Besides, since there is not restriction on the sign of $\Omega^\text{eff}_K$, then, it might happen that $\Omega_U=U/3H^2>1$. This is due to the fact that the dilaton field in the BD theory is not a standard matter field but it is a part of the gravitational field itself. This effective (dimensionless) kinetic energy density vanishes whenever: $$x=\frac{\sqrt{6}}{\omega_\textsc{bd}}\;\Rightarrow\;\dot\vphi=\frac{6}{\omega_\textsc{bd}}\,H\;\Rightarrow\;\vphi=\frac{6}{\omega_\textsc{bd}}\,\ln a,$$ or if: $$x=0\;\Rightarrow\;\dot\vphi=0\;\Rightarrow\;\vphi=const.,$$ which, provided that the matter fluid is cold dark matter, corresponds to the GR-de Sitter universe, i. e., to the $\Lambda$CDM model.\cite{peebles, lcdm} The following are useful equations which relate $\dot H/H^2$ and $\ddot\vphi/H^2$ with the phase space variables $x$, $y$ and $\xi$:

\bea &&\frac{\dot H}{H^2}=2\sqrt{6}\,x-3\omega_\textsc{bd}\,x^2-\frac{3y^2\xi}{3+2\omega_\textsc{bd}}-\frac{2+\omega_\textsc{bd}\left(1+w_m\right)}{3+2\omega_\textsc{bd}}\,3\Omega^\text{eff}_m,\nonumber\\
&&\frac{\ddot\vphi}{H^2}=-3\sqrt{6}\,x-6x^2+\frac{6y^2\xi}{3+2\omega_\textsc{bd}}+\frac{1-3w_m}{3+2\omega_\textsc{bd}}\,3\Omega^\text{eff}_m.\label{useful}\eea

Our goal will be to write the resulting system of cosmological equations (\ref{efe}), in the form of a system of autonomous ordinary differential equations (ODE-s) in terms of the variables $x$, $y$, $\xi$, of some phase space. We have\cite{quiros-prd-2015-1}:

\bea &&x'=-3x\left(1+\sqrt{6}x-\omega_\textsc{bd}x^2\right)+\frac{x+\sqrt{2/3}}{1+2\omega_\textsc{bd}/3}y^2\xi+\frac{\frac{1-3w_m}{\sqrt 6}+\left[2+\omega_\textsc{bd}(1+w_m)\right]\,x}{1+2\omega_\textsc{bd}/3}\Omega^\text{eff}_m,\nonumber\\
&&y'=y\left[3x\left(\omega_\textsc{bd}x-\frac{\xi+3}{\sqrt{6}}\right)+\frac{y^2\xi}{1+2\omega_\textsc{bd}/3}+\frac{2+\omega_\textsc{bd}\left(1+w_m\right)}{1+2\omega_\textsc{bd}/3}\Omega^\text{eff}_m\right],\nonumber\\
&&\xi'=-\sqrt{6}x\left(1-\xi\right)^2\left(\Gamma-1\right),\label{asode}\eea where $\Omega^\text{eff}_m$ is given by Eq. (\ref{friedmann-c}), and it is assumed that $\Gamma\equiv U\der^2_\vphi U/(\der_\vphi U)^2$ can be written as a function of $\xi$ \cite{ejp-2015}: $\Gamma=\Gamma(\xi)$. Hence, the properties of the dynamical system (\ref{asode}) are highly dependent on the specific functional form of the potential $U=U(\vphi)$.


\subsubsection{The dynamical system for different self-interaction potentials}\label{subsect-dsyst-pots}

In this subsection we shall write the dynamical system (\ref{asode}) for a variety of self-interaction potentials of cosmological interest. It is worth noticing that the only information on the functional form of the self-interaction potential is encoded in the definition of the parameter $\Gamma$ in Eq. (\ref{asode}). Hence, what we need is to write the latter parameter as a concrete function of the coordinate $\xi$ for given potentials.

\begin{itemize}

\item{\it The exponential potential.}

\bea U(\vphi)=M^2\,e^{k\vphi},\label{exp}\eea which, in terms of the standard BD field $\phi$ (see Eq. (\ref{vphi})), amounts to the power-law potential $V(\phi)=M^2\phi^{k+1}$ in the action (\ref{bd-action}). In Eq. (\ref{exp}), $M^2$ and $k$ are free constant parameters. In this -- the most simple -- case $$\xi=1-\frac{\der_\vphi U}{U}=1-k,$$ is a constant, so that the system of ODE-s (\ref{asode}) reduces dimensionality from 3 to 2. The fact that, for the exponential potential $\Gamma=1$, is unimportant in this case since, as said, $\xi$ is not a variable but a constant. 
\bigskip
\item{\it The combination of exponentials.}

\bea U(\vphi)=M^2\,e^{k\vphi}+N^2\,e^{m\vphi},\label{comb-exp}\eea which corresponds to the BD potential $$V(\phi)=M^2\phi^{k+1}+N^2\phi^{m+1}$$ ($M^2$, $N^2$, $k$ and $m$ are free constants), leads to the following

\bea \Gamma(\xi)=(k+m)\frac{\left(1-\frac{mk}{k+m}-\xi\right)}{\left(1-\xi\right)^2}.\label{gamma-comb-exp}\eea As a consequence the third autonomous ODE in the dynamical system (\ref{asode}) can be written as

\bea \xi'=-\sqrt{6}\,x\,\left[k+m-mk-1-(k+m-2)\xi-\xi^2\right].\label{xi-ode-comb-exp}\eea The particular case when $M^2=N^2$, $m=-k$, corresponds to the cosh potential:

\bea U(\vphi)=2M^2\cosh(k\vphi),\label{cosh}\eea for which $\Gamma(\xi)=k^2/(1-\xi)^2$, and 

\bea \xi'=-\sqrt{6}\,x\,\left[k^2-(1-\xi)^2\right].\label{xi-ode-cosh}\eea
\bigskip
\item{\it The cosh-like potentials.}

\bea U(\vphi)=M^2\cosh^k(\mu\vphi),\label{cosh-like}\eea where $M^2$, $k$ and $\mu$ are constant parameters, are also very interesting from the point of view of the cosmology\cite{matos_urena_cqg_2000, cosh-pot}. These correspond to potentials of the following kind

\bea V(\phi)=M^2 \phi\left[\cosh(\ln\phi^\mu)\right]^k,\label{cosh-like-bd}\eea in terms of the original BD field $\phi$. We have $$\xi=1-\frac{\der_\vphi U}{U}=1-k\mu\tanh(\mu\vphi),$$ so that 

\bea \Gamma(\xi)=\frac{k^2\mu^2+(k-1)(1-\xi)^2}{k(1-\xi)^2}.\label{gamma-cosh-like}\eea The resulting autonomous ODE -- third equation in (\ref{asode}) -- reads

\bea \xi'=-\frac{\sqrt{6}}{k}\,x\left[k^2\mu^2-(1-\xi)^2\right].\label{xi-ode-cosh-like}\eea Notice that, by setting $k=1$ and then replacing $k\rightarrow\mu$ one recovers the ODE (\ref{xi-ode-cosh}) for the cosh potential (\ref{cosh}).

Working in a similar way with the sinh-like potential

\bea U(\vphi)=M^2\sinh^k(\mu\vphi),\label{sinh-like}\eea we obtain: $$\xi=1-k\mu\,\text{cotanh}(\mu\vphi),$$ and the same $$\Gamma(\xi)=\frac{k^2\mu^2+(k-1)(1-\xi)^2}{k(1-\xi)^2},$$ so that the corresponding autonomous ODE is the same Eq. (\ref{xi-ode-cosh-like}) as for the cosh-like potential. The difference resides in the range of the variable $\xi$. For the cosh-like potential one has:

\bea 1-k\mu\leq\xi\leq 1+k\mu\;(-\infty<\vphi<\infty),\label{xi-range-cosh-like}\eea while, for the sinh-like one $$1+k\mu\leq\xi<\infty,$$ when $-\infty<\vphi<0$, and $$-\infty<\xi\leq 1-k\mu,$$ if $0<\vphi<\infty$. Here we have assumed that both $k$ and $\mu$ are non-negative quantities ($k\geq 0$, $\mu\geq 0$).

\end{itemize}


\subsection{Vacuum Brans-Dicke cosmology}\label{subsect-bd-vac}

A significant simplification of the dynamical equations is achieved when matter degrees of freedom are not considered. In this case, since $$\Omega^\text{eff}_m=0\;\Rightarrow\;y^2=1+\sqrt{6}x-\omega_\textsc{bd}\,x^2,$$ then the system of ODE-s (\ref{asode}) simplifies to a plane-autonomous system of ODE-s:

\bea &&x'=\left(-3x+3\frac{x+\sqrt{2/3}}{3+2\omega_\textsc{bd}}\,\xi\right)\left(1+\sqrt{6}x-\omega_\textsc{bd}x^2\right),\nonumber\\
&&\xi'=-\sqrt{6}x\left(1-\xi\right)^2\left(\Gamma-1\right).\label{x-xi-ode-vac}\eea 

In the present case one has

\bea &&\Omega_U=\frac{U}{3H^2}=y^2=1+\sqrt{6}x-\omega_\textsc{bd}x^2,\nonumber\\
&&\Omega^\text{eff}_K=x\left(\omega_\textsc{bd}x-\sqrt{6}\right)\;\Rightarrow\;\Omega^\text{eff}_K+\Omega_U=1,\label{omegas-vac}\eea where we recall that the definition of the effective (dimensionless) kinetic energy density $\Omega^\text{eff}_K$, has not the same meaning as in GR with scalar field matter: it may be a negative quantity. Here we consider non-negative self-interaction potentials $U(\vphi)\geq 0$, so that the dimensionless potential energy density $\Omega_U=y^2$, is restricted to be always non-negative: $\Omega_U=1+\sqrt{6}x-\omega_\textsc{bd}x^2\geq 0$. Otherwise, $y^2<0$, and the phase-plane would be a complex plane. Besides, we shall be interested in expanding cosmological solutions exclusively ($H\geq 0$), so that $y\geq 0$. Because of this the variable $x$ is bounded to take values within the following interval:

\bea \alpha_-\leq x\leq\alpha_+,\;\alpha_\pm=\sqrt\frac{3}{2}\left(\frac{1\pm\sqrt{1+2\omega_\textsc{bd}/3}}{\omega_\textsc{bd}}\right).\label{x-bound}\eea This means that the phase space for the vacuum BD theory $\Psi_\text{vac}$ can be defined as follows: $\Psi_\text{vac}=\left\{(x,\xi):\;\alpha_-\leq x\leq\alpha_+\right\},$ where the bounds on the variable $\xi$ --  if any -- are set by the concrete form of the self-interaction potential (see below). Another useful quantity is the deceleration parameter 

\bea &&q=-1-\frac{\dot H}{H^2}=-1-2\sqrt{6}x+3\omega_\textsc{bd}x^2+\frac{3(1+\sqrt{6}x-\omega_\textsc{bd}x^2)\xi}{3+2\omega_\textsc{bd}}.\label{dec-par-zero-m}\eea

In accordance with the results of Refs. \citen{hrycyna, hrycyna-1, hrycyna-2}, there are found four dilatonic equilibrium points, $P_i:(x_i,\xi_i)$, in the phase space $\Psi_\text{vac}$ corresponding to the dynamical system (\ref{x-xi-ode-vac}), without the specification of the function $\Gamma(\xi)$.

\begin{itemize}

\item{\it GR-de Sitter phase.} This solution corresponds to the critical point: 

\bea &&(0,0)\;\Rightarrow\;x=0\;\Rightarrow\;\vphi=\vphi_0,\;\text{and}\;y^2=1\;\Rightarrow\;3H^2=U=const.,\nonumber\eea which corresponds to accelerated expansion $q=-1$. Given that, the eigenvalues of the linearization matrix around this point depend on the concrete form of the function $\Gamma(\xi)$, $$\lambda_{1,2}=-\frac{3}{2}\left(1\pm\sqrt{1+\frac{8(1-\Gamma)}{3(3+2\omega_\textsc{bd})}}\right),$$ at first sight it appears that nothing can be said about the stability of this solution until the functional form of the self-interaction potential is specified. Notice, however, that since $\xi=0$ at this equilibrium point, this means that $U(\vphi)\propto e^\vphi$, i. e., the function $\Gamma$ is completely specified: $\Gamma=1$. As a matter of fact, the eigenvalues of the linearization matrix around $(0,0)$ are: $\lambda_1=-3,\;\lambda_2=0,$ which means that $(0,0)$ is a non-hyperbolic point.

\bigskip
\item{\it BD-de Sitter critical point.} We found another de Sitter solution: $q=-1$ $\Rightarrow\;\dot H=0$, which is associated with scaling of the effective kinetic and potential energies of the dilaton:

\bea &&P:\left(\frac{1}{\sqrt{6}(1+\omega_\textsc{bd})},1\right)\;\Rightarrow\;\frac{\Omega^\text{eff}_K}{\Omega_U}=-\frac{6+5\omega_\textsc{bd}}{12+17\omega_\textsc{bd}+6\omega^2_\textsc{bd}},\nonumber\\
&&\lambda_1=-\frac{4+3\omega_\textsc{bd}}{1+\omega_\textsc{bd}},\;\lambda_2=0,\label{dil-scaling}\eea where, as before, $\lambda_1$ and $\lambda_2$ are the eigenvalues of the linearization matrix around the critical point. We call this as BD-de Sitter critical point to differentiate it from the GR-de Sitter point. 

\bigskip
\item{\it Stiff-dilaton solution.} The effective stiff-dilaton critical points ($\Omega^\text{eff}_K=1$):

\bea &&P_\pm:\left(\alpha_\pm,1\right)\;\Rightarrow\;q_\pm=2+\sqrt{6}\,\alpha_\pm,\nonumber\\
&&\lambda^\pm_1=6\left(1+\sqrt\frac{2}{3}\,\alpha_\pm\right),\;\lambda_2=0,\label{stiff-dil-vac}\eea are also found, where the $\alpha_\pm$ are defined in Eq. (\ref{x-bound}).

\end{itemize} 

In order to make clear what the difference is between the above de Sitter solutions, let us note that the Friedmann constraint (\ref{friedmann-c}), evaluated at the BD-de Sitter point above, can be written as $$e^{-\vphi}\rho_m=3H_0^2+\frac{6+5\omega_\textsc{bd}}{6(1+\omega_\textsc{bd})^2}\,3H_0^2-U_0,$$ i. e., $e^{-\vphi}\rho_m=const.$ This means that the weakening/strengthening of the effective gravitational coupling ($G_\text{eff}\propto e^{-\vphi}$) is accompanied by a compensating growing/decreasing property of the energy density of matter $\rho_m\propto e^\vphi$, which leads to an exponential rate o expansion $a(t)\propto e^{H_0 t}$. This is to be contrasted with the GR-de Sitter solution: $3H_0^2=U_0$ $\Rightarrow\;a(t)\propto e^{\sqrt{U_0/3}\,t}$, which is obtained only for vacuum, $\rho_\text{vac}=U_0$; $\rho_m=0$.

The conclusion in Refs. \citen{hrycyna, hrycyna-1, hrycyna-2} that the obtained critical points are quite independent of the form of the function $\Gamma$, is not accurate enough. For the GR-de Sitter point, for instance, $\xi=0$, which means that $$\xi=1-\frac{\der_\vphi U}{U}=0\;\Rightarrow\;U\propto e^\vphi,$$ forcing $\Gamma=1$. For the remaining equilibrium points, $\xi=1$ $\Rightarrow\;U=const$, and $\Gamma=$undefined. This means that the equilibrium points listed above exist only for specific self-interaction potentials, but not for arbitrary potentials. Hence, contrary to the related statements in Refs. \citen{hrycyna, hrycyna-1, hrycyna-2}, the above results are not as general as they seem to be.

Given that the above critical points are all non-hyperbolic,\footnote{When the critical point under scrutiny is a non-hyperbolic point the linear analysis is not enough to get useful information on the stability of the point. In this case other tools, such as the center manifold theorem are to be invoked \cite{classic-books, classic-books-1, classic-books-2, center, center2, centre, centre-1, yoe-genly}.} resulting in a lack of information on the corresponding asymptotic properties, here we focus in the exponential potential (\ref{exp}), which includes the particular case when $$k=1\;\Rightarrow\;\xi=0\;\Rightarrow\;U(\vphi)=M^2\exp\vphi\;\Rightarrow\;\Gamma=1,$$ and the cosmological constant case $$k=0\;\Rightarrow\;\xi=1\;\Rightarrow\;U=M^2,$$ with the hope to get more precise information on the stability properties of the corresponding equilibrium configurations. For completeness we shall consider also other potentials beyond the exponential one.


\subsubsection{Exponential potential}\label{subsect-exp-pot}

In this case, since $\xi=1-k$, is a constant, the plane-autonomous system of ODE-s (\ref{x-xi-ode-vac}) simplifies to a single autonomous ODE:

\bea &&x'=-\left(\frac{\left(k+2+2\omega_\textsc{bd}\right)x-\sqrt\frac{2}{3}(1-k)}{1+2\omega_\textsc{bd}/3}\right)\left(1+\sqrt{6}x-\omega_\textsc{bd}x^2\right).\label{x-ode-vac-exp}\eea The critical points of the latter dynamical system are:

\bea x_1=\frac{\sqrt{2/3}\,(1-k)}{k+2+2\omega_\textsc{bd}},\;x_\pm=\alpha_\pm,\label{vac-exp-c-points}\eea where the $\alpha_\pm$ are given by Eq. (\ref{x-bound}). Notice that, since $x_i\neq 0$ (but for $k=1$, in which case $x_1=0$ and $q=-1$), there are not critical points associated with constant $\vphi=\vphi_0$. This means that the de Sitter phase with $\dot\vphi=0$ ($\vphi=const$), $U(\vphi)=const.$, i. e., the one which occurs in GR and which stands at the heart of the $\Lambda$CDM model, does not arise in the general case when $k\neq 1$. 

Hence, only in the particular case of the exponential potential (\ref{exp}) with $k=1$ ($\xi=0$), which corresponds to the quadratic potential in terms of the original BD variables: $V(\phi)=M^2\phi^2$, the GR-de Sitter phase is a critical point of the dynamical system (\ref{x-ode-vac-exp}). In this latter case ($k=1$) the critical points are (see Eq. (\ref{vac-exp-c-points})): $x_1=0$, $x_\pm=\alpha_\pm$. Worth noticing that $x_1=0$ corresponds to the GR--de Sitter solution $3H^2=M^2\exp\vphi_0$, meanwhile, the $x_\pm=\alpha_\pm$, correspond to the stiff-fluid (kinetic energy) dominated phase: $\Omega^\text{eff}_K=1$. While in the former case the deceleration parameter $q=-1-\dot H/H^2=-1$, in the latter case it is found to be

\bea q=2+\sqrt{6}\,\alpha_+>0.\label{dec-p-stiff}\eea

For small (linear) perturbations $\epsilon=\epsilon(\tau)$ around the critical points: $x=x_i+\epsilon$, $\epsilon\ll 1$, one has that, around the de Sitter solution: $\epsilon'=-3\epsilon$ $\Rightarrow\;\epsilon(\tau)\propto\exp(-3\tau)$, so that it is an attractor solution. Meanwhile, around the stiff-matter solutions: $$\epsilon_\pm(\tau)\propto e^{3\left(2+\sqrt{6}\,\alpha_\pm\right)\tau},$$ so that, if assume non-negative $\omega_\textsc{bd}\geq 0$, the points $x_\pm$ are always past attractors (unstable equilibrium points) since $2+\sqrt{6}\,\alpha_->0$. For negative $\omega_\textsc{bd}<0$, these points are both past attractors whenever $\omega_\textsc{bd}<-3/2$. In this latter case, for $-3/2<\omega_\textsc{bd}<0$, the point $x_+$ is a past attractor, while the point $x_-$ is a future attractor instead.


\subsubsection{Constant potential}\label{subsect-c-pot}

The constant potential $U(\vphi)=M^2$ is a particular case of the exponential (\ref{exp}), when $k=0$ ($\xi=1$ $\Rightarrow\;U=const$). In this case the autonomous ODE (\ref{x-ode-vac-exp}) simplifies:

\bea x'=\left[\frac{\sqrt{2/3}-2(1+\omega_\textsc{bd})x}{3+2\omega_\textsc{bd}}\right]\left(1+\sqrt{6}x-\omega_\textsc{bd}x^2\right).\label{x-ode-vac-cc}\eea The critical points are:

\bea x_1=\frac{1}{\sqrt{6}(1+\omega_\textsc{bd})},\;x_\pm=\alpha_\pm.\label{c-points-cc}\eea In this case, 

\bea\left.\frac{\dot H}{H^2}=-\frac{3-\sqrt{6}\omega_\textsc{bd}x}{3+2\omega_\textsc{bd}}\left[1-\sqrt{6}(1+\omega_\textsc{bd})x\right]\;\Rightarrow\;\frac{\dot H}{H^2}\right|_{x_1}=0\;\Rightarrow\;H=H_0,\label{hdot}\eea so that the point $x_1$ corresponds to BD-de Sitter expansion ($q=-1$). At $x_1$ the effective kinetic and potential energies of the dilaton scale as $$\frac{\Omega^\text{eff}_K}{\Omega_U}=-\frac{6+5\omega_\textsc{bd}}{12+17\omega_\textsc{bd}+6\omega^2_\textsc{bd}},$$ where, as mentioned before, the minus sign is not problematic since $\Omega^\text{eff}_K$ is not the kinetic energy of an actual matter field. As already shown -- see the paragraph starting below Eq. (\ref{dil-scaling}) and ending above Eq. (\ref{stiff-dil-vac}) -- this point does not correspond to a $\Lambda$CDM phase of the cosmic evolution, since, unlike in the GR case, in the BD theory the effective gravitational coupling $G_\text{eff}\propto e^{-\vphi}$ is not a constant and, besides, the de Sitter solution $H=H_0$ is obtained in the presence of ordinary matter with energy density $\rho_m\propto G^{-1}_\text{eff}$. 

Given that under a small perturbation ($\epsilon\ll 1$) around $x_1$: $$\epsilon(\tau)\propto\exp\left(-\frac{4+3\omega_\textsc{bd}}{1+\omega_\textsc{bd}}\,\tau\right),$$ this is a stable equilibrium point (future attractor) if the BD parameter $\omega_\textsc{bd}\geq 0$. In case it were a negative quantity, instead, $x_1$ were a future attractor whenever $\omega_\textsc{bd}<-4/3$ and $-1<\omega_\textsc{bd}<0$.

The critical points $x_\pm$ in Eq. (\ref{c-points-cc}), correspond to kinetic energy-dominated phases, i. e., to stiff-matter solutions $\Omega^\text{eff}_K=1$, where $q=2+\sqrt{6}\,\alpha_+>0$, and, under a small perturbation $\epsilon'=\lambda_\pm\epsilon$, $$\lambda_\pm=6\left(1+\sqrt\frac{2}{3}\,\alpha_\pm\right),$$ so that, assuming non-negative $\omega_\textsc{bd}\geq 0$, the points $x_\pm$ are always unstable (source critical points). In the case when $\omega_\textsc{bd}<0$ is a negative quantity, the point $x_-$ is unstable if $\omega_\textsc{bd}<-4/3$ (the critical point $x_+$ is always unstable).


\subsubsection{Other potentials than the exponential}\label{subsect-other-pot}

The concrete form of the dynamical system \eqref{x-xi-ode-vac} depends crucially on the function $\Gamma(\xi)$. For a combination of exponentials, for instance, one has (see Eq. \eqref{gamma-comb-exp}):

\bea &&x'=\left(-3x+3\frac{x+\sqrt{2/3}}{3+2\omega_\textsc{bd}}\,\xi\right)\left(1+\sqrt{6}x-\omega_\textsc{bd}x^2\right),\nonumber\\
&&\xi'=-\sqrt{6}x\left[k+m-km-1-(k+m-2)\,\xi-\xi^2\right].\label{ode-vac-comb-exp}\eea 

In this case (assuming that $m>k$), since 

\bea \xi=\frac{1-k+(1-m)\left(\frac{N}{M}\right)^2 e^{(m-k)\vphi}}{1+\left(\frac{N}{M}\right)^2 e^{(m-k)\vphi}},\label{xi-comb-exp}\eea as $\vphi$ undergoes $-\infty<\vphi<\infty$ $\Rightarrow\;1-m\leq\xi\leq 1-k$. Hence, the phase space where to look for equilibrium points of the dynamical system (\ref{ode-vac-comb-exp}), is the bounded compact region of the phase plane $(x,\xi)$, given by $$\Psi^\text{c.exp}_\text{vac}=\left\{(x,\xi):\alpha_-\leq x\leq\alpha_+,\;1-m\leq\xi\leq 1-k\right\},$$ where, we recall, $\alpha_\pm=\sqrt{3/2}(1\pm\sqrt{1+2\omega_\textsc{bd}/3})/\omega_\textsc{bd}$ (see Eq. \eqref{x-bound}).

In the case of the cosh and sinh-like potentials, Eq. \eqref{cosh-like} and \eqref{sinh-like} respectively, one has:

\bea &&x'=\left(-3x+3\frac{x+\sqrt{2/3}}{3+2\omega_\textsc{bd}}\,\xi\right)\left(1+\sqrt{6}x-\omega_\textsc{bd}x^2\right),\nonumber\\
&&\xi'=-\frac{\sqrt{6}}{k}\,x\left(k^2\mu^2-1+2\xi-\xi^2\right).\label{ode-vac-cosh-sinh-like}\eea 

The difference between the cosh and the sinh-like potentials is in the phase space where to look for critical points of \eqref{ode-vac-cosh-sinh-like}. For the cosh-like potentials one has that the phase space is the following bounded and compact region of the phase plane $$\Psi^\text{cosh}_\text{vac}=\left\{(x,\xi):\alpha_-\leq x\leq\alpha_+,\;1-k\mu\leq\xi\leq 1+k\mu\right\},$$ while, for the sinh-like potentials the phase space is the unbounded region $\Psi^\text{sinh}_\text{vac}=\Psi^\text{sinh-}_\text{vac}\cup\Psi^\text{sinh+}_\text{vac}$, where

\bea &&\Psi^\text{sinh-}_\text{vac}=\left\{(x,\xi):\alpha_-\leq x\leq\alpha_+,\;1+k\mu\leq\xi<\infty\right\},\nonumber\\
&&\Psi^\text{sinh+}_\text{vac}=\left\{(x,\xi):\alpha_-\leq x\leq\alpha_+,\;-\infty<\xi\leq 1-k\mu\right\}.\nonumber\eea 

A distinctive feature of the dynamical systems (\ref{ode-vac-comb-exp}) and (\ref{ode-vac-cosh-sinh-like}), is that the GR-de Sitter critical point with $x=\xi=0$, $$P_\text{dS}:\left(0,0\right)\;\Rightarrow\;H=H_0,\;\vphi=\vphi_0,$$ is shared by all of them. However, as it will be shown in section \ref{subsect-no-lcdm-phase}, this does not mean that for potentials of the kinds (\ref{comb-exp}), (\ref{cosh-like}), and (\ref{sinh-like}), with arbitrary free parameters, the $\Lambda$CDM model is an equilibrium point of the corresponding dynamical system. As a matter of fact, only for those arrangements of the free parameters which allow that the given potential approaches to the exponential $U\propto\exp\vphi$ as an asymptote, the $\Lambda$CDM model is an equilibrium configuration of the corresponding dynamical system (see the discussion in section \ref{subsect-no-lcdm-phase}).


\subsection{Brans-Dicke cosmology with matter}\label{subsect-bd-matter}

Above we have investigated the dynamical properties of the vacuum BD cosmology in the phase space. Here we shall explore the case when the field equations are sourced by pressureless dust with $w_m=0$, and for exponential potentials \eqref{exp} only. Given that, in this case, $\xi=1-k$, is a constant, the relevant phase space is a region of the phase plane $(x,y)$. The corresponding autonomous system of ODE-s results in the plane-autonomous system consisting of the first two equations in \eqref{asode}:

\bea &&x'=-3x\left(1+\sqrt{6}x-\omega_\textsc{bd}x^2\right)+\frac{3(1-k)}{3+2\omega_\textsc{bd}}\left(x+\sqrt{2/3}\right)y^2\nonumber\\
&&\;\;\;\;\;\;\;\;\;\;\;\;\;\;\;\;\;\;\;\;\;\;\;\;\;\;\;\;\;\;\;\;\;\;\;\;\;\;\;\;\;\;\;\;\;\;\;\;\;+\frac{1+\sqrt{6}\left(2+\omega_\textsc{bd}\right)\,x}{\sqrt{6}\left(3+2\omega_\textsc{bd}\right)}\,3\Omega^\text{eff}_m,\nonumber\\
&&y'=y\left[3x\left(\omega_\textsc{bd}x-\frac{4-k}{\sqrt 6}\right)+\frac{3(1-k)}{3+2\omega_\textsc{bd}}\,y^2+\frac{2+\omega_\textsc{bd}}{3+2\omega_\textsc{bd}}\,3\Omega^\text{eff}_m\right],\label{asode-m}\eea which has physically meaningful equilibrium configurations only within the phase plane: $\Psi_\text{mat}=\left\{(x,y):\;\alpha_-\leq x\leq\alpha_+,\;0\leq y\leq\sqrt{1+\sqrt{6}x-\omega_\textsc{bd}x^2}\right\},$ where we have considered that $\Omega^\text{eff}_m\geq 0$ and $y\in R^+\cup 0$. The critical points of this dynamical system are:

\bea &&P_\text{stiff}:\left(\frac{1-\sqrt{1+2\omega_\textsc{bd}/3}}{\sqrt{2/3}\omega_\textsc{bd}},0\right)\;\Rightarrow\;\Omega^\text{eff}_m=0;\nonumber\\
&&P'_\text{stiff}:\left(\frac{1+\sqrt{1+2\omega_\textsc{bd}/3}}{\sqrt{2/3}\omega_\textsc{bd}},0\right)\;\Rightarrow\;\Omega^\text{eff}_m=0;\nonumber\\
&&P_\text{sc}:\left(\frac{1}{\sqrt{6}(1+\omega_\textsc{bd})},0\right)\;\Rightarrow\;\Omega^\text{eff}_m=\frac{12+17\omega_\textsc{bd}+6\omega^2_\textsc{bd}}{6(1+\omega_\textsc{bd})^2};\nonumber\\
&&P'_\text{sc}:\left(-\frac{\sqrt{3/2}}{k+1},\frac{\sqrt{k+4+3\omega_\textsc{bd}}}{\sqrt{2}(k+1)}\right)\;\Rightarrow\;\Omega^\text{eff}_m=\frac{2k^2-3k-8-6\omega_\textsc{bd}}{2(k+1)^2},\label{c-points}\eea and

\bea &&P_*:\left(-\frac{\sqrt{2/3}(k-1)}{k+2+2\omega_\textsc{bd}},\frac{\beta}{k+2+2\omega_\textsc{bd}}\right)\;\Rightarrow\nonumber\\
&&\;\;\;\;\Omega^\text{eff}_m=\frac{12-6k-6k^2+\left(7-2k-5k^2\right)\omega_\textsc{bd}}{2(k+2+2\omega_\textsc{bd})^2},\label{c-points'}\eea where, in the last critical point we have defined the parameter: $$\beta=\sqrt{1+2\omega_\textsc{bd}/3}\sqrt{8+6\omega_\textsc{bd}-k(k-2)}.$$ The equilibrium points $P_\text{stiff}$ and $P'_\text{stiff}$ represent stiff-fluid solutions, meanwhile the remaining points represent scaling between the energy density of the dilaton and the CDM. Let us to focus into two of the above critical points: $P'_\text{sc}$ and $P_*$. As it was for vacuum BD cosmology, the de Sitter critical point does not arise unless $k=1$. In this latter case ($k=1$), for the last equilibrium point in Eq. (\ref{c-points'}) one gets: $$P_*:\left(0,1\right),\;q=-1\;(H=H_0),\;\Omega^\text{eff}_m=0,\;\lambda_{1,2}=-3,$$ where $\lambda_1$ and $\lambda_2$ are the eigenvalues of the linearization matrix around $P_*:(0,1)$. This means that, as in the vacuum case, for the exponential potential $U(\vphi)\propto\exp\vphi$ the GR-de Sitter solution is an attractor of the dynamical system \eqref{asode-m}. For the scaling point $P'_\text{sc}$, the deceleration parameter is given by $$q=\frac{k-2}{2(k+1)},$$ so that, for $k=0$, which corresponds to the constant potential $U=U_0$, the BD-de Sitter solution is obtained $$q=-1\;\Rightarrow\;a(t)\propto e^{H_0 t},\;e^{-\vphi}\rho_m=const.$$ However, since $$\Omega_m=\frac{2k^2-3k-8-6\omega_\textsc{bd}}{2(k+1)^2},$$ at $k=0$, $\Omega^\text{eff}_m=-(4+3\omega_\textsc{bd})$, is a negative quantity, unless the Brans--Dicke coupling parameter falls into the very narrow interval $-3/2<\omega_\textsc{bd}\leq-4/3$. Hence, for $k=0$, but for $-1.5<\omega_\textsc{bd}\leq-1.33$, the point $P'_\text{sc}$ does not actually belong in the phase space $\Psi_\text{mat}$.


\subsection{(Non)emergence of the $\Lambda$CDM phase from the Brans-Dicke cosmology}\label{subsect-no-lcdm-phase}

This problem has been generously discussed before in the references \citen{damour_prl_1993, hrycyna}. The conclusion on the emergence of the $\Lambda$CDM cosmology starting from the Brans-Dicke theory, seems to be supported by the existence of a de Sitter phase, which was claimed to be independent on the concrete form of the self-interaction potential of the dilaton field in Refs. \citen{hrycyna, hrycyna-1}, although in Ref. \refcite{hrycyna-2} the same authors somewhat corrected their previous claim. As we have already seen, in general -- but for the exponential potential with the unit slope: $U(\vphi)\propto e^\vphi$ -- the $\Lambda$CDM model is not an attractor of the FRW-BD cosmology. However, we think that this very important subject needs to be discussed in more length. We want to make clear that the statement on the non-universality of the GR-de Sitter equilibrium point\cite{quiros-prd-2015-1}, does not forbids the possible existence of exact de Sitter solutions for several choices of the self-interaction potential (see, for instance, Ref. \refcite{odintsov-ref}). What the statement means is that, in case such solutions existed, these would not be generic solutions but very particular (unstable) solutions instead, which are unable to represent any sensible cosmological scenario. 

It is useful to notice that de Sitter solution arises whenever $$q=-1\;\Rightarrow\;\dot H=0\;\Rightarrow\;H=H_0\;\Rightarrow\;a(t)\propto e^{H_0 t}.$$ This condition can be achieved even if $x\neq 0$. However, only when $$x=0\;\Rightarrow\;\dot\vphi=0\;\Rightarrow\;\vphi=\vphi_0,$$ the de Sitter solution can lead to the $\Lambda$CDM model, where by $\Lambda$CDM model we understand the FRW cosmology within the frame of Einstein's GR, with a cosmological constant $\Lambda$ and cold dark matter as the sources of gravity. Actually, only if $\vphi=\vphi_0$, is a constant, the action \eqref{dbd-action} -- up to a meaningless factor of $1/2$ -- is transformed into the Einstein-Hilbert action plus a matter source: $$S=\frac{1}{8\pi G_N}\int d^4x\sqrt{|g|}\left\{R-2U_0\right\}+2\int d^4x\sqrt{|g|}{\cal L}_m,$$ where $e^{\vphi_0}=1/8\pi G_N$. When ${\cal L}_m$ is the Lagrangian of CDM, the latter action is the mathematical expression of what we call as the $\Lambda$CDM cosmological model. Below we shall discuss on the (non)universality of the $\Lambda$CDM equilibrium point. In order to find related useful clues, we shall discuss first on the simpler case of the vacuum BD cosmology and then on BD cosmology with CDM.

\begin{itemize}

\item{\it Vacuum FRW-BD cosmology.} In this case the de Sitter phase arises only if assume an exponential potential of the form $$U(\vphi)\propto\exp\vphi\;\Rightarrow\;V(\phi)=M^2\phi^2,$$ which means that $\xi=0$ and $\Gamma=1$, are both completely specified, or if $\xi=1$, i. e., if $$U(\vphi)=M^2\;\Rightarrow\;V(\phi)=M^2\phi.$$ As a matter of fact, as shown in section \ref{subsect-bd-vac}, for exponential potentials of the general form: $$U(\vphi)=M^2\ e^{k\vphi}\;\Rightarrow\;V(\phi)=M^2\phi^{k+1},$$ with $k\neq 1$ and $k\neq 0$, the de Sitter critical point does not exist. In other words, speaking in terms of the original BD variables: but for the quadratic and the lineal monomials, $V(\phi)\propto\phi^2$ and $V(\phi)\propto\phi$, respectively -- also for those potentials which approach to either $\phi^2$ or $\phi$ at the stable point of the potential -- the de Sitter solution is not an equilibrium point of the corresponding dynamical system. 

Even when de Sitter solution is a critical point of (\ref{x-ode-vac-cc}), its existence, by itself, does not warrant that the $\Lambda$CDM model is approached. As an illustration, let us choose the vacuum FRW-BD cosmology driven by a constant potential (see subsection \ref{subsect-c-pot}). In this case one of the equilibrium points of the dynamical system (\ref{x-ode-vac-cc}): $$x_1=1/\sqrt{6}(1+\omega_\textsc{bd})\neq 0,$$ corresponds to the de Sitter solution since $$q=-1\;\Rightarrow\;\frac{\dot H}{H^2}=0\;\Rightarrow\;H=H_0.$$ The tricky situation here is that, although the de Sitter solution ($H=H_0$) is a critical point of the dynamical system (\ref{x-ode-vac-cc}), the $\Lambda$CDM model is not mimicked. Actually, at $x_1$, 

\bea &&x=\frac{\dot\vphi}{\sqrt{6}\,H}=\frac{1}{\sqrt{6}(1+\omega_\textsc{bd})}\;\Rightarrow\nonumber\\
&&\dot\vphi=\frac{H_0}{1+\omega_\textsc{bd}}\;\Rightarrow\;\vphi(t)=\frac{H_0\,t}{1+\omega_\textsc{bd}}+\vphi_0,\nonumber\eea i. e., the scalar field evolves linearly with the cosmic time $t$. This point corresponds to BD theory and not to GR since, while in the latter the Newton's constant $G_N$ is a true constant, in the former the effective gravitational coupling (the one measured in Cavendish-like experiments) evolves with the cosmic time: $$G_\text{eff}=\frac{4+2\omega_\textsc{bd}}{3+2\omega_\textsc{bd}}\,e^{-\vphi}\;\Rightarrow\;\frac{\dot G_\text{eff}}{G_\text{eff}}=-\frac{H_0}{1+\omega_\textsc{bd}}.$$ Taking the Hubble time to be $t_0=13.817\times 10^9$ yr (as, for instance, in Ref. \refcite{hrycyna}), i. e., the present value of the Hubble constant $H_0=7.24\times 10^{-11}$ yr$^{-1}$, one gets 

\bea \frac{\dot G_\text{eff}}{G_\text{eff}}=-\frac{1}{1+\omega_\textsc{bd}}\,7.24\times 10^{-11}\,\text{yr}^{-1}.\label{estimate}\eea 

As a consequence of the above, if consider cosmological constraints on the variability of the gravitational constant\cite{uzan-rev}, for instance the ones in Ref. \refcite{cosmo}, which uses WMAP-5yr data combined with SDSS power spectrum data: $$-1.75\times 10^{-12}\,\text{yr}^{-1}<\frac{\dot G}{G}<1.05\times 10^{-12}\,\text{yr}^{-1},$$ or the ones derived in Ref. \refcite{cosmo-1}, where the dependence of the abundances of the D, $^3$He, $^4$He, and $^7$Li upon the variation of $G$ was analyzed: $$|\dot G/G|<9\times 10^{-13}\,\text{yr}^{-1},$$ from Eq. (\ref{estimate}) one obtains the following bounds on the value of the BD coupling constant: $$\omega_\textsc{bd}>40.37\;|\;\omega_\textsc{bd}<-69.95,\;\text{and}\;\omega_\textsc{bd}>79.44\;|\;\omega_\textsc{bd}<-81.44,$$ respectively. These constraints are in acceptable agreement with the estimates of Refs. \citen{bd-coupling, aquaviva} (see, also, Ref. \refcite{chiva}).

As seen in section \ref{subsect-other-pot}, for other potentials, such as the combination of exponentials (\ref{comb-exp}), the cosh (\ref{cosh-like}) and sinh-like (\ref{sinh-like}) potentials, the GR-de Sitter solution is a critical point of the corresponding dynamical system. However, do not get confused: the above statement is not true for any arrangement of the free constants. Take, for instance, the combination of exponentials. The GR-de Sitter point $x=\xi=0$ entails that (see Eq. (\ref{xi-comb-exp})), either $k=m=1$ $\Rightarrow\;\xi=0$, or, for $m=1$, arbitrary $k$, the point is asymptotically approached as $\vphi\rightarrow\infty$ if $k<1$. In the former case ($k=m=1$) the combination of exponentials $$U(\vphi)=M^2 e^{k\vphi}+N^2 e^{m\vphi},$$ coincides with the single exponential (\ref{exp}), $U(\vphi)=(M^2+N^2)\,e^\vphi$, while in the latter case ($m=1$, $k$ arbitrary), assuming that $k<1$, the above potential tends asymptotically ($\vphi\rightarrow\infty$) to the exponential $U(\vphi)\approx N^2 e^\vphi$. For the cosh and sinh-like potentials one has: 

\bea U(\vphi)=M^2\left(e^{\mu\vphi}\pm e^{-\mu\vphi}\right)^k,\label{cosh-sinh}\eea where the ``$+$'' sign is for the cosh potential, while the ``$-$'' sign is for the sinh potential, and the $2^{-k}$ has been absorbed in the constant factor $M^2$. On the other hand, one has the following relationships (see subsection \ref{subsect-dsyst-pots}): $$\xi=1-k\mu\frac{e^{\mu\vphi}-e^{-\mu\vphi}}{e^{\mu\vphi}+e^{-\mu\vphi}},\;\xi=1-k\mu\frac{e^{\mu\vphi}+e^{-\mu\vphi}}{e^{\mu\vphi}-e^{-\mu\vphi}},$$ where the left-hand equation is for the cosh-like potential, while the right-hand one is for the sinh-like potential. Since at the GR-de Sitter point: $x=\xi=0$, then, from the above equations it follows that this critical point exists for the cosh and sinh-like potentials only if $k\mu=1$, in which case the mentioned potentials \eqref{cosh-sinh}  asymptotically approach to the exponential as $\vphi\rightarrow\infty$: $$U(\vphi)\approx M^2\,e^{k\mu\vphi}=M^2\,e^\vphi.$$

\bigskip
\item{\it FRW-BD cosmology with matter.} In the case when we consider a matter source for the BD equations of motion, the existence of a de Sitter critical point with $x=0$ $\Rightarrow\;\dot\vphi=0$ can be associated with the $\Lambda$CDM model. The autonomous system of ODE-s that can be obtained out of the cosmological FRW-BD equations of motion when these are sourced by CDM, is given by Eq. (\ref{asode-m}). The critical points of this dynamical system are depicted in \eqref{c-points}, \eqref{c-points'}. Notice that only one of them: $$P_*:\left(-\frac{\sqrt{2/3}(k-1)}{k+2+2\omega_\textsc{bd}},\frac{\beta}{k+2+2\omega_\textsc{bd}}\right),$$ where $\beta=\sqrt{1+2\omega_\textsc{bd}/3}\sqrt{8+6\omega_\textsc{bd}-k(k-2)}$, can be associated with GR-de Sitter expansion (i. e., with what we know as the $\Lambda$CDM model) in the special case when $k=1$. In this latter case $P_*:(0,1)$. Since we are considering exponential potentials of the form in Eq. \eqref{exp}, then, the GR-de Sitter equilibrium configuration is associated, exclusively, with the potential $$\frac{\der_\vphi U}{U}=k=1\;\Rightarrow\;U(\vphi)\propto e^\vphi.$$

Although in section \ref{subsect-bd-matter} we have considered only exponential potentials in FRW-BD cosmology with background dust, it is clear that the result remains the same as for the vacuum case: Only for the exponential potential with unit slope: $U(\vphi)\propto\exp\vphi$, or for potentials that approach asymptotically to $\exp\vphi$, the GR-de Sitter solution is an equilibrium configuration of the corresponding dynamical system.

\end{itemize}


\subsubsection{Final comments on the non-universality of the GR-de Sitter attractor}

The finding that only for the exponential potential with unit slope $U(\vphi)\propto\exp\vphi$, or for potentials that approach asymptotically to $\exp\vphi$, the GR-de Sitter solution is a critical point of the dynamical system \eqref{asode}, is not surprising. This result can be easily understood if perform a conformal transformation to the Einstein frame of the BD theory: $\hat g_{\mu\nu}=\Omega^2g_{\mu\nu},$ $\sqrt{|\hat g|}=\Omega^4\sqrt{|g|},$ with $\Omega^2=e^\vphi$. In this case the Jordan frame Brans-Dicke action $$S=\int d^4x\sqrt{|g|}\,e^\vphi\left[R-\omega_\textsc{bd}(\der\vphi)^2-2U\right],$$ is mapped into the Einstein's frame one\cite{faraoni-book} (see section \ref{sect-cf}): $$S=\int d^4x\sqrt{|\hat g|}\left[\hat R-\left(\omega_\textsc{bd}+\frac{3}{2}\right)\left(\hat\der\vphi\right)^2-2\hat U\right].$$ It is seen from this latter action, that the EF-BD theory is just general relativity with a self-interacting scalar field with potential $\hat U=e^{-\vphi}U$. Hence, only the exponential potential $U(\vphi)=\Lambda\,e^\vphi$, leads to general relativity plus a scalar field with a constant potential. The general relativity de Sitter state with a constant scalar field is obviously a solution. It is possible to obtain other de Sitter solutions in the Jordan frame but in such a case we need a time dependent scalar field $\vphi=\vphi(t)$, to compensate the time dependence of the Hubble parameter in the Einstein frame, so this is not the GR limit. In order to better understand the statement above, let us write the EF motion equations which are derived from the second of the above actions, in terms of the FRW metric: 

\bea &&3\hat H^2=\frac{2\omega_\textsc{bd}+3}{4}\,\dot\vphi^2+e^{-\vphi}U,\nonumber\\
&&\dot{\hat H}=-\frac{2\omega_\textsc{bd}+3}{4}\,\dot\vphi^2,\nonumber\\
&&\ddot\vphi+3\hat H\dot\vphi=\frac{2e^{-\vphi}\left(U-\der_\vphi U\right)}{2\omega_\textsc{bd}+3},\label{ef-efe}\eea respectively. Besides, the JF and the EF Hubble parameters are related by the following equation:

\bea \hat H=\frac{1}{2}\,\dot\vphi+H,\label{hubble-rel}\eea where we took into account the conformal transformation of the scale factor  $\hat a=\Omega\,a$. Notice from the second equation in (\ref{ef-efe}), that the only possibility to obtain a de Sitter solution in the Einstein's frame is that $\vphi$ be a constant ($\dot\vphi=0$). But, then, from the third equation in (\ref{ef-efe}), it follows that the self-interaction potential should be the exponential: $$U-\der_\vphi U=0\;\Rightarrow\;U=\Lambda\,e^\vphi,$$ where $\Lambda$ is an integration constant. Hence, the Friedmann equation in (\ref{ef-efe}) reads: $3\hat H^2=\Lambda$, and since $\dot\vphi=0$, the relationship (\ref{hubble-rel}) implies that in the Jordan frame we will have also a GR-de Sitter solution $H=\hat H=\sqrt{\Lambda/3}$. From the relationship (\ref{hubble-rel}) it follows, besides, that there can be other de Sitter solutions in the Jordan frame ($H=H_0$), that would require an evolving scalar field $\vphi=\vphi(t)$ which compensates the time evolution of the EF Hubble parameter: $$H_0=\hat H(t)-\frac{1}{2}\,\dot\vphi(t).$$ The JF-de Sitter solution would not be a general relativity solution since the effective gravitational coupling in the Jordan frame: $G_\text{eff}(t)\propto\exp{[-\vphi(t)]}$, would be an evolving quantity. For further details and a more exhaustive discussion on this subject see \refcite{quiros-prd-2015-1}.

The result discussed above is not exclusive of the Brans-Dicke theory. A similar result on the non-universality of the GR-de Sitter phase has been discussed in Ref. \refcite{barrow-shaw-cqg-2008}, where a STT containing a vacuum fluid and other subdominant matter stresses was investigated. It was shown that very specific conditions on the coupling function $\omega(\phi)$ are to be imposed in order for the given STT-s to have the GR-de Sitter limit. The existence of the GR-de Sitter limit has been demonstrated also in Ref. \refcite{carloni-cqg-2008} for STT-s with coupling $\propto\phi^2$ and with power-law potential: $V(\phi)=\lambda\phi^n$. Given the very specific form of the coupling function and the restricted kinds of potentials investigated in that reference, their result can be considered as another argument on the non-universality of the GR-de Sitter limit. For the study of the asymptotic dynamics of STT-s we recommend Ref. \refcite{billyard_coley_prd_1999}, where the qualitative properties of cosmological models is investigated by exploiting the formal equivalence of these theories with general relativity minimally coupled to a scalar field under a conformal transformation and field redefinition. In particular, the asymptotic behavior of spatially homogeneous cosmological models in a class of STT-s which are conformally equivalent to general relativistic Bianchi cosmologies with a scalar field and an exponential potential, was studied. Particular attention was paid to self-similar scalar-tensor cosmological models.


\subsection{Dynamics of Horndeski theories: the cubic Galileon case}\label{subsect-gal}

Inspired by the DGP model, in Ref. \refcite{nicolis_gal} it was proposed an infrared modification of gravity which is a generalization of the 4D effective theory in the DGP braneworld. The theory is invariant under the Galilean shift symmetry $\der_\mu\phi\rightarrow\der_\mu\phi+b_\mu$ in the Minkowski space-time, which keeps the equations of motion at second order. The scalar field that respects the Galilean symmetry is dubbed ``Galileon''. The model has a self-accelerating de Sitter solution with no ghost-like instability. The analysis in Ref. \refcite{nicolis_gal} is valid only for weak gravity in flat spacetime, so that the above result must change in the covariant version of the model\cite{deffayet_vikman_gal, deffayet_deser_gal, deffayet-rev, chow_gal, deffayet_prd_2011, fab_4_prl_2012, tsujikawa_lect_not}. As a matter of fact, in the covariantized theory of the Galileon the shift symmetry is not preserved, however, the equations of motion still are second order, which is primordial since the higher-derivative theories are in general plagued by the so called Ostrogradsky instability\cite{woodard_2007, ostro-theor}. Newton's gravity at short distances is recovered thanks to the Vainshtein mechanism (see subsection \ref{subsect-vain}) that is based on nonlinear field self-interactions such as $\Box\phi(\der\phi)^2$. This nonlinear effect has been employed for the brane-bending mode of the self-accelerating branch in the DGP braneworld. Galileons belong in the class of Horndeski theories (see section \ref{sect-horn}).

Here we focus in a modification of the BD theory where a piece of action containing derivatives of the BD field higher than the first one is considered. This modification is the basis of the so called ``Brans-Dicke Galileon'', formerly studied in Ref. \refcite{kazuya_gal} (see also Refs. \citen{japan_gal, chow_gal, also_gal, also_gal_1, also_gal_2}). It was demonstrated the existence of self-accelerating universe with no ghost-like instabilities on small scales if the Galileon is a BD scalar field $\phi$ with a cubic self-interaction term\cite{kazuya_gal}. The action for this model is given by:

\bea S_\text{BD}^\text{cubic}=\int d^4x\sqrt{|g|}\left[\phi R-\frac{\omega_\text{BD}}{\phi}(\der\phi)^2-2V(\phi)+\alpha^2\Box\phi\left(\frac{\der\phi}{\phi}\right)^2+2{\cal L}_m\right],\label{kazuya-action}\eea where the BD scalar field $\phi$ stands as the Galileon field, and the BD parameter $\omega_\text{BD}$ and $\alpha^2$ -- the strength of the cubic self-interaction -- are free constants. As before ${\cal L}_m$ is the Lagrangian density of the matter degrees of freedom other than the Galileon itself. The cubic interaction $\propto\Box\phi(\der\phi)^2/\phi^2$, is the unique form of interactions at cubic order that keeps the field equation for the Galileon $\phi$ of second-order \cite{nicolis_gal}. The existence of the self-accelerating universe requires a negative BD parameter $\omega_\text{BD}<0$, but, thanks to the non-linear term, small fluctuations around the solution are stable on small scales. General relativity is recovered at early times and on small scales by the cubic interaction via the Vainshtein mechanism. At late time, gravity is strongly modified and the background cosmology shows a phantom-like behaviour\cite{kazuya_gal}. As we explain below in subsection \ref{subsect-vain}, the Vainshtein mechanism is a screening mechanism due to the non-linearity of the term containing the derivatives of the scalar field. For distances from the source much smaller than the Vainshtein radius $r_\text{V}$, which depends on the source and on the parameters of the theory, the gravitational effects of the scalar field are hidden via the non-linear self-interaction so that the resulting theory is indistinguishable from general relativity\cite{chow_gal, vainsh_rev}. The influence of the scalar field becomes important only at large scales, e. g. for cosmology.

In order to perform the dynamical analysis we focus in the cubic Galileon model where the Galileon is minimally coupled to the curvature. This model is quite simpler than the BD cubic Galileon. The model of interest has the following choice ($G_5=0$):

\bea K=X-V(\phi),\;G_3=\sigma X,\;G_4=\frac{1}{2},\label{model}\eea where $\sigma=\sigma(\phi)$ is a coupling function. The resulting action reads:

\bea S=\int d^4x\frac{\sqrt{-g}}{2}\left\{R-\left[1+\sigma\Box\phi\right](\der\phi)^2-2V(\phi)\right\},\label{qbic-gal-action}\eea where the matter action piece $S_m=\int d^4x\sqrt{-g}\,{\cal L}_m$ (${\cal L}_m$ stands for the matter Lagrangian), has been omitted for simplicity but, if desired, it may be added. We shall discuss on the importance of the matter coupling in the theory \eqref{qbic-gal-action}: Thanks to the highly non-linear character of this theory, it is expected that the matter coupling sets constraints on the vacuum degrees of freedom so that the number of dynamical variables should be different from the pure vacuum case. The results we discuss here are also applicable to other modified theories of gravity. 

Given the complex form of the generalized Galileon field equations which are obtained by means of the variational principle from \eqref{qbic-gal-action}, deriving of exact cosmological solutions is by far a mammoth task. This is where the tools of the dynamical systems theory come into scene. Although the phase space dynamics of the class of models specified by the choice \eqref{model} has been investigated in detail in Ref. \refcite{genly_saridakis_jcap_2013} for a pair of choices of the coupling function $\sigma=\sigma(\phi)$ and of the potential $V(\phi)$, here we want to pay special attention to a particular case that was not investigated in that reference: the Galileon vacuum cosmology. For the more general situation when, in addition to the Galileon, the background matter is considered, in Ref. \refcite{genly_saridakis_jcap_2013} it was found that there are not any new stable late-time solutions apart from those of standard quintessence. In consequence, if one forgets about the particular kind of coupling set by the cubic term above, one may naively expect that the same result should hold true for the particular case when the standard matter degrees of freedom are removed. In agreement with our intuition, the results we shall discuss here will show quite the contrary: there is a very interesting asymptotic dynamics in the vacuum of the generalized Galileon cosmological models, which strongly departs from the asymptotic structure of standard quintessence even at late-time.

A FRW spacetime with flat spatial sections \eqref{frw-metric} is assumed. The cosmological field equations resulting from the action \eqref{qbic-gal-action}, read:

\bea &&3H^2=\rho_m+\rho_\phi,\;-2\dot H=\rho_m+p_m+\rho_\phi+p_\phi,\nonumber\\
&&\left(1+2\sigma_{,\phi}\dot\phi^2-6\sigma H\dot\phi\right)\ddot\phi+3H\dot\phi+\left(\frac{1}{2}\sigma_{,\phi\phi}\dot\phi^2-3\sigma\dot H-9\sigma H^2\right)\dot\phi^2=-V_{,\phi},\nonumber\eea where, in addition to the Galileon, a standard matter fluid with energy density $\rho_m$ and barotrotopic pressure $p_m$, is assumed. The energy density and the parametric pressure of the Galileon field are given by

\bea \rho_\phi=\frac{\dot\phi^2}{2}\left(1+\sigma_{,\phi}\dot\phi^2-6\sigma H\dot\phi\right)+V,\;p_\phi=\frac{\dot\phi^2}{2}\left(1+\sigma_{,\phi}\dot\phi^2+2\sigma\ddot\phi\right)-V.\nonumber\eea Here, for simplicity of the analysis, we choose the constant Galileon coupling case with the exponential potential:

\bea \sigma=\sigma_0\;\Rightarrow\;\sigma_{,\phi\phi}=\sigma_{,\phi}=0,\;V(\phi)=V_0\,e^{-\lambda\phi}.\label{case}\eea For definiteness we shall assume non-negative $\sigma_0\geq 0$, which is the more interesting choice (for $\sigma_0<0$, the asymptotic dynamics results in a straightforward particular case of Galileon cosmology with background matter). Under the above assumptions the cosmological Einstein's field equations read:

\bea 3H^2=\rho_m+\rho_\phi,\;-2\dot H=\rho_m+p_m+\rho_\phi+p_\phi,\label{qbic-feqs}\eea while the motion equation of the Galileon is depicted by:

\bea \left(1-6\sigma_0H\dot\phi\right)\ddot\phi+3H\dot\phi-3\sigma_0H^2\left(3+\frac{\dot H}{H^2}\right)\dot\phi^2=-V_{,\phi}.\label{qbic-kg-eq}\eea In the above equations:

\bea \rho_\phi=\frac{\dot\phi^2}{2}\left(1-6\sigma_0H\dot\phi\right)+V,\;p_\phi=\frac{\dot\phi^2}{2}\left(1+2\sigma_0\ddot\phi\right)-V.\label{qbic-rho-p}\eea Equations \eqref{qbic-feqs}, \eqref{qbic-kg-eq}, \eqref{qbic-rho-p}, are the master equations of the model.


\subsubsection{The variables of the phase space}

Our aim will be to trade the complex system of second order equations \eqref{qbic-feqs}, \eqref{qbic-kg-eq}, \eqref{qbic-rho-p}, by a system of autonomous ordinary differential equations (ODE-s). For this purpose one has to choose adequate variables of some state space. To start with one chooses the following standard, Hubble-normalized variables of the phase space\cite{copeland-wands-dsyst}:

\begin{align} x_s=\frac{\dot\phi}{\sqrt{6} H},\;y_s=\frac{\sqrt V}{\sqrt{3}H}.\label{qbic-xy-var}\end{align} In terms of these variables the Friedmann equation in (\ref{qbic-feqs}) can be written as:

\begin{align} \Omega_m=1-x_s^2-y_s^2+6\sqrt{6}\,x^3_sH^2\sigma_0,\label{qbic-friedmann-eq}\end{align} where $\Omega_i:=\rho_i/3H^2$ is the dimensionless (normalized) energy density of the $i$-th matter component. As seen from Eq. \eqref{qbic-friedmann-eq}: i) one needs yet another phase space variable to account for the factor $H^2\sigma_0$, and ii) due to the positive sign of the fourth term in the right-hand side (RHS) of Eq. (\ref{qbic-friedmann-eq}), given $x_s\geq 0$, the variables $x_s$ and $y_s$ can take arbitrary large values, while $0\leq\Omega_m\leq 1$. In consequence, we introduce the following bounded new variables of the phase space\cite{quiros_cqg_2016}:

\begin{align} x_\pm=\frac{1}{x_s\pm1},\;y=\frac{1}{y_s+1},\;z=\frac{1}{H^2\sigma_0+1},\label{qbic-n-var}\end{align} where $x_+$ is for non-negative $x_s\geq 0$ ($\dot\phi\geq 0$), while $x_-$ is for non-positive $x_s\leq 0$ ($\dot\phi\leq 0$). Besides, $0\leq x_+\leq 1$ ($-1\leq x_-\leq 0$), $0\leq y\leq 1$, and $0\leq z\leq 1$. Here we are assuming that only expanding cosmologies arise: $H\geq 0$ ($y_s\geq 0$), and that along orbits of the phase space $x_s$ does not flip sign. These assumptions are not independent of each other. Actually, at a bounce, no matter whether it is a bounce at a minimum or at a maximum size of the universe, where $\dot a=0,\;\ddot a>0$ (minimum size), or $\dot a=0,\;\ddot a<0$ (maximum size universe), since $H$ flips sign, then, necessarily $y_s\propto\sqrt{V}/H$, flips sign as well. Notice that the bounce, if present, arises at the boundary $y_s=0$ since, while $H$ flips sign $\sqrt{V}$ does not. Besides, at the bounce, simultaneously, $\dot\phi\sim H\sim 0$ and $\sqrt{V}\sim H\sim 0$ since otherwise, if assume finite $\dot\phi$ and $V$, $$x_s=\frac{\dot\phi}{\sqrt{6}H}\rightarrow\infty,\;y_s=\frac{\sqrt V}{\sqrt{3}H}\rightarrow\infty.$$

The choice of coordinates in \eqref{qbic-n-var} is specially useful in those cases where $x_s=0$, and $y_s=0$ are invariant subspaces in the ($x_s,y_s$) -- phase space. This means that orbits originated from initial conditions, say, in the quadrant $x_s\geq 0$, $y_s\geq 0$, will entirely lay in that quadrant. I. e., the orbits will not cross none of the boundaries (it could be better to say separatrices), $x_s=0$ and $y_s=0$. This is, precisely, the case for the vacuum of the generalized Galileon model \eqref{qbic-feqs}, \eqref{qbic-kg-eq}, \eqref{qbic-rho-p}. The Friedmann constraint (\ref{qbic-friedmann-eq}) can be written as:

\begin{align} \Omega_m=1-x_s^2-y_s^2-2\sqrt\frac{2}{3}\,x^3_s Q.\label{qbic-friedmann-c}\end{align} Another useful quantity is defined by,

\begin{align} Q:=-9H^2\sigma_0=9\left(\frac{z-1}{z}\right)<0.\label{Q-def}\end{align}


\subsubsection{Dynamics of the generalized Galileon cosmology with matter}

In spite of the fact that this is a particular case of the more general situation investigated in Ref. \refcite{genly_saridakis_jcap_2013}, in order to illustrate our adopted procedure, we shall discuss on the generalized Galileon cosmology in the presence of a matter fluid with energy density $\rho_m$ and barotropic pressure $p_m$\cite{quiros_cqg_2016}. For simplicity, we set $p_m=0$, so we deal with background (pressureless) dust. As said, it will be adopted the exponential potential $V(\phi)=V_0\exp(-\lambda\phi)$, and the constant Galileon coupling $\sigma=\sigma_0$, will be assumed. It is possible to trade the cosmological field equations \eqref{qbic-feqs}, \eqref{qbic-kg-eq}, \eqref{qbic-rho-p}, by the following dynamical system given in terms of the bounded variables $x_\pm$, $y$, and $z$ in \eqref{qbic-n-var}:

\begin{align} &x'_\pm=-\frac{x^2_\pm}{\sqrt{6}}\,\eta_\pm+x_\pm(1\mp x_\pm)\gamma_\pm,\nonumber\\
&y'=y(1-y)\left[\sqrt\frac{3}{2}\,\lambda\left(\frac{1\mp x_\pm}{x_\pm}\right)+\gamma_\pm\right],\nonumber\\
&z'=-2z(1-z)\gamma_\pm,\label{qbic-mat-asode}\end{align} where, for compactness of writing, we have defined:

\begin{align} &\gamma_\pm:=\left[\frac{\dot H}{H^2}\right]_\pm=-3\left\{\frac{\frac{3}{2}\,x_\pm^2\Theta_\pm(1)+2(1\mp x_\pm)^4y^2Q^2}{y^2\left[3x_\pm^4+2\sqrt{6}x_\pm^3(1\mp x_\pm)Q+2(1\mp x_\pm)^4Q^2\right]}\right.\nonumber\\
&\left.\;\;\;\;\;\;\;\;\;\;\;\;\;\;\;\;\;\;\;\;\;\;\;\;\;\;\;\;\;\;\;\;+\frac{x_\pm(1\mp x_\pm)\left[\sqrt{6}\Theta_\pm(2)-\lambda x_\pm(1\mp x_\pm)(1-y)^2\right]Q}{y^2\left[3x_\pm^4+2\sqrt{6}x_\pm^3(1\mp x_\pm)Q+2(1\mp x_\pm)^4Q^2\right]}\right\},\nonumber\\
&\eta_\pm:=\left[\frac{\ddot\phi}{H^2}\right]_\pm=-9\left\{\frac{\sqrt{6} x_\pm^3(1\mp x_\pm)y^2-\lambda x_\pm^4(1-y)^2+(1\mp x_\pm)^2\Delta_\pm Q}{y^2\left[3x_\pm^4+2\sqrt{6}x_\pm^3(1\mp x_\pm)Q+2(1\mp x_\pm)^4Q^2\right]}\right\}.\label{qbic-def-s}\end{align} We have defined also the following functions:

\begin{align} \Theta_\pm(a):=a(1\mp x_\pm)^2y^2-x^2_\pm(1-2y),\;\Delta_\pm:=x_\pm^2(1-y)^2-(1\mp2 x_\pm)y^2.\label{defs}\end{align} In \eqref{qbic-mat-asode} the comma denotes the derivative $f'=H^{-1} \dot f$, while the '$\pm$' signs account for two different branches of the dynamical system (as a matter of fact, one has two different dynamical systems). In terms of the bounded variables we can write the deceleration parameter as: $q=-(1+\dot H/H^2)=-1-\gamma_\pm$.


\begin{table}[tph!]
\tbl{Critical points $P_{c_i}:\left(x_{c_i},y_{c_i},_{c_i}\right)$ of the dynamical systems \eqref{qbic-mat-asode}, together with their main properties: existence, stability, deceleration parameter $q$, dimensionless energy density of matter $\Omega_m$, and the equation of state of the Galileon: $\omega_\phi=p_\phi/\rho_\phi$.}
{\begin{tabular}{@{}cccccc@{}} \toprule
Crit. Point& Existence & Stability & $q$ & $\Omega_m$ & $\omega_\phi$ \\ \colrule
$P^\pm_1:(\pm 1,1,0)$& always & unstable & $\frac{1}{2}$  & $1$ & undef. \\
&& (num. invest.) &&&\\
$P^\pm_2:(\pm 1,1,1)$& '' & saddle & $\frac{1}{2}$& $1$ & $1$ \\
$P^\pm_3:(\pm 1/2,1,1)$& '' & saddle & $2$ & $0$ & $1$ \\
$P^\pm_4:\left(\frac{\sqrt{6}}{\lambda\pm\sqrt{6}},\frac{\sqrt{6}}{\sqrt{6-\lambda^2}+\sqrt{6}},1\right)$& $\lambda^2< 6$ & stable if $\lambda^2<3$ & $-1+\frac{\lambda^2}{2}$ & $0$ & $-1+\frac{\lambda^2}{3}$ \\
 & & saddle if $\lambda^2>3$ & & & \\
$P^\pm_5:\left(\frac{\pm 2\lambda}{2\lambda\pm\sqrt{6}},\frac{2\lambda}{2\lambda\pm\sqrt{6}},1\right)$& $\lambda^2>3$ & stable point & $\frac{1}{2}$ & $\frac{\lambda^2-3}{\lambda^2}$ & $0$ \\
 & & spiral if $\lambda^2>\frac{24}{7}$ & & & \\ \botrule
\end{tabular}\label{tab-qbic-1}}\end{table}


\begin{table}[tph!]
\tbl{Eigenvalues of the linearization matrices around the critical points of the dynamical system \eqref{qbic-mat-asode}. We have used the following parameter definition: $\alpha:=\sqrt{-7+24/\lambda^2}$.}
{\begin{tabular}{@{}cccc@{}} \toprule
Crit. Point& $\lambda_1$ & $\lambda_2$ & $\lambda_3$ \\ \colrule
$P^\pm_1:(\pm 1,1,0)$& undef. & undef. & undef. \\
$P^\pm_2:(\pm 1,1,1)$& $-\frac{3}{2}$ & $-3$ & $\frac{3}{2}$ \\
$P^\pm_3:(\pm 1/2,1,1)$& $3\mp\sqrt\frac{3}{2}\lambda$ & $3$ & $-6$ \\
$P^\pm_4:\left(\frac{\sqrt{6}}{\lambda\pm\sqrt{6}},\frac{\sqrt{6}}{\sqrt{6-\lambda^2}+\sqrt{6}},1\right)$& $-\lambda^2$ & $-3+\frac{\lambda^2}{2}$ & $-3+\lambda^2$ \\
$P^\pm_5:\left(\frac{\pm 2\lambda}{2\lambda\pm\sqrt{6}},\frac{2\lambda}{2\lambda\pm\sqrt{6}},1\right)$& $-3$ & $-\frac{3}{4}+\alpha$ & $-\frac{3}{4}-\alpha$ \\ \botrule
\end{tabular}\label{tab-qbic-1-1}}\end{table}

The critical points of the dynamical systems \eqref{qbic-mat-asode}, together with their main properties, are summarized in TAB. \ref{tab-qbic-1}, while the eigenvalues of the linearization matrix around each one of the equilibrium points are shown in TAB. \ref{tab-qbic-1-1}. These tables reflect the fact that, the late-time asymptotics of the present Galileon model does not differ too much from the standard quintessence. This is particularly true for the late-time dynamics (points $P_4^\pm$ and $P_5^\pm$). Notice that the bigbang solution is not a global past attractor but a local one. The remaining equilibrium points coincide with those found in TAB. 1 of Ref. \refcite{copeland-wands-dsyst}: 

\begin{itemize}

\item{\it The matter dominated solution.} This corresponds to the critical point, $P_2^\pm$, which is associated with a saddle critical point. 

\bigskip
\item{\it The stiff-matter solutions.} The equilibrium points, $P_3^\pm$,are correlated with a scalar field's kinetic energy-dominated universe. In the present case these are always saddle points, while in the standard quintessence case these can be the past attractors as well. 

\bigskip
\item{\it The scalar field dominated solution.} The critical point, $P_4^\pm$, represents scaling between the scalar field's kinetic and potential energy densities. This can be either a saddle or a late-time attractor as in the quintessence model. 

\bigskip
\item{\it The matter scaling solution.} This solution is associated with the critical points, $P_5^\pm$, representing a scaling between the scalar field energy density and the energy density of matter. Whenever it exists it is a late-time attractor. It is either a focus or an spiral equilibrium point.

\end{itemize} The above results are essentially the same obtained in Ref. \refcite{genly_saridakis_jcap_2013} by means of a bit different procedure.


\subsubsection{Asymptotic dynamics of the generalized Galileon vacuum}

Apparently, the simplest case we can deal with is when the cosmic background is the vacuum ($\Omega_m=0$). In such a case the Friedmann constraint \eqref{qbic-friedmann-c} amounts to a relationship between the variables $x_s$, $y_s$ and $z$: $$Q=\frac{1}{2}\sqrt\frac{3}{2}\left(\frac{1-x_s^2-y_s^2}{x_s^3}\right),$$ so that one of these variables, say $z$: $$z=\frac{6\sqrt{6}x_s^3}{6\sqrt{6}x_s^3+x_s^2+y_s^2-1},$$ is redundant, and one ends up with a plane-autonomous system of ODE:

\begin{align}
&x'_s=\frac{1}{\sqrt 6}\frac{\ddot\phi}{H^2}-x_s\frac{\dot H}{H^2},\nonumber\\
&y'_s=-y_s\left(\sqrt\frac{3}{2}\,\lambda x_s+\frac{\dot H}{H^2}\right),\label{asode-g0-exp}
\end{align}  where

\begin{align*}
&\frac{\dot H}{H^2}=-\frac{6(1-y_s^2)(1+x_s^2-y_s^2)-3(x_s^2+y_s^2-1)(1+x_s^2-y_s^2-\sqrt{2/3}\lambda x_sy_s^2)}{4(1-y_s^2)+(x_s^2+y_s^2-1)^2},\nonumber\\
&\frac{\ddot\phi}{H^2}=-\frac{3\sqrt{6}\,x_s(x_s^2+y_s^2-1)(1+x_s^2-y_s^2)-6\sqrt{6}\,x_s(1+x_s^2-y_s^2-\sqrt{2/3}\lambda x_sy_s^2)}{4(1-y_s^2)+(x_s^2+y_s^2-1)^2}.
\end{align*}

The structure of the dynamical system \eqref{asode-g0-exp} entails that the semi-infinite planes $\{(x_s,y_s):x_s>0,\,y_s\geq 0\}$ and $\{(x_s,y_s):x_s<0,\,y_s\geq 0\}$ are invariant subspaces as well as the vertical lines $x_s=0$, $y_s\geq 0$, and $y_s=0$. Hence, the orbits originated from initial conditions in the region $\Psi^-=\{(x_s,y_s):x_s<0,\;y_s\geq 0\},$ will lay entirely in this region. The same is true for orbits in the region $\Psi^+=\{(x_s,y_s):x_s>0,\;y_s\geq 0\}.$ Furthermore, the system \eqref{asode-g0-exp} is form-invariant under the coordinate change $(x_s,y_s,\lambda)\rightarrow (-x_s,y_s,-\lambda).$ Thus, for the relevant computations we may consider to investigate just the sector $x_s\geq 0, \lambda\geq 0$. The dynamics on the sector $x_s\leq 0,\lambda\leq 0$ will be the same.  

As before we shall seek for new bounded variables so that all of the possible equilibrium points are ``visible''. Given that the vertical line $x_s=0$, is a separatrix, one may investigate the dynamics in the invariant subspaces $\Psi^-$ and $\Psi^+$, separately. Accordingly, one may relay the bounded variables defined in \eqref{qbic-n-var}. The corresponding phase space where to look for equilibrium points: $\Phi_\text{whole}=\Phi^-\cup\Phi^+$, is the union of the following bounded planes: 

\begin{align} &\Phi^+=\{(x_+,y):0\leq x_+\leq 1,\;0\leq y\leq 1\},\nonumber\\
&\Phi^-=\{(x_-,y):-1\leq x_-\leq 0,\;0\leq y\leq 1\}.\label{phipm}\end{align}

In terms of the bounded variables defined in \eqref{qbic-n-var}, the following plane-autonomous dynamical system is obtained for the Galileon vacuum with constant coupling $\sigma=\sigma_0$, and exponential potential $V=V_0\exp(-\lambda\phi)$:

\begin{align}
&x'_\pm=-\frac{x_\pm^2}{\sqrt{6}}\left[\frac{\ddot\phi}{H^2}\right]_\pm+x_\pm(1\mp x_\pm)\left[\frac{\dot H}{H^2}\right]_\pm,\nonumber\\
&y'=y(1-y)\left\{\sqrt\frac{3}{2}\lambda\left(\frac{1\mp x_\pm}{x_\pm}\right)+\left[\frac{\dot H}{H^2}\right]_\pm\right\},\label{asode-nvar}
\end{align} where

\begin{align*}
&\left[\frac{\dot H}{H^2}\right]_\pm=-3\left\{\frac{3\Theta_\pm(-1/3)\Theta_\pm(1)-\sqrt{2/3}\lambda x_\pm(1\mp x_\pm)(1-y)^2\Theta_\pm(-1)}{4x^4_\pm y^2(2y-1)+\Theta^2_\pm(-1)}\right\},\\
&\left[\frac{\ddot\phi}{H^2}\right]_\pm=3\left\{\frac{\sqrt{6}(1\mp x_\pm)\Theta_\pm(1)\left[\Theta_\pm(-1)-2x^2_\pm y^2\right]+4\lambda x^3_\pm(1\mp x_\pm)^2y^2(1-y)^2}{x_\pm\left[4x^4_\pm y^2(2y-1)+\Theta^2_\pm(-1)\right]}\right\},
\end{align*} and we have used the definition of the function $\Theta_\pm(a)$ given in Eq. (\ref{defs}). Here the '+' and '-' signs refer to two different branches, so that we have in fact two different dynamical systems: i) the one expressed in terms of the variables $x_+$, $y$ and $z$, which corresponds to the case with $\dot\phi>0$, and ii) the other expressed through $x_-$, $y$ and $z$, which corresponds to the case with $\dot\phi<0$. 

In the present case, one of the variables, say, $z$, is expressed as a function of the remaining variables $x_\pm$ and $y$ through 

\begin{align} Q_\pm=9\left(\frac{z-1}{z}\right)=\frac{x_\pm\Theta_\pm(-1)}{2\sqrt{2/3}(1\mp x_\pm)^3y^2}.\label{rel-Q}\end{align} The whole phase plane for this case is the union of the subspaces $\Phi^+$ and $\Phi^-$ defined in \eqref{phipm}: $\Phi_\text{whole}=\Phi^-\cup\Phi^+$. Its boundaries are at the edges 

\begin{align} 
&B_1:=\{(x_-,0): -1\leq x_-\leq 0\}\cup \{(x_+,0): 0\leq x_+\leq 1\},\nonumber \\ 
& B_2:=\left\{(1,y): -\infty \leq  y \leq \infty\right\},\nonumber\\
&B_3:=\{(x_-,1): -1\leq x_-\leq 0\}\cup \{(x_+,1): 0\leq x_+\leq 1\},\nonumber \\
&B_4:=\left\{(-1,y): -\infty \leq  y \leq \infty\right\},\nonumber\end{align} where $z=0$ $\Leftrightarrow$ $\sigma_0H^2\rightarrow\infty$, which means that either there is a cosmological singularity there ($H\rightarrow\infty$), or the cubic derivative interaction is decoupled from the gravitational interactions ($\sigma_0\rightarrow\infty$).


\begin{table}[tph!]
\tbl{Critical points of the dynamical system \eqref{asode-nvar} and their basic properties: existence, stability and deceleration parameter $q$.}{\begin{tabular}{@{}cccccc@{}} \toprule
Crit. Point& Existence & Stability & $q$ \\ \colrule
$P^\pm_{1v}:(\pm 1,0)$& always & saddle & $8$ \\
$P^\pm_{2v}:(\pm 1,1)$& '' & unstable & $4/5$ \\
$P^\pm_{3v}:\left(\frac{\pm\lambda}{\lambda\mp 2\sqrt{6}},0\right)$& $\pm\lambda<0$ & stable & $-4$ \\
$P_{4v}:(0,1)$& always & saddle if $\pm\lambda\geq 0$ & $-4$ \\
& & unstable if $\pm\lambda<0$ & \\
& & (num. inv.) & \\
$P^\pm_{5v}:(\pm 1/2,1)$& '' & saddle if $\pm\lambda<\sqrt{6}$ & $2$ \\
 & & stable if $\pm\lambda>\sqrt{6}$  & \\
$P^\pm_{6v}:\left(\frac{\sqrt{6}}{\lambda\pm\sqrt{6}},\frac{\sqrt{6}}{\sqrt{6-\lambda^2}+\sqrt{6}}\right)$& $\lambda^2<6$ & stable  & $-1+\frac{\lambda^2}{2}$ \\
$P^\pm_{7v}:(\pm 1,1/2)$& always & stable (num. inv.)  & $-1$ \\ \botrule
\end{tabular}\label{tab-qbic-2}}\end{table}



\begin{table}[tph!]
\tbl{Eigenvalues of the linearization matrices around the critical points of the dynamical system \eqref{asode-nvar}.}
{\begin{tabular}{@{}ccc@{}} \toprule
Crit. Point& $\lambda_1$ & $\lambda_2$ \\ \colrule
$P^\pm_{1v}:(\pm 1,0)$& $12$ & $-9$ \\
$P^\pm_{2v}:(\pm 1,1)$& $6/5$ & $9/5$ \\
$P^\pm_{3v}:\left(\frac{\pm\lambda}{\lambda\mp 2\sqrt{6}},0\right)$& $-3$ & $-12$ \\
$P_{4v}:(0,1)$& undef. & $6$ \\
$P^\pm_{5v}:(\pm 1/2,1)$& $3\mp\sqrt\frac{3}{2}\lambda$ & $-6$ \\
$P^\pm_{6v}:\left(\frac{\sqrt{6}}{\lambda\pm\sqrt{6}},\frac{\sqrt{6}}{\sqrt{6-\lambda^2}+\sqrt{6}}\right)$& $-\lambda^2$ & $-3+\frac{\lambda^2}{2}$ \\ $P^\pm_{7v}:(\pm 1,1/2)$& $0$ & $-3$ \\  \botrule
\end{tabular}\label{tab-qbic-2-1}}\end{table}


Below we discuss on the most generic solutions, including the physically relevant critical points of the dynamical system \eqref{asode-nvar} and their stability properties. These are summarized in TAB. \ref{tab-qbic-2}. In order to judge about the stability of the equilibrium points, in TAB. \ref{tab-qbic-2-1} the eigenvalues of the linearization matrix around each one of these critical points are shown. 

\begin{itemize}

\item{\it Exponential quintessence.} The equilibrium points $P_{5v}^\pm$ and $P^\pm_{6v}$, for which $z=1$, are the usual critical points found in Ref. \refcite{copeland-wands-dsyst} for the exponential quintessence model, if consider, as we do in the present section, the vacuum case (no other matter degrees of freedom than the scalar field). The points $P_{5v}^\pm$ correspond to the stiff-matter solutions, which, in the exponential quintessence case, are unstable and are expected to be relevant only at early times \cite{copeland-wands-dsyst}. In the present case we obtain a bit different result which is due to the non-vanishing Galileon coupling $\sigma=\sigma_0$: the stiff-matter solution can be stable, i. e., it can be a late-time attractor (see TAB. \ref{tab-qbic-2-1}). This is achieved if either, $\lambda>\sqrt{6}$ ('+' branch), or $\lambda<-\sqrt{6}$ ('-' branch). For either $\lambda<\sqrt{6}$ ('+' branch) or $\lambda>-\sqrt{6}$ ('-' branch), the stiff-matter solution is a saddle point in the phase space, but it can not be a source point (past attractor) as it is in the standard exponential quintessence case if $|\lambda|<\sqrt{6}$. This apparently harmless departure from the standard stability properties of the stiff-matter solution arises because the equilibrium point: $x_\pm=\pm 1/2$, $y=1$ ($z=1$), or, in terms of the standard variables $x_s$, $y_s$ in Eq. (\ref{qbic-xy-var}) (the same variables used in \cite{copeland-wands-dsyst}): $x_s=\pm 1$, $y_s=0$, is approached asymptotically not only if $\dot\phi=\pm\sqrt{6}H$, $V=0$, $\sigma_0=0$, as in the quintessence case, but also if there is a perhaps very tiny residual non-vanishing Galileon coupling $\sigma_0\neq 0$ ($\sigma_0\ll 1$): $$\dot\phi\sim H\gg V,\;\sigma_0\ll 1/H^2,\;\sigma_0\neq 0.$$ Hence, provided that $|\lambda|>\sqrt{6}$, and that the above conditions are fulfilled, the stiff-matter solution is the global attractor, meaning that the final (stable) state of the cosmic evolution is the ultra-relativistic stiff-matter stage. This behavior has not analogue in the exponential quintessence model. The critical points $P^\pm_{6v}$ have the same properties as in the exponential quintessence model \cite{copeland-wands-dsyst}. These correspond to scaling of the kinetic and potential energies of the scalar field: $$\frac{\dot\phi^2}{2V}=\frac{\lambda^2}{6-\lambda^2}.$$ Whenever they exist, they are attractors.

\bigskip
\item{\it The de Sitter solution.} Another interesting property of the present Galileon model, formerly investigated in \refcite{genly_saridakis_jcap_2013}, is that the de Sitter solution (points $P_{7v}^\pm$ in TAB. \ref{tab-qbic-2}) is a critical point of \eqref{asode-nvar}). It is a local attractor. Worth noticing that the points $P_{5v}^\pm$ and $P_{6v}$, correspond to the points $A^\pm$ and $C$ in table 1 of Ref. \refcite{genly_saridakis_jcap_2013}, respectively. The parameter $z$ is undefined in this case since, if the de Sitter point is approached along the separatrices sep$^\pm$, then $z=1$, meanwhile, for other approaching directions $z=0$. The de Sitter solution does not arise in standard exponential quintessence, unless $\lambda=0$ (constant potential case), so that its existence for any $\lambda\neq 0$ is a genuine consequence of the Galileon coupling $\sigma_0\neq 0$. As long as the former critical point exists independent on the value of the parameter $\lambda$, one might incorrectly infer that for vanishing potential, i. e., in the limit $\lambda\rightarrow\infty$, the equilibrium point $P_{7v}^\pm$ could be associated with a self-accelerating solution as in Ref. \refcite{kazuya_gal}, i. e, a de Sitter solution in which cosmic acceleration arises even in the absence of matter and for vanishing potential: $$\rho_m=p_m=V(\phi)=\dot H=0.$$ In Ref. \refcite{kazuya_gal} this kind of solution has been investigated within the context of BD theory with the cubic derivative interaction $\propto f(\phi)(\der\phi)^2\Box\phi$, so that it were not that surprising if this critical point arose in the present case, where a similar cubic interaction is being considered. However, as we shall show, the self-accelerating solution can not arise in the present case. Actually, supposing that the conditions for a self-accelerating solution are fulfilled, i. e., assuming that $\rho_m=p_m=0$, and $\dot H=0$ $\Rightarrow$ $H=H_0$, the Friedmann equation in \eqref{qbic-feqs}, amounts to the following cubic algebraic equation in $\dot\phi$: $$9\sigma_0H_0\dot\phi^3-\dot\phi^2+9H_0^2=0.$$ Any real root $\dot\phi=r_0=$const. of this equation leads to $\ddot\phi=0$, hence, the Raychaudhuri equation $$-2\dot H=\rho_\phi+p_\phi=0\;\Rightarrow\;1-3\sigma_0H_0r_0=0.$$ Exactly the same result: $3\sigma_0H_0r_0=1$, is obtained from the Klein-Gordon equation in \eqref{qbic-kg-eq}. Now, if substitute this $r_0$ back into the cubic algebraic equation above, one gets that $H^4_0=-2/27\sigma_0^2$, which can not be satisfied by any reals $H_0$ and $\sigma_0$.

\bigskip
\item{\it The bigbang solution.} The points $P_{2v}^\pm:(\pm 1,1)$ should not be confounded neither with the point $P^\pm_2$ in TAB. \ref{tab-qbic-1} ($O_1$ in Ref. \refcite{genly_saridakis_jcap_2013}), nor with $P^\pm_1$ in that table. In terms of the standard variables $x_s$, $y_s$, $P_{2v}^\pm\Rightarrow (0,0)$, which, in the case of the exponential quintessence model explored in Ref. \refcite{copeland-wands-dsyst}, coincides with $P^\pm_2$ and represents the matter-dominated solution. However, in the vacuum case, since $$\Omega_\phi=\frac{\rho_\phi}{3H^2}=1,$$ at any time, it can not represent any matter-dominance. In fact, since for $P_{2v}^\pm$, $q=4/5$, then $$\frac{\dot H}{H^2}=-\frac{9}{5}\;\Rightarrow\;H=\frac{5/9}{t-t_0}\;\Rightarrow\;a(t)\propto(t-t_0)^{5/9},$$ i. e., $P_{2v}^\pm$ is associated with a solution with a pure Galileon bigbang singularity at some initial time $t_0$ (compare with the point $P^\pm_1$ in TAB. \ref{tab-qbic-1} which is associated with a matter-dominated bigbang instead). This unstable solution, which corresponds to a source critical point in the phase space, has not analogues in the standard exponential quintessence model. It can have importance only at early stages of the cosmic evolution.

\bigskip
\item{\it The phantom solution.} One of the most interesting findings of the present investigation is the solution which is associated with the critical point $P_{3v}^\pm$. It is a stable solution (a local attractor) and represents phantom behavior. In order to illustrate the latter statement let us to choose the $P_{3v}^+$ solution, which exists only for negative $\lambda<0$. Let us set $\lambda=-\kappa$, with $\kappa>0$. At $P_{3v}^+$ we have that 

\bea x_+=\frac{\kappa}{\kappa+2\sqrt{6}}\;\Rightarrow\;\dot\phi=\frac{12}{\kappa}\,H,\;y=0\;\Rightarrow\;\frac{\sqrt{V}}{\sqrt{3}H}\rightarrow\infty,\;z=0\;\Rightarrow\;\sigma_0H^2\rightarrow\infty.\nonumber\eea From the above equations it follows that $$\phi(a)=\frac{12}{\kappa}\,\ln a+\phi_0,$$ where $\phi_0$ is an arbitrary integration constant. Additionally, since for this critical point $q=-4$ (recall that for the vacuum case $\Omega_\phi=1$): $$\frac{\dot H}{H^2}=-\frac{3}{2}\left(1+\omega_\phi\right)=3\;\Rightarrow\;\omega_\phi=-3,$$ where, by definition $\omega_\phi:=p_\phi/\rho_\phi$, the associated solution is a super-accelerating one. For this case we have that $$H(t)=\frac{1}{3(t_f-t)}\;\Rightarrow\;a(t)=\frac{a_0}{(t_f-t)^{1/3}}\;(t\leq t_f),$$ where $-3t_f$ and $\ln a_0$ are arbitrary integration constants. Given that $V\propto\exp(\kappa\phi)\propto a^{12}\propto(t_f-t)^{-4}$, then, as $t\rightarrow t_f$ asymptotically: $$H^2\sigma_0\propto(t_f-t)^{-2}\rightarrow\infty,\;\frac{\sqrt V}{H}\propto(t_f-t)^{-1}\rightarrow\infty,$$ as required. We have, also, that $$\rho_\phi(t)=3H^2(t)=\dot H(t)=\frac{1}{3(t_f-t)^2}.$$ Besides, since at $P^\pm_{3v}$, $\dot\phi=12H/\kappa$, the Friedmann equation can be written as $$V=\left(3-\frac{\alpha^2}{2}\right)H^2+3\alpha^3\sigma_0H^4,$$ where $\alpha=12/\kappa$. As seen, the self-interaction Galileon potential $V(\phi)$ asymptotically approaches to $V\propto H^4$, as required by the consistency of the phantom solution. The phantom behavior is evident from the fact that the energy density of the Galileon grows up without bounds with $t$. As seen, given that $a(t)$, $H(t)$, $\dot H(t)$, and $\rho_\phi(t)$, all blow up at $t=t_f$, i. e., in a finite time into the future, a big rip singularity \cite{odintsov, ruth} is the inevitable fate of the cosmic evolution in the present case. The point $P_{4v}$ represents super-accelerating contraction of the universe as we shall see below. In contrast to $P_{3v}^\pm$, this solution has no impact in the late-time dynamics. In this case it is required a vanishing self-interaction potential $V=0$ $\Rightarrow\;y=1$, and a finite $\dot\phi\neq 0$, and that, asymptotically, $$H\rightarrow 0\;\Rightarrow\;x_\pm\rightarrow 0,\;z\rightarrow 1.$$ As a matter of fact $$q=-4\;\Rightarrow\;\frac{\dot H}{H^2}=3\;\Rightarrow\;H(t)=-\frac{1}{3(t-t_b)}\;(t\geq t_b),$$ where the integration constant has been set $C=-3t_b$. Asymptotically, as $t\rightarrow\infty$, $H\rightarrow 0$, as required, besides $a(t)\propto(t-t_b)^{-1/3}\rightarrow 0$. Although we restricted ourselves to consider expanding cosmologies only, this point belongs in the boundary of the phase space and so, in spite of the mentioned restriction, we have taken it into consideration in our analysis.\footnote{There is yet another pair of equilibrium points of the dynamical system corresponding to the generalized Galileon vacuum, which are not found in the more general case when, in addition to the Galileon field, there is standard (pressureless) matter in the cosmic background. These are the points $P^\pm_{1v}$ in TAB. \ref{tab-qbic-2}. They correspond to a super-decelerated pace of the cosmic expansion $$H\propto\frac{1}{9(t-t_0)}\;\Rightarrow\;a(t)\propto(t-t_0)^{1/9},$$ where $9t_0$ is an arbitrary integration constant. At these points: $$\dot\phi\ll H\ll\sqrt{V},\;1\ll \sqrt{\sigma_0} H\ll(\sqrt{\sigma_0}\dot\phi)^3.$$ The points $P^\pm_{1v}$ are saddle critical points, so that the corresponding pattern of cosmic expansion can be only a transient stage of the cosmological evolution. Besides, only for e very narrow set of initial conditions the corresponding phase plane orbits approach to $P^\pm_{1v}$.}

\end{itemize}

As shown, the asymptotic structure of the vacuum Galileon model, which is a particular case of the model with matter, is not as trivial as thought. In particular the Galileons can play an important role in determining the fate of the cosmic evolution\cite{quiros_cqg_2016}. This is to be contrasted with the result for the cubic Galileon with matter (see Ref. \refcite{genly_saridakis_jcap_2013}) that the Galileons will not have impact on the late-time evolution of the universe.

As we shall discuss in subsection \ref{subsect-cosmo-vain}, the above behavior is a consequence of a kind of cosmological version of the van Dam-Veltman-Zakharov (vDVZ) discontinuity: we can not get the whole phase dynamics of the cubic Galileon vacuum in the continuous limit $\Omega_m\rightarrow 0$ of the more general dynamical system \eqref{qbic-mat-asode} corresponding to the cubic Galileon with background matter (TAB. \ref{tab-qbic-1}). The vDVZ discontinuity can be avoided if assume that the cubic self-interactions of the Galileon are somehow screened by its interactions with the background matter so that, for instance, the phantom solution that may affect the late time cosmic dynamics of the Galileon vacuum is erased from the phase space. In consequence, in the presence of matter degrees of freedom (in addition to the Galileon) the late-time dynamics of the model may be essentially the same as for standard quintessence\cite{genly_saridakis_jcap_2013}.


\section{Causality and Laplacian instability}\label{sect-c2s}

The Horndeski theories have been applied with success to describe the cosmological evolution of our Universe in different contexts \cite{kazuya_gal, japan_gal, chow_gal, also_gal, also_gal_1, also_gal_2}. An interesting subset of the Horndeski theories is composed of the so called scalar-tensor theories with a non-minimal derivative (kinetic) coupling, in particular those where the kinetic coupling is to the Einstein's tensor \cite{sushkov, saridakis-sushkov, sushkov-a, k-coup-skugoreva, matsumoto, granda, gao, germani-prl, germani}: $\propto G^{\mu\nu}\der_\mu\phi\der_\nu\phi$. The latter theory is characterized by its relative mathematical simplicity when compared with other Horndeski theories and also by its ability to account for the early (transient) inflationary stage, since it is able to explain in a unique manner both a quasi-de Sitter phase and an exit from it without any fine-tuned potential \cite{sushkov}. 

The action for the typical theory with non-minimal derivative coupling of the scalar with the Einstein's tensor: $G_{\mu\nu}\equiv R_{\mu\nu}-g_{\mu\nu}R/2$, is given by \eqref{k-coup-action} or, by the more general action:

\bea &&S=\int d^4x\frac{\sqrt{|g|}}{2}\left[R-\left(\epsilon g^{\mu\nu}-\alpha G^{\mu\nu}\right)\der_\mu\phi\der_\nu\phi-2V(\phi)\right]+S_m,\label{action-dcoup}\eea where the coupling constant $\alpha$ is a real number. The parameter $\epsilon$ can take the following values: $\epsilon=+1$ (quintessence), $\epsilon=-1$ (phantom cosmology), and $\epsilon=0$ (referred here as ``pure derivative coupling''). In the above equation $S_m$ is the action of the matter degrees of freedom other than the scalar field. 

Theories of the type \eqref{action-dcoup} have been studied in different contexts. For instance, in Ref. \refcite{rinaldi} static, spherically symmetric solutions to the gravitational field equations derived from \eqref{action-dcoup} were explored and black hole solutions with a single regular horizon were found, and their thermodynamical properties were examined (see also Ref. \refcite{k-coup-japan}). Related work regarding asymptotically locally AdS and flat black holes can be found in Refs. \citen{anabalon, cisterna_prd_2014}, while in Refs. \citen{cisterna_prd_2015, cisterna_prd_2016} the authors constructed the first neutron stars based in \eqref{action-dcoup}. The obtained construction may -- in principle -- constrain in a phenomenological way the free parameters of the model. Cosmological scenarios based in theories with kinetic coupling with the Einstein's tensor have been studied in Ref. \refcite{saridakis-sushkov} in order to examine quintessence (and phantom) models of dark energy with zero and constant self-interaction potentials. It has been shown that, in general, the universe transits from one de Sitter solution to another, depending on the coupling parameter. A variety of behaviors -- including Big Bang and Big Crunch solutions, and also cosmological bounce -- reveals the capabilities of the corresponding cosmological model. A dynamical systems analysis of the derivative coupling model with the Higgs-like potential can be found in Ref. \refcite{matsumoto}, while a similar study for the exponential potential has been performed in Ref. \refcite{huang}. It was found that, for the quintessence case, the stable fixed points are the same with and without the non-minimal derivative coupling, while for the pure derivative coupling (no standard canonical kinetic term) only the de Sitter attractor exists and the dark matter solution is unstable. Cosmology based in \eqref{action-dcoup} has been also investigated in Ref. \refcite{jinno}. The latter paper points out the existence of the Laplacian instability in the theory with kinetic coupling of the scalar field with the Einstein's tensor in the context of reheating after inflation. Particle production after inflation in the model \eqref{action-dcoup} tensor has also been studied in Ref. \refcite{jinno-others} by the same authors (see also Ref. \refcite{ema-others}).

A central aspect of the theory \eqref{action-dcoup} was investigated in Ref. \refcite{gao}, where it was found that, in the pure derivative coupling case ($\epsilon=0$), the scalar field may play the role of both dark matter and dark energy. In this case, the effective equation of state (EOS) of the scalar field $\omega_\text{eff}$ can cross the phantom divide\cite{caldwell, vikman_prd_2005, crossing-odintsov, crossing-observ, crossing-observ-1, crossing-nesseris, crossing-perivo, crossing-hu, crossing-chimento, mohseni}: $\omega_\Lambda=-1$, but this can lead to the sound speed becoming superluminal as it crosses the divide, and so is physically forbidden. The possibility of the phantom divide crossing in the model is in itself a very interesting finding, however two results we find particularly interesting in this study: i) that the crossing of the phantom divide may be linked with superluminal sound speed, and ii) that the physical limits on the sound speed are used as a basic criterion for rejection of a given cosmological model. The fact that the physical bounds on the speed of propagation of the perturbations of the field is to be taken carefully and seriously when Horndeski-type theories are under investigation, was understood also by the authors of Ref. \refcite{sup-lum-gal-1}. In that reference it was shown that, when the Dirac-Born-Infeld (DBI) Galileon is considered as a local modification to gravity, such as in the Solar system, the existing stable solutions always exhibit superluminality, casting doubt on the existence of a standard Lorentz invariant UV completion of that theory. We want to mention that there exist alternative points of view on this issue. For instance, in Refs. \citen{vikman, sup-lum-gal-2} it is shown that k-essence and Galileon theories, respectively, satisfy an analogue of Hawking's chronology protection conjecture, an argument that can be extended to include Hordenski theories in general. However, there are strong arguments that contradict such kinds of non-orthodox points of view on causality (for more on this issue see Refs. \citen{sup-lum-other, sup-lum-other-1}). In this regard we recommend the clear and pedagogical discussion on this issue given in Ref. \refcite{ellis-roy}. 

It is well known that Horndeski theories all possess some configurations with a superluminal propagation, hence, it is very important to discuss on causality within the framework of these generalizations of scalar-tensor theories. Below we shall discuss this issue in a particular case of a Horndeski theory with kinetic coupling to the Einstein's tensor and then, in subsection \ref{speed-grav-w}, we discuss on the speed of sound for the propagating scalar mode and on the speed of the gravitational waves within Horndeski theories in general. This topic is central after the detection of gravitational waves from the neutron star-neutron star merger GW170817 and the simultaneous measurement of the gamma-ray burst GRB170817A\cite{ligo}.


\subsection{Theory with kinetic coupling to the Einstein's tensor}\label{subsect-k-coup}

Given that but for the Ref. \refcite{quiros_cqg_2018}, there does not exist in the bibliography a thorough discussion on the implications for cosmology of the physical bounds on the speed of sound in the theory with the kinetic coupling to the Einstein's tensor, in this section we shall discuss on the ``$\omega_\Lambda=-1$'' barrier crossing issue in the model \eqref{action-dcoup} by paying special attention to the physical bounds on the speed of sound squared $c_s^2$. These bounds are imposed by stability and causality, two fundamental principles of classical physical theories: The squared sound speed should be non-negative $c^2_s\geq 0$ since otherwise, the cosmological model will be classically unstable against small perturbations of the background energy density, usually called as Laplacian -- also gradient -- instability. Besides, causality arguments impose that the mentioned small perturbations of the background should propagate at most at the local speed of light $c^2_s\leq 1$. It is to be mentioned that in Ref. \refcite{gao} the subject was only partially investigated -- only connection of the phantom barrier crossing with superluminality of the scalar perturbations was established -- besides only the pure derivative coupling case $\epsilon=0$ was considered in that reference. The issue was also stated but not investigated in Refs. \citen{dent, matsumoto}.

In order to implement the numeric investigation we shall explore two specific potentials: the frequently encountered in cosmological applications exponential potential \cite{huang, exp-pot-ferreira, exp-pot-ferreira-1, copeland-wands-dsyst}: $V=V_0\exp(\lambda\phi)$ and, also, the power-law potential $V=V_0\phi^{2n}$\cite{pwl-pot-peebles, pwl-pot-ratra-peebles}. The exponential potential

\bea V=V_0\,e^{\lambda\phi}\;\Rightarrow\;V'=\lambda V,\label{exp-pot}\eea where $V_0$ and $\lambda$ are real constants ($V_0\geq 0$), can be found as well in higher-order or higher-dimensional gravity theories\cite{exp-pot-origin-1, exp-pot-origin-1-barrow, wands-cqg-1994}, and in string or Kaluza-Klein type models, where the moduli fields may have effective exponential potentials\cite{exp-pot-origin-2}. Exponential potentials can also arise due to nonperturbative effects such as gaugino condensation\cite{exp-pot-origin-3}. In the present model the exponential potential has been investigated in Ref. \refcite{huang}, where a dynamical systems analysis was performed. The conclusion of the authors was that the derivative coupling to the Einstein's tensor does not modify the phase space dynamics of the quintessence\cite{copeland-wands-dsyst}. The power-law potential

\bea V=V_0\phi^{2n}\;\Rightarrow\;V'=2n\,V_0^{1/2n}V^{1-1/2n},\label{pow-law-pot}\eea where $V_0$ is a non-negative constant and $n$ is a real parameter, is also frequently found in the cosmological applications\cite{pwl-pot-peebles, pwl-pot-ratra-peebles}. In the quintessence case the inverse-power law potential exhibits the tracker behavior, a very desirable property for the quintessence if one wants to avoid the cosmic coincidence problem\cite{pwl-pot-other, pwl-pot-other-1, pwl-pot-other-2, pwl-pot-other-3}. The origin of this potential might be associated with supersymmetry considerations\cite{pwl-pot-origin, pwl-pot-origin-1}.


\begin{figure}
\includegraphics[width=4cm]{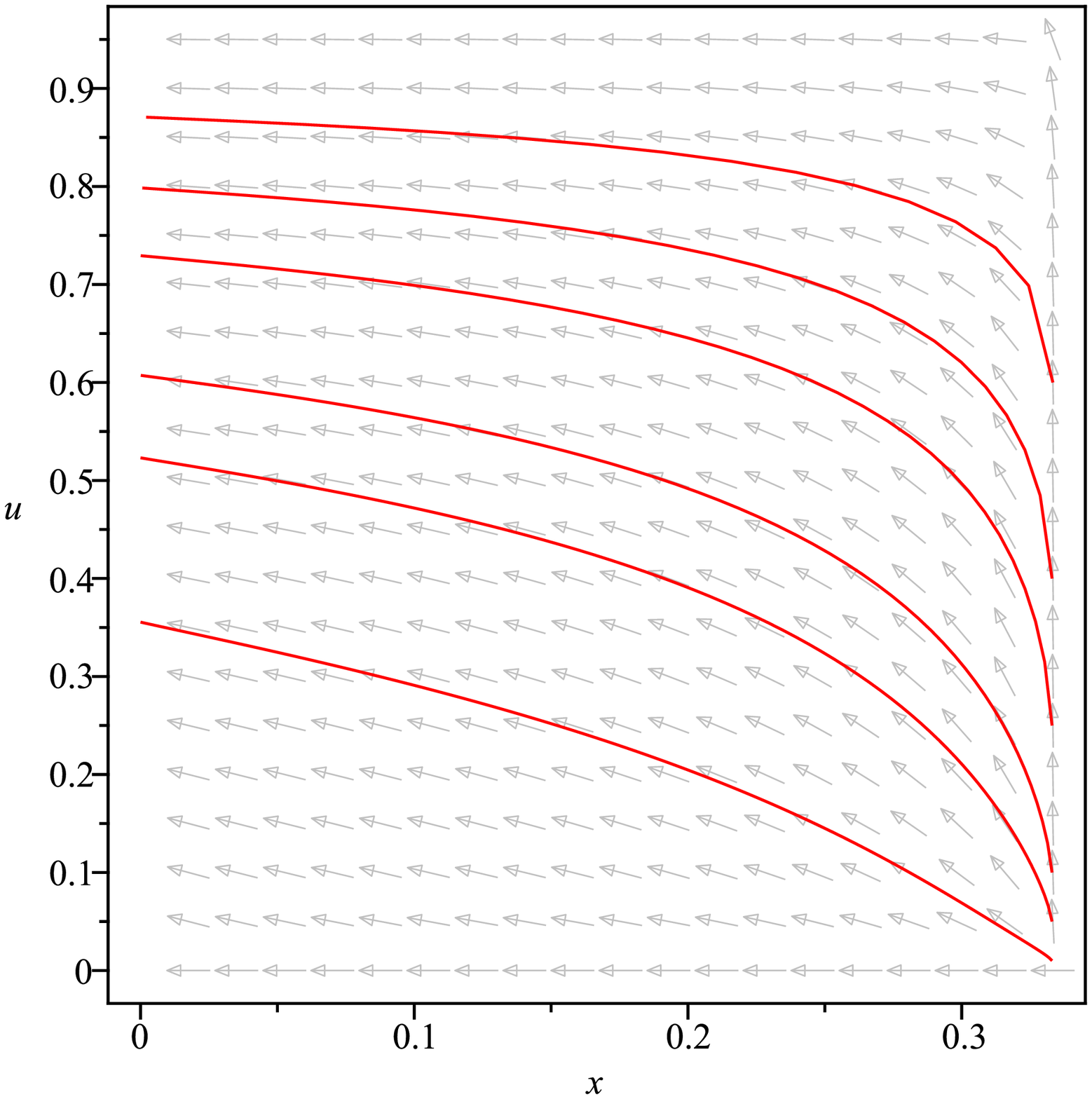}
\includegraphics[width=4.2cm]{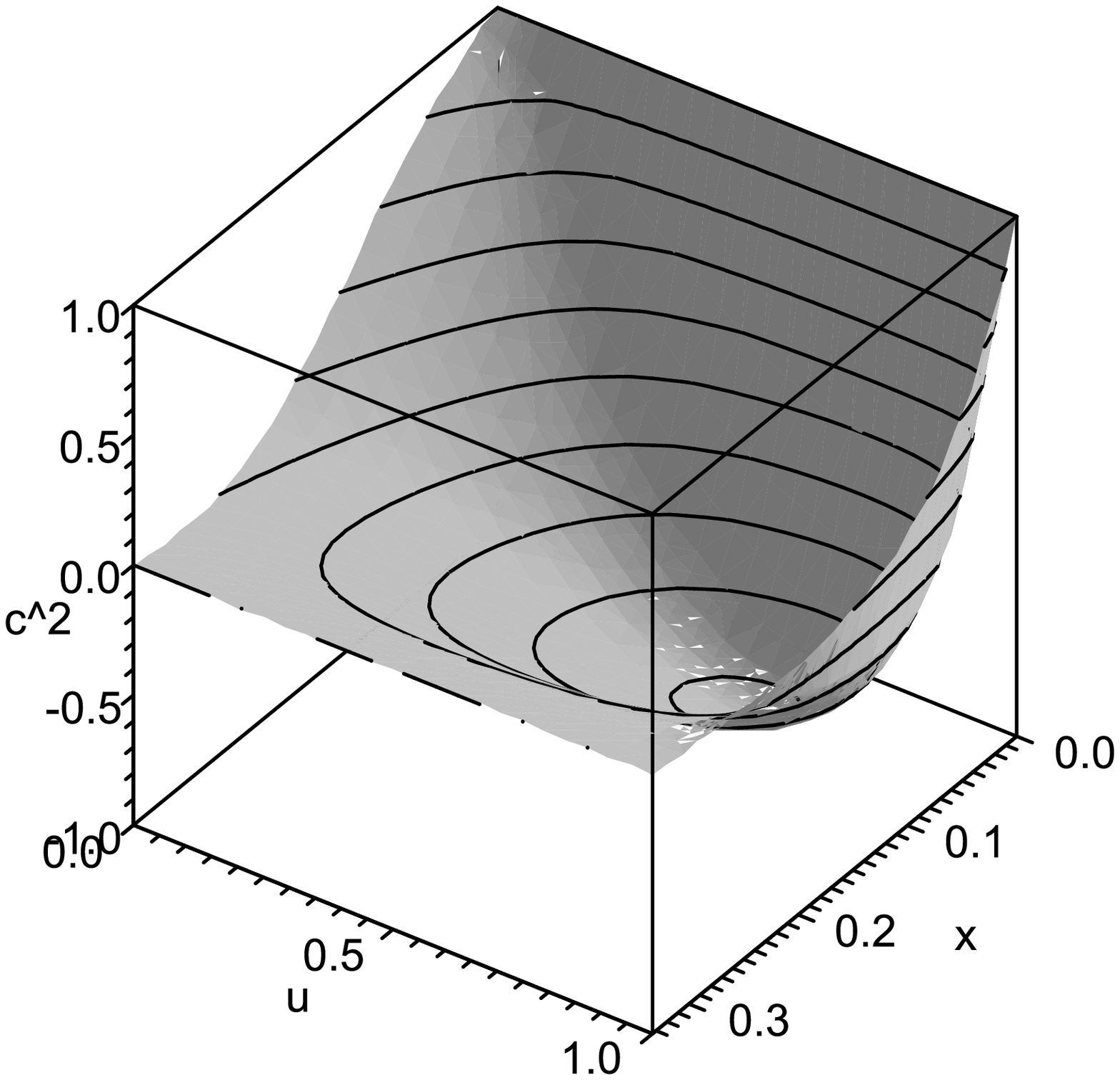}
\includegraphics[width=4.2cm]{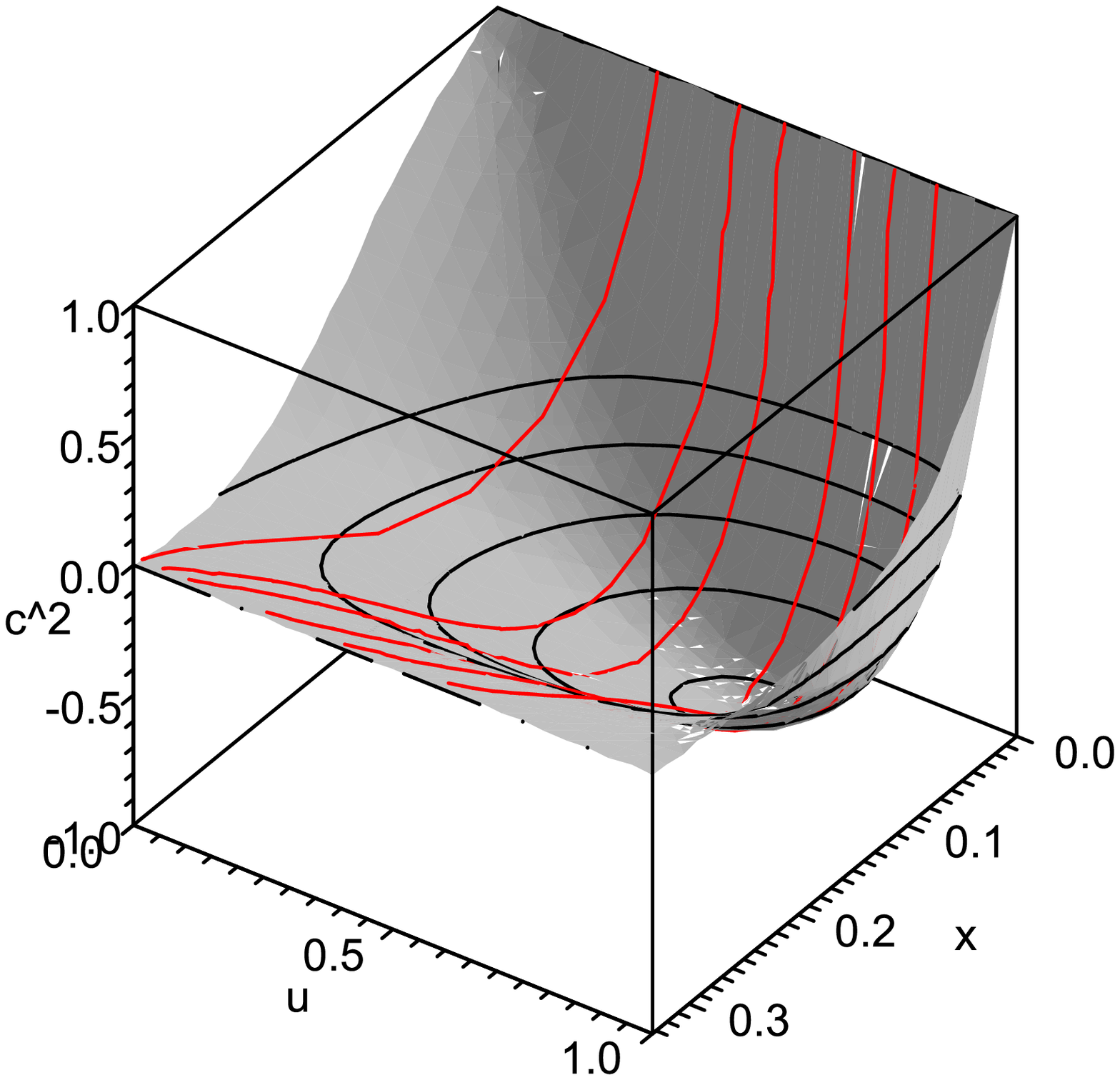}\vspace*{8pt}
\caption{The $c^2_s$-embedding schematically represented. The phase portrait of the dynamical system \eqref{ode-xu} and the plot of the surface $c^2_s=c^2_s(x,u)$ -- with contours -- in the extended (three-dimensional) phase space that is spanned by the coordinates $x$, $u$ and $c^2_s$, are shown in the left-hand and in the middle figures, respectively. In the right-hand figure the $c^2_s$-embedding diagram is drawn: the orbits (red curves) appearing in the phase portrait (left) have been embedded into the surface $c^2_s=c^2_s(x,u)$. The computations correspond to the cosmological model \eqref{action-dcoup} with positive coupling ($\alpha>0$) and for the growing exponential potential ($\lambda=5$). The contours drawn in the right-hand figure mark the region where $c^2_s<0$, i. e., where the Laplacian instability develops. The different embedded orbits correspond to whole cosmic evolutionary pathways that are associated with different sets of initial conditions. From the embedding diagram it is seen that independent on the initial conditions chosen the corresponding cosmological histories inevitably go through a stage where $c^2_s<0$, so that the classical gradient instability destroys any chance for the Universe to evolve into its present state.}\label{fig2}\end{figure}


As a qualitative support to the present discussion a geometric procedure of analysis based on the properties of the dynamical system will be used\cite{quiros_cqg_2018}. It provides a clear illustration of the failure of causality and/or of the development of Laplacian instability -- as well as of the crossing of the phantom divide -- along given phase space orbits. This procedure consists on the mapping of phase space orbits into the extended phase space, that is: the phase plane complemented with an additional dimension represented by the physical parameter of interest (the effective EOS or the squared sound speed, for instance). This is why the procedure is called as ``$P$-embedding'', where $P$ refers to the given physical parameter. In this subsection numeric computations are performed for the exponential and for the power-law potentials exclusively, including the constant and vanishing potential cases as particular cases. The embedding procedure is schematically represented in FIG. \ref{fig2}, where the $c^2_s$-embedding is illustrated for the cosmological model of interest, for the positive coupling case ($\alpha>0$) and for the monotonically growing exponential potential \eqref{exp-pot} with $\lambda=5$.

The results we shall discuss below, will show that the cosmological models based in the STT with non-minimal derivative coupling to the Einstein's tensor \eqref{action-dcoup} develop severe causality problems related with superluminal propagation of the perturbations of the scalar field. These problems are critical whenever the crossing of the phantom divide happens, however, these may arise even in the absence of the crossing. More problematic than the violations of causality in the model is the fact that it is plagued by the classical Laplacian (also gradient) instability, despite that the theory \eqref{action-dcoup} in which it is based, is free of the Ostrogradsky instability. These results confirm the inappropriateness of the kinetic coupling theories of the kind \eqref{action-dcoup}, as it has been discussed just recently in Refs. \citen{new, new-1, new-2, kobayashi_prd_2018, chagoya_arxiv, crisostomi_koyama_prd_2018, langlois_prd_2018}, on the light of the tight constraint on the difference in speed of photons and gravitons $(c^2_T-c^2)/c^2\leq 6\times 10^{-15}$ ($c_T$ is the speed of the gravitational waves) implied by the announced detection of gravitational waves from the neutron star-neutron star merger GW170817 and the simultaneous measurement of the gamma-ray burst GRB170817A\cite{ligo}.

\subsubsection{Basic equations and set up}

The main hypothesis is that the physical bounds on the speed of sound (squared) are viable criteria to reject physical theories like the one being investigated here. Other assumptions considered are the following: i) For simplicity of the discussion we shall focus in the vacuum case, i. e., in \eqref{action-dcoup} we set $S_m=0$, ii) only expanding cosmologies ($H\geq 0$) will be considered, iii) we assume non-negative energy density, i. e., non-negative self-interacting potential $V\geq 0$, and iv) only the case with $\epsilon=1$ (quintessence) will be of interest (for the pure derivative coupling case $\epsilon=0$ see Ref. \refcite{quiros_cqg_2018}). In addition, for sake of brevity, we shall discuss the case with the positive coupling $\alpha>0$ exclusively. A detailed study of the case with the negative coupling can be found also in Ref. \refcite{quiros_cqg_2018}.

As a model for the background spacetime we assume, as before, the FRW metric with flat spatial sections \eqref{frw-metric}. The cosmological field equations that can be derived from the action \eqref{action-dcoup} read:

\bea &&\;\;3H^2=\rho_\text{eff},\;-2\dot H=\rho_\text{eff}+p_\text{eff},\nonumber\\
&&\ddot\phi+3H\dot\phi=\frac{-6\alpha H\dot H\dot\phi-\der_\phi V}{\epsilon+3\alpha H^2}.\label{feqs}\eea The effective energy density and pressure of the scalar field are given by

\bea &&\rho_\text{eff}=\frac{\epsilon+9\alpha H^2}{2}\dot\phi^2+V(\phi),\label{rho}\\
&&p_\text{eff}=\frac{\epsilon-3\alpha H^2}{2}\dot\phi^2-V(\phi)-\alpha\dot\phi^2\dot H-2\alpha H\dot\phi\ddot\phi,\label{p}\eea respectively. An interesting property of the effective energy density $\rho_\text{eff}$ in \eqref{rho} and of the effective pressure $p_\text{eff}$ in \eqref{p}, is that these quantities depend not only on the scalar field matter degree of freedom $\phi$ and its derivatives $\dot\phi$ and $\ddot\phi$, but also on the curvature through $H^2$ and $\dot H$. In particular, the effective kinetic energy density of the scalar field in the right-hand-side (RHS) of the Friedmann equation above: $(\epsilon+9\alpha H^2)\dot\phi^2/2$, is contributed not only by $\dot\phi$ but also by the curvature through the squared Hubble rate. Notice that when in the above equations the non-minimal derivative coupling vanishes: $\alpha=0$, we recover the standard result of general relativity with minimally coupled scalar field matter. 

One can rewrite the Friedmann equation in the following way:

\bea 3H^2=\gamma^2\rho_\phi=\rho_\text{eff},\;\gamma=\frac{1}{\sqrt{1-3\alpha\dot\phi^2/2}},\label{3h2}\eea where $\gamma=\gamma(\dot\phi)$ is the 'boost' function and $\rho_\phi=\epsilon\dot\phi^2/2+V(\phi),$ is the standard energy density of the scalar field. Written in the latter form $\rho_\text{eff}$ is a function only of the scalar field degree of freedom $\phi$, and of its derivative $\dot\phi$ since the curvature effects are hidden in the non-canonical form of the effective energy density, i. e., in the boost function. We point out that for negative coupling ($\alpha<0$), the boost function is bounded from below and also from above: $0<\gamma\leq 1$, while for positive coupling ($\alpha>0$): $1\leq\gamma<\infty$, i. e., it is bounded from below only.

\begin{itemize}

\item{\it Non-negative coupling and upper bound on $|\dot\phi|$.} If we consider non-negative $\alpha\geq 0$, from \eqref{3h2} -- given that we consider non-negative effective energy density exclusively -- it follows that $1-3\alpha\dot\phi^2/2\geq 0$, i. e.

\bea 0\leq\dot\phi^2\leq\frac{2}{3\alpha}\;\Leftrightarrow\;-\frac{1}{3\alpha}\leq X\leq 0,\label{bounds-dphi}\eea where $X=\der_\mu\phi\der^\mu\phi/2=-\dot\phi^2/2$.\footnote{Notice that the definition of the variable $X$ in this subsection, which is the one most commonly found, differs from the one in the former sections of this review by a sign, $X\rightarrow-X$.} We want to point out here the non-conventional nature of the ``effective'' kinetic energy of the scalar field \eqref{rho} under the derivative coupling when $\alpha>0$. Actually, as just seen, the standard kinetic energy $\propto\dot\phi^2$ is bounded from above, a strange feature not arising in standard scalar-tensor theories without self-couplings. Notwithstanding, the effective kinetic energy in \eqref{rho}: $\propto(\epsilon+9\alpha H^2)\dot\phi^2,$ is not bounded due the curvature effects encoded in $H^2$. In reference \refcite{sushkov}, since in that presentation the coupling $\kappa$ is of opposite sign as compared with our $\alpha$: $\kappa=-\alpha$, the case where the standard kinetic term is bounded from above corresponds to the condition expressed by Eq. (27) in the mentioned reference (see also equations (19) and (21) of the same reference, recalling that in this review we have chosen the units where $8\pi G_N=1$, while in Ref. \refcite{sushkov}: $G_N=1$.) 

\bigskip
\item{\it New variables.} In spite of the commonly used variable $X$ (see above), in order to study both positive and negative coupling cases in a unified way, here we prefer to use the new variable:

\bea x:=\alpha\dot\phi^2/2,\label{x-var}\eea i. e. the new variable is properly the standard kinetic energy of the scalar field multiplied by the coupling constant. Hence, positive coupling entails that $x\geq 0$, while negative coupling means that $x\leq 0$. Vanishing $x=0$ means that, either the scalar field is a constant $\phi=\phi_0$, or there is not derivative coupling: $\alpha=0$. 

In the same way, in connection with the self-interaction potential term, it will be very useful to introduce the following variable:

\bea y:=\alpha V.\label{y-var}\eea For positive $\alpha$ this variable takes non-negative values: $0\leq y<\infty$, while for negative coupling ($\alpha<0$) the variable takes non-positive values instead $-\infty<y\leq 0$. We want to underline that for positive coupling ($\alpha>0$), given that $H^2$: 

\bea 3\alpha H^2=\frac{\epsilon x+y}{1-3x},\label{3h2-xy-eq}\eea should be a non-negative quantity ($H^2\geq 0$), the non-negative variable $x$ should take values in the physically meaningful interval: $0\leq x\leq 1/3$. Meanwhile, for negative coupling ($\alpha<0$) the variable $x$ is non-positive: $-\infty<x\leq 0$. As already stated above, in this subsection we shall expose the case with the positive coupling ($\alpha>0$) exclusively. For the negative coupling case see Ref. \refcite{quiros_cqg_2018}. The above variables will allow us to write the equations in a more compact manner and to make our computations independent of the specific value of the coupling constant.

\end{itemize}


\begin{figure}\begin{centering}
\includegraphics[width=5cm]{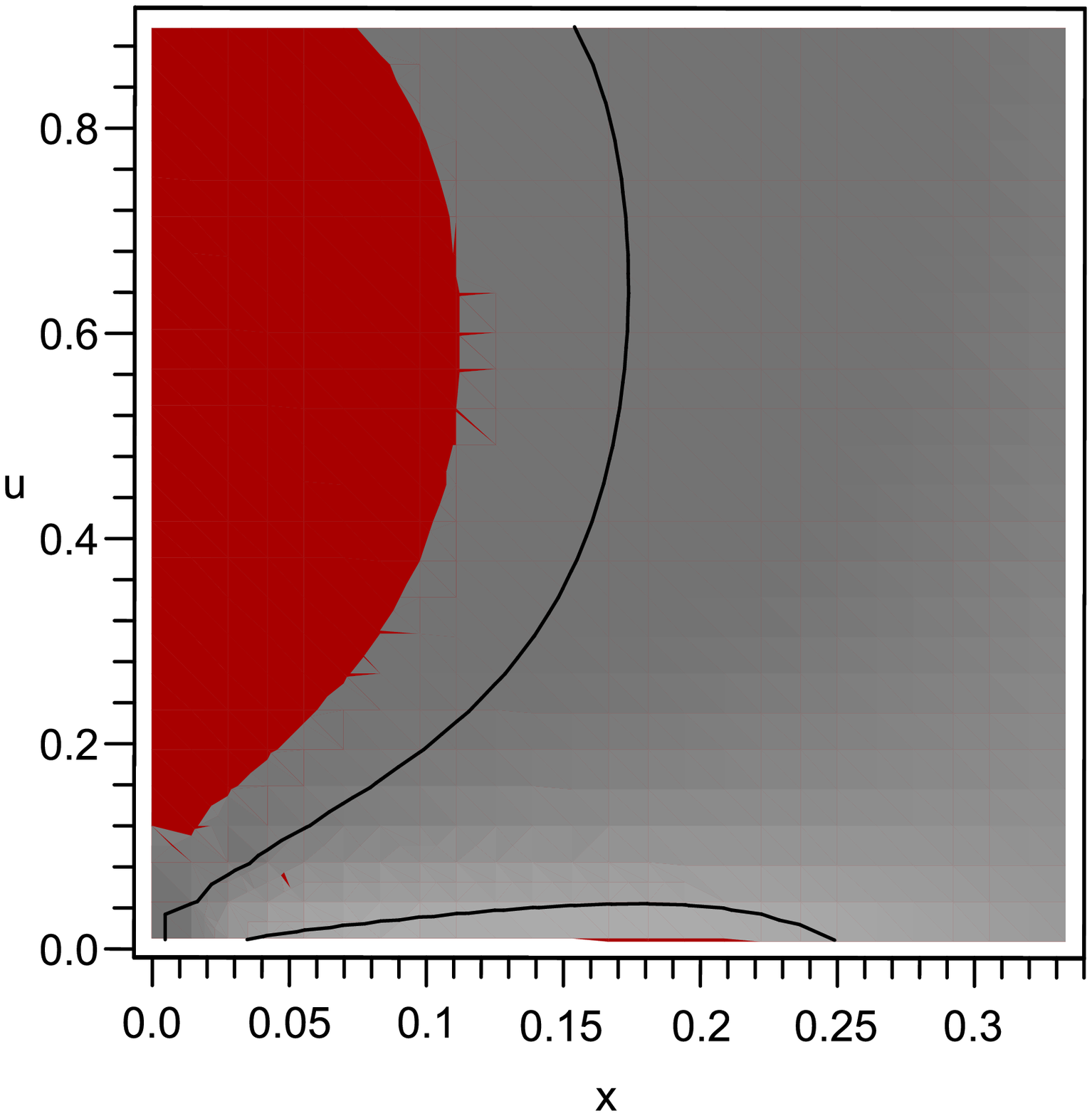}
\includegraphics[width=5cm]{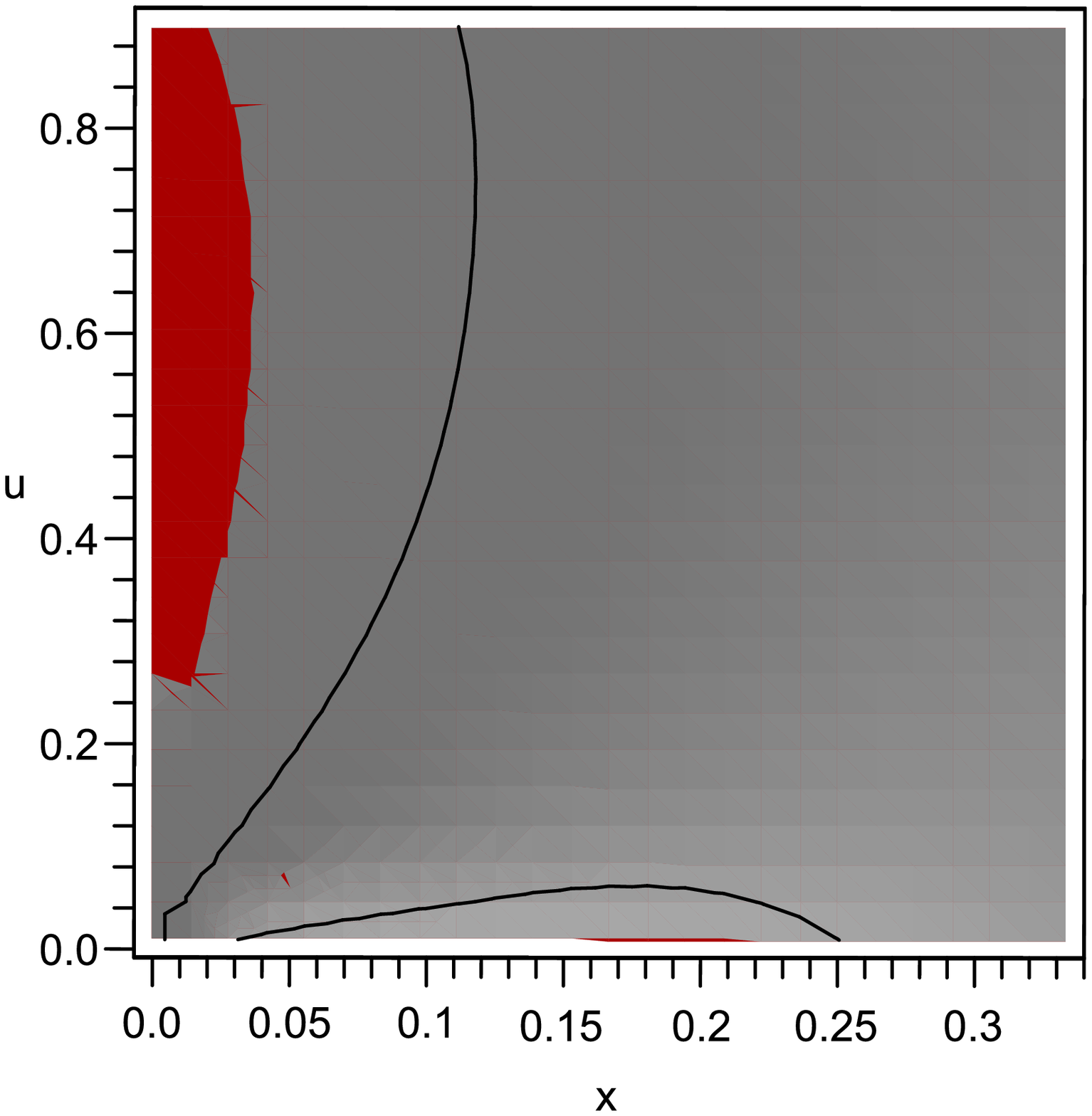}\\
\includegraphics[width=5cm]{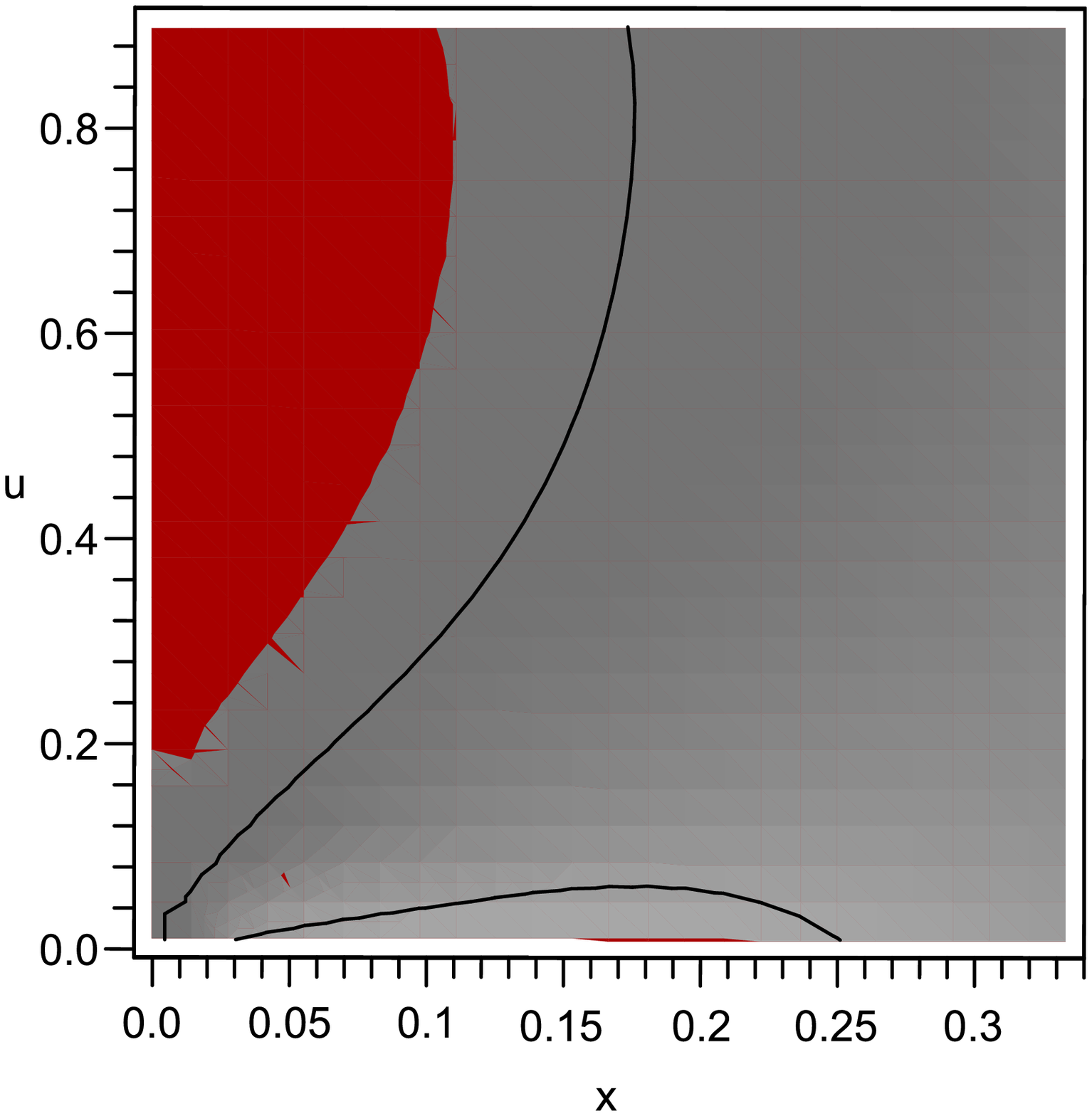}
\includegraphics[width=5cm]{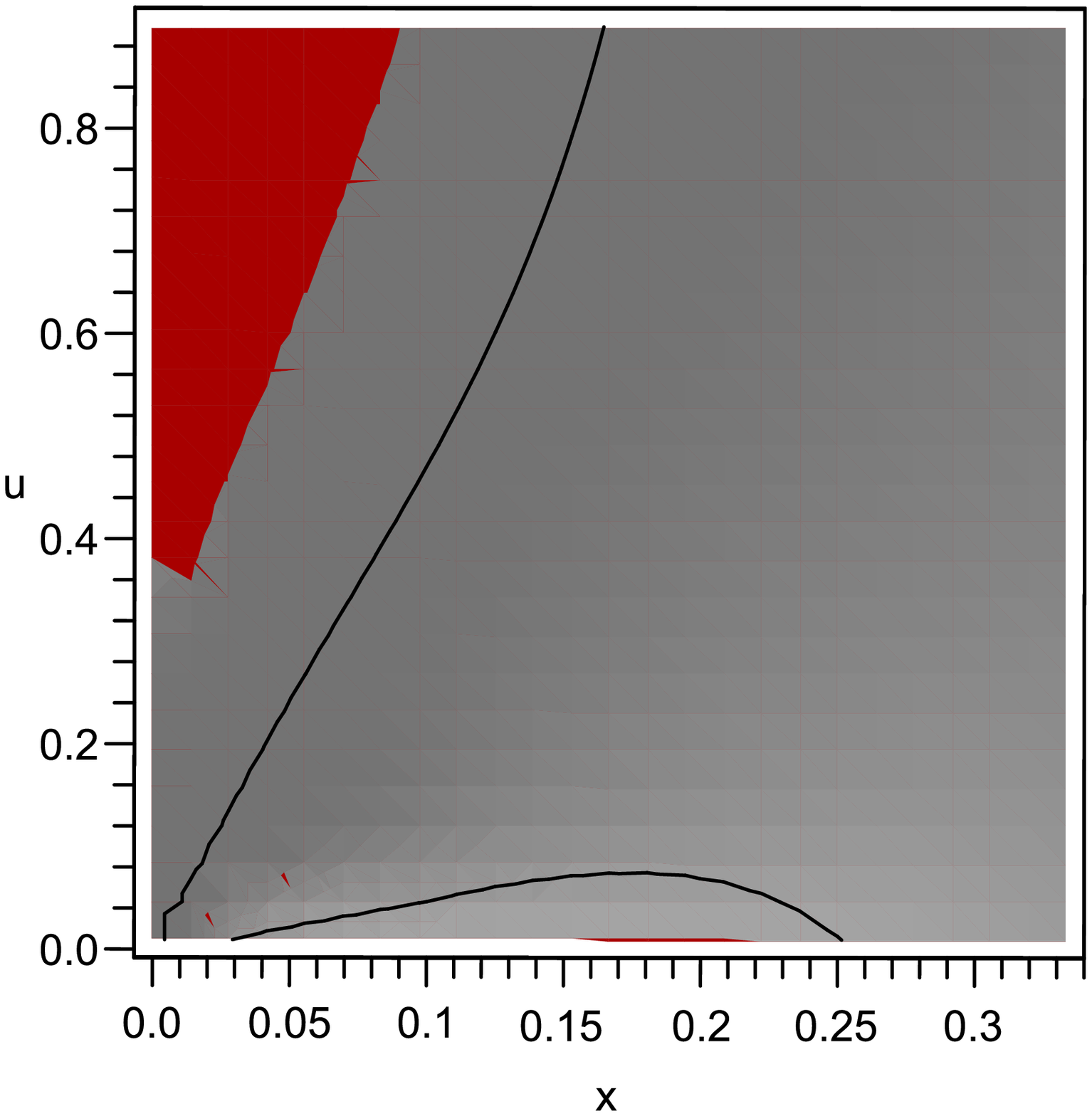}\vspace*{8pt}
\caption{Geometric representation of the bound $\omega_\text{eff}+1\geq 0$ in the $xu$-plane for positive coupling ($\alpha>0$). For illustrative purposes we have chosen two negative-slope potentials: the decaying exponential potential $V=V_0\exp{(\lambda\phi)}$ with $\lambda<0$ (top panels) and the inverse power-law potential $V=V_0\phi^{2n}$ with $n<0$ (bottom panels), for different values of the parameters $\lambda$ and $n$ respectively. In the left hand panels, from left to the right: $\lambda=-5$ and $\lambda=-2$, while in the right-hand panels: $n=-2$ and $n=-1$, respectively. Here we use the bounded variable $u=y/y+1$ ($0\leq u\leq 1$) instead of $y=\alpha V$ (the variable $x=\alpha\dot\phi^2/2$ is already bounded: $0\leq x\leq 1/3$), so that the whole phase plane $xu$ fits into a finite size box. The red-colored regions correspond to the phantom domain where $\omega_\text{eff}+1<0$. For monotonically growing potentials ($\lambda>0|n>0$) the phantom domain is not found so that the crossing is not possible.}\label{fig01}\end{centering}\end{figure}



\subsubsection{Phantom barrier crossing: General analysis}

One issue of interest when one explores cosmological models of dark energy is the possibility of crossing the so called ``phantom divide'' barrier $\omega_\Lambda=-1$\cite{caldwell, vikman_prd_2005, crossing-odintsov, crossing-observ, crossing-observ-1, crossing-nesseris, crossing-perivo, crossing-hu, crossing-chimento}. Here we shall discuss on the crossing in the theory with kinetic coupling to the Einstein's tensor\cite{mohseni}. If under the assumptions undertaken here we combine the second and third equations in \eqref{feqs}, we obtain:

\bea -2\alpha\dot H=R_1+R_2,\label{doth}\eea with (recall that $y_\phi=\alpha\der_\phi V$):

\bea R_1=\frac{2x\left[\epsilon(1-2x)+y\right](\epsilon+3y)}{(1-3x)F_\epsilon},\;R_2=\frac{2\sqrt{2x(1-3x)(\epsilon x+y)}\,y_\phi}{\sqrt{3}F_\epsilon},\label{r1-r2-def}\eea where, for compactness of writing, we have introduced the following definition:

\bea F_\epsilon\equiv F_\epsilon(x,y):=\epsilon(1-3x+6x^2)+(1+3x)y.\label{f-eps}\eea The effective EOS parameter of the scalar field is given by:

\bea \omega_\text{eff}=\frac{p_\text{eff}}{\rho_\text{eff}}=-1-\frac{2\dot H}{3H^2}=-1+\frac{R_1+R_2}{3\alpha H^2},\label{weff-def}\eea where $R_1$ and $R_2$ are given by \eqref{r1-r2-def} and, in terms of the variables $x$, $y$, the denominator $3\alpha H^2$ is given by \eqref{3h2-xy-eq}. Hence, for the effective EOS in the general case -- unspecified $\epsilon$ -- we get:

\bea \omega_\text{eff}=-1+\frac{2x(\epsilon+3y)\left[\epsilon(1-2x)+y\right]}{(\epsilon x+y)F_\epsilon}+\frac{2}{F_\epsilon}\sqrt\frac{2x(1-3x)^3}{3(\epsilon x+y)}\;y_\phi.\label{eos-master-eq}\eea 

As it can be seen from \eqref{weff-def}, the crossing of the phantom barrier is achieved only if $-2\dot H$ may change sign during the cosmic evolution. In general $-2\dot H$ is a non-negative quantity. This is specially true for the standard quintessence where in equations \eqref{feqs}, \eqref{rho} and \eqref{p} we set $\alpha=0$ and $\epsilon=1$. In this case $-2\dot H=\dot\phi^2\geq 0$, while the EOS parameter in \eqref{weff-def} can be written as

\bea \omega_\text{eff}=-1+\frac{\dot\phi^2}{3H^2},\label{quint-cross}\eea so that, given that $\dot\phi^2/3H^2$ is always non-negative, then $\omega_\text{eff}\geq -1$. In this case the phantom barrier crossing is not possible unless additional complications are considered such as, for instance: i) non-gravitational interaction of the dark energy and dark matter components\cite{n-m-int, n-m-int-pavon, n-m-int-quiros}, ii) multiple dark energy fields like in quintom models\cite{quintom-mod, quintom-mod-1, quintom-rev} or iii) extra-dimensional effects\cite{quiros_jcap_2006}. Here we shall investigate the issue within the frame of the theory \eqref{action-dcoup} where the derivatives of the scalar field are non-minimally coupled to the Einstein's tensor.

For non-negative $x$-s, i. e., for positive coupling ($\alpha>0$), the denominators of $R_1$ and of $R_2$ in \eqref{r1-r2-def} are always positive-valued. So is the numerator of the term $R_1$ which means that this term is always non-negative. Meanwhile, the sign of the numerator of the term $R_2$ is determined by the slope of the self-interaction potential: $y_\phi=\alpha\der_\phi V$. Consequently, for non-negative $0\leq x\leq 1/3$, the term $R_2$ in \eqref{doth} is the only one that may allow for the crossing of the phantom barrier. In this case two clear conclusions can be done: i) the crossing is due to the derivative coupling with strength $\alpha$, and ii) the crossing is allowed only if $\dot\phi V'=\dot V<0$, i. e., if the self-interaction potential decays with the cosmic expansion. Assuming that this is indeed the case, the competition between the positive term $R_1$ and the negative one $R_2$ during the course of the cosmic evolution is what makes possible the flip of sign of $-2\dot H=R_1+R_2$, and hence the crossing of the phantom barrier. Notice that for the constant potential $\der_\phi V=0$, as well as for the monotonically growing potentials the crossing is not possible. This is true, in particular, for the growing exponential potential: $V\propto\exp(\lambda\phi)$ with $\lambda>0$ for $\dot\phi>0$ or $\lambda<0$ for $\dot\phi<0$, and for the power-law $V\propto\phi^n$ with $n\geq 0$. 

The above results are illustrated in FIG. \ref{fig01} where a geometric representation of the quantity $\omega_\text{eff}+1$ in the $xu$-plane is shown. Here we used the new (bounded) variable: 

\bea u=\frac{y}{y+1},\;0\leq u\leq 1.\label{u-var}\eea This choice makes possible to fit the whole (semi-infinite) phase plane $xy$ into a finite size box: $\{(x,u):0\leq x\leq 1/3,0\leq u\leq 1\}$. The red-colored regions are the ones where $\omega_\text{eff}+1<0$, i. e., where the scalar field behaves like phantom matter. It is appreciated that, for negative-slope potentials (the decaying exponential and the inverse power-law in the figure), both the phantom region with $\omega_\text{eff}+1<0$ and the region where $\omega_\text{eff}+1>0$ (gray color) coexist, so that the crossing of the phantom divide is possible.


\subsubsection{Squared sound speed $c_s^2$}\label{subsect-sound-speed}

In Ref. \refcite{cartier} the authors derived the evolution equations for the most general cosmological scalar, vector and tensor perturbations in a class of non-singular cosmologies derived from higher-order corrections to the low-energy bosonic string action:

\bea {\cal L}=\frac{1}{2}f(\phi,R)-\frac{1}{2}\omega(\phi)\nabla^\mu\phi\nabla_\mu\phi-V(\phi)+{\cal L}_q,\label{cartier-lag}\eea where $f(\phi,R)$ is an algebraic function of the scalar field $\phi$ and of the curvature scalar $R$, while $\omega(\phi)$ and $V(\phi)$ are functions of the scalar field. For our purposes it is enough to consider $f(\phi,R)=R$ and $\omega(\phi)=1$. Through ${\cal L}_q$ the inclusion of higher order derivative terms is allowed:

\bea {\cal L}_q=-\frac{\lambda}{2}\xi\left[c_1R^2_\text{GB}+c_2 G^{\mu\nu}\der_\mu\phi\der_\nu\phi+c_3\Box\phi\der^\mu\phi\der_\mu\phi+c_4\left(\der^\mu\phi\der_\mu\phi\right)^2\right],\label{lag-q}\eea where $\xi=\xi(\phi)$ is a function of the scalar field, $R^2_\text{GB}\equiv R_{\mu\nu\tau\lambda}R^{\mu\nu\tau\lambda}-4R_{\mu\nu}R^{\mu\nu}+R^2$ is the Gauss-Bonnet combination, $\lambda$, $c_1,\ldots,c_4$ are constants and we have chosen the units where $\alpha'=1$. Here, without loss of generality we set $\xi=1$.

The action \eqref{qbic-gal-action} is a particular case of \eqref{cartier-lag}, so that the results of Ref. \refcite{cartier} are easily applicable to this case (see for instance Ref. \refcite{gao}). The Einstein's field equations that are derived from the Lagrangian \eqref{cartier-lag} read: 

\bea G_{\mu\nu}=T^\text{eff}_{\mu\nu}=T^{(\phi)}_{\mu\nu}+T^{(q)}_{\mu\nu},\;\Box\phi-T^{(q)}=V',\nonumber\eea where the comma stands for derivative with respect to $\phi$, $$T^{(\phi)}_{\mu\nu}=\der_\mu\phi\der_\nu\phi-\frac{1}{2}g_{\mu\nu}\left(\der^\tau\phi\der_\tau\phi\right)-g_{\mu\nu}V,$$ is the standard stress-energy tensor of a scalar field, while $$T^{(q)}_{\mu\nu}=-2\frac{\der{\cal L}_q}{\der g^{\mu\nu}}-g_{\mu\nu}{\cal L}_q,$$ and $T^{(q)}$ represent the contributions derived from the next to leading order corrections given by ${\cal L}_q$ in equation \eqref{lag-q} (equation (2) of Ref. \refcite{cartier}). These contribute towards the effective stresses and energy. 

The perturbed line-element reads\cite{cartier, hwang}:

\bea &&ds^2=-a^2(1+2\psi)d\eta^2-2a^2\left(\beta_{,i}+B_i\right)d\eta dx^i+\nonumber\\
&&\;\;\;\;\;\;\;\;\;\;\;\;\;\;\;\;\;\;\;\;\;\;\;\;a^2\left[g_{ij}(1+2\vphi)+2\gamma_{,i|j}+2C_{(i|j)}+2C_{ij}\right]dx^idx^j,\label{ds-pert}\eea where $d\eta=dt/a$, $\psi=\psi(t,{\bf x})$, $\beta=\beta(t,{\bf x})$, $\vphi=\vphi(t,{\bf x})$ and $\gamma=\gamma(t,{\bf x})$ characterize the scalar-type perturbations. The traceless modes $B_i$ and $C_i$ ($B^i_{|i}=C^i_{|i}=0$) represent the vector-type perturbations, meanwhile, $C_{ij}=C_{ij}(t,{\bf x})$ are trace free and transverse: $C^j_{i|j}=C^i_i=0$, and correspond to the tensor-type perturbations. The vertical bar denotes covariant derivative defined in terms of the space metric $g_{ij}$. Following Ref. \refcite{hwang} in Ref. \refcite{cartier} the uniform-field gauge ($\delta\phi=0$) is chosen since this gauge admits the simplest analysis. In this case each variable is replaced by its corresponding gauge-invariant combination with $\delta\phi$, for instance, for the scalar perturbation the gauge-invariant combination $$\vphi_{\delta\phi}\equiv\vphi-H\frac{\delta\phi}{\dot\phi},$$ is considered (in the uniform-field gauge $\vphi_{\delta\phi}$ is identified with $\vphi$ since $\delta\phi=0$). The second-order differential (wave) equation for the scalar-metric perturbation $\vphi_{\delta\phi}$ in closed form reads\cite{cartier}: 

\bea \frac{1}{a^3Q_s}\frac{\der}{\der t}\left(a^3Q_s\frac{\der}{\der t}\,\vphi_{\delta\phi}\right)-c^2_s\frac{\Box}{a^2}\,\vphi_{\delta\phi}=0,\label{s-waveq}\eea where $$Q_s=\frac{\dot\phi^2+\frac{3Q_a^2}{2+Q_b}+Q_c}{\left(H+\frac{Q_a}{2+Q_b}\right)^2},$$ and the squared speed of propagation of the scalar perturbation is given by

\bea c^2_s=1+\frac{(2+Q_b)Q_d+Q_aQ_e+\frac{Q_a^2Q_f}{2+Q_b}}{(2+Q_b)(\dot\phi^2+Q_c)+3Q_a^2},\label{c2s-eff}\eea with

\bea &&Q_a=\lambda\dot\phi^2\left(2c_2H+c_3\dot\phi\right),\;Q_b=\lambda c_2\dot\phi^2,\;Q_c=-3\lambda\dot\phi^2\left(c_2 H^2+2c_3H\dot\phi+2c_4\dot\phi^2\right),\nonumber\\
&&Q_d=-2\lambda\dot\phi^2\left[c_2\dot H+c_3\left(\ddot\phi-H\dot\phi\right)\right],\;Q_e=4\lambda\dot\phi\left[c_2\left(\ddot\phi-H\dot\phi\right)-c_3\dot\phi^2\right],\label{Q-s}\eea and $Q_f=2Q_b=2\lambda c_2\dot\phi^2$.

For the linearized tensor-type perturbations the following second order equation of motion is obtained\cite{cartier}: 

\bea \frac{1}{a^3Q_T}\frac{\der}{\der t}\left(a^3Q_T\frac{\der}{\der t}C^i_{\;\;j}\right)-c^2_T\frac{\Box}{a^2}C^i_{\;\;j}=\frac{1}{Q_T}\delta T^i_{\;\;j},\label{t-waveq}\eea where $\delta T^i_{\;\;j}$ includes contributions to the tensor-type energy-momentum tensor, $$Q_T=1+\frac{\lambda}{2}\,c_2\dot\phi^2,$$ and

\bea c^2_T=\frac{2-\lambda c_2\dot\phi^2}{2+\lambda c_2\dot\phi^2},\label{c2t}\eea is the squared speed of propagation of the gravitational waves perturbation. Notice that for $c^2_s>0$ and $c^2_T>0$ the wave equations \eqref{s-waveq} and \eqref{t-waveq}, respectively, are hyperbolic differential equations -- the Cauchy problem is well posed -- meanwhile for negative $c^2_s<0$ and $c^2_T<0$, these equations are elliptic so there is not propagating mode (the Cauchy problem is not well posed). In this later case a Laplacian instability develops.

For the cosmological model based in \eqref{action-dcoup} the Lagrangian \eqref{lag-q} can be written in the following way: $${\cal L}_q=\frac{3\alpha}{2}\dot\phi^2H^2,$$ where we have set $\xi=1$, $\lambda c_2=-\alpha$ (the remaining constants in \eqref{lag-q} vanish). Hence:

\bea &&Q_a=-2\alpha H\dot\phi^2,\;Q_b=-\alpha\dot\phi^2,\;Q_c=3\alpha H^2\dot\phi^2,\nonumber\\
&&Q_d=2\alpha\dot H\dot\phi^2,\;Q_e=-4\alpha\dot\phi(\ddot\phi-H\dot\phi).\label{qs}\eea For the squared speed of propagation of the gravitational waves perturbation \eqref{c2t} it is found that:

\bea c^2_T=\frac{1+\alpha\dot\phi^2/2}{1-\alpha\dot\phi^2/2},\label{c2t'}\eea where it is appreciated that, for the positive coupling $\alpha>0$, the tensor perturbations propagate superluminally. A similar result has been formerly reported in Ref. \refcite{germani} for the same model but under the slow-roll approximation, i. e., valid for primordial inflation. For negative coupling $\alpha<0$, provided that $\dot\phi^2>2/|\alpha|$ the squared sound speed of the tensor perturbations becomes negative, signaling the eventual occurrence of a Laplacian instability. For a detailed derivation of \eqref{c2s-eff} and of \eqref{c2t} within the perturbative approach we recommend Ref. \refcite{cartier}.

Equation \eqref{c2s-eff} with the substitution of the quantities \eqref{qs} will be our master equation for determining the (squared) speed of propagation of the scalar perturbations of the energy density. In terms of the field variables $x=\alpha\dot\phi^2/2$ and $y=\alpha V(\phi)$ we have that:

\bea c^2_s=1+\frac{4x[\epsilon(3-11x+6x^2)+(1-3x)y]}{3(1-x)F_\epsilon}-\frac{3(1-x)(\epsilon x+y)(\omega_\text{eff}+1)}{F_\epsilon},\label{c2s-master-eq}\eea where $\omega_\text{eff}$ is given by \eqref{eos-master-eq} and the funciton $F_\epsilon$ has been defined in \eqref{f-eps}. We have that $0\leq x\leq 1/3$ and $0\leq y<\infty$. This means that $F_\epsilon$ is always a positive function. Besides, both the numerator and the denominator in the second term in the right-hand side (RHS) of equation \eqref{c2s-master-eq} are positive quantities. The same is true for the factor $(1-x)(\epsilon x+y)/F_\epsilon$ in the third term in the RHS of the mentioned equation. Hence, while the second term always contributes towards superluminality of propagation of the scalar perturbations, the contribution of the third term depends on the sign of $\omega_\text{eff}+1$. For $\omega_\text{eff}>-1$ the superluminal contribution of the second term in the RHS of \eqref{c2s-master-eq} may be compensated by the third term. However, when $\omega_\text{eff}<-1$, both terms in the RHS of \eqref{c2s-master-eq} contribute towards superluminality of the propagation of the scalar perturbations of the energy density. This means that, whenever the crossing of the phantom divide is allowed, then $\omega_\text{eff}+1$ becomes necessarily negative during a given stage of the cosmic evolution and, consequently, causality violations are inevitable. This result is independent on the specific functional form of the self-interaction potential. 

In general, from \eqref{c2s-master-eq} it follows that whenever the condition 

\bea \frac{4x[\epsilon(3-11x+6x^2)+(1-3x)y]}{9(1-x)^2(\epsilon x+y)}>\omega_\text{eff}+1,\label{cond}\eea is fulfilled, the squared sound speed is superluminal ($c^2_s>1$). The latter condition may be satisfied only if $\omega_\text{eff}+1<0$, i. e., if $\omega_\text{eff}<-1$. For positive $\omega_\text{eff}+1>0$, the inequality \eqref{cond} is never satisfied. 

We want to point out that, although the condition $\omega_\text{eff}<-1$ boosts further superluminality of the propagation of the scalar perturbations, in general the $\omega=-1$ crossing is not required for the superluminality to arise in the present model. Actually, as seen from \eqref{c2s-master-eq}, given that the second term in the RHS of \eqref{c2s-master-eq} is always a postive quantity, superluminality arises even if $\omega_\text{eff}+1=0$.  

The potential situation where $\omega_\text{eff}+1>0$, i. e., where $\omega_\text{eff}>-1$, leads to another interesting and disturbing possibility, namely that 

\bea \omega_\text{eff}+1>\frac{(1-x)F_\epsilon}{3(1-x)^2(\epsilon x+y)}+\frac{4x[\epsilon(3-11x+6x^2)+(1-3x)y]}{9(1-x)^2(\epsilon x+y)},\label{laplac-instab}\eea that is, that $c^2_s<0$. Fulfillment of this latter bound leads to the development of the Laplacian/gradient instability. This is a classical instability associated with the uncontrolled growth of the amplitude of the scalar perturbations of the background density. 

\begin{figure}\begin{centering}
\includegraphics[width=3cm]{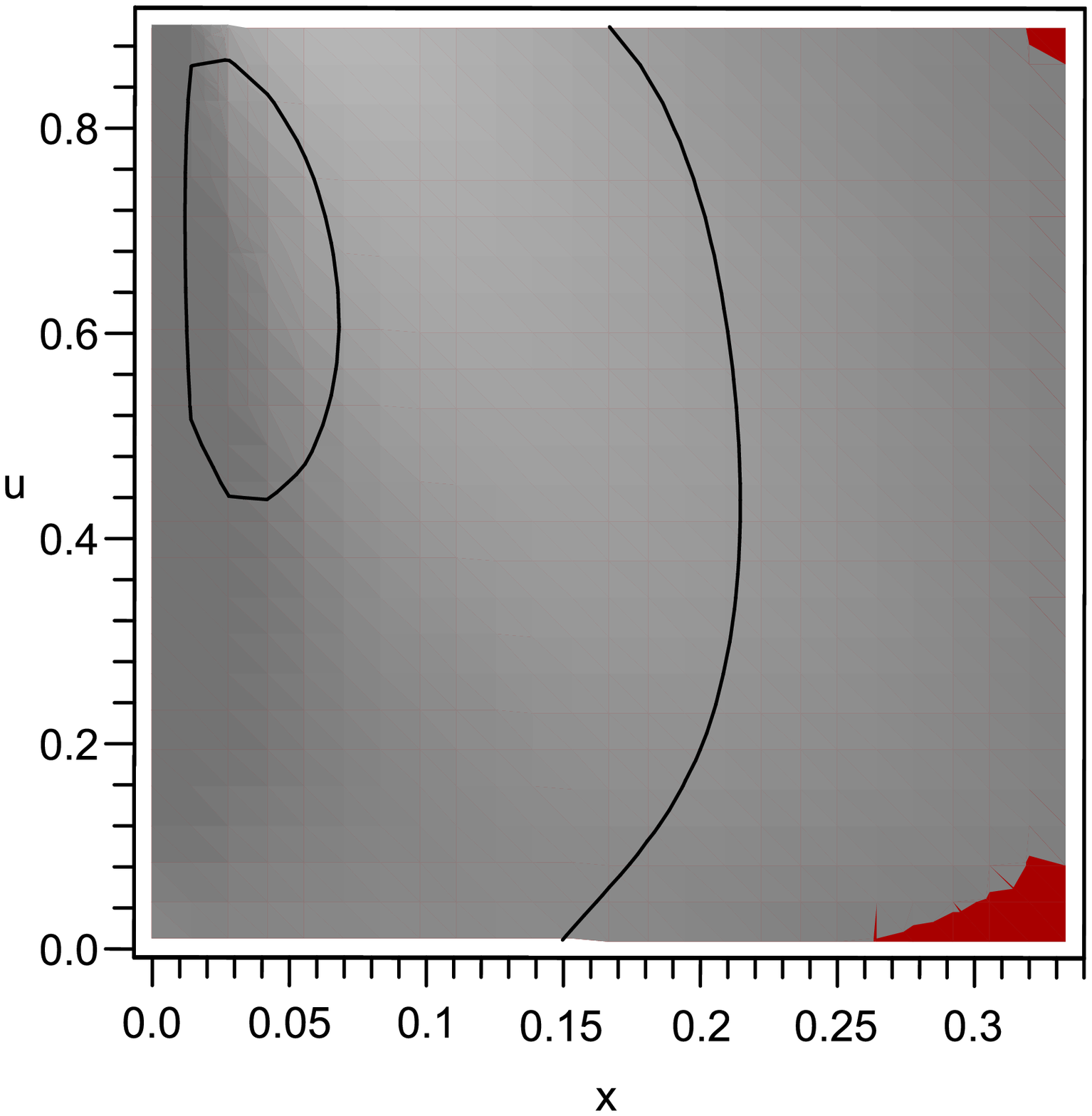}
\includegraphics[width=3cm]{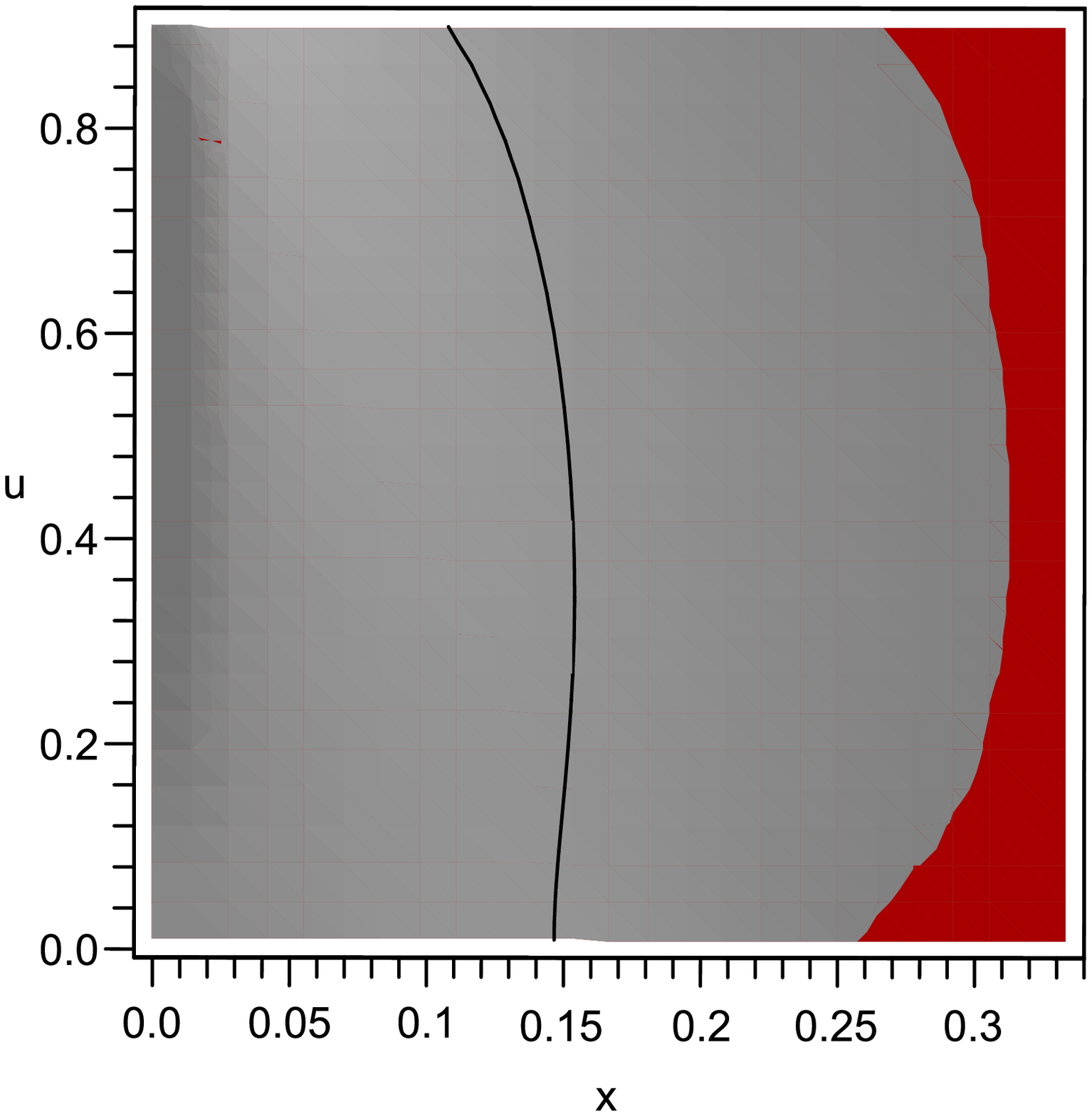}
\includegraphics[width=3cm]{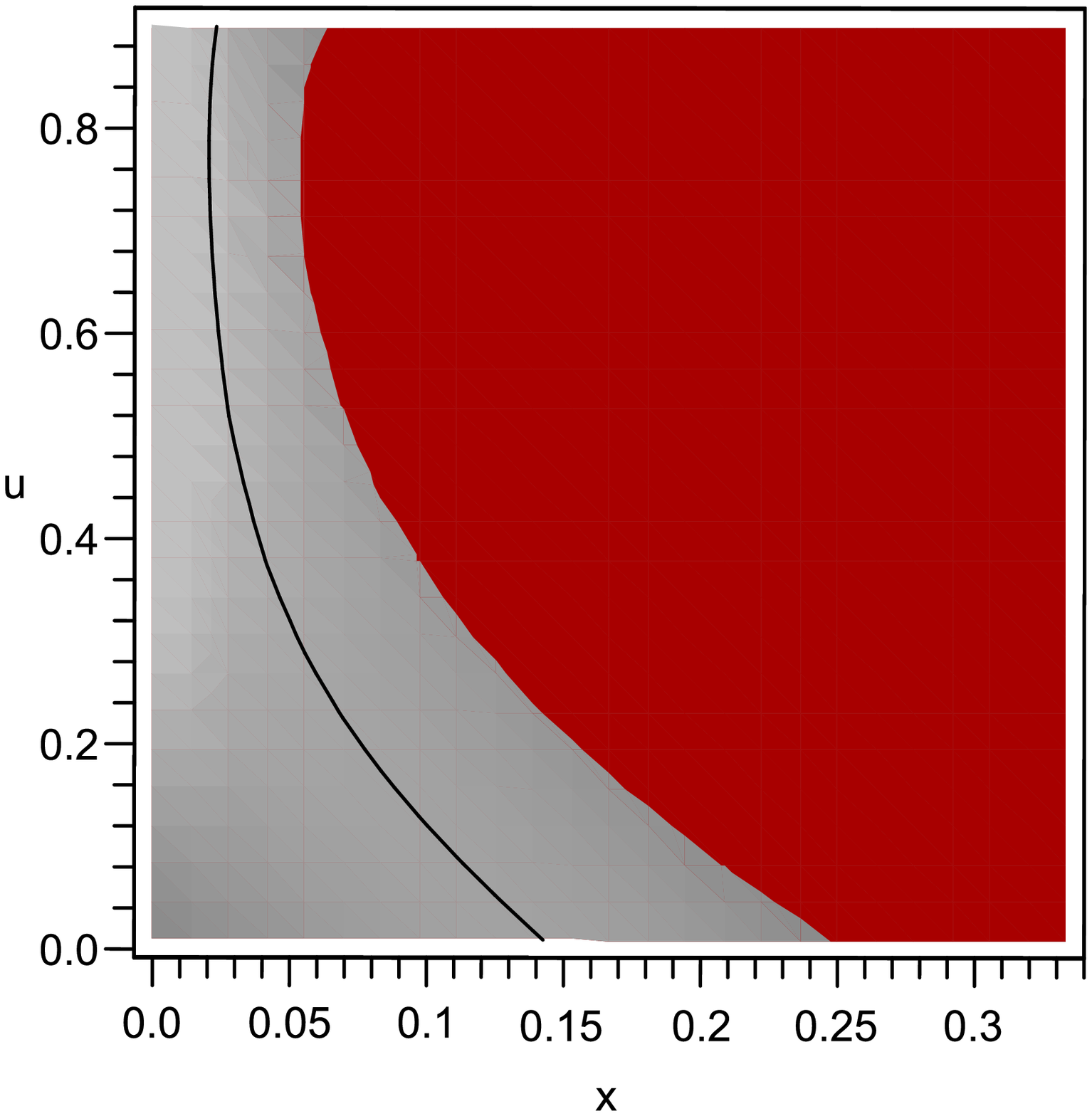}
\includegraphics[width=3cm]{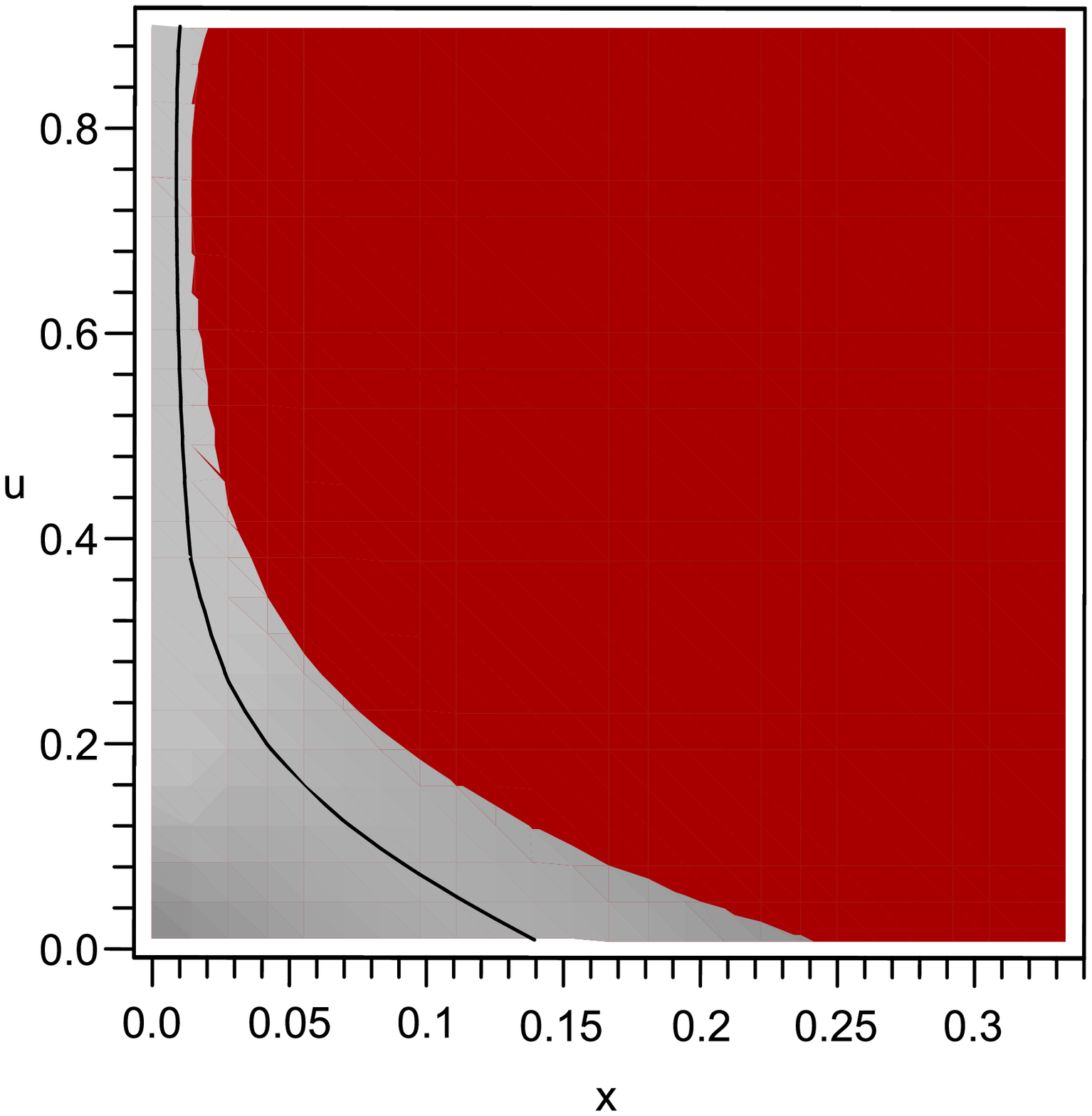}
\includegraphics[width=3cm]{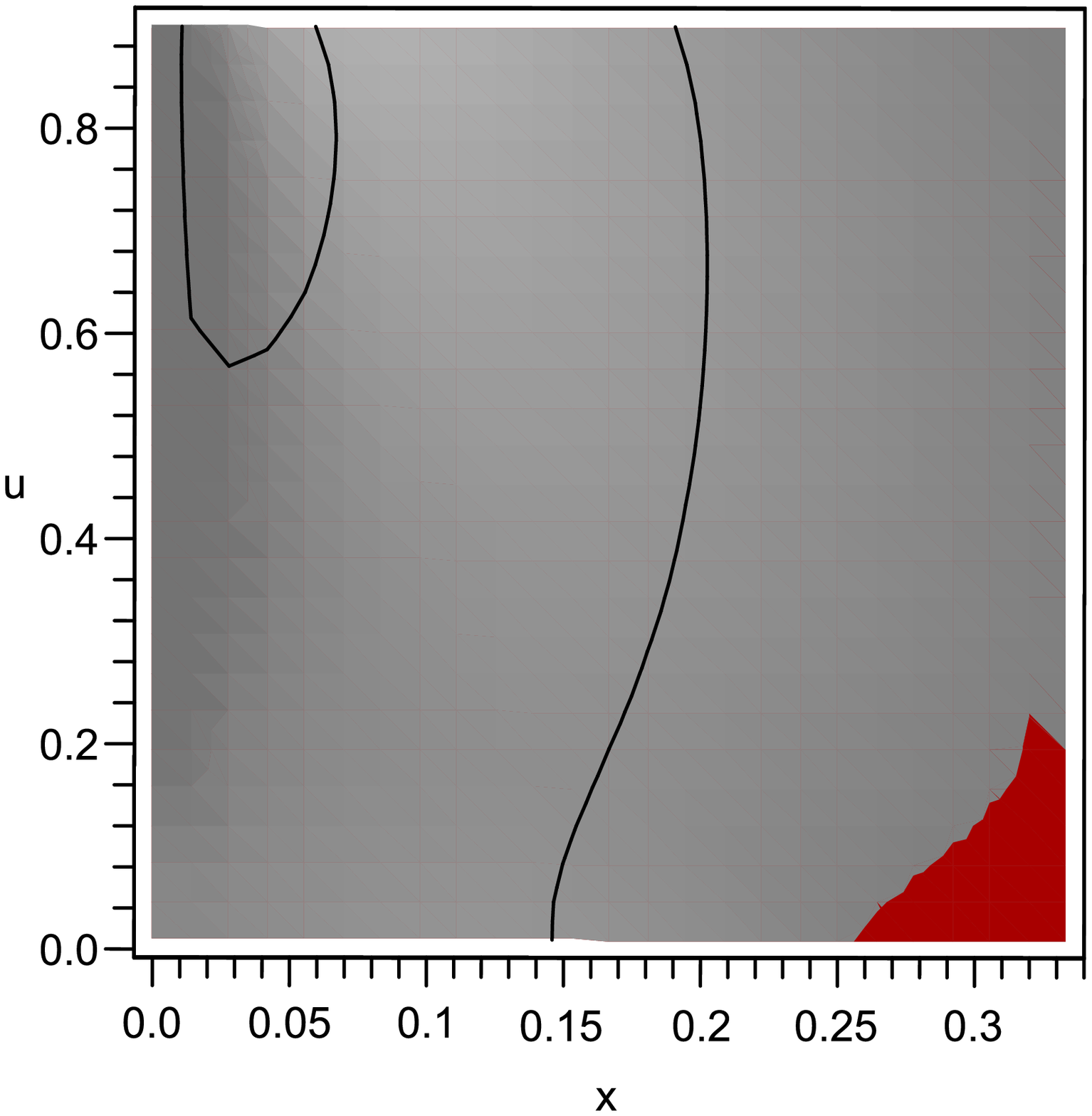}
\includegraphics[width=3cm]{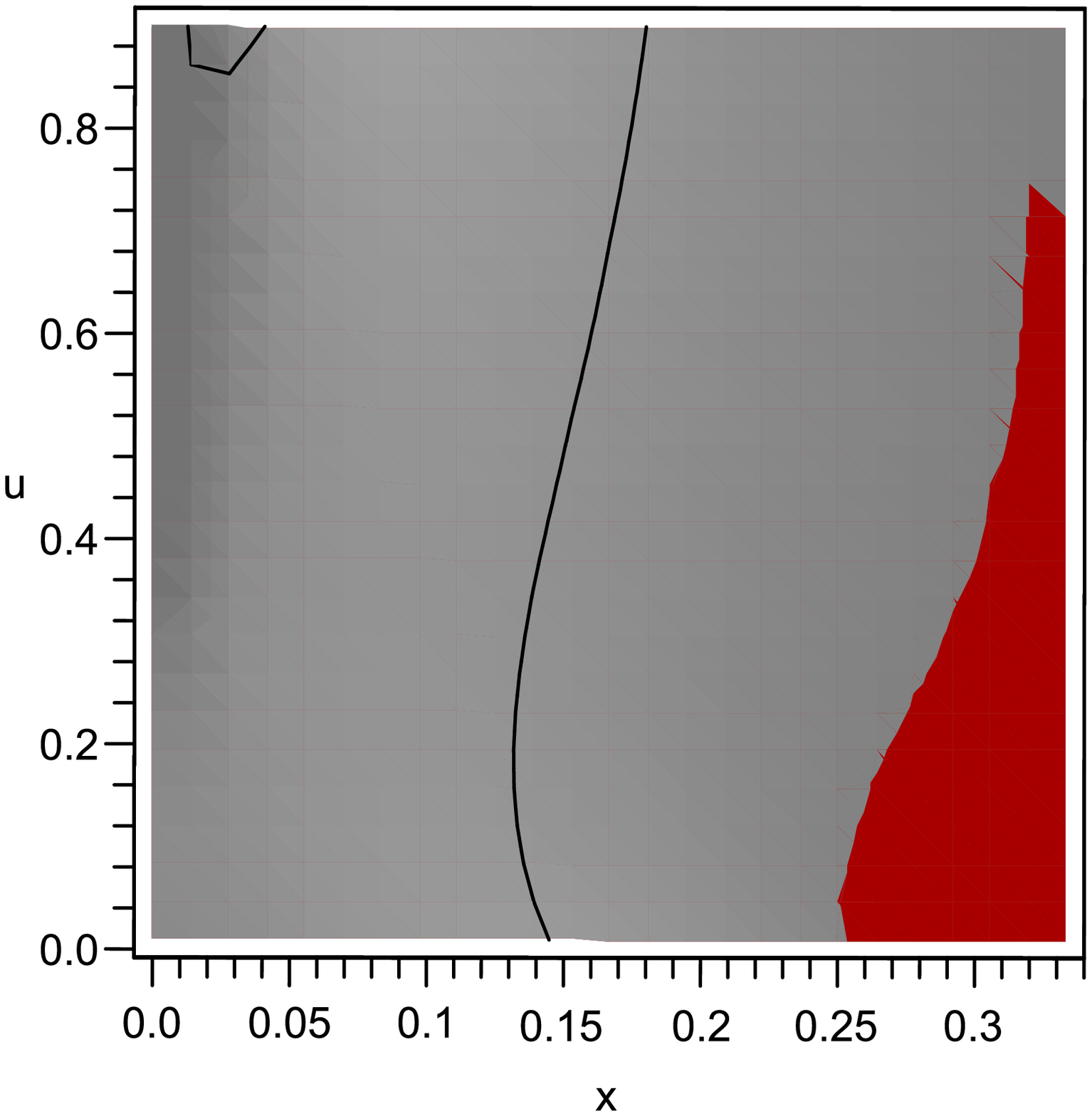}
\includegraphics[width=3cm]{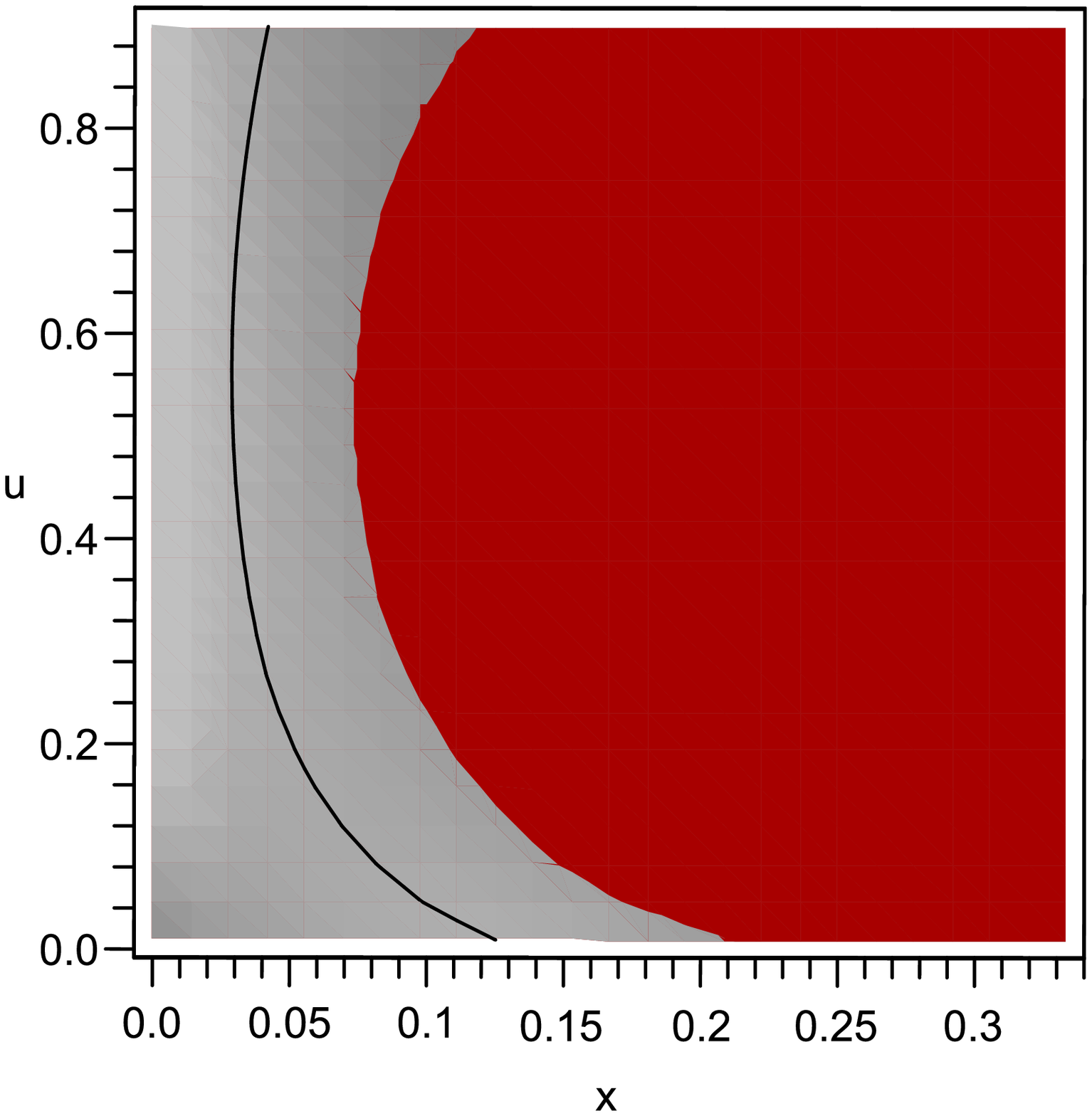}
\includegraphics[width=3cm]{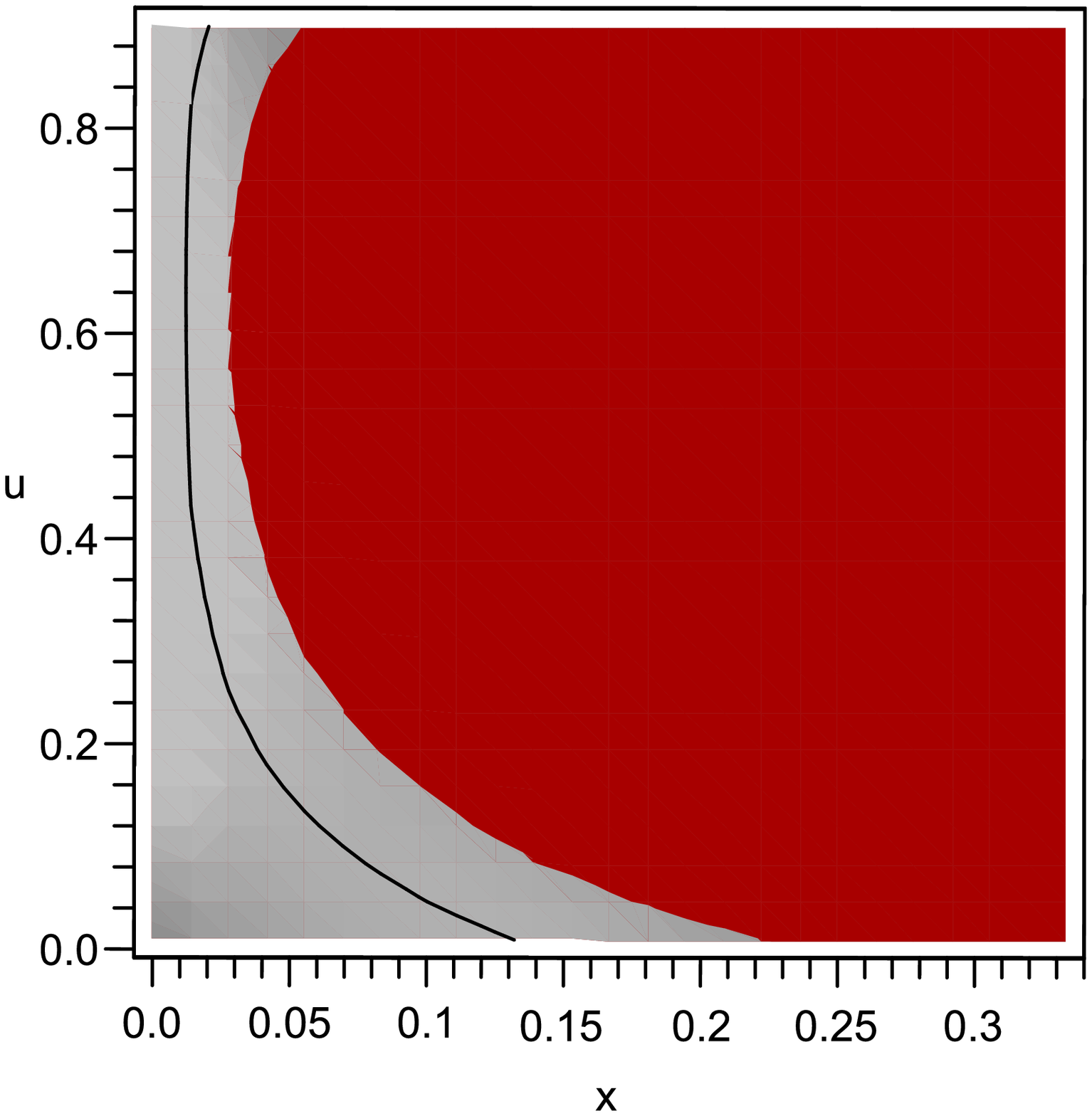}\vspace*{8pt}
\caption{Geometric representation of the bound $c^2_s\geq 0$ in the $xu$-plane for positive coupling $\alpha>0$. For illustrative purposes we consider the exponential potential $V=V_0\exp{(\lambda\phi)}$ -- top panels -- and the power-law potential $V=V_0\phi^{2n}$ -- bottom panels -- for different values of the parameters $\lambda$ and $n$ respectively. As in FIG. \ref{fig01}, here we use the bounded variables $x=\alpha\dot\phi^2/2$ ($0\leq x\leq 1/3$) and $u=y/y+1$ ($0\leq u\leq 1$) where $y=\alpha V$, so that the whole phase plane $xu$ fits into a finite size box. In the top panels, from left to the right: $\lambda=-5$, $\lambda=-2$, $\lambda=2$ and $\lambda=5$, while in the bottom panels: $n=-2$, $n=-1$, $n=1$ and $n=2$, respectively. The red-colored regions are the ones where the squared sound speed is negative ($c^2_s<0$), i. e., where the Laplacian instability eventually develops. It is seen that, although the bound $c^2_s<0$ is always met in some -- even small -- region in the $xu$-plane, for monotonically growing potentials ($\lambda>0|n>0$), i. e. for potentials that do not allow the crossing of the phantom divide, the region of the phase plane where the Laplacian instability arises is appreciably larger.}\label{fig1}\end{centering}\end{figure}

In the FIG. \ref{fig1} we have geometrically represented the bound $c^2_s\geq 0$ for the exponential \eqref{exp-pot} and for the power-law \eqref{pow-law-pot} potentials, for different values of the free parameters $\lambda$ and $n$, respectively. Meanwhile, in FIG. \ref{fig2} we have drawn the surfaces $\omega_\text{eff}=\omega_\text{eff}(x,u)$ and $c^2_s=c^2_s(x,u)$ for the growing exponential potential with $\lambda=5$. In these figures we have used the bounded coordinate in \eqref{u-var}: $$u=\frac{y}{y+1},\;0\leq u\leq 1,$$ instead of $y$ ($0\leq y<\infty$), in order to comprise the whole phase plane into a finite-size region. In the right-hand panel of FIG. \ref{fig2} different orbits of the dynamical system corresponding to the present cosmological model, have been mapped into the surface $c^2_s=c^2_s(x,u)$ in order to show geometrically, that the choice of free parameters that is not compatible with the crossing of the phantom divide -- in this case the growing exponential (positive slope) -- leads eventually to the development of the Laplacian instability. 

\begin{figure}\begin{centering}
\includegraphics[width=3.5cm]{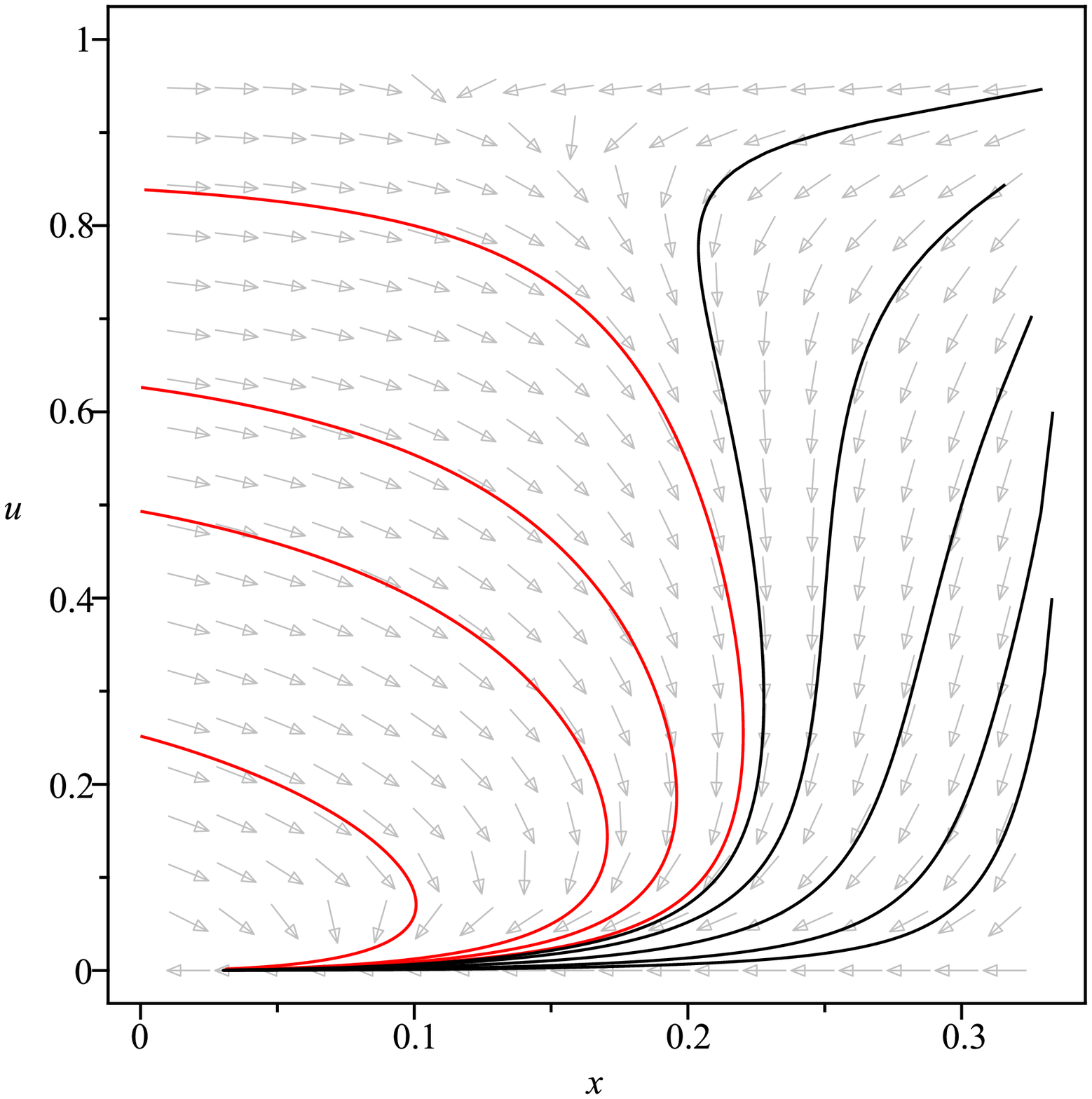}
\includegraphics[width=4.2cm]{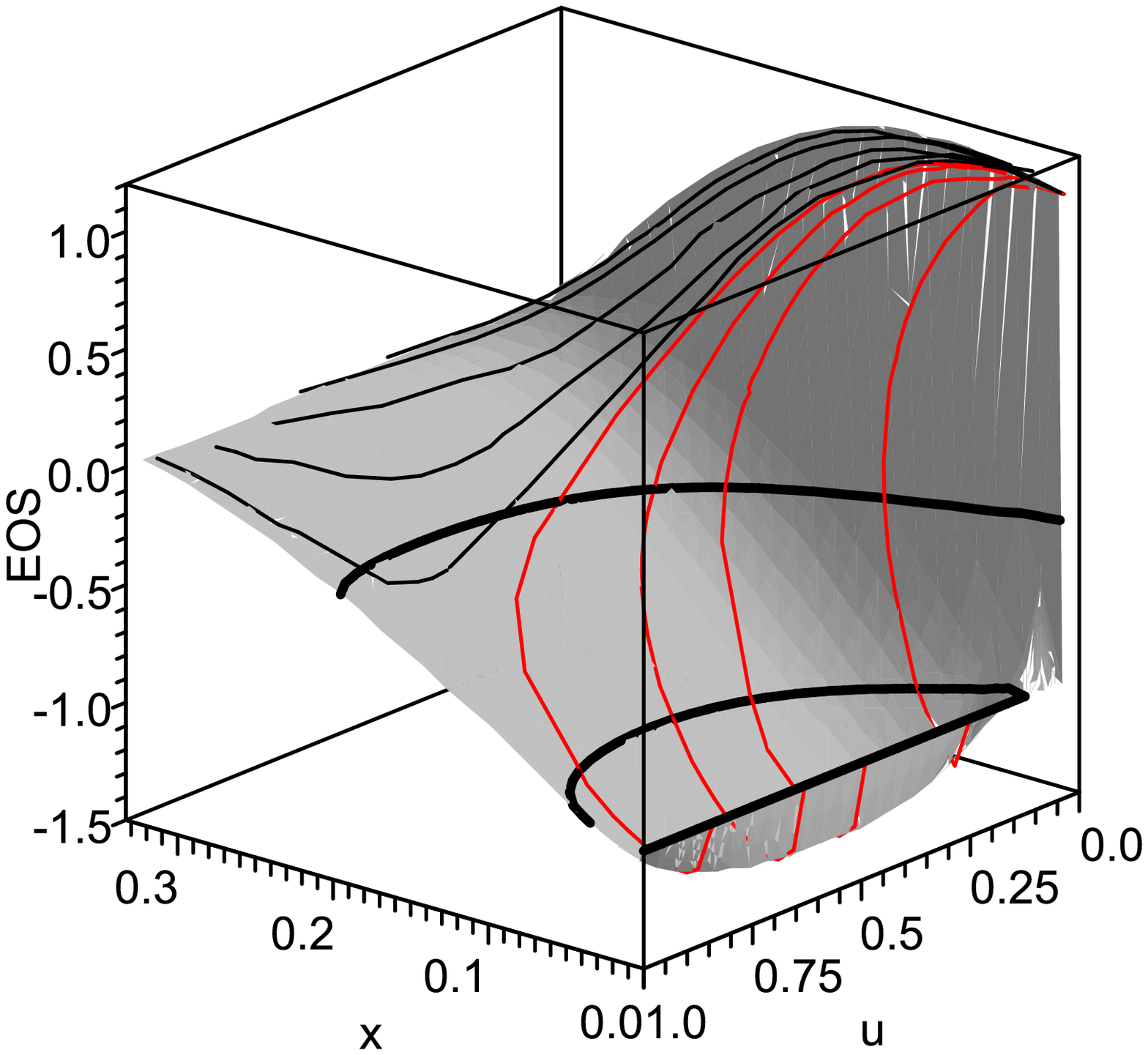}
\includegraphics[width=4.2cm]{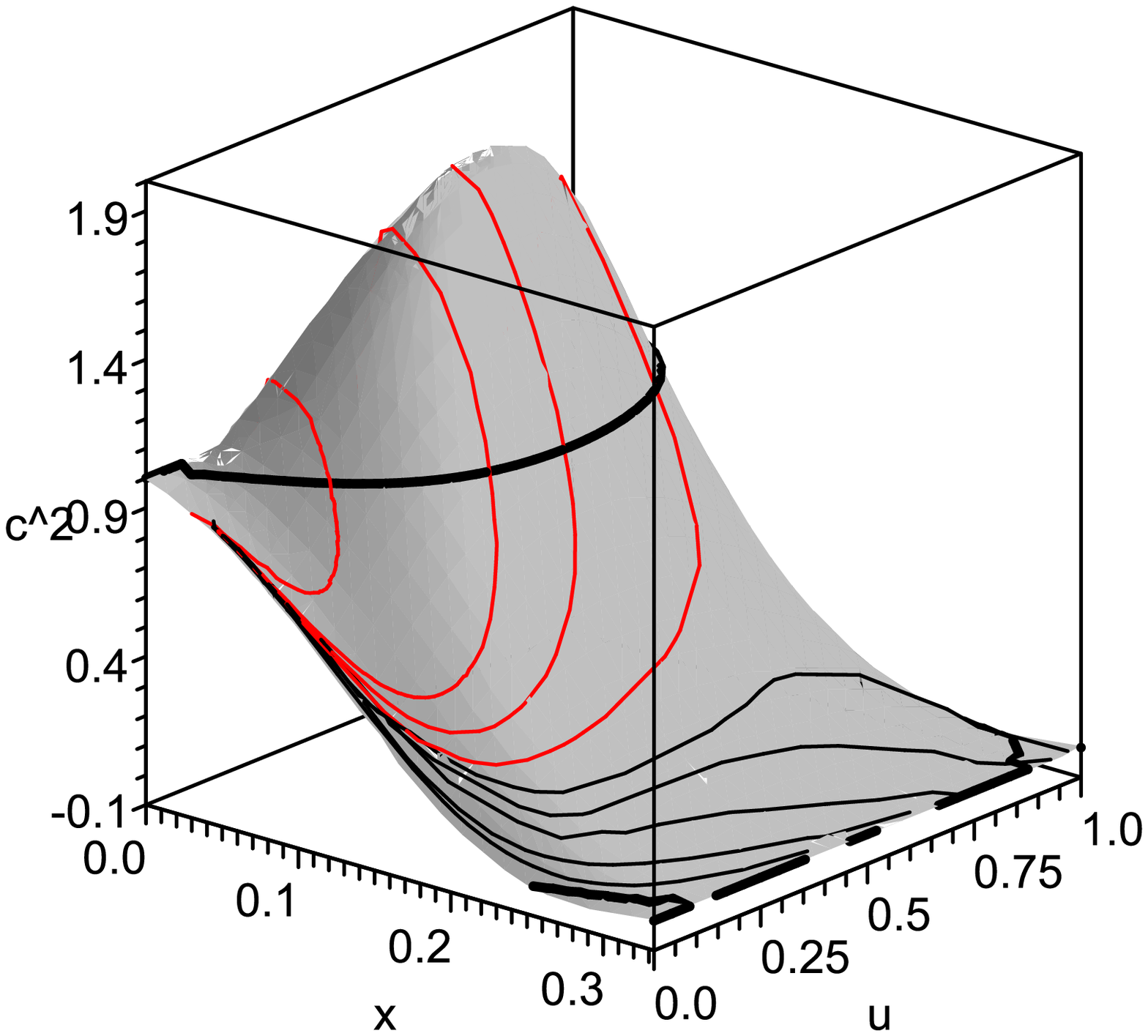}
\includegraphics[width=3.5cm]{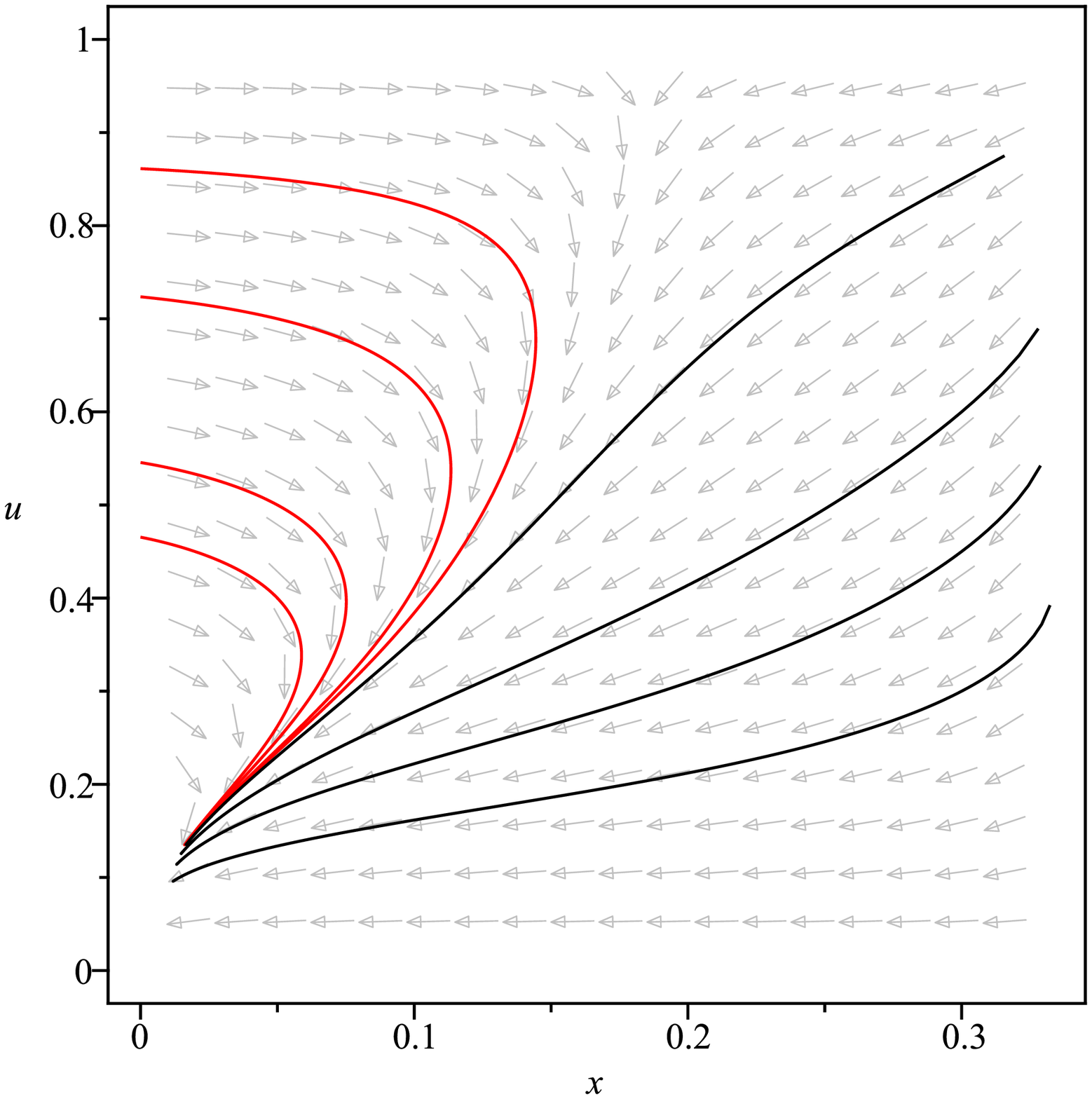}
\includegraphics[width=4.2cm]{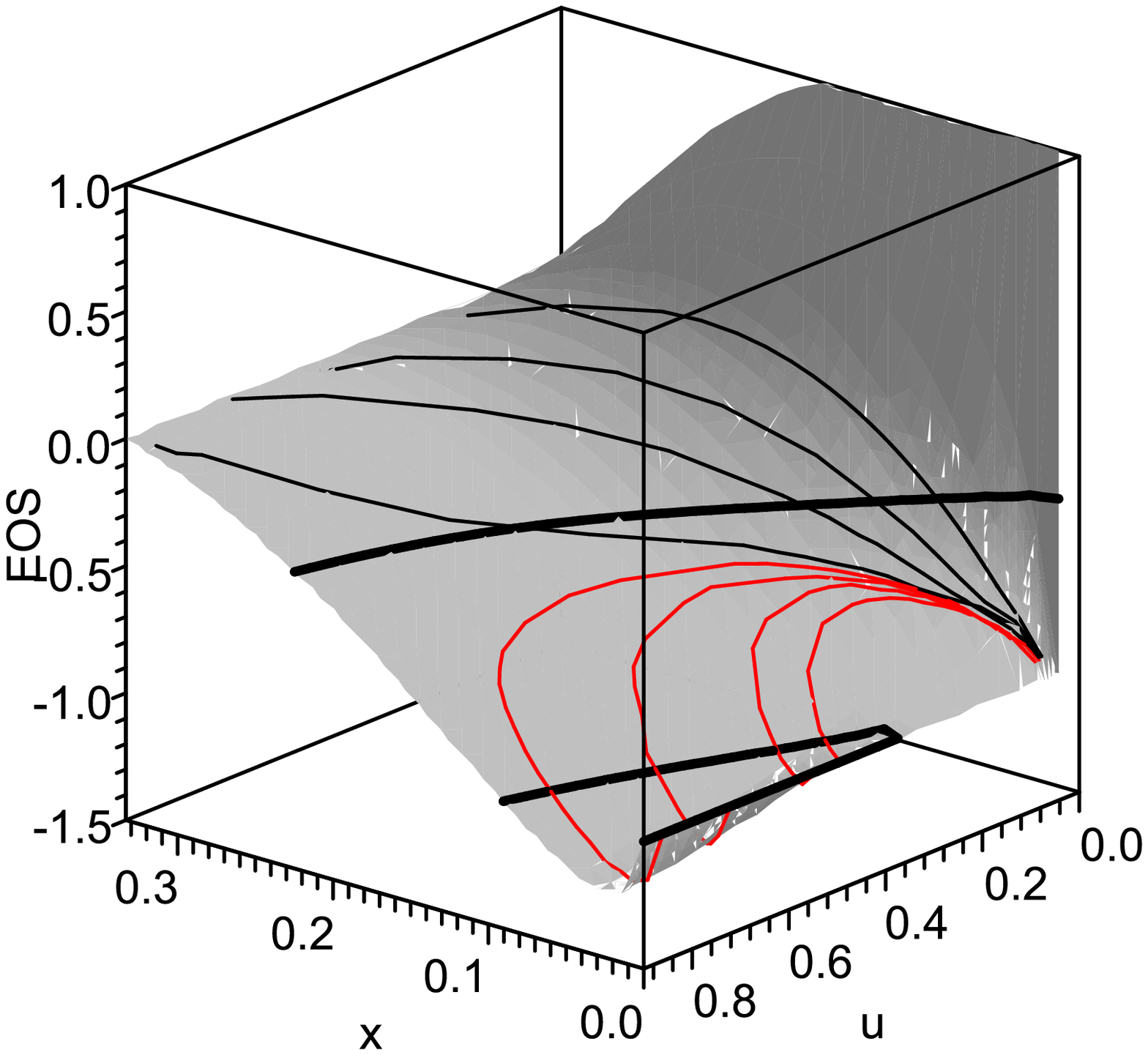}
\includegraphics[width=4.2cm]{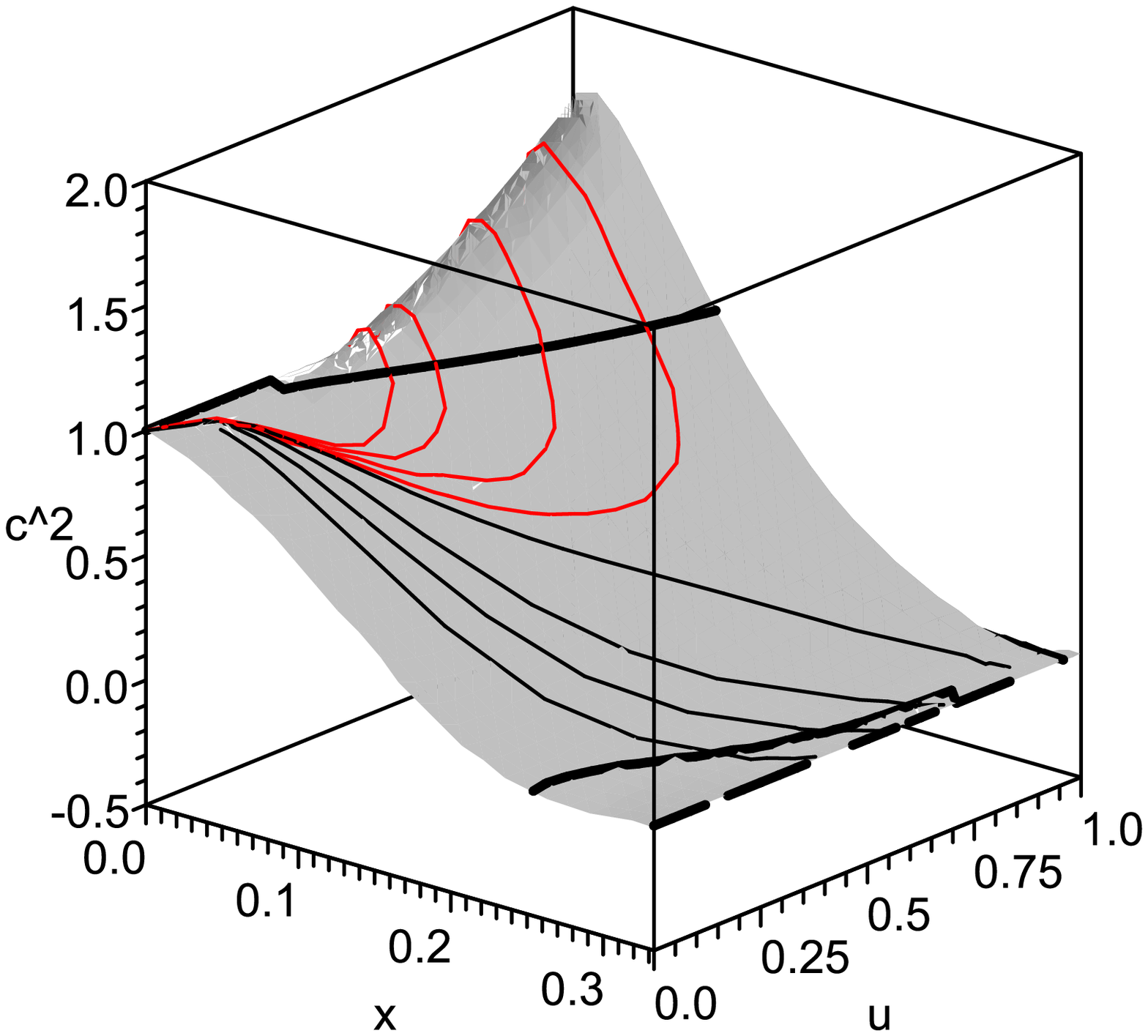}\vspace*{8pt}
\caption{Phase portraits (left) of the dynamical system \eqref{ode-xu}, EOS-embedding diagrams (middle) and $c^2_s$-embedding diagrams (right) corresponding to the cosmological model \eqref{action-dcoup} with positive coupling ($\alpha>0$). In the top panels the decaying exponential potential \eqref{exp-pot} ($\lambda=-5$) has been chosen, while in the bottom panels the inverse power-law potential \eqref{pow-law-pot} ($n=-1$) is considered. In the EOS-embedding diagrams the contours -- thick horizontal curves -- are drawn for $\omega_\text{eff}=-1/3$ (upper contour) and for $\omega_\text{eff}=-1$ (lower contour), while in the $c^2_s$-embeddings the drawn (quite irregular) contours are for $c^2_s=1$ (upper contour) and for $c^2_s=0$ (lower contour). It is seen that the red-colored orbits do the crossing of the phantom divide (middle panels) since these cross through the $\omega_\text{eff}=-1$ contour, and also violate causality since in the right-hand panels these orbits come from domains on the surface $c^2_s=c^2_s(x,u)$ that lie above the contour $c^2_s=1$, representing the local speed of light.}\label{fig3}\end{centering}\end{figure}

As already shown, the potentials that allow for the crossing of the phantom divide can lead also to causality problems. This finding is geometrically illustrated in the figure FIG. \ref{fig3}, where the EOS-embedding and $c^2_s$-embedding diagrams are shown for potentials with the negative slope: (i) decaying exponential potential \eqref{exp-pot} with $\lambda=-5$ (top panels) and (ii) inverse power-law potential \eqref{pow-law-pot} with $n=-1$ (bottom panels), respectively.


\subsubsection{Squared sound speed and the dynamical system}

In order to illustrate the main results discussed here we heavily rely on the properties of the dynamical system corresponding to the cosmological model of interest. Here we give a compact exposition of the most elementary of these properties in connection with the bounds on the squared sound speed. We want to underline that here we do not care about a detailed study of the critical points of the dynamical system and their stability. A detailed dynamical systems study of the present model can be found in Ref. \refcite{huang}. Different orbits in the given phase space will correspond to possible patterns of cosmological evolution that are sustained by the dynamical system and, consequently, by the cosmological equations \eqref{feqs}. Moreover, every possible orbit that can be generated by every possible choice of the initial conditions, represents a potential cosmic history for our universe. The critical points of the dynamical system correspond to ``outstanding'' or generic cosmological solutions of \eqref{feqs}.


Let us discuss on the asymptotic properties of the dynamical system corresponding to the cosmological equations \eqref{feqs} in the phase plane $$\psi=\{(x,y):0\leq x\leq 1/3,y\geq 0\}.$$ The second order cosmological field equations \eqref{feqs} can be traded by the following system of 2 ordinary differential equations on the variables $x$, $y$:

\bea &&x'=\frac{x[\epsilon(1-2x)+y]}{1-3x}-\frac{(1-x)(\epsilon x+y)(\omega_\text{eff}+1)}{2(1-3x)},\nonumber\\
&&y'=y_\phi\sqrt\frac{2x(\epsilon x+y)}{3(1-3x)},\label{ode-xy}\eea where the comma means derivative with respect to the time variable $d\tau=\alpha Hdt$. The problem with \eqref{ode-xy} is that the phase plane is unbounded ($0\leq y<\infty$) so that it may happen that one or several critical points of the dynamical system at infinity are unseen in a finite region of the phase plane. This is why in \eqref{u-var} we introduced the bounded variable $u=y/y+1$ ($0\leq u\leq 1$). After this choice the whole phase plane is shrunk into the phase rectangle:

\bea \psi_{\alpha>0}=\{(x,u):0\leq x\leq 1/3,0\leq u\leq 1\},\label{bound-rect}\eea and the ODE system \eqref{ode-xy} is rewritten as:

\bea &&x'=\frac{x[\epsilon(1-2x)(1-u)+u]}{(1-3x)(1-u)}-\frac{x(1-x)[\epsilon(1-u)+3u][\epsilon(1-2x)(1-u)+u]}{(1-3x)(1-u)^2F_\epsilon}\nonumber\\
&&\;\;\;\;\;\;\;\;\;\;\;\;\;\;\;\;\;\;\;\;\;\;\;\;\;\;\;\;\;\;\;\;\;\;\;\;\;\;\;\;\;\;\;\;\;-\sqrt\frac{2x(1-x)^2(1-3x)[\epsilon x(1-u)+u]}{3(1-u)}\frac{u_\phi}{(1-u)^2F_\epsilon},\nonumber\\
&&u'=u_\phi\sqrt\frac{2x[\epsilon x(1-u)+u]}{3(1-3x)(1-u)},\label{ode-xu}\eea where $$F_\epsilon=\frac{\epsilon(1-3x+6x^2)(1-u)+(1+3x)u}{1-u}.$$ In the left-hand figures in FIG. \ref{fig3} the phase portraits of the dynamical system \eqref{ode-xu} are drawn for the decaying exponential with $\lambda=-5$ (top) and for the inverse power-law with $n=-1$ (bottom), for a set of 9 and 8 different initial conditions respectively.

A crude inspection of \eqref{ode-xu} reveals that, independent of the specific functional form of the self-interaction potential, among the equilibrium configurations of the dynamical system in the phase rectangle \eqref{bound-rect}, there is a critical manifold: ${\cal M}_0=\{(0,u):0\leq u\leq 1\}.$ Equilibrium points in this manifold have different stability properties. The origin ${\cal P}_0:(0,0)$ is a stable critical point. Moreover, it is the global future attractor. The remaining points ${\cal P}_i\in{\cal M}_0$ represent unstable equilibrium configurations and can be only local sources. In the phase portraits (left-hand figures) in FIG. \ref{fig3} the red-color orbits start at local sources in ${\cal M}_0$ and end up at the global attractor ${\cal P}_0$. For points ${\cal P}_\delta:(\delta,u)$ in the neighborhood of ${\cal M}_0$, where $\delta\ll 1$ is a small parameter, we have that: $$c^2_s\approx 1-6\sqrt{2}\,\delta^{1/2}\sqrt\frac{u}{1-u}\,u_\phi,$$ where the terms $\propto\delta$ and of higher orders in the small parameter have been omitted. Hence, if we assume that $u\neq 0$ -- i. e., if exclude the global attractor at the origin -- assuming potentials with the negative slope: $$\der_\phi V<0\Rightarrow y_\phi<0\Rightarrow u_\phi<0,$$ for points in the neighborhood of the critical manifold ${\cal M}_0$, the speed of sound becomes superluminal $c^2_s>1$. This behavior is illustrated in the $c2_s$-embedding diagrams in FIG. \ref{fig3}, where it is appreciated that as the red-colored orbits leave the source points the speed of sound becomes superluminal. Notice that at the source points, as well as at the global attractor at the origin, where $\delta=0$, we have that $c^2_s=1$. For orbits that start at points to the right of the phase rectangle ($x=1/3-\delta$), it is found that there are regions in the phase plane where the squared sound speed becomes negative, signaling the development of Laplacian instability. This is illustrated in the first and second figures (from left to the right) in FIG. \ref{fig1} where the small red-colored regions in the $xu$-plane represent the domains in the phase rectangle where $c^2_s<0$. In the $c^2_s$-embedding diagrams in FIG. \ref{fig3} it is appreciated that several of the mentioned orbits (continuous black curves) indeed meet the gradient instability regions.

It has been shown in Ref. \refcite{quiros_cqg_2018} that the situation is not better for other potentials with either negative and positive slopes and for the pure derivative coupling case (including negative coupling $\alpha<0$): In all cases for a large set of initial conditions there are phase space orbits that eventually reach regions where either superluminality arises or a Laplacian instability develops.


\begin{figure}\begin{centering}
\includegraphics[width=5cm]{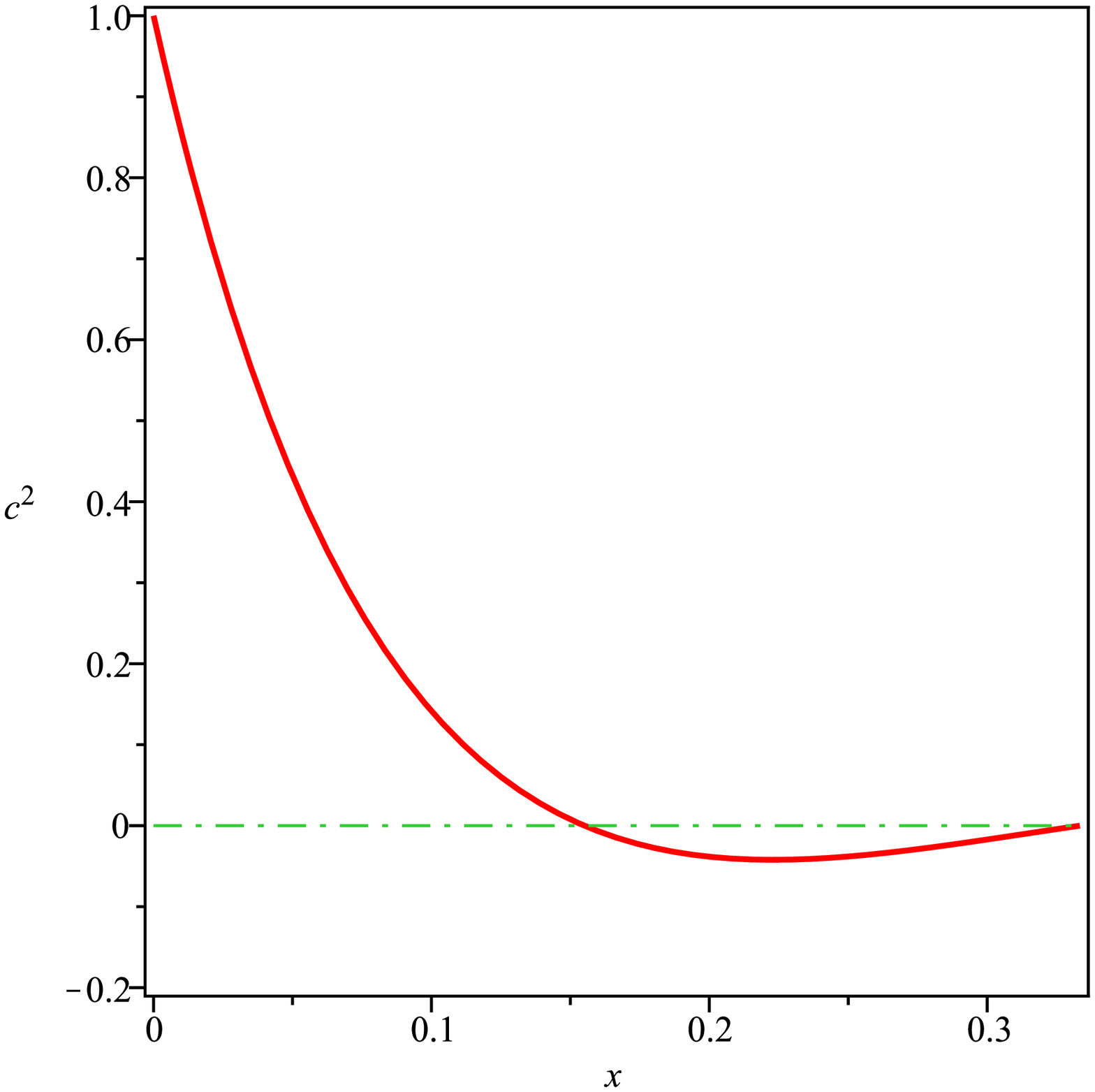}
\includegraphics[width=5cm]{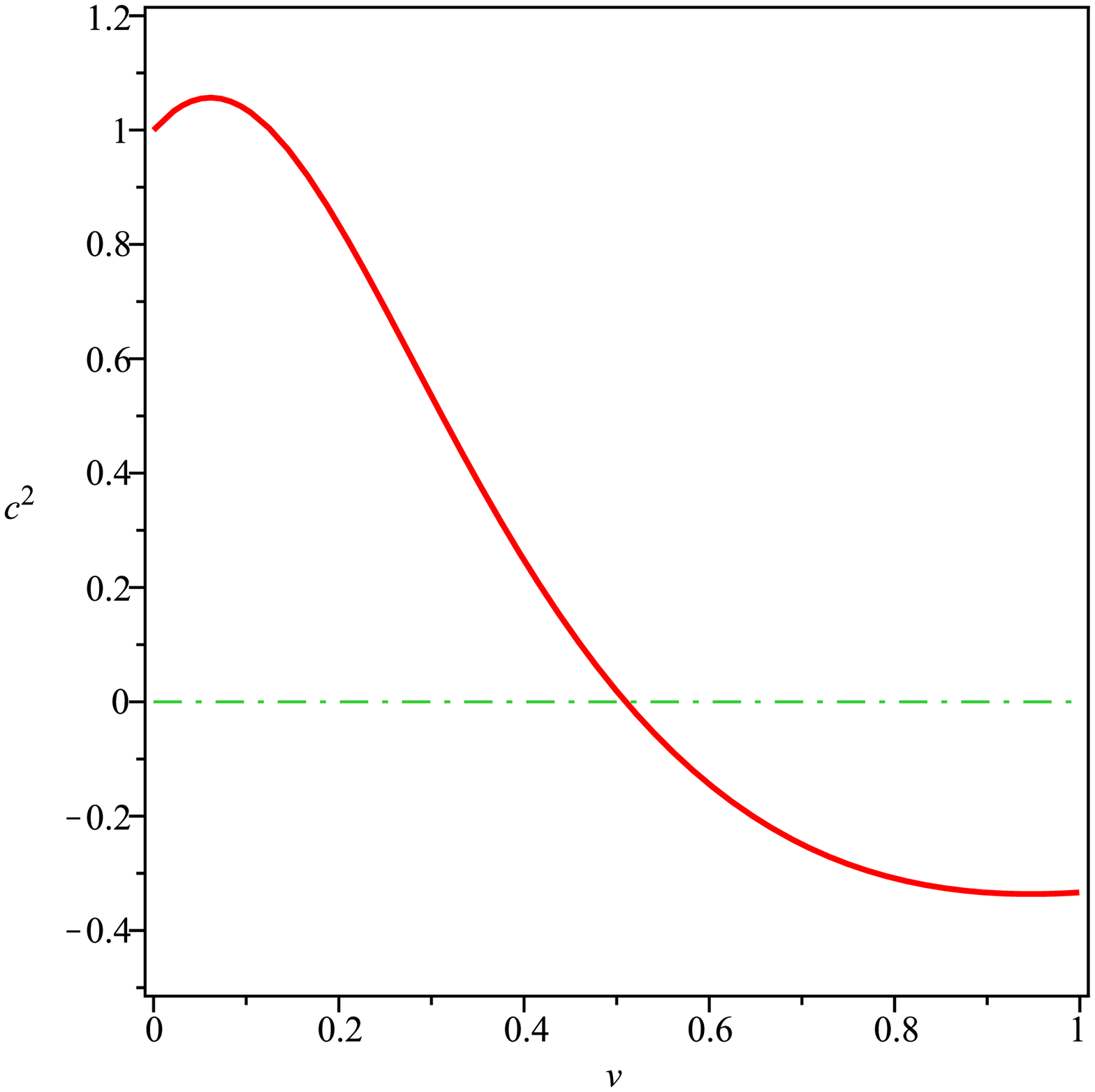}\vspace*{8pt}
\caption{Plot of $c^2_s$ vs $x$ (left) and of $c^2_s$ vs $v$ (right) for the model \eqref{action-dcoup} with the constant potential ($y_0=\alpha V_0$). The left-hand figure is for the positive coupling case $\alpha>0$ ($0\leq x\leq 1/3$), while the right-hand figure is for the negative coupling case $\alpha<0$ ($0\leq v\leq 1$). In the left we have arbitrarily set $y_0=10$, while in the right $y_0=-0.01$. The dash-dot horizontal line marks the lower bound $c^2=0$ on the squared speed of sound. It is appreciated that, independent of the sign of the coupling, there always exist an interval in the $x$/$v$-coordinate where $c^2_s<0$, meaning that a Laplacian instability eventually arises.}\label{figfin}\end{centering}\end{figure}



\subsubsection{Final remarks on causality and stability in the kinetic coupling theory}

The above results show that in general terms, without specifying the functional form of the self-interacting potential, the cosmological models based in the theory \eqref{action-dcoup} where the scalar field is kinetically coupled to the curvature, are unsatisfactory due to the occurrence of causality violations and -- what is more problematic -- of classical Laplacian instabilities, for a non-empty set of initial conditions. These results do not depend on the sign of the coupling constant $\alpha$ in \eqref{action-dcoup} as shown in Ref. \refcite{quiros_cqg_2018}. In the mentioned reference this was shown analytically and also numerically, by specifying the form of the potential. There is, however, a particular class of such models without the potential ($V=0$) and with the constant potential ($V=V_0$) that deserve separate comments since these can be treated in a fully analytical way. 

In general terms theories with the kinetic coupling of the scalar field to the Einstein's tensor -- this is true also for more general Horndeski theories -- all possess some configurations with a superluminal propagation. Besides, these theories have also the speed of propagation of the gravity waves different from the speed of light. In particular, the speed of sound for the scalar perturbations can be subluminal while, simultaneously, the speed of propagation for the gravity waves can be superluminal\cite{germani}. In Ref. \refcite{germani} this has been shown for the theory \eqref{action-dcoup} with the positive coupling and for the quartic potential during inflation. In \eqref{c2t} the squared speed of propagation of the gravity waves perturbations is given independent of the self-interaction potential: 

\bea c^2_T=\frac{1+x}{1-x}.\label{c2t-x}\eea This confirms that the speed of the gravitational waves is always superluminal if assume the positive coupling $\alpha>0$. For the negative coupling, in terms of the bounded variable $v$ ($0\leq v\leq 1$) we have that: 

\bea c^2_T=1-2v.\label{c2t-v}\eea This means that for $0\leq v\leq 1/2$ the speed of propagation of the gravitational waves meets the bounds: $0\leq c^2_T\leq 1$, meanwhile, for $v>1/2$, the squared sound speed of the tensor perturbations is a negative quantity that leads eventually to the development of a Laplacian instability.

In order to further illustrate the above results, let us to discuss in detail the constant potential case: $$V=V_0\Rightarrow y=y_0=\alpha V_0,$$ with the vanishing potential as the particular case when $y_0=0$, that can be studied analytically. We have that (for definiteness we consider $\epsilon=1$):

\bea 3\alpha H^2=\frac{x+y_0}{1-3x}.\label{3h2-y0}\eea Since for the positive coupling $0\leq x\leq 1/3$, from \eqref{3h2-y0} it follows that for $\alpha>0$ the Hubble rate is unbounded from above and bounded from below: $\sqrt{y_0/3\alpha}\leq H<\infty$. For the negative coupling $\alpha<0$ ($-\infty<x\leq 0$) the Hubble rate is bounded (in this case the constant $y_0$ should be a negative quantity as well): 

\bea &&\frac{1}{3\sqrt{-\alpha}}\leq H\leq\sqrt\frac{y_0}{3\alpha}=\sqrt\frac{V_0}{3}\;(V_0>1/|3\alpha|),\nonumber\\
&&\sqrt{V_0/3}\leq H\leq 1/3\sqrt{-\alpha},\;(V_0<1/|3\alpha|).\label{h-bound}\eea For the constant potential the dynamical system \eqref{ode-xy} reduces to a single ordinary differential equation (ODE):

\bea x'=-\frac{2x(1-2x+y_0)}{1-3x}\left[\frac{y_0+(1-3y_0)x-3x^2}{1+y_0+3(y_0-1)x+6x^2}\right].\label{ode-apos}\eea For the positive coupling ($0\leq x<1/3$), one of the critical points of the ODE \eqref{ode-apos} is at the origin $x=0$. This is a stable equilibrium point since linear perturbations $\delta$ around it ($x\rightarrow 0+\delta$) exponentially decay with the time $\tau=\alpha\ln a$: $\delta(\tau)\propto\exp(-2y_0\tau),$ or in terms of the scale factor of the Universe: $$\delta(a)\propto a^{-2\alpha y_0},$$ the perturbations decay as an inverse power-law. The above means that the cosmic dynamics ends up at the de Sitter attractor $x=0$, where $H=H_0=\sqrt{V_0/3}$. Consistently with the fact that, for the positive coupling, the late time dynamics is not modified by the kinetic coupling\cite{sushkov}, the above is the standard late time behavior expected in any scalar field model with a constant potential. For the vanishing potential the asymptotic late time dynamics corresponds to the empty static universe $H=0$, since for this particular case the origin ($x=0$) is the attractor equilibrium configuration as well: The small linear perturbations around the origin decay like $$\delta(\tau)\propto\tau^{-1}\Rightarrow\delta(a)\propto\frac{1}{\alpha\ln a}.$$ This model is plagued by the Laplacian instability as it can be seen from the top figure in FIG. \ref{figfin}, where the squared sound speed is plotted against $x$.

For the negative coupling ($-\infty<x\leq 0$) it is better to use the bounded variable $v=x/x-1$ ($0\leq v\leq 1$). In this case the autonomous ODE \eqref{ode-apos} transforms into:

\bea v'=-\frac{2v(1-v)[1+y_0+(1-y_0)v]}{1+2v}\left[\frac{y_0-(1-y_0)v-2(1+y_0)v^2}{1+y_0+(1-5y_0)v+4(1+y_0)v^2}\right].\label{ode-aneg}\eea Two of the critical points of the ODE \eqref{ode-aneg} are at the origin ($v=0$ $\Leftrightarrow x=0$), and at $v=1$ ($x\rightarrow\infty$). The dynamical equations for linear perturbations $\delta$ around these points read: $\delta'=-2y_0\delta$ and $\delta'=-\delta/3$, respectively. After integration, for perturbations around the origin $v=0$, we get that $\delta(a)\propto a^{-2\alpha y_0}$, i. e., given that both $\alpha$ and $y_0$ are negative for this case, then the perturbations decay with the cosmic expansion. Meanwhile, for perturbations around $v=1$ we get that $\delta(a)\propto a^{-\alpha/3}$ and, since $\alpha$ is negative, then the corresponding perturbation grows with the expansion of the Universe. Hence the point $v=1$ is unstable while the origin $v=0$ is the attractor. Since in this case: $$3\alpha H^2=\frac{y_0-(1+y_0)v}{1+2v},$$ in models with the constant potential (for the negative coupling) the Universe starts a the unstable de Sitter solution with $H=1/3\sqrt{-\alpha}$ and ends up its history at the late-time de Sitter solution with $$3\alpha H^2=y_0\Rightarrow H=H_0=\sqrt{V_0/3}.$$ The asymptotic de Sitter state at $v=1$: $H=1/3\sqrt{-\alpha}$, is to be associated with the primordial inflation \cite{sushkov} and the fact that it is a unstable equilibrium state warrants the natural (required) exit from the early times inflationary stage.\footnote{Transient quasi-de Sitter phases of the cosmic evolution can be found also for other potentials than the constant one.} Notice that for the above picture to make physical sense, in \eqref{h-bound} we have to choose the bottom-line bound, i. e., $V_0<1/|3\alpha|$. Otherwise the attractor would be at higher curvature than the starting point of the cosmic expansion, which is a non-sense from the point of view of the inflationary history of our Universe. 

In spite of the claims that this picture represents an appropriate description of the primordial inflation, according to \eqref{c2t-v} in the neighborhood of the inflationary equilibrium point: $v=1\mp\delta$ ($\delta\ll 1$), for the squared speed of propagation of tensor perturbations we have that: $c^2_T\approx -1\pm 2\delta$, so that the development of a Laplacian instability forbids the -- otherwise unphysical -- inflationary stage in the model. The estimated value of the coupling constant in Ref. \refcite{sushkov-a} is of about:

\bea |\alpha|\sim 10^{-74}\text{sec}^2\approx 10^{-24}\text{GeV}^{-2},\label{a-bound}\eea  where the authors chose the time at which inflation is assumed to start $t\approx 10^{-36}$sec. We may as well choose the time at which inflation is assumed to have ended: $t\approx 10^{-33}$sec. The estimated value for the coupling in this case is about 4 orders of magnitude larger: 

\bea |\alpha|\sim 10^{-70}\text{sec}^2\approx 10^{-20}\text{GeV}^{-2}.\label{a-bound'}\eea If combine the above estimates with the tight constraint on the difference in speed of photons and gravitons $|c^2_T-1|\leq 10^{-15}$ (recall that in this review we have chosen the units where $c^2=1$) implied by the announced detection of gravitational waves from the neutron star-neutron star merger GW170817 and the simultaneous measurement of the gamma-ray burst GRB170817A \cite{ligo}, since according to \eqref{c2t-x}: $$c^2_T-1=\frac{2x}{1-x}\Rightarrow 2x\leq 10^{-15},$$ we get that $\dot\phi^2\leq 10^5-10^9$GeV$^2$, i. e., $\dot\phi^2\leq 10^{-33}-10^{-29}M_\text{pl}$, where $M_\text{pl}\approx 10^{19}$GeV is the Planck mass. These estimates leave not much freedom for the scalar field to behave different from an effective cosmological constant. 

The above exposed picture is overshadowed by the stability problems associated with the scalar and tensor modes of the perturbations whose energy density grows without bound due to fact that, for these modes it may happen that $c^2_s<0$ ($c^2_T<0$). In the bottom figure in FIG. \ref{figfin} the plot of $c^2_s$ vs $v$ is drawn for $y_0=-0.01$. The conditions for the development of the Laplacian instability ($c^2_s<0$) are evident in the figure, in particular for points in the neighborhood of (including) the source equilibrium configuration that can be associated with the primordial inflation. Besides, in the neighborhood of this point we have also that $c^2_T<0$, so that the tensor modes are classically unstable as well.


\subsection{Speed of scalar and of tensor perturbations in Horndeski theories}\label{speed-grav-w}

As we have already said, an aspect of the study of the Galileon models that has gained interest recently, is related with the tight constraint on the difference in speed of photons and gravitons 

\bea |c^2_T-c^2|\leq 6\times 10^{-15}c^2,\label{const-gw}\eea where $c_T$ is the speed of the gravitational waves (recall that here $c^2=1$), implied by the announced detection of gravitational waves from the neutron star-neutron star merger GW170817 and the simultaneous measurement of the gamma-ray burst GRB170817A\cite{ligo}. Take for instance, the Horndeski-type theory with kinetic coupling of the scalar field to the Einstein's tensor: $\alpha G_{\mu\nu}\der^\mu\phi\der^\nu\phi$\cite{gao, sushkov, saridakis-sushkov, matsumoto, granda, germani-prl}. In this theory, as discussed above in subsection \ref{subsect-k-coup}, the squared speed of sound of the gravitational waves is given by\cite{germani}: 

\bea c^2_T=\frac{2+\alpha\dot\phi^2}{2-\alpha\dot\phi^2},\label{c2t-germani}\eea so that, depending on the kinetic energy of the scalar field, either a Laplacian instability develops ($\alpha\dot\phi^2>2$) or the gravitational waves may travel at superluminal speed ($\alpha\dot\phi^2<2$). Hence, in this theory the speed of propagation of the gravitational waves may substantially differ from the local speed of light thus rendering the resulting cosmological model incompatible with the above constraint \eqref{const-gw}. In contrast, as we shall see, the cubic Galileon model that is based on \eqref{qbic-gal-action}, is not constrained by the above mentioned combined detection of gravitational waves from the neutron star merger GW170817 and the simultaneous measurement of GRB170817A reported in Ref. \refcite{ligo}, since for the cubic Galileon model the speed of propagation of the tensor perturbations (gravitational waves) exactly coincides with the local speed of light: $c_T=1$.

Here we shall discuss on the speed of propagation of the tensor and scalar perturbations in Horndeski theories in general, so that the model with the kinetic coupling to the Einstein's tensor and the cubic Galileon, are particular cases. If follow the perturbative procedure of Ref. \refcite{defelice(horn_perts)_plb_2011} (see subsection \ref{subsect-horn-geff}), it can be found that the speed of propagation (squared) of the tensor perturbations in the Horndeski theories is given by the following expression\cite{g-infl-2}:

\bea c^2_T=\frac{{\cal F}_T}{{\cal G}_T}=\frac{G_4-X\left(\ddot\phi G_{5,X}+G_{5,\phi}\right)}{G_4-2XG_{4,X}-X\left(H\dot\phi G_{5,X}-G_{5,\phi}\right)},\label{c2t-horn}\eea where we have substituted the quantities ${\cal F}_T$ and ${\cal G}_T$ from \eqref{horn-perts-usef-quant}. In a similar way it can be found that the sound speed squared (speed of propagation of the scalar perturbation) is given by\cite{g-infl-2}:

\bea c^2_S=\frac{\frac{1}{a}\frac{d}{dt}\left(\frac{a}{\Theta}{\cal G}^2_T\right)-{\cal F}_T}{\frac{\Sigma}{\Theta^2}{\cal G}_T^2+3{\cal G}_T},\label{c2s-horn}\eea where the expansion $\Theta$ is to be substituted from \eqref{horn-expansion}, and 

\bea &&\Sigma:=XK_{,X}+2X^2K_{,XX}+12H\dot\phi XG_{3,X}+6H\dot\phi X^2G_{3,XX}-2XG_{3,\phi}\nonumber\\
&&\;\;\;\;\;\;\;-2X^2G_{3,\phi X}-6H^2G_4+6H^2\left(7XG_{4,X}+16X^2G_{4,XX}+4X^3G_{4,XXX}\right)\nonumber\\
&&\;\;\;\;\;\;\;-6H\dot\phi\left(G_{4,\phi}+5XG_{4,\phi X}+2X^2G_{4,\phi XX}\right)+30H^3\dot\phi XG_{5,X}+26H^3\dot\phi X^2G_{5,XX}\nonumber\\
&&\;\;\;\;\;\;\;+4H^3\dot\phi X^3G_{5,XXX}-6H^2X\left(6G_{5,\phi}+9XG_{5,\phi X}+2X^2G_{5,\phi XX}\right).\label{sigma}\eea In order for the Horndeski theories to be free of Laplacian instabilities it is required that the following bounds be jointly met: 

\bea {\cal F}_T\geq 0,\;\;{\cal G}_T>0,\;\;\frac{1}{a}\frac{d}{dt}\left(\frac{a}{\Theta}{\cal G}^2_T\right)-{\cal F}_T\geq 0,\;\;\frac{\Sigma}{\Theta^2}{\cal G}_T^2+3{\cal G}_T>0,\label{cond-ghost-free}\eea while if one requires, besides, that there would not be causality issues, the following constraints should be satisfied as well:

\bea 0\leq{\cal F}_T\leq{\cal G}_T,\;\;0\leq\frac{1}{a}\frac{d}{dt}\left(\frac{a}{\Theta}{\cal G}^2_T\right)-{\cal F}_T\leq\frac{\Sigma}{\Theta^2}{\cal G}_T^2+3{\cal G}_T.\label{cond-causal}\eea

From \eqref{c2t-horn} it immediately follows that in Horndeski theories with $G_5=0$, $G_4(\phi,X)=f(\phi)$ (i. e., the coefficient $G_4$ is not a function of $X$), the tensor perturbations propagate at the speed of light: $c_T^2=1$. General relativity with a scalar field (the basis for quintessence and also for K-essence models), Brans-Dicke and NMC theories, and also the cubic Galileon, are examples of theories that belong in this class. For the BD cubic Galileon\cite{kazuya_gal}, for instance, $G_4=\phi$, while ,$G_4=1/2$, for the cubic Galileon model given by \eqref{qbic-gal-action}. In both cases $G_4\neq f(X)$ and $G_5=0$. Hence, no matter which model for the cubic Galileon to choose, the speed squared of the gravitational waves coincides with the speed of light. Meanwhile, for the kinetic coupling theory \eqref{k-coup-action}, since, $G_4=1/2$, $G_5=-\alpha\phi/2$, then the speed squared of the tensor perturbations \eqref{c2t-germani}; $$c_T^2=\frac{1+\alpha X}{1-\alpha X},$$ may vary from point to point in spacetime, thus making the theory cosmologically highly improbable due to the tight constraints on $c_T$ coming from the GW170817, GRB170817A events. For the covariant Galileon\cite{nicolis_gal, deffayet_vikman_gal, deffayet_deser_gal, deffayet_prd_2011, fab_4_prl_2012, deffayet-rev}, since, $$K=-2c_2,\;G_3=-2c_3 X,\;G_4=\frac{1}{2}-4c_4 X^2,\;G_5=-4c_5 X^2,$$ the speed squared of the tensor perturbations is given by: $$c_T^2=\frac{1-8c_4X^2+16c_5X^2\ddot\phi}{1+24c_4X^2+16c_5X^2H\dot\phi},$$ so that there can be similar problems of causality and Laplacian instabilities may arise as well.

The fact that the speed of the tensor perturbations coincides with the speed of light does not mean that the sound speed squared (the speed of the scalar perturbations) equals that of light also. So that, even in this case, one has to take care about possible causality issue and/or gradient instability. Take, for instance, the cubic Galileon model depicted by the action \eqref{qbic-gal-action}. This model is specified by the following choice of functions in the Horndeski Lagrangians: $K=X-V(\phi)$, $G_3=\sigma X$ (for simplicity we choose $\sigma=$ constant), $G_4=1/2$, $G_5=0$. As mentioned in the above paragraph, for this choice $c_T^2=1$, i. e., the speed squared of the tensor perturbations equals that of light. However, according to \eqref{c2s-horn}: $$c_S^2=\frac{-\dot H+\sigma X\left(3\ddot\phi+H\dot\phi-2\sigma X^2\right)}{X\left[1+6\sigma\left(H\dot\phi+\sigma X^2\right)\right]}.$$ In dependence of the sign of the derivatives of the Hubble parameter and of the scalar field, this expression for the sound speed squared can take even negative values, signaling the occurrence of a gradient instability. Another example can be the K-essence models where $K=K(\phi,X)$, $G_4=1/2$, and $G_3=G_5=0$. In this case equation \eqref{c2s-horn} can be written in the following form: $$c^2_S=-\frac{\dot H}{XK_{,X}+2X^2K_{,XX}},$$ so that nothing forbids the occurrence of gradient instabilities and/or of causality issues. Recall that for both models: cubic Galileon and K-essence, the speed of the tensor perturbations $c^2_T=1$.


\section{Screening mechanisms}\label{sect-screen}

Although the search for the elusive scalar field in terrestrial experiments seems to have given a positive result with the discovery of the Higgs particle\cite{higgs_prl_1964, higgs_phys_lett_1964, englert_prl_1964, guralnik_prl_1964, higgs_discov_1, higgs_discov_2, higgs_discov_3, higgs_discov_4}, the search for the scalar-field as a co-carrier of the gravitational interactions in local (solar system) experiments, as well as in the cosmological context, has been less successful. A possible explanation of the elusiveness of the gravitational scalar-field relies on the so called screening mechanisms, that allow to hide it from the reach of the local experiments. There are known several screening mechanisms within the STT, however, in this section we shall explore only two of the most efficient ways which scalar fields may have found to evade local searches: i) the chameleon and ii) the Vainshtein screening mechanisms.


\subsection{Chameleon fields}\label{subsect-cham}

Although many aspects of BD theory have been well-explored in the past \cite{fujii_book_2004, faraoni-book}, other aspects have been cleared up just recently. Thanks to the chameleon effect \cite{cham, cham-1, cham-khoury, cham-wei, cham-tamaki, cham-mota, cham-olive, cosmo-cham, cham-5-force, cham-rad, cham-fdr, cham-bd-1, cham-bd, bisabr_ass_2014, rev-cham}, for instance, it was just recently understood that the experimental bounds on the BD coupling parameter $\omega_\textsc{bd}$, which were set up through experiments in the solar system, might not apply in the large cosmological scales if consider BD theory with a potential. According to the chameleon effect, the effective mass of the scalar field $m_\phi$, depends on the background energy density of the environment: In the large cosmological scales where the background energy density is of the order of the critical density $\rho_\text{crit}\sim 10^{-31}$ g/cm$^3$, the effective mass is very small $m_\phi\sim H_0\sim 10^{-33}$ eV, so that the scalar field has impact in the cosmological dynamics. Meanwhile, in the solar system, where the averaged energy density of the environment is huge compared with $\rho_\text{crit}$, the effective mass is large $m_\phi>1$ mm$^{-1}$ ($m_\phi> 10^{-3}$ eV), so that the Yukawa--like contribution of the scalar field to the gravitational interaction $\propto e^{-m_\phi r}/r$, is short-ranged, leading to an effective screening of the scalar field in the solar system. Below we shall briefly sketch the mathematics of chameleon effect. We shall heavily rely on Refs. \citen{cham, cham-1, cham-wei}. The commonly studied mathematical model is based in the following action:

\bea S_\text{cham}=\int d^4x\sqrt{|g|}\left[\frac{1}{2}R-\frac{1}{2}(\der\phi)^2-V(\phi)\right]+\int d^4x\sqrt{|g|}{\cal L}_m(\psi^{(i)},g^{(i)}_{\mu\nu}),\label{khoury-action}\eea where the $\psi^{(i)}$ account for the different matter degrees of freedom. The chameleon $\phi$ interacts directly with the matter degrees of freedom through conformal couplings. This is realized by allowing the different matter species $\psi^{(i)}$ to couple to different conformal metrics $g^{(i)}_{\mu\nu}$;

\bea g^{(i)}_{\mu\nu}=e^{2\beta_i\phi}g_{\mu\nu},\label{cham-coup}\eea where $g_{\mu\nu}$ is the ``Einstein's frame metric'' -- in the discussion on the conformal transformations in section \ref{sect-cf} it will be clear why we put the quotation marks --  and the $\beta_i$-s are dimensionless constants of order unity\cite{cham, cham-1}. The KG motion equation derived from \eqref{khoury-action} reads:

\bea \Box\phi=\der_\phi V-\sum_i\beta_i e^{4\beta_i\phi}g^{\mu\nu}_{(i)}T^{(i)}_{\mu\nu},\label{khoury-kg-eq}\eea where $T^{(i)}_{\mu\nu}$ is the stress-energy tensor of the $i$-th matter species. For non-relativistic dust-like matter $g^{\mu\nu}_{(i)}T^{(i)}_{\mu\nu}=-\rho_m^{(i)}$ (the density of the $i$-th matter species). A non-trivial assumption of the chameleon model of Refs. \citen{cham, cham-1} is that the matter density that has to be considered in the motion equations is not $\rho_m^{(i)}$ but its EF counterpart, $\hat\rho_m^{(i)}=\rho_m^{(i)}\exp{(3\beta_i\phi)}$, which is the one that is conserved in the Einstein's frame. Hence, the KG motion equation for the chameleon becomes:

\bea \Box\phi=\der_\phi V+\sum_i\beta_i e^{\beta_i\phi}\hat\rho_m^{(i)}.\label{khoury-kg-eq'}\eea A look at this equation shows that the dynamics of the chameleon is governed not by $V(\phi)$ alone, but by the effective potential, 

\bea V_\text{eff}(\phi,\hat\rho_m^{(i)})=V(\phi)+\sum_i\hat\rho_m^{(i)}\,e^{\beta_i\phi},\label{khoury-eff-pot}\eea which is a function not only of the scalar field, but also depends on the matter density of the environment. The main idea behind the chameleon effect is that the effective potential, $V_\text{eff}(\phi)$, is a minimum at some $\phi_\text{min}$ even if $V(\phi)$ is monotonic. Actually, whenever, $V(\phi)$, is monotonically decreasing function and $\beta_i>0$, or $V(\phi)$ is monotonically increasing while $\beta_i<0$, the effective potential is a minimum at a value of the field $\phi_\text{min}$, which satisfies; $$\der_\phi V_\text{eff}(\phi_\text{min})=\der_\phi V(\phi_\text{min})+\sum_i\beta_i\hat\rho^{(i)}_m e^{\beta_i\phi_\text{min}}=0.$$ The mass of small fluctuations about $\phi_\text{min}$:

\bea m^2_\text{eff}=\der_\phi^2 V_\text{eff}(\phi_\text{min})=\der^2_\phi V(\phi_\text{min})+\sum_i\beta^2_i\hat\rho^{(i)}_m e^{\beta_i\phi_\text{min}},\label{khoury-mass}\eea is a function of the environmental matter density as well. Hence, the scalar field acquires a mass which depends on the local matter density. The denser the environment, the more massive the chameleon is. The effective mass, in turn, determines the reach of the Yukawa-type interaction, $\phi\propto e^{-m_\text{eff}r}/r$. The larger the effective mass the weaker the fifth-force associated with the chameleon field is, i. e., the faster the interaction decays with the distance. It is expected that the effective mass of the scalar field is sufficiently large on Earth so as to evade current constraints on violation of the Einstein's equivalence principle (EEP) and on fifth-force\cite{will-lrr-2014}.

\subsubsection{Thin-shell effect}

The thin-shell effect is appreciable only for large bodies. Its physical basis may be understood if assume that the effective mass of the chameleon fluctuations about the minimum of the effective potential inside the large body is large enough, so that, but for a thin-shell below the surface of the body, the chameleon field is effectively screened. Hence, it is the contribution coming from this thin shell the one that counts. 

In order to sketch the mathematical basis for the thin-shell effect let us assume a single matter species so that we drop the index '$i$'. We restrict our discussion to the static, spherically symmetric case. Consider a spherically symmetric compact body of radius $R$, with homogeneous density $\hat\rho$ and mass $M=4\pi\hat\rho R^3/3$. Ignoring back-reaction effects the KG motion equation \eqref{khoury-kg-eq'} becomes: 

\bea \frac{d^2\phi}{dr^2}+\frac{2}{r}\frac{d\phi}{dr}=\der_\phi V+\beta\hat\rho(r)\,e^{\beta\phi}.\label{cham-eom}\eea Appropriate boundary conditions: $$\left.\frac{d\phi}{dr}\right|_{r=0}=0,\;\;\lim_{r\rightarrow\infty}\phi=\phi_\text{out},$$ and continuity of $\phi$ and of its derivative $d\phi/dr$ at the boundary of the body, $r=R$, are required. Here $\phi_\text{in}$, $\phi_\text{out}$, are the values of the field that minimize the effective potential inside and outside of the compact object, respectively. Hence, $m_\text{in}$ and $m_\text{out}$ are the masses of the fluctuations of the field about $\phi_\text{in}$ and $\phi_\text{out}$, respectively. The approximate solution of \eqref{cham-eom} outside of the compact object reads: $$\phi(r)\simeq-\left(\frac{3\beta}{4\pi}\right)\left(\frac{\Delta R}{R}\right)\frac{M\;e^{-m_\text{out}(r-R)}}{r}+\phi_\text{out},$$ with $$\frac{\Delta R}{R}\simeq\frac{8\pi(\phi_\text{out}-\phi_\text{in})}{6\beta M/R},$$ where $\Delta R$ is the thickness of the thin shell. The $\phi$-profile outside large objects is suppressed by a factor $\Delta R/R\ll 1$. For details of the computations and for estimates see Refs. \citen{cham, cham-1}.


\subsection{Brans-Dicke chameleon}\label{subsect-bd-cham}

There is one aspect of the chameleon effect that we want to discuss in detail. If take a look at the existing bibliography on this subject -- see the above subsection \ref{subsect-cham} -- one immediately finds that this effect is almost exclusively described in the Einstein frame, where the chameleon is minimally coupled to the curvature scalar, at the cost of being non-minimally coupled to the matter sector of the action. This means that there is a fifth-force effect that deviates particles' paths from the geodesics of the gravitational metric or, in other words: in terms of the gravitational metric the matter stress-energy tensor is not conserved. In this case one expects that the chameleon effect may screen the fifth-force from local experiments. In what concerns to BD gravity theory the chameleon effect has been investigated by including, in addition to the usual NMC of the scalar field with the curvature, also a non-minimal coupling between the BD field and the matter Lagrangian\cite{cham-bd-1, cham-bd, bisabr_ass_2014}, so that there is a fifth-force effect that is to be screened by the chameleon. 

But, what about standard (Jordan frame) BD theory with a self-interacting scalar field? In this case, given that the matter fields interact with the BD field only gravitationally, there is not any fifth-force to be screened. This is why, in the first place the Brans-Dicke theory is a metric theory of gravity\cite{will-lrr-2014}, while theories where the matter degrees of freedom are non-minimally coupled to the scalar field are not metric. Why then would one search for a screening mechanism like the chameleon effect in this case? The answer is straightforward: Because one needs to weaken the very stringent constraint on the BD coupling parameter $\omega_\text{BD}>4\times 10^4$ -- the only free parameter of BD theory -- if the Brans-Dicke theory of gravity is to be differentiated from GR. A detailed description of the chameleon effect in the JF can be found only in Ref. \refcite{quiros_prd_2015}, where the BD chameleon is minimally coupled to the matter sector as it is customary in BD theories. Since the chameleon effect is apparent in the density dependence of the dilaton's mass, we think that the absence of appropriate discussion in the JF, is due to the unconventional way which the self-interaction potential of the dilaton arises in the corresponding Klein-Gordon equation that governs its dynamics. 

Here we shall focus in the description of the chameleon effect in the Jordan frame of the BD theory with a potential\cite{quiros_prd_2015}. It will be shown that in a cosmological context, provided that the effective chameleon potential has a minimum within a region of constant matter density, the GR-de Sitter solution can be, at most, either a local attractor or a saddle point of the BD theory within that region. In contrast, as it has been shown in Refs. \citen{hrycyna, hrycyna-1, hrycyna-2, quiros-prd-2015-1} by means of the tools of the dynamical systems theory, in a cosmological setting the GR-de Sitter solution can be a global attractor of the BD theory exclusively for the quadratic potential: $V(\phi)=M^2\phi^2$, or for any BD potential that asymptotes to the quadratic one, $V(\phi)\rightarrow M^2\phi^2$ (see section \ref{sect-dsyst-bd}). There are found several works on the de Sitter (inflationary) solutions within the scalar-tensor theory in the bibliography, in particular within the BD theory but, just for illustration here we mention those in Refs. \citen{barrow-ref, barrow-ref-1, odintsov-ref}.

We assume the BD theory\cite{bd-1961, brans_phd_thesis, dicke-1962} with the potential\cite{fujii_book_2004, faraoni-book}, to dictate the dynamics of gravity and matter. In the Jordan frame\footnote{Sometimes it is convenient to rescale the BD scalar field and, consequently, the self-interaction potential: $$\phi=e^\vphi,\;V(\phi)=e^\vphi\,U(\vphi),$$ so that, the action \eqref{bd-action} is transformed into the string frame BD action: $$S_\textsc{sf}^\vphi=\int d^4x\sqrt{|g|}e^\vphi\left\{R-\omega_\textsc{bd}(\der\vphi)^2-2U+2e^{-\vphi}{\cal L}_m\right\}.$$} it is depicted by the action \eqref{bd-action} or by the corresponding equations of motion \eqref{bd-feq}, \eqref{kgbd-eq}:

\bea &&G_{\mu\nu}=\frac{1}{\phi}\,T^{(m)}_{\mu\nu}+\frac{\omega_\textsc{bd}}{\phi^2}\left[\der_\mu\phi\der_\nu\phi-\frac{1}{2}g_{\mu\nu}\left(\der\phi\right)^2\right]-g_{\mu\nu}\frac{V}{\phi}+\frac{1}{\phi}\left(\nabla_\mu\der_\nu\phi-g_{\mu\nu}\Box\phi\right),\nonumber\\
&&\Box\phi=\frac{2}{3+2\omega_\textsc{bd}}\left(\phi\der_\phi V-2V+\frac{1}{2}\,T^{(m)}\right),\label{bd-mot-eq}\eea where, as already pointed out, $\Box\equiv g^{\mu\nu}\nabla_\mu\nabla_\nu$, is the D'Alembertian operator, and $$T^{(m)}_{\mu\nu}=-\frac{2}{\sqrt{|g|}}\frac{\der\left(\sqrt{|g|}\,{\cal L}_{m}\right)}{\der g^{\mu\nu}},$$ is the conserved stress-energy tensor of the matter degrees of freedom $$\nabla^\mu T^{(m)}_{\mu\nu}=0.$$


\subsubsection{The Klein-Gordon equation and the mass of the scalar field}

The mass (squared) of the BD scalar field can be computed with the help of the following equation\cite{mass}: 

\bea m_\phi^2=\frac{2}{3+2\omega_\textsc{bd}}\left[\phi \der^2_\phi V(\phi)-\der_\phi V(\phi)\right].\label{bd-mass}\eea This mass is the one which is associated with a Yukawa-like term $\phi(r)\propto\exp(-m_\phi r)/r$, when the Klein-Gordon equation in \eqref{bd-mot-eq} is considered in the weak-field, slow-motion regime, and in the spherically symmetric case. For completeness of our exposition, here we shall explain the main reasoning line behind this result\cite{mass}.

In general for a scalar field which satisfies the standard KG equation 

\bea \Box\phi=\der_\phi V_\text{eff}+S,\label{kg-s-eq}\eea where $V_\text{eff}=V_\text{eff}(\phi)$ is the effective self-interaction potential of the scalar field $\phi$, and $S$ is a source term ($S$ does not depend on $\phi$), the effective mass squared of the scalar field is defined by $m^2_\phi=\der^2_\phi V_\text{eff}$. The problem with this definition is that the BD scalar field does not satisfy the usual KG equation, but the one in Eq. (\ref{bd-mot-eq}), where the self-interaction potential of the BD field appears in an unconventional way. In order to fix this problem, one notices that, if introduce the effective potential\cite{mass}

\bea V_\text{eff}(\phi)=\frac{2}{3+2\omega_\textsc{bd}}\left[\phi V(\phi)-3\int d\phi V(\phi)\right],\label{eff-pot}\eea so that $$\der_\phi V_\text{eff}=\frac{2}{3+2\omega_\textsc{bd}}\left[\phi\der_\phi V(\phi)-2V(\phi)\right],$$ then, the KG equation in \eqref{bd-mot-eq} can be rewritten in the more conventional way: 

\bea \Box\phi=\der_\phi V_\text{eff}+\frac{1}{3+2\omega_\textsc{bd}}\,T^{(m)},\label{kg-source}\eea where the second term in the right-hand side (RHS) of this equation, is the source term which does not depend explicitly on the $\phi$-field.

What one usually calls as an effective mass, is a concept that is linked with the oscillations of the field around the minimum of the effective potential, which propagate in spacetime. These oscillations, or excitations, are the ones that carry energy-momentum and, if required, may be quantized. For simplicity consider the vacuum case $T^{(m)}_{\mu\nu}=0$ of Eq. (\ref{kg-source}). Let us assume next that $V_\text{eff}$ is a minimum at some $\phi_*$. Given that the equation (\ref{kg-source}) is non--linear, one may consider small deviations around $\phi_*$: $\phi=\phi_*+\delta\phi$ ($\delta\phi\ll 1$), then, up to terms linear in the deviation, one gets: $$\der_\phi V_\text{eff}(\phi)\approx\der_\phi V_\text{eff}(\phi_*)+\der^2_\phi V_\text{eff}(\phi_*)\,\delta\phi+...=m^2_*\delta\phi,$$ where $m^2_*\equiv \der^2_\phi V_\text{eff}(\phi_*)$, is the effective mass of the scalar field perturbations. Working in a flat Minkowski background (in spherical coordinates) $$ds^2=-dt^2+dr^2+r^2\left(d\theta^2+\sin^2\theta d\phi^2\right),$$ which amounts to ignoring the curvature effects and the backreaction of the scalar field perturbations on the metric, and imposing separation of variables, the perturbations $$\delta\vphi=\delta\vphi(t,r)=\sum_n e^{-i\omega_n t}\psi_n(r),$$ where $\omega_n$ is the angular frequency of the oscillations of the $n$-th excitation of the field, obey the Helmholtz equation: 

\bea \frac{d^2\psi_n}{dr^2}+\frac{2}{r}\frac{d\psi_n}{dr}+{\bf k}_n^2\psi_n=0,\label{helmholtz-eq}\eea where ${\bf k}_n^2=\omega^2_n-m^2_*$, is the wave--number squared. Eq. (\ref{helmholtz-eq}) is solved by the spherical waves: $$\psi_n(r)=C_n\frac{e^{i|{\bf k}_n|r}}{r},$$ where the $C_n$--s are integration constants. For $|\omega_n|<m_*$, the Eq. (\ref{helmholtz-eq}) has the Yukawa--type solution:

\bea \psi_n(r)=C_n\frac{e^{-\sqrt{m^2_*-\omega_n^2}\,r}}{r}.\label{yukawa-sol}\eea It is understood that the modes with the lowest energies $E_n=\hbar\,\omega_n=\omega_n\ll m_*$, are the ones which are more easily excited in the small oscillations approximation, and, hence, are the prevailing ones. In particular, $\omega_0\ll m_*$, so that

\bea \psi(r)\sim\psi_0(r)=C\frac{e^{-m_*\,r}}{r}.\label{0-mode}\eea These lowest order modes are the ones with the shortest effective Compton wave length $\lambda_*\approx m^{-1}_*$, and are the ones which decide the range of the Yukawa-type interaction, i. e., these are the ones which decide the effective screening of the $\phi$--field.

In what follows, we ignore the oscillations in time by assuming the static situation. This amounts to ignoring all of the higher-energy excitations of the field. This assumption bears no consequences for the qualitative analysis. However, once a friction term $\propto\dot\phi$ arises, for instance in a cosmological context, the oscillations of the field in time around the minimum, are necessarily to be considered.


\subsubsection{The field-theoretical mass}

The above analysis suggests that one may introduce a general definition of the mass (squared) of the BD scalar field $m^2_\phi\rightarrow m^2_*=\der^2_\phi V_\text{eff}(\phi_*)$, $$m^2_\phi=\der^2_\phi V_\text{eff}(\phi)=\frac{2}{3+2\omega_\textsc{bd}}\left[\phi \der^2_\phi V(\phi)-\der_\phi V(\phi)\right],$$ which is just Eq. \eqref{bd-mass}. What if the effective potential $V_\text{eff}$ has no minimums at all? The quartic potential $V(\phi)=\lambda\phi^4$, for instance, has a minimum at $\phi=0$, however, the corresponding effective potential: 

\bea V_\text{eff}(\phi)=\frac{4\lambda\,\phi^5}{5(3+2\omega_\textsc{bd})},\label{eff-q-pot}\eea has no minimums. In this case, our understanding of what an effective mass means, might have no meaning at all. In particular, the screened Coulomb--type potential (the mentioned Yukawa-like solution), being the most relevant physical manifestation of a massive propagator, might not arise. The corresponding ``mass'' in Eq. \eqref{bd-mass} would be just a useful field theoretical construction with the dimensions of mass, no more.

In spite of this, following the most widespread point of view, here we shall consider that, even away from the minimum of the effective potential, the field parameter $m^2_\phi$ given by Eq. \eqref{bd-mass}, represents the effective mass (squared) of the scalar field. In order to differentiate the mentioned field theoretic parameter from an actual effective mass (the one with consequences for fifth-force experiments), we shall call the latter as ``effective mass'', while the former as ``effective field-theoretical mass''.


\subsubsection{The chameleon mass}\label{cham-mass-sec}

Due to the chameleon mechanism, the screening effect may arise even if the effective potential $V_\text{eff}$ does not develop minimums. As we shall see, all what one needs is to include the source term in the RHS of the KG equation (\ref{kg-s-eq}), within a redefined effective potential, which we shall call effective chameleonic potential: 

\bea V_\text{ch}=V_\text{eff}+\phi S\;\Rightarrow\;\Box\phi=\der_\phi V_\text{ch}=\der_\phi V_\text{eff}+S.\label{cham-pot-idea}\eea 

Usually the chameleon effect is discussed, exclusively, in the Einstein frame formulation of the Brans-Dicke theory, where the scalar field couples directly with the matter degrees of freedom\cite{cham, cham-1, cham-khoury, cham-wei, cham-tamaki, cham-mota, cham-olive, cosmo-cham, cham-5-force, cham-rad, cham-fdr, cham-bd-1, cham-bd, bisabr_ass_2014, rev-cham}. Due to the non-trivial way which the self-interaction potential enters in the KG equation, the discussion of the chameleon effect in the Jordan frame formulation of BD theory seems more obscure than in the Einstein frame.

Below we shall show that, regardless of the unconventional form of the potential in the KGBD equation in \eqref{bd-mot-eq}, the chameleon effect can be discussed in the Jordan frame as well, if introduce the following definition of the effective chameleon potential:

\bea V_\text{ch}(\phi)=V_\text{eff}(\phi)+\frac{\phi T^{(m)}}{3+2\omega_\text{BD}}=\frac{2\phi V(\phi)-6\int d\phi V(\phi)+\phi T^{(m)}}{3+2\omega_\textsc{bd}},\label{bd-cham-pot}\eea so that 

\bea \der_\phi V_\text{ch}=\frac{2}{3+2\omega_\textsc{bd}}\left(\phi\der_\phi V-2V+\frac{1}{2}\,T^{(m)}\right),\label{der-bd-cham-pot}\eea coincides with the RHS of the BDKG equation in \eqref{bd-mot-eq}, and the latter can be written in the form of the conventional KG equation without a source: $\Box\phi=\der_\phi V_\text{ch}.$ 

As in the standard case, the effective mass squared: $m^2_{\phi_*}=\der^2_\phi V_\text{ch}(\phi_*),$ may be defined for the small perturbations of the BD scalar field around the minimum $\phi_*$ of the chameleon potential $V_\text{ch}$. Actually, under the assumption of spherical symmetry, given that $V_\text{ch}$ is a minimum at some $\phi_*$, if follow the procedure explained above, in the weak-field and low-velocity limit (basically the case when the curvature effects and the back-reaction on the metric are ignored): $$\frac{d^2(\delta\phi)}{dr^2}+\frac{2}{r}\frac{d(\delta\phi)}{dr}=m^2_{\phi_*}\delta\phi,$$ where $$m^2_{\phi_*}=\der^2_\phi V_\text{ch}(\phi_*),$$ is the effective mass of the perturbations around the minimum of the chameleon potential. 

We recall that, although when dealing with the chameleon effect we care only about the spatial deviation about the minimum $\delta\phi(r)$, as a means to linearize the BDKG equation, in general these deviations are also time-dependent so that we have time-dependent perturbations around the minimum of the chameleon potential. These may be viewed as periodic oscillations of the BD field about the minimum, and the resulting effective mass can be interpreted as the mass of the corresponding scalar excitations propagating in a flat background.

Solving for the above Helmholtz equation, one has for $\phi(r)=\phi_*+\delta\phi(r)$, the following solution:$$\phi(r)=\phi_*+C_1\frac{e^{-m_{\phi_*} r}}{r}+C_2\frac{e^{m_{\phi_*} r}}{r},$$ where $C_1$ and $C_2$ are integration constants which we can determine through the boundary conditions. If assume, for instance, that $\phi(r)$ tends asymptotically to a constant value $\phi_\infty$: $$\lim_{r\rightarrow\infty}\phi(r)=\phi_\infty\;\Rightarrow\;\phi(r)=\phi_\infty+C_1\frac{e^{-m_{\phi_*} r}}{r}.$$ 

Depending on the physical situation at hand, other boundary conditions are required in order to fix the remaining constants $C_1$ and $C_2$. For instance, if assume regularity of the solution at the origin $r=0$, then $C_1=-C_2$, $$\phi(r)=\phi_*+C_2\frac{\sinh(m_{\phi_*} r)}{r},$$ so that $\phi(0)=\phi_*+C_2 m_*$, etc. The interesting thing here is that the effective chameleon mass $m_{\phi_*}=m_{\phi_*}(\rho)$, is a function of the surrounding density $\rho$. This property of the effective mass of the BD scalar field perturbations is what is called, primarily, as the chameleon effect. Actually, the BD chameleon effect is related with the fact that the density of matter $\rho$ in the argument of the effective mass: $m_{\phi_*}(\rho)$, is the density measured by co-moving observers (with four-velocity $\delta^\mu_0$) in the JFBD theory: $\rho=T^{(m)}_{\mu\nu}\delta^\mu_0\delta^\nu_0$. This is, besides, the density of matter that is conserved in the JF (also in the SF) formulation of the Brans-Dicke theory. 

This is to be contrasted with the original chameleon effect of Refs. \citen{cham, cham-1}, where the physically meaningful matter density $\rho_i$ is not the one measured by co-moving observers in the EF, i. e., this is not the conserved one in this frame, neither in the conformal one, but a density which does not depend on the dilaton. Actually, in Refs. \citen{cham, cham-1} the matter fields couple to the conformal metric $g^{(i)}_{\mu\nu}=\exp(2\beta_i\phi/M_\textsc{pl})\,g_{\mu\nu}$, while the density of the non-relativistic fluid measured by EF co-moving observers is denoted by $\tilde\rho_i$. It is assumed that what matters is the $\phi$-independent density $\rho_i=\tilde\rho_i\exp(3\beta_i\phi/M_\textsc{pl})$, which is the one conserved in the EF. While this choice may not be unique, in the Jordan frame of the Brans-Dicke theory (the same for the SF), one does not have this ambiguity: the matter density measured by JF(SF) co-moving observers $\rho$, is the one conserved in the Jordan/string frames and, additionally, it does not depend on the BD-field.


\subsubsection{The Brans-Dicke chameleon: examples}\label{eje-sec}

Let us illustrate how the chameleon effect arises in the Jordan frame of BD theory, by exploring a pair of examples.

\begin{itemize}

\item{\it The quartic potential.}

In the first place let us choose the example with the quartic potential\cite{cham-khoury}: 

\bea V(\phi)=\lambda\phi^4,\label{q-pot}\eea where we assume that the free parameter $\lambda\geq 0$, is a non-negative constant. In this case, as said, the effective potential \eqref{eff-q-pot}: $V_\text{eff}(\phi)\propto\phi^5$, does not develop minimums. Yet the corresponding chameleon potential \eqref{bd-cham-pot}:

\bea V_\text{ch}(\phi)=\frac{4}{3+2\omega_\textsc{bd}}\left[\frac{\lambda}{5}\,\phi^5+\frac{T^{(m)}}{4}\,\phi\right]=\frac{4}{3+2\omega_\textsc{bd}}\left[\frac{\lambda}{5}\,\phi^5-\frac{\rho}{4}\,\phi\right],\label{bd-cham-pot-q}\eea where we have assumed a homogeneous, pressureless dust background: $T^{(m)}=-\rho$, can have a minimum if $\omega_\textsc{bd}\geq-3/2$. Actually, at the value $$\phi_*=\left(\frac{\rho}{4\lambda}\right)^{1/4},$$ the derivatives of the above chameleon potential $$\der_\phi V_\text{ch}(\phi_*)=0,\;\der^2_\phi V_\text{ch}(\phi_*)=\frac{16\lambda}{3+2\omega_\textsc{bd}}\left(\frac{\rho}{4\lambda}\right)^{3/4}.$$ Hence, provided that $\omega_\textsc{bd}\geq-3/2$, since $\der^2_\phi V_\text{ch}(\phi_*)>0$, the chameleon potential $V_\text{ch}$ in Eq. (\ref{bd-cham-pot-q}), is a minimum at $\phi_*$. In this case we can identify a physically meaningful effective mass of the BD field:

\bea m^2_{\phi_*}=\der^2_\phi V_\text{ch}(\phi_*)=\frac{16\lambda}{3+2\omega_\textsc{bd}}\left(\frac{\rho}{4\lambda}\right)^{3/4}.\label{phys-bd-mass}\eea In general, $\rho$ can be a function of the spacetime point $\rho=\rho(x)$, however, in most applications the function $\rho(x)$ is assumed piece--wise constant. For instance, one may imagine an spherical spatial region of radius $R$, filled with a static fluid with homogeneous and isotropic constant density $\rho_0$, and surrounded by a fluid with a different (also homogeneous) constant density $\rho_\infty$, so that: \[\rho(r)=\left\{\begin{array}{ll} \rho_0 & \mbox{if $r\leq R$};\\ \rho_\infty & \mbox{if $r\gg R$}.\end{array} \right.\] In such a case the effective mass $m_{\phi_*}$ of the BD scalar field would be one for modes propagating inside the spherical region $m_{\phi_0}$, and another different value $m_{\phi_\infty}$, for scalar modes propagating outside of (far from) the spherical region: $$m_{\phi_0}=\sqrt{\der^2_\phi V_\text{ch}(\phi_0)},\;m_{\phi_\infty}=\sqrt{\der^2_\phi V_\text{ch}(\phi_\infty)}.$$ For the quartic potential, in particular, one would have that: $$m_{\phi_0}=\frac{2(4\lambda)^{1/8}\,\rho^{3/8}_0}{\sqrt{3+2\omega_\textsc{bd}}},\;m_{\phi_\infty}=\frac{2(4\lambda)^{1/8}\,\rho^{3/8}_\infty}{\sqrt{3+2\omega_\textsc{bd}}},$$ inside and outside of the spherical region of radius $R$, respectively. In case the the gravitational configuration of matter were given by a point--dependent density profile $\rho=\rho(x)$, such as, for instance, in a cosmological context where $\rho=\rho(t)$ ($t$ is the cosmic time), the effective chameleon mass $m_{\phi_*}$ were point-dependent as well. However, as it is well known, the masses of point particles in the JF/SF formulations of the BD theory, are constants by definition. Otherwise, these particles would not follow geodesics of the JF/SF metric. In general, coexistence of particles of constant mass and particle excitations with point-dependent mass, bring about problems with the equivalence principle. Besides, if $\rho=\rho(x)$, then, the resulting effective mass will be a field-theoretical construction which has an anomalous behavior under the conformal transformation of the metric. In order to evade any possible discussion on the equivalence principle, or on the anomalous behavior of the effective field-theoretical chameleon mass under the conformal transformations of the metric, here we adopt the most widespread handling of the chameleon effect, and we assume that the density of matter has piece-wise constant profile in the sense explained above. 

\bigskip
\item{\it The quadratic monomial: the massless BD field.}

One peculiar note about the effective chameleon potentials: if look at equation \eqref{bd-cham-pot}, one sees that the quadratic potential $V(\phi)=M^2\phi^2$, plays a singular role. Actually, if substitute this potential into \eqref{bd-cham-pot}, one obtains that the resulting chameleon potential $$V_\text{ch}(\phi)=\frac{\phi\,T^{(m)}}{3+2\omega_\textsc{bd}},$$ does not have a minimum. Besides, the second derivative vanishes: $\der^2_\phi V_\text{ch}=0$. This means that the quadratic potential does not generate the chameleon effect. The same is true for any potential that asymptotes to $\phi^2$, for instance $V(\phi)\propto\cosh(\lambda\phi)-1$. Even the effective field-theoretical mass squared \eqref{bd-mass} vanishes for the quadratic monomial. Hence, since the chameleon effect does not work, the scalar field can not be screened from solar systems experiments. This looks like a bad news since, as shown in Ref. \citen{hrycyna, hrycyna-1, hrycyna-2, quiros-prd-2015-1} -- see section \ref{sect-dsyst-bd} -- only for the quadratic monomial, or for potentials that asymptote to it, the BD theory has the $\Lambda$CDM solution as a global attractor. In the case of a standard scalar field $\sigma$ whose dynamics is governed by the usual Klein-Gordon equation $\Box\sigma=\der_\sigma V(\sigma)$, the quadratic monomial $V(\sigma)\propto\sigma^2$, is also a singular potential in the sense that it is the only potential for which the BDKG equation is a linear differential equation, i. e., the superposition principle is satisfied.

\end{itemize}


\subsubsection{The quartic potential: estimates}\label{estimate-sec}

Notice that, similar to the chameleon effect arising in the Einstein frame of the BD theory\cite{cham, cham-1}, the mass squared of the BD field given by Eq. \eqref{phys-bd-mass}, i. e., the Jordan frame mass -- the one that determines the range of the Yukawa-like correction\cite{sotiriou} -- depends on the background energy density $m_{\phi_*}\propto\rho^{3/8}$. As it can be seen, this dependence of the mass of the scalar field on the ambient energy density improves the one in Ref. \refcite{cham-khoury}: $m_\phi\propto\rho^{1/3}$, just by a fraction. 

In order to make estimates, let us write $$m_{\phi_*}=\frac{4^{5/8}\lambda^{1/8}}{\sqrt{3+2\omega_\textsc{bd}}}\left(\frac{\rho}{M^4_\textsc{pl}}\right)^{3/8}M_\textsc{pl},$$ or in ``user-friendly'' units (using the terminology of Ref. \refcite{cham-khoury}): 

\bea m_{\phi_*}[\text{mm}^{-1}]\approx\frac{10\lambda^{1/8}}{\sqrt{3+2\omega_\textsc{bd}}}\left(\rho[\text{g/cm}^3]\right)^{3/8}.\label{estimate-eq}\eea Let us assume that the scalar field is immersed in the earth atmosphere with mean density $\rho^\text{atm}\approx 10^{-3}$ g/cm$^3$, then, provided that the millimeter range screening\cite{cham-khoury}: $(m_{\phi_*}^\text{atm})^{-1}\sim 1$ mm, is undertaken, from Eq. \eqref{estimate-eq} it follows that $$\omega_\textsc{bd}\approx 1.6\,\lambda^{1/4}-1.5,$$ so that, if consider, for instance, that $\lambda\sim 1$, one gets that $\omega_\textsc{bd}\approx 0.1$ can be of order unity or smaller. What this means is that the BD theory may describe the gravitational phenomena with a coupling constant of order unity and, yet, the chameleon potential \eqref{bd-cham-pot-q} may effectively screen the BD field from experiments that look for violation of the Newton's law, for distances above the millimeter.

The next question is whether the above potential can be a good candidate for cosmology as well. The ratio of the mass of the BD field measured at large cosmological scales, to the scalar field mass estimated in earth's atmosphere: 

\bea \frac{m_{\phi_*}^\text{cosm}}{m_{\phi_*}^\text{atm}}=\left(\frac{\rho^\text{crit}}{\rho^\text{atm}}\right)^{3/8}\approx 3\times 10^{-11},\label{estimate'}\eea where we have taken into account that the critical energy density of the universe $\rho^\text{crit}\sim 10^{-31}$ g/cm$^3$. If consider the millimeter-range screening above, $(m^\text{atm}_{\phi_*})^{-1}\approx 1$mm $\Rightarrow\;m^\text{atm}_{\phi_*}\approx 10^{-4}$ eV, then the estimated mass of the cosmological BD scalar field $$m_{\phi_*}^\text{cosm}\approx 3\times 10^{-11} m_{\phi_*}^\text{atm}\approx 3\times 10^{-15}\;\text{eV},$$ is by some 18 orders of magnitude heavier than the expected value $m_\phi^\text{cosm}\sim H_0\sim 10^{-33}$ eV. Hence, if assume that the BD scalar field with a fixed potential $V(\phi)=\lambda\phi^4$, is effectively screened from solar system experimentation, the BD field would not have cosmological implications. The ``reconciliation'' between terrestrial and cosmological bounds, at once, can be achieved by power-law potentials leading to BD chameleon mass, $m_{\phi_*}\propto(\rho)^{k/2}$, with the power $k\approx 29/14\approx 2.071$, or higher. Of course, the reconciliation is natural if, for instance, $m_{\phi_*}\propto\exp\rho$.

\subsubsection{Thin-shell effect}

Our estimates above are unsatisfactory in many aspects. First of all, a lot of simplification has been made for sake of transparency of our analysis. For instance, the well known thin-shell effect\cite{cham, cham-1}, which arises due to the non-linearity of the BD scalar field, and which is significant for large bodies, has not been considered in our analysis (for a detailed exposition of the thin-shell effect we recommend Ref. \refcite{cham-khoury}). Nevertheless, even if take into account the thin-shell effect, the physical implications of the huge difference between the cosmic and terrestrial mass scales: $m_{\phi_*}^\text{cosm}/m_{\phi_*}^\text{atm}\sim 10^{-11}$, can not be erased by the thin-shell mediated weakening of the effective coupling of the BD field to the surrounding matter. Actually, the additional contribution of the chameleon BD field to the Newtonian gravitational potential energy of a given mass $M_b$, is expressed by $$\Delta U^*_N\propto-\beta^{*2}_\text{eff}M_b\frac{e^{-r/\lambda_\text{eff}}}{r},$$ where $\beta^*_\text{eff}$ is the effective coupling of the chameleon field to the surrounding matter, and $\lambda_\text{eff}=m_{\phi_*}^{-1}$, is its effective Compton length. We have that 

\bea \lambda_{\phi_*}^\text{atm}\approx 10^{18}\lambda_\phi^\text{atm},\label{lambda-ratio}\eea where $\lambda_\phi^\text{atm}\sim 1$mm, is the Compton length of the chameleon field which is consistent with the experiments on fifth-force, while $\lambda_{\phi_*}^\text{atm}$ is the effective range of the scalar field mediated interaction, computed with the potential $V(\phi)\propto\phi^4$, under the assumption that the cosmological bound $\lambda_{\phi_*}^\text{cosm}\sim 10^{26}$m, is met: $$\lambda^\text{atm}_{\phi_*}\sim 10^{-11}\lambda^\text{cosm}_{\phi_*}\sim 10^{15}\,\text{m}=10^{18}\,\text{mm}.$$ Then, requiring that $$\frac{\Delta U_N^*}{\Delta U_N}=\left(\frac{\beta^*_\text{eff}}{\beta_\text{eff}}\right)^2\frac{e^{r/\lambda_\phi^\text{atm}}}{e^{r/\lambda_{\phi_*}^\text{atm}}}\approx 1,$$ for the given potential $V(\phi)\propto\phi^4$, the expected weakening of the effective coupling of the chameleon BD field to the surrounding matter, is an unnaturally large effect: $\left(\beta^*_\text{eff}\right)^2\sim\exp\left(-10^{20}\right)\beta_\text{eff}^2,$ where we have assumed that $\beta_\text{eff}\sim 1$. In order to obtain the above estimate, we have arbitrarily set the distance from the source of gravity $r\approx 10^2\lambda_{\phi_*}^\text{atm}$ and the Eq. \eqref{lambda-ratio} has been considered.

The above results are true, in general, for power-law potentials of arbitrary power: $V(\phi)\propto\phi^\lambda$. In this latter case, for the mass squared of the BD field, one gets: $m^2\propto\rho^{(\lambda-1)/\lambda}$, where, as $\lambda\rightarrow\infty$, $(\lambda-1)/\lambda\rightarrow 1$. This means that the latter power can never exceed unity $(\lambda-1)/\lambda\leq 1$. Recall that, a necessary requirement, when the power-law potential $V(\phi)\propto\phi^\lambda$ is allowed to explain cosmological and terrestrial bounds at once, amounts to: $(\lambda-1)/\lambda>2.071$. 

Our conclusion is that, in general, terrestrial and solar system bounds on the mass of the BD scalar field, and bounds of cosmological origin, are difficult to reconcile through a single chameleon potential.


\subsection{The Vainshtein mechanism}\label{subsect-vain}

The relatively recent increase of interest in the Vainshtein screening effect\cite{vainshtein, vainsh-deff-gabad, khoury-rev} is mainly due to the study of the DGP brane model\cite{dgp_plb_2000, deffayet(dgp)_prd_2002, luty_porrati_rattazzi_jhep_2003, nicolis_rattazzi_jhep_2004}. In a certain decoupling limit of the DGP theory\cite{luty_porrati_rattazzi_jhep_2003, nicolis_rattazzi_jhep_2004} (here we do not use the units system where $M_\text{Pl}=1$): $M_\text{Pl}\rightarrow\infty,$ $M_5\rightarrow\infty,$ while keeping the strong-coupling scale $(M_\text{Pl}/r_c^2)^{1/3}$ fixed -- here $M_5$ is the 5D Planck mass and $r_c\equiv M^2_\text{Pl}/2M^3_5$ is the crossover scale which separates 5D and 4D regimes -- the resulting theory is local on the brane, and describes a self-interacting scalar field coupled to weak-field gravity in 4D:

\bea {\cal L}_\text{dec}=-\frac{M^2_\text{Pl}}{4}h^{\mu\nu}G^L_{\mu\nu}-3(\der\phi)^2-\frac{r^2_c}{M_\text{Pl}}(\der\phi)^2\Box\phi+\frac{1}{2}h^{\mu\nu}T^{(m)}_{\mu\nu}+\frac{\phi}{M_\text{Pl}}T^{(m)},\label{dgp-decoup-lag}\eea where $\phi$ accounts for the brane-bending mode (longitudinal graviton), $G^L_{\mu\nu}$ is the linearized Einstein's tensor \eqref{lin-etensor} and $T^{(m)}=\eta^{\mu\nu}T^{(m)}_{\mu\nu}$, is the trace of the stress-energy tensor of matter. The above action is invariant under the Galilean shift symmetry, $\der_\mu\phi\rightarrow\der_\mu\phi+c_\mu$. In regions of high energy density the non-linearities in the equations of motion for $\phi$ dominate, which results in its decoupling\cite{chow_gal} from the remaining degrees of freedom, so that the resulting theory is general relativity. This screening mechanism relies on the higher order derivatives, as opposed to the chameleon effect that relies in the density dependence of the effective potential. 

Nonlinear interactions of $\phi$ are important near an astrophysical source and, as said, result in the decoupling of this helicity-0 mode from the source. The characteristic scale below which $\phi$ is strongly coupled, is called as ``Vainshtein radius'' and is given by, $r_V=(r^2_c r_S)^{1/3}$, where $r_S$ is the Schwarzschild radius of the source. The decoupling limit corresponds to the formal limit, $r_S\rightarrow 0$, with $r_V$ held fixed.

In order to understand how the strong interactions of the Galileon $\phi$ lead to its decoupling near the source, following Ref. \refcite{chow_gal}, let us write the motion equation for the Galileon \eqref{dgp-decoup-lag}: 

\bea \der_\mu j^\mu=-\frac{T^{(m)}}{2M_\text{Pl}},\label{gal-eom}\eea where the ``current'' is defined as: $$j_\mu=3\der_\mu\phi+\frac{r^2_c}{M_\text{Pl}}\der_\mu\phi\Box\phi-\frac{r^2_c}{M_\text{Pl}}\der_\mu(\der\phi)^2.$$ In the static, spherically symmetric case, assuming a point mass with $T^{(m)}=-M\delta^3(r)$, the motion equation \eqref{gal-eom} can be written in the form of the following algebraic equation on $d\phi/dr$: $$\left(\frac{d\phi}{dr}\right)^2+\frac{3M_\text{Pl}r}{2r^2_c}\left(\frac{d\phi}{dr}\right)-\frac{M^2_\text{Pl}r_S}{4rr_c^2}=0.$$ The roots of this algebraic equation are: $$\frac{1}{M_\text{Pl}}\left(\frac{d\phi}{dr}\right)_\pm=\frac{3rr_S}{4r_V^3}\left[-1\pm\sqrt{1+\frac{4r^3_V}{9r^3}}\right].$$ We keep the '$+$' branch of the solution only (the ghost-free normal branch of the DGP) because it satisfies, $d\phi/dr\rightarrow 0$ as $r\rightarrow\infty$. For the self-accelerating branch, instead, $d\phi/dr\rightarrow-\infty$ as $r\rightarrow\infty$. The ratio of the Galileon-mediated force, $|\vec{F}_\phi|=M^{-1}_\text{Pl}d\phi/dr$, to the Newtonian gravitational force: $$|\vec{F}_N|=\frac{d\Phi}{dr}=\frac{G_N M}{r^2}=\frac{r_S}{2r^2},$$ reads:

\bea \frac{|\vec{F}_\phi|}{|\vec{F}_N|}=\frac{3r^3}{2r^3_V}\left[-1+\sqrt{1+\frac{4r^3_V}{9r^3}}\right].\label{force-ratio}\eea It immediately follows that, deep inside the Vainshtein radius: $r\ll r_V$, $$\frac{|\vec{F}_\phi|}{|\vec{F}_N|}\simeq\left(\frac{r}{r_V}\right)^{3/2}\ll 1,$$ so that the Galileon-mediated force is clearly suppressed.


\subsubsection{Cosmological analogue of the Vainshtein screening}\label{subsect-cosmo-vain}

Here we shall discuss on a cosmological analogue of the Vainshtein mechanism. This was explored for the first time in Ref. \refcite{chow_gal}. Then, in Ref. \refcite{kazuya_gal} the cubic Brans-Dicke Galileon model was explored where, again, a cosmological version of the Vainshtein screening was found. A similar cubic Galileon, but this time minimally coupled to the curvature, was explored in Ref. \refcite{quiros_cqg_2016} for the exponential potential, and then in Ref. \refcite{quiros-cqg-2018-1} for potentials beyond the exponential one. It was shown that a kind of vDVZ discontinuity arises: It is not possible to recover all of the vacuum solutions of the theory by going to the continuous limit when the matter density vanishes. It was then conjectured that the non-linear (cubic derivative) interactions of the Galileon with the background matter ``screens'' the relevant vacuum solutions\cite{quiros-cqg-2018-1}, which seems like another cosmological version of the Vainshtein mechanism. Below we shall briefly discuss on the above mentioned cosmological realizations of this screening mechanism.

\begin{itemize}

\item{\it Cubic Brans-Dicke Galileon.} As already mentioned, in Ref. \refcite{kazuya_gal} a covariant Brans-Dicke Galileon model exhibiting the self-accelerating solution was proposed that was free of ghostlike instabilities. The key feature in the model was the cubic self-interaction term of the form $f(\phi)\Box\phi(\der\phi)^2$. This is the unique form of interactions at cubic order yielding a second-order motion equation for the Galileon field. A related cubic Galileon model given by the Lagrangian $${\cal L}=-3(\der\phi)^2-\frac{1}{\Lambda^3}\Box\phi(\der\phi)^2+\frac{g}{M_\text{Pl}}\phi\,T^{(m)},$$ where $g\sim{\cal O}(1)$ for gravitational strength coupling, $M_\text{Pl}$ is the Planck scale, $\Lambda$ is the strong-coupling of the theory and $T^{(m)}\equiv g^{\mu\nu}T^{(m)}_{\mu\nu}$ is the trace of the stress-energy tensor of matter, provides the simplest non-trivial theory exhibiting the Vainshtein screening mechanism\cite{vainsh_rev, vainshtein, vainsh-deff-gabad, khoury-rev}. A similar Lagrangian is found in Ref. \refcite{chow_gal}. The Vainshtein mechanism relies on the cubic self-interaction term $\Box\phi(\der\phi)^2/\Lambda^3$ becoming large compared to the kinetic term $(\der\phi)^2$ near massive objects. The above cubic Galileon Lagrangian belongs in the wider class of the so called Horndeski theories\cite{horndeski_gal} that represent the generalization of STT to include higher-derivative terms. The motion equations for Horndeski theories are second order, thus warranting the absence of the Ostrogradsky instability. In a flat FRW cosmological background the motion equations for the cubic BD Galileon that are derived from the action \eqref{kazuya-action}, with $V(\phi)=\Lambda\phi$ and $\alpha^2=M^{-2}$, read\cite{kazuya_gal}:

\bea &&3H^2+3P\left(1-\frac{P^2}{M^2}\right)H=\frac{\rho_m}{\phi}+\frac{\omega_\text{BD}}{2}P^2+\frac{P^4}{M^2}+\Lambda,\nonumber\\
&&-2\dot H=\frac{\rho_m+p_m}{\phi}+\left(\omega_\text{BD}+1+\frac{P^2}{M^2}\right)P^2-HP\left(1-\frac{3P^2}{M^2}\right)+\dot P\left(1-\frac{P^2}{M^2}\right),\nonumber\\
&&\left(3+2\omega_\text{BD}+\frac{12HP}{M^2}+\frac{5P^2}{M^2}\right)\left(\dot P+P^2\right)+\left(3+2\omega_\text{BD}-\frac{P^2}{M^2}\right)3HP\nonumber\\
&&\;\;\;\;\;\;\;\;\;\;\;\;\;\;\;\;\;\;\;\;\;\;\;\;\;\;\;\;\;\;\;\;\;\;\;\;\;+\frac{6P^2}{M^2}\left(\dot H+3H^2\right)-\frac{4P^4}{M^2}=2\Lambda+\frac{\rho_m-3p_m}{\phi},\label{kazuya-feqs}\eea where $P\equiv\dot\phi/\phi$ and $\rho_m$ ($p_m$) is the matter density (barotropic pressure) of the background matter fluid. If combine the first two equations above, and set the conditions for self-accelerating solutions: $\Lambda=\rho_m=p_m=0$ and $\dot H=\dot P=0$, one gets\cite{kazuya_gal}: $$\left(\frac{P}{H}\right)^2+\frac{4}{2+\omega_\text{BD}}\left(\frac{P}{H}\right)+\frac{6}{2+\omega_\text{BD}}=0,$$ whose roots are: $$Z_\pm\equiv\left(\frac{P}{H}\right)_\pm=\frac{2}{2+\omega_\text{BD}}\left(-1\pm\sqrt{-\frac{3\omega_\text{BD}+4}{2}}\right).$$ These solutions exist if $\omega_\text{BD}<-4/3$. For this self-accelerating solution the Friedmann equation can be written as, $$H^2=M^2\frac{3\left(1+Z_--\omega_\text{BD}Z^2_-/6\right)}{Z^3_-(3+Z_-)},$$ where we choose the minus branch of the solution since $H^2$ has to be positive. In order to describe the acceleration today, $M$ should be fine-tuned: $M\sim H_0$ ($H_0$ is the present value of the Hubble parameter). At high energies/large curvature, when the nonlinear terms in the field equations \eqref{kazuya-feqs} dominate over the linear term, we get that, $P=M/\sqrt{3}\sim H_0\ll H$. This is equivalent to the formal limit $P\rightarrow 0$ in the field equations, which corresponds to the GR limit. This is how the cosmological Vainshtein screening mechanism takes place in this model.

\bigskip
\item{\it Cubic Galileon minimally coupled to the curvature.} A simplified cubic Galileon model of cosmological interest is given by the action \eqref{qbic-gal-action}. This scenario has been probed to be very interesting and has been studied in detail in the literature. In Refs. \citen{genly_saridakis_jcap_2013} a dynamical systems study of the model \eqref{qbic-gal-action} with the inclusion of the background matter (${\cal L}_m$) was developed for a pair of self-interaction potentials, showing that the cubic self-interaction of the Galileon has no impact in the late-time cosmic dynamics. A very interesting result was obtained in Ref. \refcite{quiros_cqg_2016} (see the discussion in subsection \ref{subsect-gal}) for the cubic Galileon \eqref{qbic-gal-action} with the exponential self-interaction potential: When the matter degrees of freedom other than the Galileon itself are removed, i. e., when the vacuum Galileon action \eqref{qbic-gal-action} with ${\cal L}_m=0$ is considered, the late time dynamics can be indeed modified by the presence of a phantom attractor associated with super-accelerated expansion, a result that has no analogue in the case when the matter Lagrangian ${\cal L}_m$ is considered. Hence it results that one can not recover the cubic Galileon vacuum dynamics continuously from the more general case with the inclusion of matter by setting to zero the matter energy density (and the pressure). This is a kind of cosmological vDVZ discontinuity\cite{vdvz, vdvz-1, vdvz-2}, as in similar cases in the bibliography, can be evaded by means of the cosmological analogue of the Vainshtein screening mechanism\cite{kazuya_gal, chow_gal, vainsh_rev, vainshtein, vainsh-deff-gabad, khoury-rev} that is triggered by the cubic term in \eqref{qbic-gal-action}: $\sigma\Box\phi(\der\phi)^2$. It happens that, as the effective (phantom-like) energy density grows up with the cosmic expansion, the cubic self-interaction term dominates the dynamics of the expansion, leading to the decoupling of the Galileon from the remaining degrees of freedom, and to the eventual recovering of general relativity. This is how, in the presence of background matter, the cosmological Vainshtein screening mechanism prevents the occurrence of a big-rip singularity a finite time into the future in the present cubic Galileon model. In other words, in the presence of standard matter, the phantom attractor arising in the vacuum case is erased from the phase space by means of the Vainshtein-like screening, a fact that is consistent with the result of Ref. \refcite{genly_saridakis_jcap_2013} that the late-time dynamics of the model \eqref{qbic-gal-action} with the presence of background matter is basically the same as for the standard quintessence (see also Ref. \refcite{quiros_cqg_2016}). The above result was obtained in Ref. \refcite{quiros_cqg_2016} for a particular choice of the self-interaction potential: the exponential potential, while in Ref. \refcite{quiros-cqg-2018-1} it was shown that the above kind of cosmological vDVZ discontinuity, and its resolution through the cosmological version of the Vainshtein screening effect, is independent of the specific form of the potential $V$. Here we summarize the result of Ref. \refcite{quiros-cqg-2018-1}.

The solutions that arise only in the vacuum case, but that are not found when background matter is considered, are the following:

\begin{enumerate}

\item{\bf The phantom solution.} The solution associated with the critical point, $P_{6v}^\pm$ in Ref. \refcite{quiros-cqg-2018-1}, which is either a stable critical point (a local attractor) or a saddle point, depending on the parameters values. As shown in Ref. \refcite{quiros-cqg-2018-1}, for this critical point $q=-4$ (it is a super-accelerated solution), then: $\dot H=3H^2$, so that $$H(t)=\frac{1}{3(t_f-t)}\;\Rightarrow\;a(t)=\frac{a_0}{(t_f-t)^{1/3}},$$ where $-3t_f$ and $\ln a_0$ are arbitrary integration constants, and $t\leq t_f$. Besides, for the effective energy density we have that: $$\rho_\text{eff}(t)=3H^2(t)=\dot H(t)=\frac{1}{3(t_f-t)^2},$$ where the phantom behavior is evident from the fact that the energy density of the cubic Galileon grows up with $t$ without bounds. Given that $a(t)$, $H(t)$, $\dot H(t)$, and $\rho_\text{eff}(t)$, all blow up at $t=t_f$, i. e., in a finite time into the future, a big rip singularity \cite{odintsov, ruth} may be the inevitable fate of the cosmic evolution. 

\bigskip
\item{\bf Super-accelerated contraction.} The equilibrium points $P_{7v}$ and $P_{8v}$ in Ref. \refcite{quiros-cqg-2018-1} are associated with super-accelerated contraction of the universe and correspond to unstable nodes in the phase-space. These super-accelerated solutions have no impact in the late-time dynamics. The only difference between them is that in  $P_{7v}$ the super-accelerated contraction is fueled by the Galileon with (asymptotically) constant potential, meanwhile, in $P_{8v}$ the contraction is driven by the pure kinetic energy of the Galileon (vanishing potential).

\end{enumerate} None of the solutions $P^\pm_{6v}$, $P_{7v}$ and $P_{8v}$ are found if the Galileon vacuum is filled with standard matter degrees of freedom. The phantom solution, $P^\pm_{6v}$, is perhaps the most distinctive feature of the complexity of the cubic Galileon vacuum (the super-accelerated solutions $P_{7v}$ and $P_{8v}$ are not of importance for our analysis since these correspond to contracting universe). This may have implications for the late-time asymptotics and, hence, may be of importance for the future destiny of our universe. The fact that the addition of standard matter degrees of freedom, say dust-like dark matter, screens this vacuum effect is a very interesting example of the physical role of the cubic self-interaction of the Galileon $\propto(\Box\phi)(\der\phi)^2$, that is intimately linked with the cosmological Vainshtein screening mechanism\cite{vainsh_rev, vainshtein, vainsh-deff-gabad, khoury-rev}. This cosmological screening effect is similar to the cosmological versions of the Vainshtein mechanism explained in Ref. \refcite{chow_gal} (see also Ref. \refcite{kazuya_gal}) that operates at high energies when the non-linear terms in the equations of motion dominate over the linear one, thus leading to the recovery of general relativity. In the case of the phantom vacuum solution $P^\pm_{6v}$, since it is related with a big rip singularity where $$\dot H\sim H^2\sim a^6\sim\rho_\text{eff}\rightarrow\infty,$$ what happens is that, in the presence of background matter, at late times the universe enters a high energy regime where the cubic term dominates. This results in that the Galileon decouples from the other matter degrees of freedom, so that we are left effectively with general relativity\cite{chow_gal}. 

For the super-accelerated contracting solutions $P_{7v}$ and $P_{8v}$, the explanation of the screening effect is a bit different since in this case the effective energy density (the cubic Galileon's energy density) dilutes with the contraction while the matter energy density grows up with the cosmic time: $\rho_m\sim a^{-3}\propto t$. In order to expose our reasoning line in this case, let us rewrite the Friedmann equation in the following convenient way: 

\bea H^2+\sigma_0\dot\phi^3H=\frac{1}{3}\left(\rho_m+\rho_\phi\right),\label{dgp-fried}\eea where, as before, $\rho_\phi=\dot\phi^2/2+V$. Written in this form the Friedmann equation for the cubic Galileon resembles the one for the DGP braneworld \cite{dgp_plb_2000, deffayet(dgp)_prd_2002, luty_porrati_rattazzi_jhep_2003, nicolis_rattazzi_jhep_2004, roy-rev}: $$H^2\pm\frac{1}{r_c}H=\frac{1}{3}\rho_m,$$ where the crossover scale $r_c=G_{(5)}/2G_{(4)}$ is half of the ratio between the 5D and 4D gravitational couplings. If compare this latter equation with \eqref{dgp-fried} one can identify the cubic term $\sigma_0\dot\phi^3$ with the inverse of certain ``crossover'' scale: $r_*=(\sigma_0\dot\phi^3)^{-1}$. In correspondence one may also identify a Vainshtein radius: $r_V=(r_g r_*^2)^{1/3}$, within which the non-linear cubic interaction becomes important. In this latter relationship $r_g$ is the Schwarzschild radius of the universe that is roughly the Hubble scale $r_g\sim H^{-1}$. An acceptable estimate for $r_*$ would be that $r_*\sim H^{-1}$ also. Hence, since for the super-accelerated solutions $H=-1/3(t+C_0)$ -- see the former section -- where $C_0$ is an integration constant which, for simplicity, may be set to zero, then the Vainshtein radius grows up with the cosmic time $$r_V\sim |H|^{-1}\propto t,$$ while the physical distances go like: $d_\text{phys}=ra(t)\propto t^{-1/3}$. As the contraction proceeds, eventually, there will be a regime where the physical distances start becoming smaller than the Vainshtein radius: $d_\text{phys}\lesssim r_V$, so that the non-linear (cubic) self-interaction of the Galileon becomes dominating. This leads to the decoupling of the Galileon interactions which results in the recovering of general relativity. Although in this demonstration we have assumed the estimate $r_*\sim H^{-1}$, we can see that even without the assumption of any estimate, since in general for $\dot\phi>0$ the crossover scale $r_*=(\sigma_0\dot\phi^3)^{-1}$ decays with the cosmic time, the above conclusion is always true.

\end{itemize}


\subsubsection{Vainshtein screening in Horndeski and beyond Horndeski theories}

An effective theory of the Vainshtein mechanism was developed in Ref. \refcite{niz_kazuya_vainsh} for Horndeski theories in general. The
effective theory is described by a generalization of the Galileon Lagrangian, which is used to study the stability of spherically symmetric configurations exhibiting the Vainshtein effect. A clear exposition of the Vainshtein screening effect in a cosmological background in Horndeski theories can be found, for instance, in Refs. \citen{kimura_vainsh, dima_prd_2018}.

The new non-linear interactions beyond Horndeski -- see subsection \ref{subsect-beyond-horn} -- change the behavior of the gravitational potentials inside of matter overdensities in a fundamental way: The strength of the gravitational interaction depends not only on the enclosed mass but also on the local matter energy density. As a result $\Phi$ and $\psi$ no longer coincide, implying that GR is not recovered inside the source. This indicates a breakdown of the Vainshtein screening mechanism in beyond Horndeski theories\cite{bhorn(vainsh)} which means, in turn, that these theories may be in conflict with local (Solar system) experiments.


\subsection{Hybrid metric-Palatini gravity}\label{subsect-hybrid}

Before we end up this section we want to briefly comment on an interesting possibility that, although not being properly a screening effect, allows the elusive scalar field to pass the Solar system observational constraints even if it is very light. The resulting theory renders possible the existence of a long-range scalar field which can modify the cosmological -- and galactic -- dynamics, but leaves the Solar system unaffected. This kind of theories are known under the name 'hybrid gravity'\cite{harko-hybrid, capoz-hybrid-1, capoz-hybrid-2}. The action for the hybrid metric-Palatini reads:

\bea S=\frac{1}{2}\int d^4x\sqrt{|g|}\left[R+f({\cal R})\right]+S_m,\label{hybrid-action}\eea where $S_m$ is the matter piece of action, $R$ is the standard Ricci scalar, while ${\cal R}=g^{\mu\nu}{\cal R}_{\mu\nu}$ is the Palatini curvature scalar, with ${\cal R}_{\mu\nu}$ defined in terms of an independent connection $\hat\Gamma^\alpha_{\mu\nu}$ (see below). The above action can be transformed into a dynamically equivalent scalar-tensor theory by introducing an auxiliary field $\psi$ such that $$S=\frac{1}{2}\int d^4x\sqrt{|g|}\left[R+f(\psi)-\frac{df}{d\psi}\left({\cal R}-\psi\right)\right]+S_m.$$ By rearranging the terms and redefining: $\phi\equiv df/d\psi$, $2V(\phi)=\psi df/d\psi-f(\psi)$, the above action becomes $$S=\frac{1}{2}\int d^4x\sqrt{|g|}\left[R+\phi{\cal R}-2V\right]+S_m.$$ After manipulating the equations of motion that are derived from this action by varying with respect to the metric, to the scalar field $\phi$ and to the independent connection, the following relationship between the Ricci scalar and the Palatini curvature scalar, is obtained\cite{harko-hybrid}:

\bea {\cal R}_{\mu\nu}=R_{\mu\nu}+\frac{3}{2\phi^2}\der_\mu\phi\der_\nu\phi-\frac{1}{\phi}\left(\nabla_\mu\nabla_\nu+\frac{1}{2}g_{\mu\nu}\Box\right)\phi.\label{ricci-palatini-rel}\eea Hence, the independent connection turns out to be the Christoffel symbols of the conformal metric $\hat g_{\mu\nu}=\phi g_{\mu\nu}$.

By taking into account \eqref{ricci-palatini-rel}, the action for hybrid gravity can be recast into the following scalar-tensor action:

\bea S=\frac{1}{2}\int d^4x\sqrt{|g|}\left[\left(1+\phi\right)R+\frac{3}{2\phi}(\der\phi)^2-2V(\phi)\right]+S_m.\label{stt-hybrid-action}\eea This action is similar to that of the BD theory with $\omega_\text{BD}=-3/2$, but for the non-minimal coupling to the curvature which, for the BD theory reads: $\phi R$, instead of the above $(1+\phi)R$. This subtle distinction makes the whole difference. To start with, the BD theory with the anomalous coupling, $\omega_\text{BD}=-3/2$, can be consistently coupled only to traceless matter, i. e., to radiation-like matter fields (see below in subsection \ref{subsect-conf-inv-ex}). This problem is absent in the present theory, which can be clearly seen from the derived motion equations:

\bea &&(1+\phi)G_{\mu\nu}=T^{(m)}_{\mu\nu}-\frac{3}{2\phi}\left[\der_\mu\phi\der_\nu\phi-\frac{1}{2}g_{\mu\nu}(\der\phi)^2\right]-Vg_{\mu\nu}+(\nabla_\mu\nabla_\nu-g_{\mu\nu}\Box)\phi,\nonumber\\
&&-\Box\phi+\frac{1}{2\phi}(\der\phi)^2=\frac{\phi}{3}T^{(m)}-\frac{2\phi}{3}\left[2V-(1+\phi)\frac{dV}{d\phi}\right].\label{hybrid-moteq}\eea In the weak-field limit, far from the sources (we are considering spherical symmetry), the scalar field behaves like (here, temporarily, we return tu usual units $1\rightarrow 8\pi G_N$): $$\phi(r)\approx\phi_0+\frac{2G_NM\phi_0}{3r}\,e^{-m_0 r},$$ where $\phi_0$ is the amplitude of the background value and (here the prime denotes derivative with respect to the scalar field): $$m_0^2\equiv\frac{2}{3}\left[2V-V'-\phi(1+\phi)V''\right]|_{\phi=\phi_0},$$ is the effective mass of the scalar field. The following expressions for the effective (Cavendish-like) Newton's constant $G_\text{eff}$, and for the post-Newtonian parameter $\gamma$, are obtained\cite{capoz-hybrid-2}:

\bea G_\text{eff}=\frac{G_N}{1+\phi_0}\left[1-\left(\frac{\phi_0}{3}\right)e^{-m_0 r}\right],\;\;\gamma=\frac{3+\phi_0 e^{-m_0 r}}{3-\phi_0 e^{-m_0 r}}.\label{eff-post-n-pars}\eea It is seen from these expressions that, for small $\phi_0\ll 1$, $G_\text{eff}\approx G_N$ and $\gamma\approx 1$, independent of the magnitude of the effective mass, $m_0$, of the scalar field. This means that even a very light scalar field can pass the experimental checks in the Solar system. This result is to be contrasted with the one obtained in the metric formulation of the $f(R)$ theories or, for instance, with the chameleon effect, where a heavy mass $m_0$ is required in order to evade the tight constraints coming from Solar system experiments.


\section{The conformal frames conundrum}\label{sect-cf}

The so called conformal frames' issue is, perhaps, one of the oldest controversies concerning scalar-tensor theories of gravity\cite{dicke-1962, faraoni-book, faraoni_rev_1997, faraoni_ijtp_1999, faraoni_prd_2007, sarkar_mpla_2007, deruelle_veiled_2011, deruelle_nordstrom_2011, quiros_grg_2013, sotiriou_etall_ijmpd_2008}. The controversy may be stated in the following way: Under conformal transformations of the metric the given STT may be formulated in a -- in principle infinite -- set of mathematically equivalent field variables, called as conformal frames. Among these the Jordan frame (JF) and the Einstein's frame (EF) are the most outstanding. The following related questions are the core of the ``conformal transformation's issue''.

\begin{enumerate}

\item Are the different conformal frames not only mathematically equivalent but, also, physically equivalent?

\item If the answer to the former question were negative, then: which one of the conformal frames is the physical one, i. e., the one in terms of whose field variables to interpret the physical consequences of the theory? 

\end{enumerate} The controversy originates from the lack of consensus among different researchers -- also among the different points of view of the same researcher along its research history -- regarding their answer to the above questions. There are even very clever classifications of the different works -- of different authors and of the same author -- on this issue\cite{faraoni_rev_1997}. That the controversy has not been resolved yet is clear from the amount of yearly work on the issue where there is no agreement on the correct answer to these questions\cite{saal_cqg_2016, indios_consrvd_prd_2018, thermod_prd_2018, ct-1, ct-2, ct(fresh-view)-3, paliatha-2, ct(quant-equiv)-4, ct(quant-equiv)-5, ct(quant-equiv)-6, ct-ineq-nojiri, ct-ineq-brisc, ct-ineq-capoz, ct-ineq-brooker, ct-ineq-baha, ct(inequiv)-7, ct(inequiv)-8, ct(inequiv)-9, quiros-arxiv-2018}. In this section we shall discuss on the conformal frames' issue from the classical point of view exclusively. For a related discussion based on quantum arguments we recommend Refs. \citen{ct(quant-equiv)-4, ct(quant-equiv)-5, ct(quant-equiv)-6, ct(inequiv)-7, ct(inequiv)-8} and references therein.


\subsection{Conformal transformations of the metric}

A conformal transformation of the metric:

\bea g_{\mu\nu}\rightarrow\Omega^{-2}g_{\mu\nu}\;\left(g^{\mu\nu}\rightarrow\Omega^2g^{\mu\nu}\right),\;\sqrt{|g|}\rightarrow\Omega^{-4}\sqrt{|g|},\label{conf-t-eq}\eea where $\Omega^2=\Omega^2(x)$ is the conformal factor, is a point-dependent rescaling of the metric that -- besides angles -- preserves the causal structure of the spacetime. Here we underline that the above conformal transformation of the metric is not to be confounded with conformal transformations implying simultaneous coordinate rescalings which are properly diffeomorphisms. Under \eqref{conf-t-eq} the affine connection coefficients (the Christoffel symbols) transform like:

\bea \{^\sigma_{\mu\nu}\}\rightarrow\{^\sigma_{\mu\nu}\}-\delta^\sigma_\mu\nabla_\nu\left(\ln\Omega\right)-\delta^\sigma_\nu\nabla_\mu\left(\ln\Omega\right)+g_{\mu\nu}\nabla^\sigma\left(\ln\Omega\right).\label{aff-conf-t-eq}\eea Under \eqref{conf-t-eq}, \eqref{aff-conf-t-eq}, the components of the Ricci tensor and the curvature scalar transform in the following form:

\bea R_{\mu\nu}&\rightarrow & R_{\mu\nu}+2\nabla_\mu\left(\ln\Omega\right)\nabla_\nu\left(\ln\Omega\right)-2g_{\mu\nu}\left(\nabla\ln\Omega\right)^2+2\nabla_\mu\nabla_\nu\left(\ln\Omega\right)+g_{\mu\nu}\Box\ln\Omega,\nonumber\\
R&\rightarrow &\Omega^2\left[R+6\Box\ln\Omega-6\left(\nabla\ln\Omega\right)^2\right].\label{ricci-conf-t-eq}\eea Taking into account the above transformation laws we can write the corresponding transformation law for the Einstein's tensor:

\bea &&G_{\mu\nu}\rightarrow G_{\mu\nu}+2\nabla_\mu\left(\ln\Omega\right)\nabla_\nu\left(\ln\Omega\right)+g_{\mu\nu}\left(\nabla\ln\Omega\right)^2\nonumber\\
&&\;\;\;\;\;\;\;\;\;\;\;\;\;\;\;\;\;\;\;\;\;\;\;\;\;\;\;\;\;\;\;\;+2\nabla_\mu\nabla_\nu\left(\ln\Omega\right)-2g_{\mu\nu}\Box\ln\Omega.\label{etensor-conf-t-eq}\eea  The transformation law for the D'Alembertian of a scalar field $\psi$ under \eqref{conf-t-eq} reads:

\bea \Box\psi\rightarrow\Omega^2\left[\Box\psi-2\nabla^\sigma\left(\ln\Omega\right)\nabla_\sigma\psi\right].\label{box-conf-t-eq}\eea

For the extrinsic curvature $K_{\mu\nu}=h^\sigma_\mu h^\lambda_\nu\nabla_\sigma n_\lambda$ ($K=h^{\mu\nu}K_{\mu\nu}$), of a 3D hypersurface ortogonal to the unit vector $n_\mu$ with the metric $h_{\mu\nu}=g_{\mu\nu}\pm n_\mu n_\nu$, induced on it, the transformation under \eqref{conf-t-eq} reads:

\bea K_{\mu\nu}\rightarrow\Omega^{-1}\left(K_{\mu\nu}-h_{\mu\nu}n^\sigma\der_\sigma\Omega\right)\;\Rightarrow\;K\rightarrow\Omega^{-1}\left(K-3n^\sigma\der_\sigma\Omega\right).\label{ct-ext-curv}\eea


\subsection{Jordan frame and Einstein frame formulations of BD theory}\label{subsect-bd-jf-ef}

Under the conformal transformation of the metric \eqref{conf-t-eq} with $\Omega^2=\phi$, together with the rescaling of the BD scalar field: $\phi\rightarrow\exp\vphi$, the Jordan frame BD action \eqref{bd-action}:

\bea S^\text{JF}_\text{BD}=\int_{\cal M} d^4x\sqrt{|g|}\left[\phi R-\frac{\omega_\text{BD}}{\phi}(\der\phi)^2-2V(\phi)+2{\cal L}_m\right],\label{jf-bd-action}\eea is transformed into the Einstein frame:

\bea S^\text{EF}_\text{BD}=\int d^4x\sqrt{|g|}\left[R-\left(\omega_\text{BD}+\frac{3}{2}\right)(\der\vphi)^2-2V(\vphi)+2e^{-2\vphi}{\cal L}_m\right],\label{ef-bd-action}\eea where, under \eqref{conf-t-eq}: $V(\phi)\rightarrow e^{2\vphi}V(\vphi)$. If consider the boundary term in \eqref{jf-bd-action}, then under the conformal transformation \eqref{conf-t-eq}, it is transformed into its EF counterpart in \eqref{ef-bd-action}: $$2\int_{\der\cal M} d^3x\sqrt{|h|}\phi K\rightarrow 2\int_{\der\cal M} d^3x\sqrt{|h|}K,$$ where $h$ is the determinant of the metric $h_{\mu\nu}$ induced on the boundary and $K=h^{\mu\nu}K_{\mu\nu}$ is its extrinsic curvature scalar. In the latter transformation law for the boundary term it has been taken into account that terms coming from $6\phi\Omega^2\Box(\ln\Omega)$ in the EF action \eqref{ef-bd-action} compensate the terms $-6n^\mu\der_\mu\Omega$ in the EF boundary action (see Appendix A of Ref. \refcite{copeland-wands-rev}).

Under \eqref{conf-t-eq} the stress-energy tensor of matter transforms in the following way:

\bea T^{(m)}_{\mu\nu}\rightarrow\Omega^{-2}T^{(m)}_{\mu\nu}\;\;\Rightarrow\,\;T^{(m)}=\Omega^{-4}T^{(m)},\label{conf-t-set}\eea while the conservation equation that takes place in the JF transforms into non-conservation equation in the EF:

\bea \nabla^\mu T^{(m)}_{\mu\nu}=0\;\;\rightarrow\;\;\nabla^\mu T^{(m)}_{\mu\nu}=-\frac{\der_\nu\Omega}{\Omega}\,T^{(m)}.\label{conf-t-cons-eq}\eea 

The latter transformation property of the conservation equation is reminiscent of the transformation of the geodesics of the metric under \eqref{conf-t-eq}. Actually, under the conformal transformation of the metric the JF geodesic equation (the same as in GR):

\bea \frac{d^2x^\mu}{d\tau^2}+\left\{^{\;\mu}_{\sigma\lambda}\right\}\frac{dx^\sigma}{d\tau}\frac{dx^\lambda}{d\tau}=0,\label{jf-geod-eq}\eea where $\tau$ is an affine parameter along the geodesic, is transformed into the following EF equation of motion,

\bea \frac{d^2x^\mu}{d\tau^2}+\left\{^{\;\mu}_{\sigma\lambda}\right\}\frac{dx^\sigma}{d\tau}\frac{dx^\lambda}{d\tau}=2\frac{dx^\sigma}{d\tau}\frac{dx^\mu}{d\tau}\der_\sigma\left(\ln\Omega\right)-g_{\sigma\lambda}\frac{dx^\sigma}{d\tau}\frac{dx^\lambda}{d\tau}\der^\mu\left(\ln\Omega\right).\label{ef-geod-eq}\eea It can be shown that under the affine reparametrization: $d\tau\rightarrow f^{-1}(\Omega)d\tau$, with $f(\Omega)=\Omega$, the first term in the RHS of \eqref{ef-geod-eq} can be eliminated, 

\bea \frac{d^2x^\mu}{d\tau^2}+\left\{^{\;\mu}_{\sigma\lambda}\right\}\frac{dx^\sigma}{d\tau}\frac{dx^\lambda}{d\tau}=-g_{\sigma\lambda}\frac{dx^\sigma}{d\tau}\frac{dx^\lambda}{d\tau}\der^\mu\left(\ln\Omega\right).\label{ef-geod-eq'}\eea However, the second term in the RHS of \eqref{ef-geod-eq} can not be eliminated by any affine transformation whatsoever\cite{quiros_grg_2013}. What this means is that in the EFBD there is a univerdsal fifth-force, $$f^\mu_\text{fifth}=-g_{\sigma\lambda}\frac{dx^\sigma}{d\tau}\frac{dx^\lambda}{d\tau}\der^\mu\left(\ln\Omega\right),$$ that deviates the motion of a given particle from being geodesic. A distinctive feature of this fifth-force effect is that it acts only on particles with the mass. For massless particles like the photons, gravitons, etc., that move at the speed of light, $g_{\sigma\lambda}dx^\sigma dx^\lambda=0$, so that $f^\mu_\text{fifth}=0$, i. e., massless particles move along geodesics of the metric. The same conclusion is evident from the continuity (non-conservation) equation in the right-hand of \eqref{conf-t-cons-eq}, where, as seen, since for a fluid of massless particles, $T^{(m)}=0$, then the conservation equation is preserved under the conformal transformation \eqref{conf-t-eq}.


\subsection{The Einstein frame vs the Jordan frame}\label{subsect-jf-ef}

While both formulations of the BD theory discussed above: JF and EF, are in a relationship of mathematical equivalence through \eqref{conf-t-eq}, their physical equivalence may be, at least, questionable. But, before we start the discussion, we must agree on what to regard as `physical equivalence' of the different conformal frames since, otherwise, the discussion is meaningless. 

When one thinks on physical equivalence one of the first examples that comes to one's mind is the theory of general relativity. The physical equivalence of the different coordinate frames in which the GR laws -- expressed through the action principle and the derived equations of motion -- can be formulated is sustained by the invariance of these laws under general coordinate transformations. This leads naturally to the existence of a set of measurable quantities: the invariants of the geometry such as the line element, the curvature scalar and other quantities that are not transformed by the general coordinate transformations. Another example can be the gauge theories, where the gauge symmetry warrants that the theory can be formulated in a set of infinitely many physically equivalent gauges. In this case the quantities that have the physical meaning, i. e., those that are connected with measurable quantities, are gauge-invariant. As before, the guiding principle that supports the physical equivalence of the different gauges is the underlying symmetry. Take as a very simple example the electromagnetic gauge theory of a Fermion field $\psi(x)$, that is given by the following Lagrangian:

\bea {\cal L}_\text{gauge}=\bar\psi(x)\left(iD-m\right)\psi(x)-\frac{1}{4}F_{\mu\nu}F^{\mu\nu},\label{gauge-lag}\eea where the gauge derivative $D\equiv\gamma^\mu(\der_\mu-ig A_\mu)$ ($\gamma^\mu$ are the Dirac gamma-matrices while $A_\mu$ are the electromagnetic potentials) and $F_{\mu\nu}\equiv\der_\nu A_\mu-\der_\mu A_\nu$. The above Lagrangian is invariant under the following gauge transformations: $$\psi(x)\rightarrow e^{i\alpha(x)}\psi,\;\;\bar\psi(x)\rightarrow e^{-i\alpha(x)}\bar\psi,\;\;A_\mu\rightarrow A_\mu+\frac{1}{g}\der_\mu\alpha.$$ Quantities that are invariant under the above transformations, such as, for instance those $\propto \bar\psi\psi$ (and related), or $F_{\mu\nu}F^{\mu\nu}$, are the ones that have the physical meaning. The above procedure can be straightforwardly generalized to a collection of Fermion fields and of gauge fields in the electro-weak theory, for instance.

By analogy, one may expect that physical equivalence of the conformal frames should be linked with conformal invariance of the laws of physics, in particular, of the gravitational laws. Actually, following the spirit of the above examples: coordinate invariance of the laws of gravity in GR and gauge invariance of the laws of electromagnetism, one should require the action and the field equations of the theory -- representing the physical laws -- to be invariant under \eqref{conf-t-eq}. Then one may search for quantities that are not transformed by the conformal transformations of the metric, and regard them as the measurable quantities of the theory. This is the natural way in which one may think regarding invariance of the physical laws under conformal transformations.


\subsubsection{An example of a conformal invariant STT}\label{subsect-conf-inv-ex}

An example of a STT theory which is invariant under the conformal transformation of the metric plus a redefinition of the scalar field, known as Weyl rescalings: 

\bea g_{\mu\nu}\rightarrow\Omega^{-2}g_{\mu\nu},\;\;\vphi\rightarrow\vphi+2\ln\Omega,\label{weyl-t}\eea is given by the following vacuum action\cite{quiros_grg_2013}: 

\bea S=\int d^4x\sqrt{|g|} e^\vphi\left[R+\frac{3}{2}(\der\vphi)^2\right],\label{conf-inv-action}\eea and the corresponding field equations,

\bea &&G_{\mu\nu}=-\frac{1}{2}\left[\der_\mu\vphi\der_\nu\vphi+\frac{1}{2}g_{\mu\nu}(\der\vphi)^2\right]+\left(\nabla_\mu\nabla_\nu-g_{\mu\nu}\Box\right)\vphi,\nonumber\\
&&\Box\vphi+\frac{1}{2}(\der\vphi)^2-\frac{1}{3}R=0.\label{conf-inv-feqs}\eea The above equations (including the action itself) are not transformed by \eqref{weyl-t}. Besides, the motion equation for the scalar field is not independent from the Einstein's equation, since the trace of the first equation in \eqref{conf-inv-feqs} coincides with the second (KG) equation. This means that there is not an independent equation that governs the dynamics of $\vphi$, so that, he scalar field in this theory -- as in any other conformal invariant theory -- is not a dynamical degree of freedom. This is an inevitable consequence of conformal invariance. Actually, besides the four degrees of freedom to make diffeomorphisms, there is one more degree of freedom to make conformal transformations, so that we can set the scalar field $\vphi$ to any function we want. This entails that, instead of a given metric tensor $g_{\mu\nu}$, one has a whole class of conformal metrics, $\Omega^2g_{\mu\nu}$, through which one can geometrically interpret the physical (gravitational) phenomena, i. e., one has at our disposal a class of equivalent geometrical `realizations' of given physical laws. These different geometrical realizations is what we call as different (equivalent) representations of the theory.

The problem with the theory \eqref{conf-inv-action} is that, excluding traceless matter, the remaining matter degrees of freedom can not be consistently coupled to the theory since, as noticed before, the trace of the Einstein's equation for vacuum coincides with the Klein-Gordon equation. This fact would be immediately understood by the reader if we would mention from the start that the action above is nothing but Brans-Dicke theory with the special value of the coupling constant\cite{deser} $\omega_\text{BD}=-3/2$. Regardless of this, it serves as an example of a theory that really embodies the conformal invariance of the laws of physics invoked in Refs. \citen{faraoni_prd_2007, dicke-1962} and related work.


\subsection{The controversy}

What does physical equivalence of JF and EF representations of BD theory would entail after all? In order to put this issue into context let us notice first that, while the JFBD theory \eqref{jf-bd-action} is a STT of gravity in the sense that the gravitational interactions are carried by the metric field of geometric origin, together with the non-geometric BD scalar field, the EFBD theory \eqref{ef-bd-action} is a purely geometric theory of gravity indistinguishable from general relativity but for the presence of an additional non-gravitational universal interaction (fifth-force) between the scalar and the remaining matter fields through the interaction term $\propto e^{-2\vphi}{\cal L}_m$ in the action. This is seen from comparison of the geodesic equations in the Jordan frame \eqref{jf-geod-eq} with the corresponding EF motion equations \eqref{ef-geod-eq}, that can not be reduced to a geodesic by any redefinition whatsoever of the affine parameter. Hence, apparently, JFBD and EFBD are to be regarded as different theories and not as physically equivalent representations of a same theory. 

In spite of the apparent clarity of the above argument, we recommend to read the discussion on this subject in Ref. \refcite{sotiriou_etall_ijmpd_2008} (see specially sections 3, 4 and 5 therein) where a different perspective is presented. In that reference one may find interesting arguments that are shared by many researchers. Given that many aspects of the controversy on the physical equivalence of the different conformal representations are reflected in the discussion in that reference, below we cite several selected statements made therein, with the hope that these can help us to understand the issue: 

\begin{itemize}

\item ``The freedom of having an arbitrary conformal factor is due to the fact that the EEP does not forbid a conformal rescaling in order to arrive at special-relativistic expressions of the physical laws in the local freely falling frame.''

\item ``It should be stressed that all conformal metrics $\phi g_{\mu\nu}$, $\phi$ being the conformal factor, can be used to write down the equations or the action of the theory.''

\item ``As pointed out ... any metric theory can perfectly well be given a representation that appears to violate the metric postulates (recall, for instance, that $g_{\mu\nu}$ is a member of a family of conformal metrics and that there is no {\it a priori} reason why this metric should be used to write down the field equations)''

\item ``... many misconceptions arise when a theory is identified with one of its representations and other representations are implicitly treated as different theories.''

\item ``... the arbitrariness that inevitably exists in choosing the physical variables is bound to affect the representation.''

\item ``Thus, there will be representations in which it will be obvious that a certain principle is satisfied and others in which it will be more intricate to see that. However, it is clear that the theory is one and the same and that the axioms or principles are independent of the representation.''

\item ``This situation is very similar to a gauge theory in which one must be careful to derive only gauge-independent results. Every gauge is an admissible ``representation'' of the theory, but only gauge-invariant quantities should be computed for comparison with experiment. In the case of scalar-tensor gravity, however, it is not clear what a ``gauge'' is and how to identify the analog of ``gauge-independent'' quantities.''

\end{itemize} 

What could be missing in the arguments exposed in the above listed statements? The statements in the first three items above, for instance, are all related with the existence of a class of conformal metrics, $\Omega^2g_{\mu\nu}$. But, the BD theory like any other STT is not invariant under the conformal transformations so that this equivalence class is not well-suited to this theory.\footnote{One should not be confused by the argument usually found in the bibliography that the gravitational part of the BD action: $$S_\text{BD}=\int d^4x\sqrt{|g|}\,e^\vphi\left[R-\omega(\der\vphi)^2\right],$$ where we have rescaled the BD field, $\vphi\rightarrow\ln\phi$, and the subindex ``BD'' in the coupling constant $\omega$, has been omitted, is invariant under the transformation \eqref{weyl-t} that includes a conformal transformation of the metric \eqref{conf-t-eq}, plus a transformation of the coupling constant, $$\omega\rightarrow\omega-2\der_\vphi\ln\Omega\left(1-\der_\vphi\ln\Omega\right)(2\omega+3).$$ One should notice first that, actually, the BD action above is form-invariant under the mentioned transformations. However, these imply that in general a constant value of the coupling constant in the Jordan frame is transformed into a function of $\vphi$ in the conformal frame, or, if $\der_\vphi\ln\Omega=\alpha$, is a constant, one constant value of the coupling constant is mapped into a different constant value in the conformal frame. This, in turn, has implications for the measured value of the gravitational constant \eqref{bd-eff-g}, so that one has actually two different theories: BD theory with different values -- even different behaviors -- of the coupling ``constant''. Notice also, that for the special value $\omega=-3/2$, the coupling is not transformed under the above transformations. This has been properly noticed in the ending paragraph of the former subsection \ref{subsect-conf-inv-ex}.} As a counterexample to this, in the former subsection \ref{subsect-conf-inv-ex} we presented the action (and the corresponding motion equations) of a theory that really embodies the conformal invariance, so that the existence of an equivalence class of conformal metrics, $\Omega^2g_{\mu\nu}$, is a natural consequence (see also in the next section \ref{sect-scale-inv} where we discuss on existing scale-invariant theories of gravity). A necessary requirement for the existence of this equivalence class is that the scalar field is not determined by a motion equation, i. e., it should be non-dynamical. This is, precisely, the price to pay for conformal invariance, since then, in addition to the four degrees of freedom to make coordinate transformations one has an additional degree of freedom to make conformal transformations of the metric. Contrary to this, in the BD theory -- like in any other STT -- there is always a motion equation that governs the dynamics of the scalar field. This means that, by solving the motion equations one is able to determine not only the spacetime metric, but also the scalar field, so that there is not any freedom in choosing $\phi$. Hence, the missing argument in the reasoning line displayed by the above listed statements is the need for conformal invariance of the equations of the theory (the action plus the derived motion equations) in order to accommodate an equivalence class of conformal metrics: BD and STT-s in general are not conformal invariant theories so that these are not well suited to embody physical equivalence of the conformally related formulations. 

In the example in subsection \ref{subsect-conf-inv-ex}, for instance, the action \eqref{conf-inv-action} and the derived motion equations \eqref{conf-inv-feqs} are invariant under the conformal transformation (toguether with a scalar field redefinition) in \eqref{weyl-t}. In order to explain the implications of conformal invariance of the gravitational laws in this example, let us assume that the pair, $(g^{(0)}_{\mu\nu},\vphi_{(0)})$, where $g^{(0)}_{\mu\nu}=g^{(0)}_{\mu\nu}(x)$ and $\vphi_{(0)}=\vphi_{(0)}(x)$ are point-dependent functions, accounts for any given ``starting'' representation of the theory \eqref{conf-inv-action}. By means of the conformal transformation \eqref{weyl-t} with a chosen specific function $\Omega^2_{(k)}=f_{(k)}(\vphi_{(0)})$ ($f_{(k)}$ is a positive continuous function), from this starting representation of the theory, a new representation is obtained: $$g^{(k)}_{\mu\nu}=\Omega^2_{(k)}g^{(0)}_{\mu\nu},\;\;\vphi_{(k)}=\vphi_{(0)}-2\ln\Omega_{(k)}.$$ Since both pairs, $(g^{(0)}_{\mu\nu},\vphi_{(0)})$, and $(g^{(k)}_{\mu\nu},\vphi_{(k)})$, obey the same equations of motion \eqref{conf-inv-feqs} (these obey also, of course, the same action principle \eqref{conf-inv-action}), we can say that these pairs amount to two different but physically equivalent representations of the same theory. Here one may define conformal invariant quantities that can be related to the measurables of the theory as, for instance, $d\tau^2_\text{inv}:=e^\vphi d\tau^2$ ($d\tau$ is the coordinate-invariant time interval).

We shall not discuss on the arguments displayed in the items 4, 5 and 6 above, since these are highly dependent on the assumed definition of what to understand by a representation of a theory and, hence, we shall inevitably end up discussing on semantics issues that have nothing to do with the core of the conformal frames' controversy. However, the statement in the last (seventh) item needs of some discussion. This statement contradicts the viewpoint on physical equivalence of the conformal frames advocated in Ref. \refcite{sotiriou_etall_ijmpd_2008}. According to the authors, the situation on physical equivalence of the conformal frames should be compared with a gauge theory -- as we have discussed before in the example given by the Lagrangian \eqref{gauge-lag} -- where, although every gauge is an admissible representation of the theory, only gauge-invariant quantities are of relevance for purposes of comparison with the experiment. The authors themselves recognize in a sentence of this last statement that, in the case of scalar-tensor theories ``it is not clear what a gauge is and how to identify the analog of gauge-independent quantities''.\footnote{In Ref. \citen{jarv-1, jarv-2} a formalism was developed that allows to construct the invariants that are to be linked with measurable quantities in the STT. In this regard we want to make a comment: Attaching physical (measurable) meaning to gauge invariant quantities in a theory that is not itself gauge invariant, makes sense only as an additional postulate, so that one ends up dealing with a completely different theory (not a STT in the standard sense).} In view of our above arguments, the lack of clarity in what to understand by a gauge and what to identify by gauge-independent quantities within the framework of the STT-s, in connection with conformal transformations, is due to the fact that these theories do not embody conformal invariance being the necessary requirement for the existence of physically equivalent conformal representations.

Let us comment on other arguments that are also found in the bibliography on conformal frames. In Ref. \refcite{dicke-1962}, for instance, the physical equivalence of JF and EF formulations of BD theory is assumed by allowing the units of time, length, mass and the derived quantities to scale with appropriate powers of the conformal factor $\Omega$ (see, also, Ref. \refcite{faraoni_prd_2007}). According to the argument given in these references, physics must be invariant under the choice of the units, i. e., under the rescaling of the units of length, time and mass. The logic consequence is that, since physics is invariant under a change of units, it is invariant under a conformal transformation, provided that the units of length, time, and mass are scaled\cite{faraoni_prd_2007}. In this regard such concepts like ``EF with running units'' is encountered. The main idea behind this latter concept is that, what really matters when measurements are concerned, is the ratio of the quantity being measured, for instance the mass $m_p$ of a given particle of the SMP, to the unit of measurement (say, the energy $m_A$ of some emission line of some atom): $m_p/m_A$, and since the ratio is a dimensionless quantity, it is not transformed by the conformal transformations of the metric \eqref{conf-t-eq}, so that the measured value is the same in the JF and in the EF (or in any of the conformally related frames). No matter how `natural' such kind of argument could seem, the fact is that conformal transformations of the metric are about point-dependent rescalings of the metric tensor with the spacetime coincidences, i. e., the coordinates, held fixed. In other words, the conformal transformations do not affect the measurements. Hence the above mentioned argument is a redundancy. But what is more confusing is how, according to the above reasoning line, the conformal transformation between the JF and the EF of the BD theory can be reconciled with the assumed conformal invariance of the physical laws. Recall that, according to our adopted view point that is shared by the existing approaches to physical equivalence as linked with a symmetry of the equations of motion (as in GR and in the gauge theories), conformal invariance of the gravitational laws governed by some STT theory, inevitably requires the action and the derived motion equations to be unchanged by the conformal transformation \eqref{conf-t-eq} (see the example of a conformal invariant STT above), but this is not the case of BD theory which is transformed from the JF to the EF. The mere existence of the different conformal frames is an evidence of the lack of conformal invariance of the laws of gravity that are governed by the BD theory of gravity and the more general STT-s. 

Here we have avoided the semantic issue on what to consider a theory and what a representation of a theory, by following the mainstream of thinking regarding physical equivalence of different representations of a theory: Physical equivalence of different representations of a theory is regarded as a concept intimately linked with invariance of the laws of physics -- represented by an action principle and the derived motion equations -- under given transformations; be it coordinate, gauge or conformal transformations.


\subsection{When the JF and EF representations are regarded as non-equivalent formulations}

Up to this point in our discussion we have considered the controversy that arises when the different conformal frames are considered as physically equivalent representations of a same theory. In this subsection we shall briefly expose the point of view according to which the different conformal frames are regarded as physically non-equivalent representations, i. e., here we shall be concerned with the second of the questions that stand at the core of the conformal frames' controversy stated at the beginning of this section. For instance, in the Refs. \citen{faraoni_ijtp_1999, sarkar_mpla_2007, ct-ineq-nojiri, ct-ineq-brisc, ct-ineq-capoz, ct-ineq-brooker, ct-ineq-baha, ct(inequiv)-7, ct(inequiv)-8, ct(inequiv)-9} the physical equivalence of the JF and EF conformal frames is challenged both classically\cite{faraoni_ijtp_1999, sarkar_mpla_2007, ct-ineq-nojiri, ct-ineq-brisc, ct-ineq-capoz, ct-ineq-brooker, ct-ineq-baha} and at the quantum level\cite{ct(inequiv)-7, ct(inequiv)-8, ct(inequiv)-9}. In Ref. \refcite{faraoni_ijtp_1999} an example based on gravitational waves is explored in order to clarify the issue. It is seemingly demonstrated therein that the EF is the better suited frame to describe the physical phenomena. It has been shown in Ref. \refcite{sarkar_mpla_2007} that the gravitational deflection of light to second order accuracy may observationally distinguish the two conformally related frames of the BD theory. Meanwhile in Refs. \citen{ct-ineq-nojiri, ct-ineq-brisc, ct-ineq-capoz, ct-ineq-brooker, ct-ineq-baha}, by means of the equivalence between the $f(R)$ and STT theories, the physical non-equivalence of the JF and EF frames is demonstrated. The non-equivalence of these formulations of the BD theory from the physical standpoint has been investigated also in Refs. \citen{kaloper_prd_1998, quiros_prd_2000, quiros_prd_2000_1} in what regards to the spacetime singularities. 

Here, as before, if appropriate care is not taken about involved concepts, the discussion may go on to a semantic issue. First, what means that the different conformal frames are physically non-equivalent? After all, when one compares two different frames, even when these are related by a mathematical relationship of equivalence, as long as the physical laws are not invariant under the equivalence relationship, what one is comparing is two different theories with their own set of measurable quantities. Hence, it is natural to get different predictions for a given quantity when computed in terms of the measurable quantities of one or another frame. In this regard, looking for evidence on the non-equivalence of the different conformal frames amounts to looking for evidence in favor of one or the other theoretical framework, no more. This is precisely the point of view we have exposed in the former subsections.

A different thing is to search for a physical conformal frame among the conformally related ones. This would be a task inevitably doomed to failure. Actually, if the conformally related frames are physically equivalent, then all (or none) of them are physical. If they are not physically equivalent, then the different frames represent actually different theories: for instance JFBD is a metric STT while EFBD is GR supplemented with an additional non-gravitational universal fifth-force, i. e., a non-metric theory. In this case what matters is not whether the theory is physical or not but whether the theory's predictions meet or not the experimental evidence. Nevertheless one founds statements like this (here we do not cite any particular work since this kind of statement is generalized among researchers that are not particularly familiar with the conformal transformation's issue): ``... the matter is coupled to the conformal metric $\Omega^2g_{\mu\nu}$ (the physical metric) and not to the gravitational metric $g_{\mu\nu}$.'' It is not difficult to understand that, in such cases when one may differentiate the gravitational metric from a metric to which the matter is coupled -- which in such kind of statement means that the latter is the metric in terms of which the stress-energy tensor of matter is conserved -- what one has is not a STT, nor even GR, but a bimetric theory of gravity. That this is not usually recognized is just an indication of the lack of understanding of what a conformal transformation of the metric really entails for the STT-s.


\subsection{Weyl symmetry and the geometrical aspect of the conformal transformations issue}\label{subsect-weyl}

A less known aspect of the conformal transformations of STT, in particular of BD theory, has been explored in Ref. \refcite{quiros_grg_2013} (see also Refs. \citen{quiros_prd_2000, quiros_prd_2000_1, quiros_npb_2002, scholz, sc_inv_scholz_1, sc_inv_scholz_2, romero_cqg_2012, romero_prd_2014, almeida_prd_2014, lobo_epjc_2015, pucheu_prd_2016, pucheau_barreto_prd_2016, romero_jmp_2018, romero_cqg_2018}). It is linked with the geometrical face of the conformal transformations \eqref{conf-t-eq}. According to the line of reasoning in Ref. \refcite{quiros_grg_2013}, under the conformal transformation \eqref{conf-t-eq}, the transformation of the Christoffel symbols of the metric $g_{\mu\nu}$ in \eqref{aff-conf-t-eq} may be interpreted in the following alternative way:

\bea \{^\sigma_{\mu\nu}\}\rightarrow\Gamma^\sigma_{\mu\nu}\equiv\{^\sigma_{\mu\nu}\}+\frac{1}{2}\left(\delta^\sigma_\mu\der_\nu\vphi+\delta^\sigma_\nu\der_\mu\vphi-g_{\mu\nu}\der^\sigma\vphi\right),\label{alt-way}\eea where $\Gamma^\sigma_{\mu\nu}$ are the affine connection of a Weyl-integrable manifold (WIM) and the Weyl gauge scalar $\vphi\equiv\ln\Omega^2$ is identified with the logarithm of the conformal factor. This is a particular case in the class of the more general Weyl geometries\cite{weyl}. For a compact introduction to the Weyl geometry, including the Weyl-integrable case, see section 2 of Ref. \refcite{quiros_grg_2013}. The metricity condition of the WIM (the supra-index $(w)$ means that given quantities and operators are defined in terms of the affine connection $\Gamma^\sigma_{\mu\nu}$); $\nabla^{(w)}_\sigma g_{\mu\nu}=-\der_\sigma\vphi g_{\mu\nu},$ is not transformed by the conformal transformation \eqref{weyl-t} that is complemented with a redefinition of the gauge scalar. Usually these transformations are called as ``Weyl rescalings''. The affine connection of the WIM, is also invariant under the Weyl rescalings and so are the WIM Ricci tensor $R^{(w)}_{\mu\nu}$ and the related Einstein's tensor: $G^{(w)}_{\mu\nu}$, and the geodesic of the Weyl-integrable manifold: $$\frac{d^2x^\mu}{d\tau^2}+\Gamma^\mu_{\sigma\lambda}\frac{dx^\sigma}{d\tau}\frac{dx^\lambda}{d\tau}-\frac{1}{2}\der_\sigma\vphi\frac{dx^\sigma}{d\tau}\frac{dx^\lambda}{d\tau}=0,$$ or after an appropriate redefinition of the affine parameter: $d\tau\rightarrow e^{-\vphi/2}d\tau$; $$\frac{d^2x^\mu}{d\tau^2}+\Gamma^\mu_{\sigma\lambda}\frac{dx^\sigma}{d\tau}\frac{dx^\lambda}{d\tau}=0.$$ It is then proposed in Ref. \refcite{quiros_grg_2013} that the geometrical structure better suited to address conformal invariance or invariance under Weyl rescalings is not (pseudo)Riemann geometry -- as it is implicitly assumed when the issue is discussed in the bibliography -- but Weyl-integrable geometry instead. In this context, if assume a WIM as the geometrical structure to be associated with the action \eqref{conf-inv-action}, the conformal invariance of the latter theory is complemented with the conformal invariance of the associated geometrical background. 

Even if forget about conformal invariance, as in the case of the BD theory and the STT-s, the issue of the conformal frames acquires a new dimension if assume the alternative way \eqref{alt-way}. Actually, in this case a conformal transformation from the JF to the EF takes us from (pseudo)Riemannian manifold into a WIM. Hence, from the start it is not required to compare these frames since these are associated with different geometrical structures. For a more detailed discussion on this new aspect of the conformal transformation's issue we recommend Ref. \refcite{quiros_grg_2013}.


\subsection{Disformal transformations}\label{subsect-disf}

A quarter century ago in Ref. \refcite{disf-t-beken} the question was stated on whether the conformal transformation of the kind \eqref{conf-t-eq} is the most general relation between two geometries allowed by physics? The author studied this question by supposing that the physical geometry on which matter dynamics take place could be Finslerian rather than just Riemannian. By asking for validity of the weak equivalence principle and avoiding causality issues, the conclusion was reached that the Finsler geometry has to reduce to a Riemann geometry whose metric - the physical metric - is related to the gravitational metric by a generalization of the conformal transformation called as ``disformal transformations''\cite{disf-t-beken, disf-t-bruneton, disf-t-appleby, disf-t-bettoni, disf-t-kim, disf-t-rua, disf-t-arroja, disf-t-tsuji, disf-t-achour}. These are given by the following equation:

\bea g_{\mu\nu}\rightarrow\bar g_{\mu\nu}=A(\phi,X)g_{\mu\nu}+B(\phi,X)\der_\mu\phi\der_\nu\phi,\label{disf-t-eq}\eea where, as before, $X\equiv-(\der\phi)^2/2$, stands for the kinetic term. Given that the disformal functions $A$ and $B$ depend not only on $\phi$, but also on its kinetic energy, it is implicit a dependence on the metric hidden in $X$. The disformal metric can have, depending on the sign of $B$, light cones wider or narrower than those of the metric\cite{disf-t-beken, disf-t-bruneton}. The above disformal transformation must be invertible, with inverse\cite{disf-t-bettoni}:

\bea \bar g^{\mu\nu}=A^{-1}g^{\mu\nu}-\frac{B/A}{A+2BX}\der^\mu\phi\der^\nu\phi,\label{disf-t-eq'}\eea with invertible volume element: $\sqrt{|\bar g|}=A^2\sqrt{1+2XB/A}\sqrt{|g|}.$ As stated in Ref. \refcite{disf-t-bettoni}, disformal transformations have for the Horndeski action \eqref{horn-action} a role very similar to that of conformal transformations for the STT. A special case of the disformal transformations,

\bea g_{\mu\nu}\rightarrow\bar g_{\mu\nu}=A(\phi)g_{\mu\nu}+B(\phi)\der_\mu\phi\der_\nu\phi,\label{disf-t-spec}\eea where the disformal functions depend only on the scalar field, preserves second-order field equations, warranting the Horndeski action \eqref{horn-action} to be formally invariant under \eqref{disf-t-spec}. In this case the effect of the disformal transformation \eqref{disf-t-spec} can be recast into appropriate renormalization of the functions $K(\phi,X)$, $G_i(\phi,X)$ in the Horndeski action (see the appendix C of Ref. \refcite{disf-t-bettoni} for details). 

Only as an illustration here we shall display how the ``Einstein's frame'' cubic Galileon action (here we use the units' system $M^2_\text{Pl}=(8\pi G_N)^{-1}$ with $M_\text{Pl}\simeq 1.22\times 10^{19}$ GeV):

\bea &&S=\int d^4x\sqrt{|g|}\left[\frac{M^2_\text{Pl}}{2}R-\frac{c_2}{2}(\der\phi)^2-\frac{c_3}{M^3}(\der\phi)^2\Box\phi\right.\nonumber\\
&&\;\;\;\;\;\;\;\;\;\;\;\;\;\;\;\;\;\;\;\;\;\;\;\;\;\;\;\;\;\left.+{\cal L}_m-\frac{c_G}{M_\text{Pl}M^3}T^{(m)}_{\mu\nu}\der^\mu\phi\der^\nu\phi-\frac{c_0}{M_\text{Pl}}\phi T\right],\label{cov-gal-action}\eea is transformed into the ``Jordan frame'' one by a disformal transformation of the kind \eqref{disf-t-spec}. The demonstration for this particular case was given in Ref. \refcite{disf-t-appleby} through performing the transformations in the weak field limit, then absorbing like terms by renormalizing the constants $c_i$-s, and, finally, by promoting the obtained actions to their full, non-linear counterparts. The JF action is the following:

\bea &&S=\int d^4x\sqrt{|g|}\left[\left(1-\frac{2c_0}{M_\text{Pl}}\phi\right)\frac{M^2_\text{Pl}}{2}R-\frac{c_2}{2}(\der\phi)^2-\frac{c_3}{M^3}(\der\phi)^2\Box\phi\right.\nonumber\\
&&\;\;\;\;\;\;\;\;\;\;\;\;\;\;\;\;\;\;\;\;\;\;\;\;\;\;\;\;\;\;\;\;\;\;\;\;\;\;\;\;\;\;\;\;\;\;\;\;\;\;\;\;\;\;\;\;\;\left.-\frac{M_\text{Pl} c_G}{M^3}\,G_{\mu\nu}\der^\mu\phi\der^\nu\phi+{\cal L}_m\right].\label{cov-gal-disf-action}\eea

Since the issue of the disformal transformations is relatively contemporary, to date no questions have arisen in what regards to the physical equivalence of the disformal frames. Besides, the existing works on the cubic Galileon -- and related models -- have been performed, almost exclusively, in the JF.


\section{Scale-invariant theories of gravity}\label{sect-scale-inv}

One of the cornerstones of the present models of the fundamental interactions is the existence of a number of symmetries that are shared by the laws of nature\cite{weinberg_salam_phys_rev_1962, higgs_prl_1964, higgs_phys_lett_1964, englert_prl_1964, guralnik_prl_1964, coleman_weinberg_prd_1973, coleman_jackiw_prd_1974, susskind_prd_1979, zee_prl_1979}. As the cosmic expansion proceeds several of these symmtries break down to generate the Universe we see today. Scale invariance is one of the symmetries that has played an important role in the building of the unified interactions. It is required for the renormalization procedure to work appropriately at very short distances\cite{smolin_npb_1979, zee_prd_1981, cheng_prl_1988, goldberger_prl_2008}. In what regards gravitation theories scale invariance has been also investigated from different perspectives\cite{smolin_npb_1979, zee_prd_1981, cheng_prl_1988, goldberger_prl_2008, sc_inv_indios, sc_inv_indios_1, sc_inv_shapo, sc_inv_percacci, sc_inv_prester, sc_inv_quiros, sc_inv_padilla, sc_inv_bars, sc_inv_bars_1, sc_inv_bars_2, sc_inv_bars_3, sc_inv_bars_4, sc_inv_carrasco, sc_inv_quiros_1, sc_inv_jackiw, sc_inv_alpha, sc_inv_alpha_1, sc_inv_alpha_2, sc_inv_alpha_3, sc_inv_alpha_4, salvio-1, sc_inv_ghorbani, sc_inv_farz, sc_inv_vanzo, sc_inv_alvarez, sc_inv_khoze, sc_inv_karananas, sc_inv_tambalo, sc_inv_kannike, javier-1, javier-2,sc_inv_ferreira, sc_inv_ferreira_1, sc_inv_ferreira_2, sc_inv_maeder, sc_inv_maeder_1, sc_inv_myung}. In Ref. \refcite{weyl} the first serious attempt to create a scale-invariant theory of gravity (and of electromagnetism) was made. Due to an unobserved broadening of the atomic spectral lines this attempt had a very short history\cite{sc_inv_scholz_1, sc_inv_scholz_2, sc_inv_perlick, sc_inv_novello}. A scale-invariant extension of general relativity based on Weyl's geometry is explored in Ref. \refcite{smolin_npb_1979}. If the theory contains a Higgs phase then, at large distances this phase reduces to GR. In Ref. \refcite{zee_prd_1981} it has been shown that gravity may arise as consequence of dynamical symmetry breaking in a scale -- also gauge -- invariant world. A quantum field theory of electroweak and gravitational interactions with local scale invariance and local $SU(2)\times U(1)$ gauge invariance is proposed in Ref. \refcite{cheng_prl_1988}. The requirement of local scale invariance leads to the existence of Weyl's vector meson which absorbs the Higgs particle remaining in the SMP.

In general any theory of gravity can be made Weyl-invariant by introducing a dilaton. In Ref. \refcite{sc_inv_percacci} it is shown how to construct renormalization group equations for such kind of theories, while in Ref. \citen{sc_inv_odintsov, sc_inv_odintsov_1, sc_inv_odintsov_2, sc_inv_odintsov_3} it has been shown that scale invariance is very much related with the effect of asymptotic conformal invariance, where quantum field theory predicts that theory becomes effectively conformal invariant. In Ref. \refcite{sc_inv_padilla} the authors present the most general actions of a single scalar field and two scalar fields coupled to gravity, consistent with second order field equations in four dimensions (4D), possessing local scale invariance. It has been shown that Weyl-invariant dilaton gravity provides a description of black holes without classical spacetime singularities\cite{sc_inv_prester}. The singularities appear due to ill-behavior of gauge fixing conditions, one example being the gauge in which the theory is classically equivalent to GR. In Refs. \citen{sc_inv_bars, sc_inv_bars_1, sc_inv_bars_2, sc_inv_bars_3, sc_inv_bars_4} it is shown how to lift a generic non-scale invariant action in Einstein frame into a Weyl-invariant theory and a new general form for Lagrangians consistent with Weyl symmetry is presented. Advantages of such a conformally invariant formulation of particle physics and gravity include the possibility of constructing geodesically complete cosmologies\cite{sc_inv_bars_3}. In this regard see critical comments in Refs. \citen{sc_inv_carrasco, sc_inv_quiros_1, sc_inv_jackiw} and the reply in Ref. \refcite{sc_inv_bars_4}. In Refs. \citen{sc_inv_alpha} a new class of chaotic inflation models with spontaneously broken conformal invariance has been developed. In this vein a broad class of multi-field inflationary models with spontaneously broken conformal invariance is described in Ref. \refcite{sc_inv_alpha_2}, while generalized versions of these models where the inflaton has a non-minimal coupling to gravity with $\xi<0$, different from its conformal value $\xi=-1/6$, are investigated in Ref. \refcite{sc_inv_alpha_3}.

In order to discuss on scale invariance of the gravitational laws it is useful to write the following prototype action\cite{deser, smolin_npb_1979}:

\bea S=\int d^4x\sqrt{|g|}\left[\frac{\phi^2}{12}\,R+\frac{1}{2}(\der\phi)^2\pm\frac{\lambda}{12}\,\phi^4\right].\label{deser-action}\eea Since, under the Weyl rescalings (also called as scale transformations in this review):

\bea g_{\mu\nu}\rightarrow\Omega^{-2}g_{\mu\nu},\;\phi\rightarrow\Omega\,\phi,\label{scale-t}\eea the combination $\sqrt{|g|}[\phi^2R+6(\der\phi)^2]$ is kept unchanged -- as well as the scalar density $\sqrt{|g|}\phi^4$ -- then the action (\ref{deser-action}) is invariant under \eqref{scale-t}. Any scalar field which appears in the gravitational action the way $\phi$ does, is said to be conformally coupled to gravity. Hence, for instance, the following action\cite{sc_inv_prester, sc_inv_bars, sc_inv_bars_1, sc_inv_bars_2, sc_inv_bars_3, sc_inv_bars_4, sc_inv_carrasco, sc_inv_quiros_1, sc_inv_jackiw, sc_inv_alpha, sc_inv_alpha_1, sc_inv_alpha_2, sc_inv_alpha_3, sc_inv_alpha_4}:

\bea S=\int d^4x\sqrt{|g|}\left[\frac{\left(\phi^2-\sigma^2\right)}{12}\,R+\frac{1}{2}(\der\phi)^2-\frac{1}{2}(\der\sigma)^2\right],\label{bars-action}\eea is also invariant under \eqref{scale-t} since both $\phi$ and $\sigma$ are conformally coupled to gravity, provided that the additional scalar field $\sigma$ transforms in the same way as $\phi$: $\sigma\rightarrow\Omega\,\sigma$. For the coupling $\propto (\phi^2-\sigma^2)^{-1}$ to be positive and the theory Weyl-invariant, the scalar $\vphi$ must have a wrong sign kinetic energy -- just like in \eqref{deser-action} -- potentially making it a ghost. However, the local Weyl gauge symmetry compensates, thus ensuring the theory is unitary\cite{sc_inv_bars, sc_inv_bars_1, sc_inv_bars_2, sc_inv_bars_3, sc_inv_bars_4}.

In the model of Refs. \citen{sc_inv_bars_3, sc_inv_bars_4}, in order to get geodesically complete spacetimes it is required that not only the field $\vphi$, but also the doublet Higgs field $H$ be a set of conformally coupled scalars consistent with $SU(2)\times U(1)$. The corresponding Weyl-invariant action that describes the coupling of gravity and the standard model reads: $S=\int d^4x\sqrt{|g|}\left({\cal L}+{\cal L}_\text{SMP}\right)$, where ${\cal L}_\text{SMP}$ is the Lagrangian of the standard model of particles and,

\bea {\cal L}=\frac{1}{12}\left(\phi^2-2H^\dag H\right)R+\frac{1}{2}(\der\phi)^2-|DH|^2-\frac{\lambda}{4}\left(H^\dag H-\alpha^2\phi^2\right)^2-\frac{\lambda'}{4}\phi^4.\label{higgs-bars-action}\eea In the above equation we have used the notation $|DH|^2\equiv g^{\mu\nu}D_\mu H^\dag D_\nu H$, where $D_\mu$ stands for the gauge-covariant derivative. The above action is invariant under the Weyl rescalings \eqref{scale-t} plus the following rescaling of the remaining fields in ${\cal L}$ and in ${\cal L}_\text{SMP}$: $$H\rightarrow\Omega H,\;\;\psi\rightarrow\Omega^{3/2}\psi,\;\;A^a_\mu\rightarrow A^a_\mu,$$ where $\psi$ stands for the fermion fields for quarks or leptons, and the $A^a_\mu$ are the gauge fields for the photon, gluons, $W^\pm$ and $Z^0$ bosons. In this theory the only scale is generated by gauge fixing $\phi$ to a constant: $\phi(x)\rightarrow\phi_0$. All dimensionful parameters emerge from this single source: $$\frac{1}{16\pi G_N}=\frac{\phi_0^2}{12},\;\;\frac{\Lambda}{16\pi G_N}=\lambda'\phi_0^4,\;\;H^\dag_0 H_0=\alpha^2\phi_0^2=\frac{v^2}{2}.$$

In Refs. \citen{sc_inv_alpha, sc_inv_alpha_1, sc_inv_alpha_2, sc_inv_alpha_3, sc_inv_alpha_3}, a new class of conformally invariant theories which allow inflation, even if the scalar potential is very steep in terms of the original conformal variables, was explored. In order to understand how the cosmological attractor arises in these theories let us to investigate a toy model\cite{sc_inv_alpha_4} given by the action \eqref{bars-action} supplemented with the potential term $$-\int d^4x\sqrt{|g|}\frac{\lambda}{36}\left(\phi^2-\sigma^2\right)^2.$$ In addition to the invariance under Weyl rescalings, $$g_{\mu\nu}\rightarrow\Omega^{-2}g_{\mu\nu},\;\;(\phi,\sigma)\rightarrow\Omega(\phi,\sigma),$$ it has a global $SO(1,1)$ symmetry with respect to a boost between the two fields $\phi$, $\sigma$, preserving the value of $\phi^2-\sigma^2$, which resembles the Lorentz symmetry of the theory of special relativity. Since, $\phi^2-\sigma^2>0$ in order to have gravity rather than antigravity, $\phi$ represents a cutoff for the possible values of $\sigma$. Notice, however, that $\phi$ is not a physical degree of freedom since it may be gauged away, for instance, by fixing the gauge: $\phi^2-\sigma^2=6$, so that $$\phi=\sqrt{6}\cosh(\vphi/\sqrt{6}),\,\;\sigma=\sqrt{6}\sinh(\vphi/\sqrt{6}).$$ Under this choice of the gauge, the starting action is transformed into the following GR action with a canonical (minimally coupled) scalar field and a cosmological constant: $$S=\int d^4x\sqrt{|g|}\left[\frac{1}{2}R-\frac{1}{2}(\der\vphi)^2-\lambda\right].$$ The potential term $\propto(\phi^2-\sigma^2)^2$ is the ``placeholder'' for what becomes a cosmological constant. The main idea developed in Refs. \citen{sc_inv_alpha, sc_inv_alpha_1, sc_inv_alpha_2, sc_inv_alpha_3, sc_inv_alpha_4} is that one can construct a class of inflationary models by locally modifying the would be cosmological constant (the placeholder): $$-\int d^4x\sqrt{|g|}\frac{\lambda}{36}\left(\phi^2-\sigma^2\right)^2\rightarrow-\int d^4x\sqrt{|g|}\frac{1}{36}f^2(\sigma/\phi)\left(\phi^2-\sigma^2\right)^2,$$ where $f^2(\sigma/\phi)$ is an arbitrary function of the ratio $\sigma/\phi$. Through $f^2(\sigma/\phi)$ one deforms the starting $SO(1,1)$ symmetry. In the gauge $\phi^2-\sigma^2=6$, one gets: $$S=\int d^4x\sqrt{|g|}\left[\frac{1}{2}R-\frac{1}{2}(\der\vphi)^2-f^2(\tanh\vphi/\sqrt{6})\right].$$ Hence, since asymptotically $\tanh\vphi/\sqrt{6}\rightarrow\pm 1$, i. e., $f^2(\tanh\vphi/\sqrt{6})\rightarrow$ const, the system in the large $\vphi$-limit evolves asymptotically towards its critical point where the $SO(1,1)$ symmetry is restored.


\subsection{Anomalous matter coupling in scale invariant theories of gravity}

The motion equations that can be derived from \eqref{deser-action} (here for simplicity we drop the potential term $\propto\phi^4$) read:

\bea &&\phi^2G_{\mu\nu}=-4\der_\mu\phi\der_\nu\phi+g_{\mu\nu}(\der\phi)^2+2\phi\left(\nabla_\mu\nabla_\nu-g_{\mu\nu}\Box\right)\phi,\nonumber\\
&&\Box\phi-\frac{1}{6}\,R\phi=0.\label{deser-feqs}\eea The interesting thing is that the trace of the first equation above (the Einstein's equation) exactly coincides with the KG motion equation that is derived from \eqref{deser-action}. Hence, if add minimally coupled matter with stress-energy tensor $T_{\mu\nu}^{(m)}$ (it would appear with a factor of $6$ in the RHS of the first equation), the trace of the Einstein's equation then would yield: $$\Box\phi-\frac{1}{6}\,R\phi=\frac{1}{\phi}T^{(m)},$$ while the KG equation continues being given by the second equation in \eqref{deser-feqs}. Hence, only traceless matter: $T^{(m)}=0$, can be consistently coupled in the above scale invariant theory of gravity, unless one allows the matter to couple non-minimally\cite{sc_inv_quiros}. This result is easily understood if note that under the scalar field replacement, $\phi\rightarrow\phi^2$, up to the factor of $1/12$, the action \eqref{deser-action} is just BD theory with the special (anomalous) value $\omega_\text{BD}=-3/2$ of the coupling constant. This problem on the anomalous coupling of matter in the theory \eqref{deser-action} is usually misunderstood, or just evaded.


\subsection{The (forgotten) geometrical aspect of scale invariance}

As shown in Ref. \refcite{sc_inv_jackiw}, scale invariance of the theory \eqref{deser-action} -- the same for \eqref{bars-action} -- does not have any dynamical role since its associated Noether symmetry current vanishes. Another (perhaps related) aspect of the scale invariance of gravity theories that is not discussed as frequently as desired, is related with its geometrical aspect: Scale invariance - invariance under Weyl rescalings \eqref{scale-t} -- of the action \eqref{deser-action} (the same for \eqref{bars-action}), is meaningless until a geometrical background is specified. Here by geometrical background we do not understand just a metric but a whole geometrical set up. I. e., a set of geometrical laws that define a geometrical structure, for instance, (pseudo)Riemann geometry, or Weyl geometry, etc. Usually it is implicitly assumed that the background geometry is (pseudo)Riemann, but, in what regards scale invariance, this implicit choice has its own drawbacks. As demonstrated in subsection \ref{subsect-bd-jf-ef}, under a conformal transformation of the metric in \eqref{scale-t} the geodesics of the metric are transformed into non-geodesics in the conformal frame. Hence, assuming the action \eqref{deser-action} to be defined on a pseudo-Riemann manifold, means that, while the gravitational laws -- represented by the equations of motion derived from that action -- are indeed invariant under the Weyl rescalings \eqref{scale-t}, the geodesics of the metric are transformed into non-geodesics paths, which means, in turn, that there exists an anomalous fifth-force effect in one of the conformally related representations given that it is absent in the other one. This effect, by itself, invalidates the assumed Weyl invariance of the laws of gravity in the theory \eqref{deser-action} and/or in \eqref{bars-action}, since the ``gauge'' field $\phi$ becomes into a dynamical degree of freedom, that is incompatible with scale invariance. 

There is, however, a way out of this problem. Actually, let us assume that the background geometry is Weyl-integrable\footnote{For a compact introduction to Weyl geometry in general and to Weyl-integrable geometry as its ``healthy'' particular case, see section 2 of Ref. \refcite{quiros_grg_2013}.} instead of pseudo-Riemann (see subsection \ref{subsect-weyl} of the former section). In this case the affine connection of the geometry does not coincide with the Christoffel symbols of the metric but are given, instead, by the $\Gamma^\sigma_{\mu\nu}$ defined in \eqref{alt-way}. Under the Weyl rescalings \eqref{weyl-t}: $$g_{\mu\nu}\rightarrow\Omega^{-2}g_{\mu\nu},\;\;\vphi\rightarrow\vphi+2\ln\Omega,$$ the above affine connection is not transformed; $\Gamma^\sigma_{\mu\nu}\rightarrow\Gamma^\sigma_{\mu\nu}$, while (quantities with supra-index $(w)$ are defined in terms of the affine connection $\Gamma^\sigma_{\mu\nu}$): $$R^{(w)}_{\mu\nu}\rightarrow R^{(w)}_{\mu\nu},\;\;R^{(w)}\rightarrow\Omega^2R^{(w)}\Rightarrow G^{(w)}_{\mu\nu}\rightarrow G^{(w)}_{\mu\nu},$$ where $G^{(w)}_{\mu\nu}$ stands for the Weyl-integrable Einstein's tensor. Hence, thinking along these lines, one may replace the BD action with the anomalous value $\omega_\text{BD}=-3/2$ of the coupling constant, by its Weyl-integrable (also Weyl-invariant) counterpart\cite{sc_inv_quiros}: $$\int d^4x\sqrt{|g|}e^\vphi\left[R+\frac{3}{2}(\der\vphi)^2\right]\rightarrow\int d^4x\sqrt{|g|}e^\vphi R^{(w)}.$$ In this latter theory the scalar field $\vphi$ is a gauge field that has not any dynamical content since it can be safely gauged away without any physical consequences. The derived Weyl-integrable Einstein's equations\cite{sc_inv_quiros}: $$G^{(w)}_{\mu\nu}=\frac{e^{-\vphi}}{M^2_\text{Pl}}T^{(m)}_{\mu\nu},$$ are not only Weyl-invariant, but also admit coupling of matter degrees of freedom other than the radiation.

As a matter of fact, one of the first Weyl-invariant theories of gravity with the SMP minimally coupled was proposed in Ref. \refcite{smolin_npb_1979} and then a related model was proposed in Ref. \refcite{cheng_prl_1988}. In these proposals the Weyl geometry was assumed so that, instead of a Weyl-gauge scalar as above, a Weyl-gauge vector played the role of the gravitational gauge field. Let us briefly comment on the second, more modern proposal\cite{cheng_prl_1988}. In this case the non-minimally coupled scalar field $\phi$ is identified with the Higgs gauge boson in the unitary gauge $H^T=(0,h)/\sqrt{2}$:

\bea &&S=\int d^4x\sqrt{|g|}\left[\frac{\xi|h|^2}{2}\,R^{(w)}-\frac{1}{2}|Dh|^2-\frac{\lambda}{4}(|h|^2-v_0^2)^2\right.\nonumber\\
&&\left.\;\;\;\;\;\;\;\;\;\;\;\;\;\;\;\;\;\;-\frac{1}{4}\left(H_{\mu\nu}H^{\mu\nu}+W^k_{\mu\nu}W^{\mu\nu}_k+B_{\mu\nu}B^{\mu\nu}\right)\right],\label{cheng-action}\eea where $$|h|^2\equiv h^\dag h=h^2,\;|Dh|^2\equiv g^{\mu\nu}(D_\mu h)^\dag(D_\nu h),$$ $W^k_{\mu\nu}$ and $B_{\mu\nu}$ are the field strengths of the $SU(2)$ and $U(1)$ bosons respectively (see the appendix), and $\xi$ is the non-minimal coupling parameter. In the theory \eqref{cheng-action} the electroweak symmetry breaking potential not only allows for generation of masses of the gauge bosons (and fermions) but, also, generates the Planck mass $M_\text{Pl}=\sqrt\xi\,v_0$, where $v_0\approx 246$ GeV, and $\xi\sim 10^{32}-10^{34}$ is too large to meet the observational constraints. Before breakdown of scale symmetry, the action \eqref{cheng-action} is invariant under \eqref{scale-t} plus the Higgs field rescaling $h\rightarrow\Omega\,h.$ The gauge covariant derivative of the Higgs field in \eqref{cheng-action} is defined as $D_\mu h:=(D^*_\mu-w_\mu/2)h,$ where $$D^*_\mu h\equiv\left(\der_\mu+\frac{i}{2}\,g W^k_\mu\sigma^k+\frac{i}{2}\,g'B_\mu\right)h,$$ is the gauge covariant derivative in the standard EW theory, with $W^k_\mu=(W^\pm_\mu,W^0_\mu)$ - the SU(2) bosons, $B_\mu$ - the U(1) boson, $\sigma^k$ - the Pauli matrices, and ($g,g'$) - the gauge couplings. The requirement of local scale invariance of this theory leads to the existence of Weyl's vector meson which absorbs the Higgs particle remaining in the Weinberg-Salam model.


\section{Conclusion}\label{sect-conclu}

In this review we have discussed on several of the less known aspects of scalar-tensor theories of gravity, nonetheless we have included also much of the well known aspects of these theories in order for the review to be self-contained. Among the less known issues on STT-s we may mention the screening mechanisms. The modern history of these mechanisms that allow the elusive scalar to hide itself from local (Solar system) experiments started no long ago -- as compared with the long history of the scalar fields themselves -- when the chameleon screening was proposed in Ref. \refcite{cham}. Then, some 43 year after the Brans-Dicke theory was proposed\cite{bd-1961}, we knew that the stringent constraints coming from Solar-system experiments\cite{will-lrr-2014} may be evaded by the self-interaction effects of the scalar field in the presence of background matter with high enough density. Even more recently\cite{nicolis_gal, deffayet_vikman_gal, deffayet_deser_gal}, it was rediscovered\cite{horndeski_gal} that the scalar-tensor theories may be generalized to include higher order derivatives of the scalar field, while maintaining the motion equations at second order. This last discovery has opened the way to a much better understanding of what to regard as a scalar-tensor theory of gravity. We also learned that there can be other less trivial screening mechanisms to hide the scalar field from being detected by direct experimentation. The Vainshtein screening\cite{vainshtein, vainsh-deff-gabad, khoury-rev, kazuya_gal, quiros-cqg-2018-1, niz_kazuya_vainsh, kimura_vainsh, dima_prd_2018, bhorn(vainsh)
}, in particular, is originated by the non-linear effects of the higher derivatives contributions in the presence of background matter. When these effects start dominating -- typically much below the Vainshtein radius -- the derivative interactions decouple from the rest of the degrees of freedom, thus rendering the STT to become into GR. The higher-derivatives generalizations of STT, also known as Horndeski theories, come not without drawbacks. Among these we may mention causality and stability issues related with the speed of propagation of the tensor and scalar perturbations in these theories. Even if the Horndeski theories are free of the Ostrogradsky instability\cite{woodard_2007, ostro-theor} -- related with motion equations containing derivatives higher than second order -- nothing forbids these theories from having gradient instabilities associated with negative squared sound speed.

Although further study of Horndeski -- and beyond Horndeski\cite{bhorn-1, bhorn-2, bhorn-3, bhorn-4, bhorn-5, langlois_noui_jcap_2016_1} -- theories may hide surprises, there are other -- not as contemporary -- issues, whose exploration may lead us to a much better understanding not only of STT, but also of the fundamental interactions of nature\cite{weinberg_salam_phys_rev_1962, higgs_prl_1964, higgs_phys_lett_1964, englert_prl_1964, guralnik_prl_1964, coleman_weinberg_prd_1973, coleman_jackiw_prd_1974, susskind_prd_1979, zee_prl_1979, copeland-wands-rev}. In this regard, the conformal transformation's issue\cite{faraoni-book, faraoni_rev_1997, faraoni_ijtp_1999, faraoni_prd_2007, sarkar_mpla_2007, deruelle_veiled_2011, deruelle_nordstrom_2011, quiros_grg_2013, sotiriou_etall_ijmpd_2008, saal_cqg_2016, indios_consrvd_prd_2018, thermod_prd_2018, ct-1, ct-2, ct(fresh-view)-3, paliatha-2, ct(quant-equiv)-4, ct(quant-equiv)-5, ct(quant-equiv)-6, ct-ineq-nojiri, ct-ineq-brisc, ct-ineq-capoz, ct-ineq-brooker, ct-ineq-baha, ct(inequiv)-7, ct(inequiv)-8, ct(inequiv)-9} and the very much related subject of scale-invariant (also Weyl-invariant) theories of gravity and of the fundamental interactions\cite{smolin_npb_1979, zee_prd_1981, cheng_prl_1988, goldberger_prl_2008, sc_inv_indios, sc_inv_indios_1, sc_inv_shapo, sc_inv_percacci, sc_inv_prester, sc_inv_quiros, sc_inv_padilla, sc_inv_bars, sc_inv_bars_1, sc_inv_bars_2, sc_inv_bars_3, sc_inv_bars_4, sc_inv_carrasco, sc_inv_quiros_1, sc_inv_jackiw, sc_inv_alpha, sc_inv_alpha_1, sc_inv_alpha_2, sc_inv_alpha_3, sc_inv_alpha_4, salvio-1, sc_inv_ghorbani, sc_inv_farz, sc_inv_vanzo, sc_inv_alvarez, sc_inv_khoze, sc_inv_karananas, sc_inv_tambalo, sc_inv_kannike, javier-1, javier-2, sc_inv_ferreira, sc_inv_ferreira_1, sc_inv_ferreira_2, sc_inv_maeder, sc_inv_maeder_1, sc_inv_myung}, still lacks deeper and more systematic investigation. In particular, the geometrical aspect of the issue has not been explored in all its deepness. Could it be that, after all, the geometry of our world would not be Riemannian but some other geometrical structure? This may be an interesting question that has not been yet seriously settled.

These and other interesting issues have been covered in the present review with the hope that young researchers could become interested and start wondering whether several of the most profound questions of fundamental physics could be answered by searching within the framework of scalar tensor theories of gravity.


\section*{Acknowledgments}

The author thanks Elias C. Vanegas for the invitation to write this review. This occasion reminds the author that some 16 years ago a project for a book covering similar subjects, in co-authorship with Carl H Brans, was planned and then was abandoned due to the appearance of the book of Ref. \refcite{fujii_book_2004}. This frustrated attempt, however, marked the beginning of a long-lasting friendship with Carl Brans\cite{brans_homenag}. I would like to express him my most sincere gratitude for the inspiration and for his advise. The author also acknowledges his colleagues (and co-authors in many works) T. Matos, R. Maartens, R. Lazkoz, L.P. Chimento, L.A. Ure\~na, R. Garc\'ia-Salcedo, T. Gonz\'alez, G. Leon, U. Nucamendi, R. De Arcia and F.A. Horta-Rangel for their encourage and support, and for so many fruitful discussions. The long-lasting support of the author's research activity by the CONACyT is also acknowledged.


\end{document}